\documentclass[twocolumn,traditabstract,longauth]{aa}
\usepackage{graphicx,amsmath}
\usepackage{epsf}
\usepackage{txfonts}
\usepackage{url}
\usepackage[breaklinks, colorlinks, citecolor=blue]{hyperref} 

\usepackage{natbib}
\usepackage{upgreek}

\usepackage{fixltx2e}

\def\setsymbol#1#2{\expandafter\def\csname #1\endcsname{#2}}
\def\getsymbol#1{\csname #1\endcsname}

\def\Planck{\textit{Planck}}

\def\allearlypapers{\nocite{planck2011-1.1, planck2011-1.3, planck2011-1.4, planck2011-1.5, planck2011-1.6, planck2011-1.7, planck2011-1.10, planck2011-1.10sup, planck2011-5.1a, planck2011-5.1b, planck2011-5.2a, planck2011-5.2b, planck2011-5.2c, planck2011-6.1, planck2011-6.2, planck2011-6.3a, planck2011-6.4a, planck2011-6.4b, planck2011-6.6, planck2011-7.0, planck2011-7.2, planck2011-7.3, planck2011-7.7a, planck2011-7.7b, planck2011-7.12, planck2011-7.13}}

\def\all2013resultspapers{\nocite{planck2013-p01, planck2013-p02, planck2013-p02a, planck2013-p02d, planck2013-p02b, planck2013-p03, planck2013-p03c, planck2013-p03f, planck2013-p03d, planck2013-p03e, planck2013-p01a, planck2013-p06, planck2013-p03a, planck2013-pip88, planck2013-p08, planck2013-p11, planck2013-p12, planck2013-p13, planck2013-p14, planck2013-p15, planck2013-p05b, planck2013-p17, planck2013-p09, planck2013-p09a, planck2013-p20, planck2013-p19, planck2013-pipaberration, planck2013-p05, planck2013-p05a, planck2013-pip56, planck2013-p06b}}

\newbox\tablebox    \newdimen\tablewidth
\def\leaderfil{\leaders\hbox to 5pt{\hss.\hss}\hfil}
\def\endPlancktable{\tablewidth=\columnwidth 
    $$\hss\copy\tablebox\hss$$
    \vskip-\lastskip\vskip -2pt}
\def\endPlancktablewide{\tablewidth=\textwidth 
    $$\hss\copy\tablebox\hss$$
    \vskip-\lastskip\vskip -2pt}
\def\tablenote#1 #2\par{\begingroup \parindent=0.8em
    \abovedisplayshortskip=0pt\belowdisplayshortskip=0pt
    \noindent
    $$\hss\vbox{\hsize\tablewidth \hangindent=\parindent \hangafter=1 \noindent
    \hbox to \parindent{$^#1$\hss}\strut#2\strut\par}\hss$$
    \endgroup}
\def\doubleline{\vskip 3pt\hrule \vskip 1.5pt \hrule \vskip 5pt}

\def\L2{\ifmmode L_2\else $L_2$\fi}

\def\DeltaT{\ifmmode \Delta T\else $\Delta T$\fi}
\def\deltat{\ifmmode \Delta t\else $\Delta t$\fi}
\def\fknee{\ifmmode f_{\rm knee}\else $f_{\rm knee}$\fi}
\def\Fmax{\ifmmode F_{\rm max}\else $F_{\rm max}$\fi}
\def\solar{\ifmmode{\rm M}_{\mathord\odot}\else${\rm M}_{\mathord\odot}$\fi}
\def\Msolar{\ifmmode{\rm M}_{\mathord\odot}\else${\rm M}_{\mathord\odot}$\fi}
\def\Lsolar{\ifmmode{\rm L}_{\mathord\odot}\else${\rm L}_{\mathord\odot}$\fi}

\def\inv{\ifmmode^{-1}\else$^{-1}$\fi}
\def\mo{\ifmmode^{-1}\else$^{-1}$\fi}
\def\sup#1{\ifmmode ^{\rm #1}\else $^{\rm #1}$\fi}
\def\expo#1{\ifmmode \times 10^{#1}\else $\times 10^{#1}$\fi}
\def\,{\thinspace}
\def\lsim{\mathrel{\raise .4ex\hbox{\rlap{$<$}\lower 1.2ex\hbox{$\sim$}}}}
\def\gsim{\mathrel{\raise .4ex\hbox{\rlap{$>$}\lower 1.2ex\hbox{$\sim$}}}}

\def\simprop{\mathrel{\raise .4ex\hbox{\rlap{$\propto$}\lower 1.2ex\hbox{$\sim$}}}}
\def\deg{\ifmmode^\circ\else$^\circ$\fi}
\def\pdeg{\ifmmode $\setbox0=\hbox{$^{\circ}$}\rlap{\hskip.11\wd0 .}$^{\circ}
          \else \setbox0=\hbox{$^{\circ}$}\rlap{\hskip.11\wd0 .}$^{\circ}$\fi}
\def\arcs{\ifmmode {^{\scriptstyle\prime\prime}}
          \else $^{\scriptstyle\prime\prime}$\fi}
\def\arcm{\ifmmode {^{\scriptstyle\prime}}
          \else $^{\scriptstyle\prime}$\fi}
\newdimen\sa  \newdimen\sb
\def\parcs{\sa=.07em \sb=.03em
     \ifmmode \hbox{\rlap{.}}^{\scriptstyle\prime\kern -\sb\prime}\hbox{\kern -\sa}
     \else \rlap{.}$^{\scriptstyle\prime\kern -\sb\prime}$\kern -\sa\fi}
\def\parcm{\sa=.08em \sb=.03em
     \ifmmode \hbox{\rlap{.}\kern\sa}^{\scriptstyle\prime}\hbox{\kern-\sb}
     \else \rlap{.}\kern\sa$^{\scriptstyle\prime}$\kern-\sb\fi}
\def\ra[#1 #2 #3.#4]{#1\sup{h}#2\sup{m}#3\sup{s}\llap.#4}
\def\dec[#1 #2 #3.#4]{#1\deg#2\arcm#3\arcs\llap.#4}
\def\deco[#1 #2 #3]{#1\deg#2\arcm#3\arcs}
\def\rra[#1 #2]{#1\sup{h}#2\sup{m}}

\def\dots{\relax\ifmmode \ldots\else $\ldots$\fi}
\def\WHzsr{\ifmmode $W\,Hz\mo\,sr\mo$\else W\,Hz\mo\,sr\mo\fi}
\def\mHz{\ifmmode $\,mHz$\else \,mHz\fi}
\def\GHz{\ifmmode $\,GHz$\else \,GHz\fi}
\def\mKs{\ifmmode $\,mK\,s$^{1/2}\else \,mK\,s$^{1/2}$\fi}
\def\muKs{\ifmmode \,\mu$K\,s$^{1/2}\else \,$\mu$K\,s$^{1/2}$\fi}
\def\muKRJs{\ifmmode \,\mu$K$_{\rm RJ}$\,s$^{1/2}\else \,$\mu$K$_{\rm RJ}$\,s$^{1/2}$\fi}
\def\muKHz{\ifmmode \,\mu$K\,Hz$^{-1/2}\else \,$\mu$K\,Hz$^{-1/2}$\fi}
\def\MJysr{\ifmmode \,$MJy\,sr\mo$\else \,MJy\,sr\mo\fi}
\def\MJysrmK{\ifmmode \,$MJy\,sr\mo$\,mK$_{\rm CMB}\mo\else \,MJy\,sr\mo\,mK$_{\rm CMB}\mo$\fi}
\def\microns{\ifmmode \,\mu$m$\else \,$\mu$m\fi}
\def\micron{\microns}
\def\muK{\ifmmode \,\mu$K$\else \,$\mu$\hbox{K}\fi}
\def\microK{\ifmmode \,\mu$K$\else \,$\mu$\hbox{K}\fi}
\def\muW{\ifmmode \,\mu$W$\else \,$\mu$\hbox{W}\fi}
\def\kms{\ifmmode $\,km\,s$^{-1}\else \,km\,s$^{-1}$\fi}
\def\kmsMpc{\ifmmode $\,\kms\,Mpc\mo$\else \,\kms\,Mpc\mo\fi}
\setsymbol{LFI:center:frequency:70GHz:units}{70.3\,GHz}
\setsymbol{LFI:center:frequency:44GHz:units}{44.1\,GHz}
\setsymbol{LFI:center:frequency:30GHz:units}{28.5\,GHz}

\setsymbol{LFI:center:frequency:70GHz}{70.3}
\setsymbol{LFI:center:frequency:44GHz}{44.1}
\setsymbol{LFI:center:frequency:30GHz}{28.5}

\setsymbol{LFI:center:frequency:LFI18:Rad:M:units}{71.7\GHz}
\setsymbol{LFI:center:frequency:LFI19:Rad:M:units}{67.5\GHz}
\setsymbol{LFI:center:frequency:LFI20:Rad:M:units}{69.2\GHz}
\setsymbol{LFI:center:frequency:LFI21:Rad:M:units}{70.4\GHz}
\setsymbol{LFI:center:frequency:LFI22:Rad:M:units}{71.5\GHz}
\setsymbol{LFI:center:frequency:LFI23:Rad:M:units}{70.8\GHz}
\setsymbol{LFI:center:frequency:LFI24:Rad:M:units}{44.4\GHz}
\setsymbol{LFI:center:frequency:LFI25:Rad:M:units}{44.0\GHz}
\setsymbol{LFI:center:frequency:LFI26:Rad:M:units}{43.9\GHz}
\setsymbol{LFI:center:frequency:LFI27:Rad:M:units}{28.3\GHz}
\setsymbol{LFI:center:frequency:LFI28:Rad:M:units}{28.8\GHz}
\setsymbol{LFI:center:frequency:LFI18:Rad:S:units}{70.1\GHz}
\setsymbol{LFI:center:frequency:LFI19:Rad:S:units}{69.6\GHz}
\setsymbol{LFI:center:frequency:LFI20:Rad:S:units}{69.5\GHz}
\setsymbol{LFI:center:frequency:LFI21:Rad:S:units}{69.5\GHz}
\setsymbol{LFI:center:frequency:LFI22:Rad:S:units}{72.8\GHz}
\setsymbol{LFI:center:frequency:LFI23:Rad:S:units}{71.3\GHz}
\setsymbol{LFI:center:frequency:LFI24:Rad:S:units}{44.1\GHz}
\setsymbol{LFI:center:frequency:LFI25:Rad:S:units}{44.1\GHz}
\setsymbol{LFI:center:frequency:LFI26:Rad:S:units}{44.1\GHz}
\setsymbol{LFI:center:frequency:LFI27:Rad:S:units}{28.5\GHz}
\setsymbol{LFI:center:frequency:LFI28:Rad:S:units}{28.2\GHz}

\setsymbol{LFI:center:frequency:LFI18:Rad:M}{71.7}
\setsymbol{LFI:center:frequency:LFI19:Rad:M}{67.5}
\setsymbol{LFI:center:frequency:LFI20:Rad:M}{69.2}
\setsymbol{LFI:center:frequency:LFI21:Rad:M}{70.4}
\setsymbol{LFI:center:frequency:LFI22:Rad:M}{71.5}
\setsymbol{LFI:center:frequency:LFI23:Rad:M}{70.8}
\setsymbol{LFI:center:frequency:LFI24:Rad:M}{44.4}
\setsymbol{LFI:center:frequency:LFI25:Rad:M}{44.0}
\setsymbol{LFI:center:frequency:LFI26:Rad:M}{43.9}
\setsymbol{LFI:center:frequency:LFI27:Rad:M}{28.3}
\setsymbol{LFI:center:frequency:LFI28:Rad:M}{28.8}
\setsymbol{LFI:center:frequency:LFI18:Rad:S}{70.1}
\setsymbol{LFI:center:frequency:LFI19:Rad:S}{69.6}
\setsymbol{LFI:center:frequency:LFI20:Rad:S}{69.5}
\setsymbol{LFI:center:frequency:LFI21:Rad:S}{69.5}
\setsymbol{LFI:center:frequency:LFI22:Rad:S}{72.8}
\setsymbol{LFI:center:frequency:LFI23:Rad:S}{71.3}
\setsymbol{LFI:center:frequency:LFI24:Rad:S}{44.1}
\setsymbol{LFI:center:frequency:LFI25:Rad:S}{44.1}
\setsymbol{LFI:center:frequency:LFI26:Rad:S}{44.1}
\setsymbol{LFI:center:frequency:LFI27:Rad:S}{28.5}
\setsymbol{LFI:center:frequency:LFI28:Rad:S}{28.2}

\setsymbol{LFI:white:noise:sensitivity:70GHz:units}{134.7\muKs}
\setsymbol{LFI:white:noise:sensitivity:44GHz:units}{164.7\muKs}
\setsymbol{LFI:white:noise:sensitivity:30GHz:units}{143.4\muKs}

\setsymbol{LFI:white:noise:sensitivity:70GHz}{134.7}
\setsymbol{LFI:white:noise:sensitivity:44GHz}{164.7}
\setsymbol{LFI:white:noise:sensitivity:30GHz}{143.4}

\setsymbol{LFI:white:noise:sensitivity:LFI18:Rad:M:units}{512.0\muKs}
\setsymbol{LFI:white:noise:sensitivity:LFI19:Rad:M:units}{581.4\muKs}
\setsymbol{LFI:white:noise:sensitivity:LFI20:Rad:M:units}{590.8\muKs}
\setsymbol{LFI:white:noise:sensitivity:LFI21:Rad:M:units}{455.2\muKs}
\setsymbol{LFI:white:noise:sensitivity:LFI22:Rad:M:units}{492.0\muKs}
\setsymbol{LFI:white:noise:sensitivity:LFI23:Rad:M:units}{507.7\muKs}
\setsymbol{LFI:white:noise:sensitivity:LFI24:Rad:M:units}{462.2\muKs}
\setsymbol{LFI:white:noise:sensitivity:LFI25:Rad:M:units}{413.6\muKs}
\setsymbol{LFI:white:noise:sensitivity:LFI26:Rad:M:units}{478.6\muKs}
\setsymbol{LFI:white:noise:sensitivity:LFI27:Rad:M:units}{277.7\muKs}
\setsymbol{LFI:white:noise:sensitivity:LFI28:Rad:M:units}{312.3\muKs}
\setsymbol{LFI:white:noise:sensitivity:LFI18:Rad:S:units}{465.7\muKs}
\setsymbol{LFI:white:noise:sensitivity:LFI19:Rad:S:units}{555.6\muKs}
\setsymbol{LFI:white:noise:sensitivity:LFI20:Rad:S:units}{623.2\muKs}
\setsymbol{LFI:white:noise:sensitivity:LFI21:Rad:S:units}{564.1\muKs}
\setsymbol{LFI:white:noise:sensitivity:LFI22:Rad:S:units}{534.4\muKs}
\setsymbol{LFI:white:noise:sensitivity:LFI23:Rad:S:units}{542.4\muKs}
\setsymbol{LFI:white:noise:sensitivity:LFI24:Rad:S:units}{399.2\muKs}
\setsymbol{LFI:white:noise:sensitivity:LFI25:Rad:S:units}{392.6\muKs}
\setsymbol{LFI:white:noise:sensitivity:LFI26:Rad:S:units}{418.6\muKs}
\setsymbol{LFI:white:noise:sensitivity:LFI27:Rad:S:units}{302.9\muKs}
\setsymbol{LFI:white:noise:sensitivity:LFI28:Rad:S:units}{285.3\muKs}

\setsymbol{LFI:white:noise:sensitivity:LFI18:Rad:M}{512.0}
\setsymbol{LFI:white:noise:sensitivity:LFI19:Rad:M}{581.4}
\setsymbol{LFI:white:noise:sensitivity:LFI20:Rad:M}{590.8}
\setsymbol{LFI:white:noise:sensitivity:LFI21:Rad:M}{455.2}
\setsymbol{LFI:white:noise:sensitivity:LFI22:Rad:M}{492.0}
\setsymbol{LFI:white:noise:sensitivity:LFI23:Rad:M}{507.7}
\setsymbol{LFI:white:noise:sensitivity:LFI24:Rad:M}{462.2}
\setsymbol{LFI:white:noise:sensitivity:LFI25:Rad:M}{413.6}
\setsymbol{LFI:white:noise:sensitivity:LFI26:Rad:M}{478.6}
\setsymbol{LFI:white:noise:sensitivity:LFI27:Rad:M}{277.7}
\setsymbol{LFI:white:noise:sensitivity:LFI28:Rad:M}{312.3}
\setsymbol{LFI:white:noise:sensitivity:LFI18:Rad:S}{465.7}
\setsymbol{LFI:white:noise:sensitivity:LFI19:Rad:S}{555.6}
\setsymbol{LFI:white:noise:sensitivity:LFI20:Rad:S}{623.2}
\setsymbol{LFI:white:noise:sensitivity:LFI21:Rad:S}{564.1}
\setsymbol{LFI:white:noise:sensitivity:LFI22:Rad:S}{534.4}
\setsymbol{LFI:white:noise:sensitivity:LFI23:Rad:S}{542.4}
\setsymbol{LFI:white:noise:sensitivity:LFI24:Rad:S}{399.2}
\setsymbol{LFI:white:noise:sensitivity:LFI25:Rad:S}{392.6}
\setsymbol{LFI:white:noise:sensitivity:LFI26:Rad:S}{418.6}
\setsymbol{LFI:white:noise:sensitivity:LFI27:Rad:S}{302.9}
\setsymbol{LFI:white:noise:sensitivity:LFI28:Rad:S}{285.3}

\setsymbol{LFI:knee:frequency:70GHz:units}{29.5\mHz}
\setsymbol{LFI:knee:frequency:44GHz:units}{56.2\mHz}
\setsymbol{LFI:knee:frequency:30GHz:units}{113.7\mHz}

\setsymbol{LFI:knee:frequency:70GHz}{29.5}
\setsymbol{LFI:knee:frequency:44GHz}{56.2}
\setsymbol{LFI:knee:frequency:30GHz}{113.7}

\setsymbol{LFI:knee:frequency:LFI18:Rad:M:units}{16.3\mHz}
\setsymbol{LFI:knee:frequency:LFI19:Rad:M:units}{15.1\mHz}
\setsymbol{LFI:knee:frequency:LFI20:Rad:M:units}{18.7\mHz}
\setsymbol{LFI:knee:frequency:LFI21:Rad:M:units}{37.2\mHz}
\setsymbol{LFI:knee:frequency:LFI22:Rad:M:units}{12.7\mHz}
\setsymbol{LFI:knee:frequency:LFI23:Rad:M:units}{34.6\mHz}
\setsymbol{LFI:knee:frequency:LFI24:Rad:M:units}{46.2\mHz}
\setsymbol{LFI:knee:frequency:LFI25:Rad:M:units}{24.9\mHz}
\setsymbol{LFI:knee:frequency:LFI26:Rad:M:units}{67.6\mHz}
\setsymbol{LFI:knee:frequency:LFI27:Rad:M:units}{187.4\mHz}
\setsymbol{LFI:knee:frequency:LFI28:Rad:M:units}{122.2\mHz}
\setsymbol{LFI:knee:frequency:LFI18:Rad:S:units}{17.7\mHz}
\setsymbol{LFI:knee:frequency:LFI19:Rad:S:units}{22.0\mHz}
\setsymbol{LFI:knee:frequency:LFI20:Rad:S:units}{8.7\mHz}
\setsymbol{LFI:knee:frequency:LFI21:Rad:S:units}{25.9\mHz}
\setsymbol{LFI:knee:frequency:LFI22:Rad:S:units}{15.8\mHz}
\setsymbol{LFI:knee:frequency:LFI23:Rad:S:units}{129.8\mHz}
\setsymbol{LFI:knee:frequency:LFI24:Rad:S:units}{100.9\mHz}
\setsymbol{LFI:knee:frequency:LFI25:Rad:S:units}{38.9\mHz}
\setsymbol{LFI:knee:frequency:LFI26:Rad:S:units}{58.9\mHz}
\setsymbol{LFI:knee:frequency:LFI27:Rad:S:units}{104.4\mHz}
\setsymbol{LFI:knee:frequency:LFI28:Rad:S:units}{40.7\mHz}

\setsymbol{LFI:knee:frequency:LFI18:Rad:M}{16.3}
\setsymbol{LFI:knee:frequency:LFI19:Rad:M}{15.1}
\setsymbol{LFI:knee:frequency:LFI20:Rad:M}{18.7}
\setsymbol{LFI:knee:frequency:LFI21:Rad:M}{37.2}
\setsymbol{LFI:knee:frequency:LFI22:Rad:M}{12.7}
\setsymbol{LFI:knee:frequency:LFI23:Rad:M}{34.6}
\setsymbol{LFI:knee:frequency:LFI24:Rad:M}{46.2}
\setsymbol{LFI:knee:frequency:LFI25:Rad:M}{24.9}
\setsymbol{LFI:knee:frequency:LFI26:Rad:M}{67.6}
\setsymbol{LFI:knee:frequency:LFI27:Rad:M}{187.4}
\setsymbol{LFI:knee:frequency:LFI28:Rad:M}{122.2}
\setsymbol{LFI:knee:frequency:LFI18:Rad:S}{17.7}
\setsymbol{LFI:knee:frequency:LFI19:Rad:S}{22.0}
\setsymbol{LFI:knee:frequency:LFI20:Rad:S}{8.7}
\setsymbol{LFI:knee:frequency:LFI21:Rad:S}{25.9}
\setsymbol{LFI:knee:frequency:LFI22:Rad:S}{15.8}
\setsymbol{LFI:knee:frequency:LFI23:Rad:S}{129.8}
\setsymbol{LFI:knee:frequency:LFI24:Rad:S}{100.9}
\setsymbol{LFI:knee:frequency:LFI25:Rad:S}{38.9}
\setsymbol{LFI:knee:frequency:LFI26:Rad:S}{58.9}
\setsymbol{LFI:knee:frequency:LFI27:Rad:S}{104.4}
\setsymbol{LFI:knee:frequency:LFI28:Rad:S}{40.7}

\setsymbol{LFI:slope:70GHz:units}{$-1.03$\mHz}
\setsymbol{LFI:slope:44GHz:units}{$-0.89$\mHz}
\setsymbol{LFI:slope:30GHz:units}{$-0.87$\mHz}

\setsymbol{LFI:slope:70GHz}{$-1.03$}
\setsymbol{LFI:slope:44GHz}{$-0.89$}
\setsymbol{LFI:slope:30GHz}{$-0.87$}

\setsymbol{LFI:slope:LFI18:Rad:M:units}{$-1.04$\mHz}
\setsymbol{LFI:slope:LFI19:Rad:M:units}{$-1.09$\mHz}
\setsymbol{LFI:slope:LFI20:Rad:M:units}{$-0.69$\mHz}
\setsymbol{LFI:slope:LFI21:Rad:M:units}{$-1.56$\mHz}
\setsymbol{LFI:slope:LFI22:Rad:M:units}{$-1.01$\mHz}
\setsymbol{LFI:slope:LFI23:Rad:M:units}{$-0.96$\mHz}
\setsymbol{LFI:slope:LFI24:Rad:M:units}{$-0.83$\mHz}
\setsymbol{LFI:slope:LFI25:Rad:M:units}{$-0.91$\mHz}
\setsymbol{LFI:slope:LFI26:Rad:M:units}{$-0.95$\mHz}
\setsymbol{LFI:slope:LFI27:Rad:M:units}{$-0.87$\mHz}
\setsymbol{LFI:slope:LFI28:Rad:M:units}{$-0.88$\mHz}
\setsymbol{LFI:slope:LFI18:Rad:S:units}{$-1.15$\mHz}
\setsymbol{LFI:slope:LFI19:Rad:S:units}{$-1.00$\mHz}
\setsymbol{LFI:slope:LFI20:Rad:S:units}{$-0.95$\mHz}
\setsymbol{LFI:slope:LFI21:Rad:S:units}{$-0.92$\mHz}
\setsymbol{LFI:slope:LFI22:Rad:S:units}{$-1.01$\mHz}
\setsymbol{LFI:slope:LFI23:Rad:S:units}{$-0.95$\mHz}
\setsymbol{LFI:slope:LFI24:Rad:S:units}{$-0.73$\mHz}
\setsymbol{LFI:slope:LFI25:Rad:S:units}{$-1.16$\mHz}
\setsymbol{LFI:slope:LFI26:Rad:S:units}{$-0.79$\mHz}
\setsymbol{LFI:slope:LFI27:Rad:S:units}{$-0.82$\mHz}
\setsymbol{LFI:slope:LFI28:Rad:S:units}{$-0.91$\mHz}

\setsymbol{LFI:slope:LFI18:Rad:M}{$-1.04$}
\setsymbol{LFI:slope:LFI19:Rad:M}{$-1.09$}
\setsymbol{LFI:slope:LFI20:Rad:M}{$-0.69$}
\setsymbol{LFI:slope:LFI21:Rad:M}{$-1.56$}
\setsymbol{LFI:slope:LFI22:Rad:M}{$-1.01$}
\setsymbol{LFI:slope:LFI23:Rad:M}{$-0.96$}
\setsymbol{LFI:slope:LFI24:Rad:M}{$-0.83$}
\setsymbol{LFI:slope:LFI25:Rad:M}{$-0.91$}
\setsymbol{LFI:slope:LFI26:Rad:M}{$-0.95$}
\setsymbol{LFI:slope:LFI27:Rad:M}{$-0.87$}
\setsymbol{LFI:slope:LFI28:Rad:M}{$-0.88$}
\setsymbol{LFI:slope:LFI18:Rad:S}{$-1.15$}
\setsymbol{LFI:slope:LFI19:Rad:S}{$-1.00$}
\setsymbol{LFI:slope:LFI20:Rad:S}{$-0.95$}
\setsymbol{LFI:slope:LFI21:Rad:S}{$-0.92$}
\setsymbol{LFI:slope:LFI22:Rad:S}{$-1.01$}
\setsymbol{LFI:slope:LFI23:Rad:S}{$-0.95$}
\setsymbol{LFI:slope:LFI24:Rad:S}{$-0.73$}
\setsymbol{LFI:slope:LFI25:Rad:S}{$-1.16$}
\setsymbol{LFI:slope:LFI26:Rad:S}{$-0.79$}
\setsymbol{LFI:slope:LFI27:Rad:S}{$-0.82$}
\setsymbol{LFI:slope:LFI28:Rad:S}{$-0.91$}

\setsymbol{LFI:FWHM:70GHz:units}{13\parcm01}
\setsymbol{LFI:FWHM:44GHz:units}{27\parcm92}
\setsymbol{LFI:FWHM:30GHz:units}{32\parcm65}

\setsymbol{LFI:FWHM:70GHz}{13.01}
\setsymbol{LFI:FWHM:44GHz}{27.92}
\setsymbol{LFI:FWHM:30GHz}{32.65}

\setsymbol{LFI:FWHM:LFI18:units}{13\parcm39}
\setsymbol{LFI:FWHM:LFI19:units}{13\parcm01}
\setsymbol{LFI:FWHM:LFI20:units}{12\parcm75}
\setsymbol{LFI:FWHM:LFI21:units}{12\parcm74}
\setsymbol{LFI:FWHM:LFI22:units}{12\parcm87}
\setsymbol{LFI:FWHM:LFI23:units}{13\parcm27}
\setsymbol{LFI:FWHM:LFI24:units}{22\parcm98}
\setsymbol{LFI:FWHM:LFI25:units}{30\parcm46}
\setsymbol{LFI:FWHM:LFI26:units}{30\parcm31}
\setsymbol{LFI:FWHM:LFI27:units}{32\parcm65}
\setsymbol{LFI:FWHM:LFI28:units}{32\parcm66}

\setsymbol{LFI:FWHM:LFI18}{13.39}
\setsymbol{LFI:FWHM:LFI19}{13.01}
\setsymbol{LFI:FWHM:LFI20}{12.75}
\setsymbol{LFI:FWHM:LFI21}{12.74}
\setsymbol{LFI:FWHM:LFI22}{12.87}
\setsymbol{LFI:FWHM:LFI23}{13.27}
\setsymbol{LFI:FWHM:LFI24}{22.98}
\setsymbol{LFI:FWHM:LFI25}{30.46}
\setsymbol{LFI:FWHM:LFI26}{30.31}
\setsymbol{LFI:FWHM:LFI27}{32.65}
\setsymbol{LFI:FWHM:LFI28}{32.66}

\setsymbol{LFI:FWHM:uncertainty:LFI18:units}{0.170\arcm}
\setsymbol{LFI:FWHM:uncertainty:LFI19:units}{0.174\arcm}
\setsymbol{LFI:FWHM:uncertainty:LFI20:units}{0.170\arcm}
\setsymbol{LFI:FWHM:uncertainty:LFI21:units}{0.156\arcm}
\setsymbol{LFI:FWHM:uncertainty:LFI22:units}{0.164\arcm}
\setsymbol{LFI:FWHM:uncertainty:LFI23:units}{0.171\arcm}
\setsymbol{LFI:FWHM:uncertainty:LFI24:units}{0.652\arcm}
\setsymbol{LFI:FWHM:uncertainty:LFI25:units}{1.075\arcm}
\setsymbol{LFI:FWHM:uncertainty:LFI26:units}{1.131\arcm}
\setsymbol{LFI:FWHM:uncertainty:LFI27:units}{1.266\arcm}
\setsymbol{LFI:FWHM:uncertainty:LFI28:units}{1.287\arcm}

\setsymbol{LFI:FWHM:uncertainty:LFI18}{0.170}
\setsymbol{LFI:FWHM:uncertainty:LFI19}{0.174}
\setsymbol{LFI:FWHM:uncertainty:LFI20}{0.170}
\setsymbol{LFI:FWHM:uncertainty:LFI21}{0.156}
\setsymbol{LFI:FWHM:uncertainty:LFI22}{0.164}
\setsymbol{LFI:FWHM:uncertainty:LFI23}{0.171}
\setsymbol{LFI:FWHM:uncertainty:LFI24}{0.652}
\setsymbol{LFI:FWHM:uncertainty:LFI25}{1.075}
\setsymbol{LFI:FWHM:uncertainty:LFI26}{1.131}
\setsymbol{LFI:FWHM:uncertainty:LFI27}{1.266}
\setsymbol{LFI:FWHM:uncertainty:LFI28}{1.287}

\setsymbol{HFI:center:frequency:100GHz:units}{100\,GHz}
\setsymbol{HFI:center:frequency:143GHz:units}{143\,GHz}
\setsymbol{HFI:center:frequency:217GHz:units}{217\,GHz}
\setsymbol{HFI:center:frequency:353GHz:units}{353\,GHz}
\setsymbol{HFI:center:frequency:545GHz:units}{545\,GHz}
\setsymbol{HFI:center:frequency:857GHz:units}{857\,GHz}

\setsymbol{HFI:center:frequency:100GHz}{100}
\setsymbol{HFI:center:frequency:143GHz}{143}
\setsymbol{HFI:center:frequency:217GHz}{217}
\setsymbol{HFI:center:frequency:353GHz}{353}
\setsymbol{HFI:center:frequency:545GHz}{545}
\setsymbol{HFI:center:frequency:857GHz}{857}

\setsymbol{HFI:Ndetectors:100GHz}{8}
\setsymbol{HFI:Ndetectors:143GHz}{11}
\setsymbol{HFI:Ndetectors:217GHz}{12}
\setsymbol{HFI:Ndetectors:353GHz}{12}
\setsymbol{HFI:Ndetectors:545GHz}{3}
\setsymbol{HFI:Ndetectors:857GHz}{4}

\setsymbol{HFI:FWHM:Maps:100GHz:units}{9\parcm88}
\setsymbol{HFI:FWHM:Maps:143GHz:units}{7\parcm18}
\setsymbol{HFI:FWHM:Maps:217GHz:units}{4\parcm87}
\setsymbol{HFI:FWHM:Maps:353GHz:units}{4\parcm65}
\setsymbol{HFI:FWHM:Maps:545GHz:units}{4\parcm72}
\setsymbol{HFI:FWHM:Maps:857GHz:units}{4\parcm39}
\setsymbol{HFI:FWHM:Maps:100GHz}{9.88}
\setsymbol{HFI:FWHM:Maps:143GHz}{7.18}
\setsymbol{HFI:FWHM:Maps:217GHz}{4.87}
\setsymbol{HFI:FWHM:Maps:353GHz}{4.65}
\setsymbol{HFI:FWHM:Maps:545GHz}{4.72}
\setsymbol{HFI:FWHM:Maps:857GHz}{4.39}

\setsymbol{HFI:beam:ellipticity:Maps:100GHz}{1.15}
\setsymbol{HFI:beam:ellipticity:Maps:143GHz}{1.01}
\setsymbol{HFI:beam:ellipticity:Maps:217GHz}{1.06}
\setsymbol{HFI:beam:ellipticity:Maps:353GHz}{1.05}
\setsymbol{HFI:beam:ellipticity:Maps:545GHz}{1.14}
\setsymbol{HFI:beam:ellipticity:Maps:857GHz}{1.19}

\setsymbol{HFI:FWHM:Mars:100GHz:units}{9\parcm37}
\setsymbol{HFI:FWHM:Mars:143GHz:units}{7\parcm04}
\setsymbol{HFI:FWHM:Mars:217GHz:units}{4\parcm68}
\setsymbol{HFI:FWHM:Mars:353GHz:units}{4\parcm43}
\setsymbol{HFI:FWHM:Mars:545GHz:units}{3\parcm80}
\setsymbol{HFI:FWHM:Mars:857GHz:units}{3\parcm67}

\setsymbol{HFI:FWHM:Mars:100GHz}{9.37}
\setsymbol{HFI:FWHM:Mars:143GHz}{7.04}
\setsymbol{HFI:FWHM:Mars:217GHz}{4.68}
\setsymbol{HFI:FWHM:Mars:353GHz}{4.43}
\setsymbol{HFI:FWHM:Mars:545GHz}{3.80}
\setsymbol{HFI:FWHM:Mars:857GHz}{3.67}

\setsymbol{HFI:beam:ellipticity:Mars:100GHz}{1.18}
\setsymbol{HFI:beam:ellipticity:Mars:143GHz}{1.03}
\setsymbol{HFI:beam:ellipticity:Mars:217GHz}{1.14}
\setsymbol{HFI:beam:ellipticity:Mars:353GHz}{1.09}
\setsymbol{HFI:beam:ellipticity:Mars:545GHz}{1.25}
\setsymbol{HFI:beam:ellipticity:Mars:857GHz}{1.03}

\setsymbol{HFI:CMB:relative:calibration:100GHz}{$\lsim 1\%$}
\setsymbol{HFI:CMB:relative:calibration:143GHz}{$\lsim 1\%$}
\setsymbol{HFI:CMB:relative:calibration:217GHz}{$\lsim 1\%$}
\setsymbol{HFI:CMB:relative:calibration:353GHz}{$\lsim 1\%$}
\setsymbol{HFI:CMB:relative:calibration:545GHz}{}
\setsymbol{HFI:CMB:relative:calibration:857GHz}{}

\setsymbol{HFI:CMB:absolute:calibration:100GHz}{$\lsim 2\%$}
\setsymbol{HFI:CMB:absolute:calibration:143GHz}{$\lsim 2\%$}
\setsymbol{HFI:CMB:absolute:calibration:217GHz}{$\lsim 2\%$}
\setsymbol{HFI:CMB:absolute:calibration:353GHz}{$\lsim 2\%$}
\setsymbol{HFI:CMB:absolute:calibration:545GHz}{}
\setsymbol{HFI:CMB:absolute:calibration:857GHz}{}

\setsymbol{HFI:FIRAS:gain:calibration:accuracy:statistical:100GHz}{}
\setsymbol{HFI:FIRAS:gain:calibration:accuracy:statistical:143GHz}{}
\setsymbol{HFI:FIRAS:gain:calibration:accuracy:statistical:217GHz}{}
\setsymbol{HFI:FIRAS:gain:calibration:accuracy:statistical:353GHz}{2.5\%}
\setsymbol{HFI:FIRAS:gain:calibration:accuracy:statistical:545GHz}{1\%}
\setsymbol{HFI:FIRAS:gain:calibration:accuracy:statistical:857GHz}{0.5\%}

\setsymbol{HFI:FIRAS:gain:calibration:accuracy:systematic:100GHz}{}
\setsymbol{HFI:FIRAS:gain:calibration:accuracy:systematic:143GHz}{}
\setsymbol{HFI:FIRAS:gain:calibration:accuracy:systematic:217GHz}{}
\setsymbol{HFI:FIRAS:gain:calibration:accuracy:systematic:353GHz}{}
\setsymbol{HFI:FIRAS:gain:calibration:accuracy:systematic:545GHz}{7\%}
\setsymbol{HFI:FIRAS:gain:calibration:accuracy:systematic:857GHz}{7\%}

\setsymbol{HFI:FIRAS:zero:point:accuracy:100GHz:units}{0.8\MJysr}
\setsymbol{HFI:FIRAS:zero:point:accuracy:143GHz:units}{}
\setsymbol{HFI:FIRAS:zero:point:accuracy:217GHz:units}{}
\setsymbol{HFI:FIRAS:zero:point:accuracy:353GHz:units}{1.4\MJysr}
\setsymbol{HFI:FIRAS:zero:point:accuracy:545GHz:units}{2.2\MJysr}
\setsymbol{HFI:FIRAS:zero:point:accuracy:857GHz:units}{1.7\MJysr}

\setsymbol{HFI:FIRAS:zero:point:accuracy:100GHz}{0.8}
\setsymbol{HFI:FIRAS:zero:point:accuracy:143GHz}{}
\setsymbol{HFI:FIRAS:zero:point:accuracy:217GHz}{}
\setsymbol{HFI:FIRAS:zero:point:accuracy:353GHz}{1.4}
\setsymbol{HFI:FIRAS:zero:point:accuracy:545GHz}{2.2}
\setsymbol{HFI:FIRAS:zero:point:accuracy:857GHz}{1.7}

\setsymbol{HFI:unit:conversion:100GHz:units}{0.2415\MJysrmK}
\setsymbol{HFI:unit:conversion:143GHz:units}{0.3694\MJysrmK}
\setsymbol{HFI:unit:conversion:217GHz:units}{0.4811\MJysrmK}
\setsymbol{HFI:unit:conversion:353GHz:units}{0.2883\MJysrmK}
\setsymbol{HFI:unit:conversion:545GHz:units}{0.05826\MJysrmK}
\setsymbol{HFI:unit:conversion:857GHz:units}{0.002238\MJysrmK}

\setsymbol{HFI:unit:conversion:100GHz}{0.2415}
\setsymbol{HFI:unit:conversion:143GHz}{0.3694}
\setsymbol{HFI:unit:conversion:217GHz}{0.4811}
\setsymbol{HFI:unit:conversion:353GHz}{0.2883}
\setsymbol{HFI:unit:conversion:545GHz}{0.05826}
\setsymbol{HFI:unit:conversion:857GHz}{0.002238}

\setsymbol{HFI:colour:correction:alpha=-2:V1.01:100GHz}{0.9893}
\setsymbol{HFI:colour:correction:alpha=-2:V1.01:143GHz}{0.9759}
\setsymbol{HFI:colour:correction:alpha=-2:V1.01:217GHz}{1.0007}
\setsymbol{HFI:colour:correction:alpha=-2:V1.01:353GHz}{1.0028}
\setsymbol{HFI:colour:correction:alpha=-2:V1.01:545GHz}{1.0019}
\setsymbol{HFI:colour:correction:alpha=-2:V1.01:857GHz}{0.9889}

\setsymbol{HFI:colour:correction:alpha=0:V1.01:100GHz}{1.0008}
\setsymbol{HFI:colour:correction:alpha=0:V1.01:143GHz}{1.0148}
\setsymbol{HFI:colour:correction:alpha=0:V1.01:217GHz}{0.9909}
\setsymbol{HFI:colour:correction:alpha=0:V1.01:353GHz}{0.9888}
\setsymbol{HFI:colour:correction:alpha=0:V1.01:545GHz}{0.9878}
\setsymbol{HFI:colour:correction:alpha=0:V1.01:857GHz}{1.0014}

\providecommand{\sorthelp}[1]{}

\newcommand{\ha}{H$\alpha$} 
\newcommand{\hi}{\ion{H}{i}} 
\newcommand{\hii}{\ion{H}{ii}}

\def\vhel{\ifmmode{V_{{\rm HEL}}}\else{$V_{{\rm HEL}}$}\fi}
\def\vsys{\ifmmode{V_{\rm sys}}\else{$V_{\rm sys}$}\fi}
\def\kms{\ifmmode{\,{\rm km\,s}^{-1}}\else{km\,s$^{-1}$}\fi}
\def\vlsr{\ifmmode{v_{\rm lsr}}\else{$v_{\rm lsr}$}\fi}
\def\ltsim{\ifmmode\stackrel{<}{_{\sim}}\else$\stackrel{<}{_{\sim}}$\fi}
\def\gtsim{\ifmmode\stackrel{>}{_{\sim}}\else$\stackrel{>}{_{\sim}}$\fi}

\newcommand{\planck}{\Planck} 

\newcommand{\IRAS}{\textit{IRAS\/}}

\newcommand{\WMAP}{\textit{WMAP\/}}
\newcommand{\COBE}{\textit{COBE\/}}
\newcommand{\DIRBE}{\COBE-DIRBE}

\newcommand{\Fermi}{\textit{Fermi\/}}
\newcommand{\MHz}{\,MHz}

\newcommand{\GeV}{\,GeV}
\newcommand{\kpc}{\,kpc}
\newcommand{\pc}{\,pc}
\newcommand{\um}{\,\micron}

\newcommand{\uK}{\,$\mu$K}
\newcommand{\healpix}{HEALPix}
\newcommand{\pion}{$\pi_0$}

\begin{document}
\author{\small
Planck Collaboration:
P.~A.~R.~Ade\inst{80}
\and
N.~Aghanim\inst{54}
\and
M.~I.~R.~Alves\inst{54}
\and
M.~Arnaud\inst{66}
\and
M.~Ashdown\inst{63, 6}
\and
F.~Atrio-Barandela\inst{17}
\and
J.~Aumont\inst{54}
\and
C.~Baccigalupi\inst{79}
\and
A.~J.~Banday\inst{83, 10}
\and
R.~B.~Barreiro\inst{60}
% ...
}
\institute{\small
APC, AstroParticule et Cosmologie, Universit\'{e} Paris Diderot, CNRS/IN2P3, CEA/lrfu, Observatoire de Paris, Sorbonne Paris Cit\'{e}, 10, rue Alice Domon et L\'{e}onie Duquet, 75205 Paris Cedex 13, France\\
\and
Aalto University Mets\"{a}hovi Radio Observatory and Dept of Radio Science and Engineering, P.O. Box 13000, FI-00076 AALTO, Finland\\
\and
African Institute for Mathematical Sciences, 6-8 Melrose Road, Muizenberg, Cape Town, South Africa\\
\and
Agenzia Spaziale Italiana Science Data Center, Via del Politecnico snc, 00133, Roma, Italy\\
\and
Agenzia Spaziale Italiana, Viale Liegi 26, Roma, Italy\\
\and
Astrophysics Group, Cavendish Laboratory, University of Cambridge, J J Thomson Avenue, Cambridge CB3 0HE, U.K.\\
\and
Astrophysics \& Cosmology Research Unit, School of Mathematics, Statistics \& Computer Science, University of KwaZulu-Natal, Westville Campus, Private Bag X54001, Durban 4000, South Africa\\
\and
Atacama Large Millimeter/submillimeter Array, ALMA Santiago Central Offices, Alonso de Cordova 3107, Vitacura, Casilla 763 0355, Santiago, Chile\\
\and
CITA, University of Toronto, 60 St. George St., Toronto, ON M5S 3H8, Canada\\
\and
CNRS, IRAP, 9 Av. colonel Roche, BP 44346, F-31028 Toulouse cedex 4, France\\
\and
California Institute of Technology, Pasadena, California, U.S.A.\\
\and
Computational Cosmology Center, Lawrence Berkeley National Laboratory, Berkeley, California, U.S.A.\\
\and
Consejo Superior de Investigaciones Cient\'{\i}ficas (CSIC), Madrid, Spain\\
\and
DSM/Irfu/SPP, CEA-Saclay, F-91191 Gif-sur-Yvette Cedex, France\\
\and
DTU Space, National Space Institute, Technical University of Denmark, Elektrovej 327, DK-2800 Kgs. Lyngby, Denmark\\
\and
D\'{e}partement de Physique Th\'{e}orique, Universit\'{e} de Gen\`{e}ve, 24, Quai E. Ansermet,1211 Gen\`{e}ve 4, Switzerland\\
\and
Departamento de F\'{\i}sica Fundamental, Facultad de Ciencias, Universidad de Salamanca, 37008 Salamanca, Spain\\
\and
Departamento de F\'{\i}sica, Universidad de Oviedo, Avda. Calvo Sotelo s/n, Oviedo, Spain\\
\and
Department of Astrophysics/IMAPP, Radboud University Nijmegen, P.O. Box 9010, 6500 GL Nijmegen, The Netherlands\\
\and
Department of Electrical Engineering and Computer Sciences, University of California, Berkeley, California, U.S.A.\\
\and
Department of Physics \& Astronomy, University of British Columbia, 6224 Agricultural Road, Vancouver, British Columbia, Canada\\
\and
Department of Physics and Astronomy, Dana and David Dornsife College of Letter, Arts and Sciences, University of Southern California, Los Angeles, CA 90089, U.S.A.\\
\and
Department of Physics and Astronomy, University College London, London WC1E 6BT, U.K.\\
\and
Department of Physics, Florida State University, Keen Physics Building, 77 Chieftan Way, Tallahassee, Florida, U.S.A.\\
\and
Department of Physics, Gustaf H\"{a}llstr\"{o}min katu 2a, University of Helsinki, Helsinki, Finland\\
\and
Department of Physics, Princeton University, Princeton, New Jersey, U.S.A.\\
\and
Department of Physics, University of California, Santa Barbara, California, U.S.A.\\
\and
Department of Physics, University of Illinois at Urbana-Champaign, 1110 West Green Street, Urbana, Illinois, U.S.A.\\
\and
Dipartimento di Fisica e Astronomia G. Galilei, Universit\`{a} degli Studi di Padova, via Marzolo 8, 35131 Padova, Italy\\
\and
Dipartimento di Fisica e Scienze della Terra, Universit\`{a} di Ferrara, Via Saragat 1, 44122 Ferrara, Italy\\
\and
Dipartimento di Fisica, Universit\`{a} La Sapienza, P. le A. Moro 2, Roma, Italy\\
\and
Dipartimento di Fisica, Universit\`{a} degli Studi di Milano, Via Celoria, 16, Milano, Italy\\
\and
Dipartimento di Fisica, Universit\`{a} degli Studi di Trieste, via A. Valerio 2, Trieste, Italy\\
\and
Dipartimento di Fisica, Universit\`{a} di Roma Tor Vergata, Via della Ricerca Scientifica, 1, Roma, Italy\\
\and
Discovery Center, Niels Bohr Institute, Blegdamsvej 17, Copenhagen, Denmark\\
\and
Dpto. Astrof\'{i}sica, Universidad de La Laguna (ULL), E-38206 La Laguna, Tenerife, Spain\\
\and
European Southern Observatory, ESO Vitacura, Alonso de Cordova 3107, Vitacura, Casilla 19001, Santiago, Chile\\
\and
European Space Agency, ESAC, Planck Science Office, Camino bajo del Castillo, s/n, Urbanizaci\'{o}n Villafranca del Castillo, Villanueva de la Ca\~{n}ada, Madrid, Spain\\
\and
European Space Agency, ESTEC, Keplerlaan 1, 2201 AZ Noordwijk, The Netherlands\\
\and
Helsinki Institute of Physics, Gustaf H\"{a}llstr\"{o}min katu 2, University of Helsinki, Helsinki, Finland\\
\and
INAF - Osservatorio Astronomico di Padova, Vicolo dell'Osservatorio 5, Padova, Italy\\
\and
INAF - Osservatorio Astronomico di Roma, via di Frascati 33, Monte Porzio Catone, Italy\\
\and
INAF - Osservatorio Astronomico di Trieste, Via G.B. Tiepolo 11, Trieste, Italy\\
\and
INAF Istituto di Radioastronomia, Via P. Gobetti 101, 40129 Bologna, Italy\\
\and
INAF/IASF Bologna, Via Gobetti 101, Bologna, Italy\\
\and
INAF/IASF Milano, Via E. Bassini 15, Milano, Italy\\
\and
INFN, Sezione di Bologna, Via Irnerio 46, I-40126, Bologna, Italy\\
\and
INFN, Sezione di Roma 1, Universit\`{a} di Roma Sapienza, Piazzale Aldo Moro 2, 00185, Roma, Italy\\
\and
INFN/National Institute for Nuclear Physics, Via Valerio 2, I-34127 Trieste, Italy\\
\and
IPAG: Institut de Plan\'{e}tologie et d'Astrophysique de Grenoble, Universit\'{e} Joseph Fourier, Grenoble 1 / CNRS-INSU, UMR 5274, Grenoble, F-38041, France\\
\and
IUCAA, Post Bag 4, Ganeshkhind, Pune University Campus, Pune 411 007, India\\
\and
Imperial College London, Astrophysics group, Blackett Laboratory, Prince Consort Road, London, SW7 2AZ, U.K.\\
\and
Infrared Processing and Analysis Center, California Institute of Technology, Pasadena, CA 91125, U.S.A.\\
\and
Institut d'Astrophysique Spatiale, CNRS (UMR8617) Universit\'{e} Paris-Sud 11, B\^{a}timent 121, Orsay, France\\
\and
Institut d'Astrophysique de Paris, CNRS (UMR7095), 98 bis Boulevard Arago, F-75014, Paris, France\\
\and
Institute for Space Sciences, Bucharest-Magurale, Romania\\
\and
Institute of Astronomy, University of Cambridge, Madingley Road, Cambridge CB3 0HA, U.K.\\
\and
Institute of Theoretical Astrophysics, University of Oslo, Blindern, Oslo, Norway\\
\and
Instituto de Astrof\'{\i}sica de Canarias, C/V\'{\i}a L\'{a}ctea s/n, La Laguna, Tenerife, Spain\\
\and
Instituto de F\'{\i}sica de Cantabria (CSIC-Universidad de Cantabria), Avda. de los Castros s/n, Santander, Spain\\
\and
Jet Propulsion Laboratory, California Institute of Technology, 4800 Oak Grove Drive, Pasadena, California, U.S.A.\\
\and
Jodrell Bank Centre for Astrophysics, Alan Turing Building, School of Physics and Astronomy, The University of Manchester, Oxford Road, Manchester, M13 9PL, U.K.\\
\and
Kavli Institute for Cosmology Cambridge, Madingley Road, Cambridge, CB3 0HA, U.K.\\
\and
LAL, Universit\'{e} Paris-Sud, CNRS/IN2P3, Orsay, France\\
\and
LERMA, CNRS, Observatoire de Paris, 61 Avenue de l'Observatoire, Paris, France\\
\and
Laboratoire AIM, IRFU/Service d'Astrophysique - CEA/DSM - CNRS - Universit\'{e} Paris Diderot, B\^{a}t. 709, CEA-Saclay, F-91191 Gif-sur-Yvette Cedex, France\\
\and
Laboratoire Traitement et Communication de l'Information, CNRS (UMR 5141) and T\'{e}l\'{e}com ParisTech, 46 rue Barrault F-75634 Paris Cedex 13, France\\
\and
Laboratoire de Physique Subatomique et de Cosmologie, Universit\'{e} Joseph Fourier Grenoble I, CNRS/IN2P3, Institut National Polytechnique de Grenoble, 53 rue des Martyrs, 38026 Grenoble cedex, France\\
\and
Laboratoire de Physique Th\'{e}orique, Universit\'{e} Paris-Sud 11 \& CNRS, B\^{a}timent 210, 91405 Orsay, France\\
\and
Lawrence Berkeley National Laboratory, Berkeley, California, U.S.A.\\
\and
Max-Planck-Institut f\"{u}r Astrophysik, Karl-Schwarzschild-Str. 1, 85741 Garching, Germany\\
\and
Max-Planck-Institut f\"{u}r Extraterrestrische Physik, Giessenbachstra{\ss}e, 85748 Garching, Germany\\
\and
Max-Planck-Institut f\"{u}r Radioastronomie, Auf dem H\"{u}gel 69, 53121 Bonn, Germany\\
\and
MilliLab, VTT Technical Research Centre of Finland, Tietotie 3, Espoo, Finland\\
\and
National University of Ireland, Department of Experimental Physics, Maynooth, Co. Kildare, Ireland\\
\and
Niels Bohr Institute, Blegdamsvej 17, Copenhagen, Denmark\\
\and
Observational Cosmology, Mail Stop 367-17, California Institute of Technology, Pasadena, CA, 91125, U.S.A.\\
\and
Optical Science Laboratory, University College London, Gower Street, London, U.K.\\
\and
SISSA, Astrophysics Sector, via Bonomea 265, 34136, Trieste, Italy\\
\and
School of Physics and Astronomy, Cardiff University, Queens Buildings, The Parade, Cardiff, CF24 3AA, U.K.\\
\and
Special Astrophysical Observatory, Russian Academy of Sciences, Nizhnij Arkhyz, Zelenchukskiy region, Karachai-Cherkessian Republic, 369167, Russia\\
\and
UPMC Univ Paris 06, UMR7095, 98 bis Boulevard Arago, F-75014, Paris, France\\
\and
Universit\'{e} de Toulouse, UPS-OMP, IRAP, F-31028 Toulouse cedex 4, France\\
% ...
}

\title{\emph{Planck} intermediate results. XXIII. Galactic plane emission components derived from \emph{Planck} with ancillary data}

\abstract{
\Planck\ data when combined with ancillary data provide a unique opportunity to separate the diffuse emission components of the inner Galaxy. The purpose of the paper is to elucidate the morphology of the various emission components in the strong star-formation region lying inside the solar radius and to clarify the relationship between the various components. The region of the Galactic plane covered is $l=300\deg \rightarrow 0\deg \rightarrow 60\deg$ where star-formation is highest and the emission is strong enough to make meaningful component separation. The latitude widths in this longitude range lie between 1\deg\ and 2\deg, which correspond to FWHM $z$-widths of 100--200\pc\ at a typical distance of 6\kpc. The four emission components studied here are synchrotron, free-free, anomalous microwave emission (AME), and thermal (vibrational) dust emission. These components are identified by constructing spectral energy distributions (SEDs) at positions along the Galactic plane using the wide frequency coverage of \Planck\ (\mbox{28.4--857\GHz}) in combination with low-frequency radio data at 0.408--2.3\GHz\ plus \WMAP\ data at 23--94\GHz, along with far-infrared (FIR) data from \DIRBE\ and \IRAS. The free-free component is determined from radio recombination line (RRL) data. AME is found to be comparable in brightness to the free-free emission on the Galactic plane in the frequency range 20--40\GHz\ with a width in latitude similar to that of the thermal dust; it comprises $45\pm1$\,\% of the total 28.4\GHz\ emission in the longitude range $l=300\deg \rightarrow 0\deg\rightarrow 60\deg$. The free-free component is the narrowest, reflecting the fact that it is produced by current star-formation as traced by the narrow distribution of OB stars. It is the dominant emission on the plane between 60 and 100\GHz. RRLs from this ionized gas are used to assess its distance, leading to a free-free $z$-width of FWHM $\approx100$\,pc. The narrow synchrotron component has a low-frequency brightness spectral index $\beta_{\rm synch}\approx-2.7$ that is similar to the broad synchrotron component indicating that they are both populated by the cosmic ray electrons of the same spectral index. The width of this narrow synchrotron component is significantly larger than that of the other three components, suggesting that it is generated in an assembly of older supernova remnants that have expanded to sizes of order 150\pc\ in $3\times10^5$\,yr; pulsars of a similar age have a similar spread in latitude. The thermal dust is identified in the SEDs with average parameters of $T_\mathrm{dust}=20.4\pm0.4$\,K, $\beta_\mathrm{FIR}=1.94\pm0.03$ ($>353$\,GHz), and $\beta_\mathrm{mm}=1.67\pm0.02$ ($<353$\,GHz). The latitude distributions of gamma-rays, CO, and the emission in high-frequency \Planck\ bands have similar widths, showing that they are all indicators of the total gaseous matter on the plane in the inner Galaxy.
}


\titlerunning{Galactic plane emission from \planck\ with ancillary data}

\authorrunning{Planck Collaboration}
\maketitle
%
\allearlypapers 

\section{Introduction}
\label{sec:intro}

At radio, millimetre, and far-infrared (FIR) wavelengths the emission from the interstellar medium (ISM) is generated by four distinct mechanisms. The three long-recognized components are the free-free emission from ionized gas \citep[e.g.,][]{D3:2003}, the synchrotron emission from relativistic electrons spiralling in the Galactic magnetic field \citep{Rybicki:1979}, and thermal (vibrational) emission from large dust grains \citep{Desert:1990}. Another component has been identified, namely the anomalous microwave emission (AME) correlated with dust and emitting in the frequency range 15--50\GHz. This emission has a peaked spectrum centred at 20--30\GHz\ and is believed to be electric dipole radiation from small spinning dust grains \citep{Draine:1998a, Draine:1998b}. A new flat-synchrotron component, the microwave haze, associated with gamma-ray emission, has also been identified at intermediate latitudes \citep{Dobler:2010,planck2012-IX}, but is not of concern here.

The aim of this paper is to study the inner Galaxy where the four emission components at radio and millimetre wavelengths peak in a narrow latitude band. In the past the components have been difficult to separate because their morphologies are similar, which is not the case at higher latitudes where component separation techniques have been successful \citep{gold2010,planck2013-XII}. The region $l=300\deg \rightarrow 0\deg \rightarrow 60\deg$, $|b|<2\deg$\ is the most active star formation area in the Galaxy. The latitude- or $z$-width of a component reflects its recent history. The most recent (current) star formation may be traced in the distribution of the dense CO and CS molecular clouds. The free-free emission traces both the current OB stars in individual \hii\ regions and the diffuse emission produced by the sum of all the ionizing stars. The narrow distribution of synchrotron emission traces the supernova remnants (SNRs) as they expand and become a diffuse background over their $10^5$--$10^6$\,yr lifetime. The latitude distribution of pulsars, which are also born in SN events, is also relevant to the discussion. The total gaseous mass distribution in latitude can be inferred from the thermal dust emission \mbox{\citep{planck2011-7.0}} and the gamma-ray \pion-decay data \citep{Grenier:2005}.  We will show that it is possible to separate the narrow latitude distribution from the broad distribution in each of the four mechanisms, in emission from the Gould Belt system in which the Sun lies, in the local spiral arm, and in the Galactic halo.
 
There are a number of methods available for separating the four emission components at intermediate and high Galactic latitudes. These methods assume that the components have  different distributions on the sky, and require templates for each component. The 408\MHz\ map of \citet{Haslam:1982} is used for the synchrotron emission template, and the \ha\ maps (WHAM, \citealp{Reynolds:1998}; SHASSA, \citealp{Gaustad:2001}; VTSS, \citealp{Dennison:1998}) are used for the free-free template. The FIR maps (\IRAS, \citealp{Miville-Deschenes:2005}; \DIRBE, \citealp{Hauser:1998}) are used as templates for the dust-correlated AME and the thermal dust. However, on the Galactic plane the four components have similar narrow latitude distributions and cannot be separated morphologically. Nevertheless, with a large frequency coverage it is possible to construct comprehensive spectral energy distributions (SEDs) along the Galactic plane and fit spectral models to them. This is where \planck\footnote{\planck\ (\url{http://www.esa.int/Planck}) is a project of the European Space Agency -- ESA -- with instruments provided by two scientific Consortia funded by ESA member states (in particular the lead countries: France and Italy) with contributions from NASA (USA), and telescope reflectors provided in a collaboration between ESA and a scientific Consortium led and funded by Denmark.} \citep{tauber2010a, planck2011-1.1} with its wide frequency coverage from 28.4 to 857\GHz\ combined with radio and FIR data can throw new light on the four Galactic plane components.

We will now summarize what is known about each of the components as determined at intermediate latitudes, which may act as a guide to what may be expected on the Galactic plane.

The Galactic synchrotron emission is generated by relativistic cosmic ray electrons (CRE) spiralling in the Galactic magnetic field. They have their origin in supernova shocks following the collapse of massive stars. The relativistic electrons eventually escape from the expanding remnants into the thin disk and subsequently into the Galactic halo. The high-energy component of the cosmic rays escapes from the supernova site and collides with interstellar gas to produce gamma-rays by \pion-decay. Neutron stars (pulsars) can also be produced in the stellar collapse; they are a good indicator of the Galactic supernova rate. The spectral index of the synchrotron power spectrum, $\alpha$, is directly related to the spectral index, $\delta$, of the CRE energy spectrum by the relation $\alpha = 0.5\times(\delta - 1)$ \citep{Rohlfs:2004}. The equivalent relation for the synchrotron brightness spectral index is $\beta_{\rm synch} = 0.5\times(\delta + 3)$  (where $T_{\rm b} \propto\nu^{\beta}$). Energy loss by radiation steepens the electron energy spectrum and accordingly the synchrotron spectrum also steepens at higher frequencies. The synchrotron brightness temperature spectral index at frequencies below $\sim 1$\GHz\ at intermediate latitudes is found to lie in the range $-2.5$ to $-2.7$ \citep{Lawson:1987, Broadbent:1989, Platania1998, Platania2003}. However it steepens significantly by 23\GHz\  to values of $-2.9$ to $-3.2$ as measured from 408\MHz\ \citep{Banday:2003, Davies:2006, gold2010,Ghosh:2011}. The steepening must occur somewhere in the range 2--20\GHz\ \citep{dunkley2009,Peel:2011,Strong:2011}. At \WMAP\ frequencies the spectral index is $-3.0$  at intermediate latitudes \citep{Banday:2003, Davies:2006, Kogut:2011}.

The free-free emission arises  from the ionized component of the ISM. This ionization is produced by OB stars recently formed in high-density regions; some of the emission comes from compact \hii\ regions and some is from the diffuse medium excited by the interstellar radiation field. The spectral shape of the free-free emission is well defined \citep{Bennett:1992,D3:2003,Draine2011Book}. The brightness temperature spectral index is $\beta_{\rm ff} \approx -2.10$ at 1\GHz, and steepens to $\beta_{\rm ff} \approx -2.14$ at 100\GHz. \ha\ data provide a good template for free-free emission at intermediate and high Galactic latitudes, although even here a correction is required to account for absorption by interstellar dust, which varies depending on the distributions of dust and ionized ISM along the line of sight. A fraction of the \ha\ emission may be scattered from the dust and would not have associated radio emission \citep{Mattila:2007, Witt:2010, Brandt2012}. This fraction lies in the range 14\,\% for low-density dust regions to 50\,\% in cirrus-dominated regions. A more direct determination of the free-free emission in dust-embedded \hii\ features at intermediate Galactic latitudes can be made using radio recombination lines (RRLs). On the Galactic plane this is the only direct method of determining the free-free emission (\citealp{Hart:1976}; \citealp{Lockman:1976}; \citealp{Alves2010, Alves2012}). In order to convert the RRL measurements into free-free brightness temperature ($T_\mathrm{b}$), it is necessary to know the electron temperature, $T_{\rm e}$ (as $T_\mathrm{b} \propto T_\mathrm{e}^{1.1}$). The fall in $T_{\rm e}$ towards the Galactic centre seen in individual \hii\ regions \citep{Shaver:1983, Paladini:2004} also applies to the diffuse free-free emission \citep{Alves2012}. The \planck\ 70.4 and 100\GHz\ channels are dominated by free-free emission on the Galactic plane. We will compare the estimates of the free-free from the component separation methods with the direct RRL determination for the longitude range $l = 20\deg$--$44\deg$.

Anomalous microwave emission (AME) has been observed in numerous experiments operating in the frequency range 10--60\GHz\ and peaks at 20--40\GHz\ \citep{Kogut:1996,deOliveira-costa:1999, Banday:2003, Lagache:2003, Finkbeiner:2004, Davies:2006, Miville-Deschenes:2008, Ysard:2010a, gold2010}. It is closely correlated with thermal emission from dust grains. Electric dipole radiation from small, rapidly rotating dust grains is thought to be responsible. Theoretical models predict a spectrum peaking typically in the range 10--60\GHz, depending on the properties of the dust grains and their environment \citep{Draine:1998a, Draine:1998b, Ali-Haimoud:2009,Dobler:2009, Ysard:2010a,Hoang2011}. The most compelling evidence for spinning dust grains comes from dedicated observations covering frequencies ranging from the gigahertz radio bands (e.g., COSMOSOMAS, Cosmic Background Imager, Very Small Array) through the millimetre bands of \WMAP\ and \planck\ to the FIR of \IRAS\ and \DIRBE. The best-studied examples at present are the Perseus molecular cloud \citep{Watson:2005,Tibbs2010,planck2011-7.2} and the $\rho$~Ophiuchi photo-dissociation region \citep{Casassus:2008,planck2011-7.2}; in these cases the peak is at $ 25$--30\GHz. AME has also been detected in dust clouds associated with \hii\ regions although with a range of emissivities \citep{Todorovic:2010, planck2011-7.2,planck2013-XII, planck2013-XV}. We will show that the narrow distribution of the total AME emission on the the Galactic plane arises from a range of physical environments that include the diffuse ionized medium, \hii\ regions, molecular clouds, and dust clouds with a spread of grain properties.

The spectrum of thermal dust is comprised of a cooler large-grain component and a warmer small-grain component \citep{Desert:1990,planck2011-7.12}. The large grains are responsible for the Rayleigh-Jeans emission tail measured by \planck\ while the small grains are responsible for the emission seen at near-IR wavelengths ($\sim5$--40\um) and also produce AME \citep{planck2011-7.3}. The near-IR band also includes emission from polycyclic aromatic hydrocarbons (PAHs) that may also contribute to AME \citep{Tielens:2008,Ysard:2010a,Ysard:2010b}. The PAH fraction is greatest in regions of highest gas and molecular surface density \citep{Sandstrom:2010}. Characterizing the Rayleigh-Jeans dust emission in the \planck\ regime is critical for the present study since it extends into the free-free, synchrotron, and AME domains. It is generally modelled in terms of a dust temperature, $T_\mathrm{dust}$, and a spectral index, $\beta_{\rm dust}$, by fitting over a broad spectral range. This is discussed in more detail in \citet{planck2011-7.0} and \citet{planck2011-7.12}. Conditions on the Galactic plane could be rather different from those at intermediate and high latitudes where the median value of $T_\mathrm{dust}$ is $19.7\pm1.4$\,K and $\beta_{\rm dust} = 1.62\pm0.10$ \citep{planck2013-p06b}.

Sections 2 and 3 describe the \Planck\ and ancillary data used in this paper. The analysis methods for component separation are discussed in Sect.~4; these include {\tt FastMEM} and SEDs. Section 5 describes the separation of the narrow and broad latitude distributions in each emission component and quantifies the narrow component in terms of a Gaussian amplitude and FWHM latitude width. Section 6 gives a discussion of the results of the component separation and uses the relative latitude widths to help understand the physical processes that lead to the different observed widths. The conclusions are presented in Sect.~7.


\section{\emph{Planck} data}
\label{sec:planckdata}

\Planck\ \citep{tauber2010a, planck2011-1.1} is the third generation space mission to measure the anisotropy of the cosmic microwave background (CMB).  It observed the sky in nine frequency bands covering 28.4--857\,GHz with high sensitivity and with angular resolution from 31\arcm\ to 5\arcm.  The Low Frequency Instrument (LFI; \citealt{Mandolesi2010, Bersanelli2010, planck2011-1.4}) covered the 28.4, 44.1, and 70.4\,GHz bands with amplifiers cooled to 20\,\hbox{K}.  The High Frequency Instrument (HFI; \citealt{Lamarre2010, planck2011-1.5}) covered the 100, 143, 217, 353, 545, and 857\,GHz bands with bolometers cooled to 0.1\,\hbox{K}.  Polarization was measured in all but the highest two bands \citep{Leahy2010, Rosset2010}.  A combination of radiative cooling and three mechanical coolers produced the temperatures needed for the detectors and optics \citep{planck2011-1.3}.  Two data processing centers (DPCs) check and calibrate the data and make maps of the sky \citep{planck2011-1.7, planck2011-1.6}.  \Planck's sensitivity, angular resolution, and frequency coverage make it a powerful instrument for Galactic and extragalactic astrophysics as well as cosmology.  Early astrophysics results are given in Planck Collaboration VIII--XXVI 2011, based on data taken between 13~August 2009 and 7~June 2010. Intermediate astrophysics results are now being presented in a series of papers based on data taken between 13~August 2009 and 27~November 2010 \citep{planck2012-IX,planck2013-XII,planck2013-XIV,planck2013-XV} and in the astrophysics papers from the 2013 cosmology results \citep{planck2013-p01,planck2013-p06b,planck2013-p03a}.

In this paper we use \planck\ data from the 2013 product release \citep{planck2013-p01} that are available from the \Planck\ Legacy Archive.\footnote{\url{http://www.sciops.esa.int/index.php?project=planck\&page=Planck\_Legacy\_Archive}} The data were acquired by \Planck\ during its ``nominal'' operations period from 13~August 2009 to 27 November 2010. Specifically, we use the nine temperature maps at nominal frequencies of 28.4, 44.1, 70.4, 100, 143, 217, 353, 545, and 857\GHz\ (see Table~\ref{tab:data}).

We assume Gaussian beams and use the average beamwidths given in \citet{planck2013-p01}. We note that, due to the smoothing applied here, details of the beam, such as the exact shape of the beam and its variation across the sky, do not significantly affect the results. Furthermore, extended emission is less sensitive to the details of the beam. We convert from CMB thermodynamic units to Rayleigh-Jeans brightness temperature units using the standard conversion factors given in \citet{planck2013-p28}. Colour corrections to account for the finite bandpass at each frequency are applied during the modelling of each spectrum; these are typically 1--3\,\% for LFI and 10--15\,\% for HFI. The \Planck\ bands centred at 100, 217, and 353\GHz\ are contaminated by Galactic CO lines. When determining SEDs for thermal dust we apply a CO correction (see Sect.~\ref{sec:co}).

\begin{table*}[tb]
\begingroup
\newdimen\tblskip \tblskip=5pt
\caption{ Summary of the data.}
 \label{tab:data}
\nointerlineskip
\vskip -3mm
\footnotesize
\setbox\tablebox=\vbox{
 \newdimen\digitwidth 
 \setbox0=\hbox{\rm 0} 
 \digitwidth=\wd0 
 \catcode`*=\active 
 \def*{\kern\digitwidth}
 \newdimen\signwidth 
 \setbox0=\hbox{+} 
 \signwidth=\wd0 
 \catcode`!=\active 
 \def!{\kern\signwidth}
 \halign{\hbox to 2in{#\leaderfil}\tabskip 0.5em&
 \hfil#\hfil&
 \hfil#\hfil&
 \hfil#\hfil&
 \hfil#\hfil&
 \hfil#\hfil\tabskip 0pt\cr
 \noalign{\doubleline\vskip 2pt}
 \omit\hfil Telescope/\hfil& Frequency & Angular
 resolution&Coverage&Notes \cr
 \omit\hfil survey \hfil& [GHz]& [arcmin]& & \cr
\noalign{\vskip 4pt\hrule\vskip 6pt}
JB/Eff/Parkes\tablefootmark{a} & \phantom{0000}0.408 & $51\parcm0$ & Full-sky & NCSA desourced/destriped version\tablefootmark{b} \cr
Stockert/Villa-Elisa\tablefootmark{c} & \phantom{0000}1.420 & $35\parcm4$ & Full-sky & Courtesy of W. Reich \cr
HartRAO\tablefootmark{d} & \phantom{0000}2.326 & $20\parcm0$ & Southern sky & Courtesy of J. Jonas (priv. comm.) \cr
\WMAP\ 9-yr\tablefootmark{e} & \phantom{000}22.8 & $51\parcm3$ & Full-sky & $1^{\circ}$ smoothed version\tablefootmark{b} \cr
\planck\ LFI\tablefootmark{f} & \phantom{000}28.4 & $32\parcm3 $ & Full-sky & \cr
\WMAP\ 9-yr\tablefootmark{e} & \phantom{000}33.0 & $39\parcm1$ & Full-sky & $1^{\circ}$ smoothed version\tablefootmark{b} \cr
\WMAP\ 9-yr\tablefootmark{e} & \phantom{000}40.7 & $30\parcm8$ & Full-sky & $1^{\circ}$ smoothed version\tablefootmark{b} \cr
\planck\ LFI\tablefootmark{f} & \phantom{000}44.1 & $27\parcm1$ & Full-sky & \cr
\WMAP\ 9-yr\tablefootmark{e} & \phantom{000}60.7 & $21\parcm1$ & Full-sky & $1^{\circ}$ smoothed version\tablefootmark{b} \cr
\planck\ LFI\tablefootmark{f} & \phantom{000}70.4 & $13\parcm3$ & Full-sky & \cr
\WMAP\ 9-yr\tablefootmark{e} & \phantom{000}93.5 & $14\parcm8$ & Full-sky & $1^{\circ}$ smoothed version\tablefootmark{b} \cr
\planck\ HFI\tablefootmark{f} & \phantom{00}100 & $*9\parcm7$ & Full-sky & \cr
\planck\ HFI\tablefootmark{f} & \phantom{00}143 & $*7\parcm3$ & Full-sky & \cr
\planck\ HFI\tablefootmark{f} & \phantom{00}217 & $*5\parcm0$ & Full-sky & \cr
\planck\ HFI\tablefootmark{f} & \phantom{00}353 & $*4\parcm8$ & Full-sky & \cr
\planck\ HFI\tablefootmark{f} & \phantom{00}545 & $*4\parcm7$ & Full-sky & \cr
\planck\ HFI\tablefootmark{f} & \phantom{00}857 & $*4\parcm3$ & Full-sky & \cr
\DIRBE\tablefootmark{g} & \phantom{0}1249 & $37\parcm1$ & Full-sky & LAMBDA website\tablefootmark{b} \cr
\DIRBE\tablefootmark{g} & \phantom{0}2141 & $38\parcm0$ & Full-sky & LAMBDA website\tablefootmark{b} \cr
\DIRBE\tablefootmark{g} & \phantom{0}2997 & $38\parcm6$ & Full-sky & LAMBDA website\tablefootmark{b} \cr
\IRAS\ (IRIS)\tablefootmark{h} & \phantom{0}2997 & $*4\parcm7$ & Near-full sky & \cr
\IRAS\ (IRIS)\tablefootmark{h} & 11988 & $*3\parcm6$ & Near-full sky & \cr
\IRAS\ (IRIS)\tablefootmark{h} & 24975 & $*3\parcm5$ & Near-full sky & \cr
\planck\ HFI\tablefootmark{i} & CO & $*9\parcm7$ & Full-sky & Type 1 MILCA version\cr
Parkes\tablefootmark{j} & RRL & $\sim 15\arcm$ & $l=20\deg$--$44\deg$, $|b|\leq 4\deg$ & \cr
\textit{Fermi}\tablefootmark{k} & Gamma-ray & $\sim 88\arcm$ & Full-sky & \citet{Dobler:2010}\tablefootmark{l} \cr
Arecibo/ATNF/GB-NRAO/JB\tablefootmark{m} & Pulsars & \ldots & Full-sky & ATNF Pulsar Catalogue\tablefootmark{n}\cr
Leiden/ Dwingeloo\tablefootmark{o} & \hi\     & $\sim 30$--$60\arcm$ & Full-sky & LAMBDA website\tablefootmark{b}\cr
\noalign{\vskip 3pt\hrule\vskip 4pt}
}}
\endPlancktablewide
\tablefoot{\tablefoottext{a}{\citet{Haslam:1982}}~\tablefoottext{b}{\url{http://lambda.gsfc.nasa.gov/}}~\tablefoottext{c}{\citet{Reich:1982}}~\tablefoottext{d}{\citet{Jonas:1998}}~\tablefoottext{e}{\citet{Bennett2013}}~\tablefoottext{f}{\citet{planck2013-p01}}~\tablefoottext{g}{\citet{Hauser:1998}}~\tablefoottext{h}{\citet{Miville-Deschenes:2005}}~\tablefoottext{i}{\citet{planck2013-p03a}}~\tablefoottext{j}{\citet{Alves2012}}~\tablefoottext{k}{\citet{Atwood:2009}}~\tablefoottext{l}{\url{http://astrometry.fas.harvard.edu/skymaps/fermi/}}~\tablefoottext{m}{\citet{Manchester:2005}}~\tablefoottext{n}{\url{http://www.atnf.csiro.au/research/pulsar/psrcat}} ~\tablefoottext{o}{\citet{Hartmann:1997,Dickey:1990}} }
\endgroup
\end{table*}

\section{Ancillary data}
\label{sec:ancdata}

\begin{figure*}
\centering
\includegraphics[angle=90,scale=0.34]{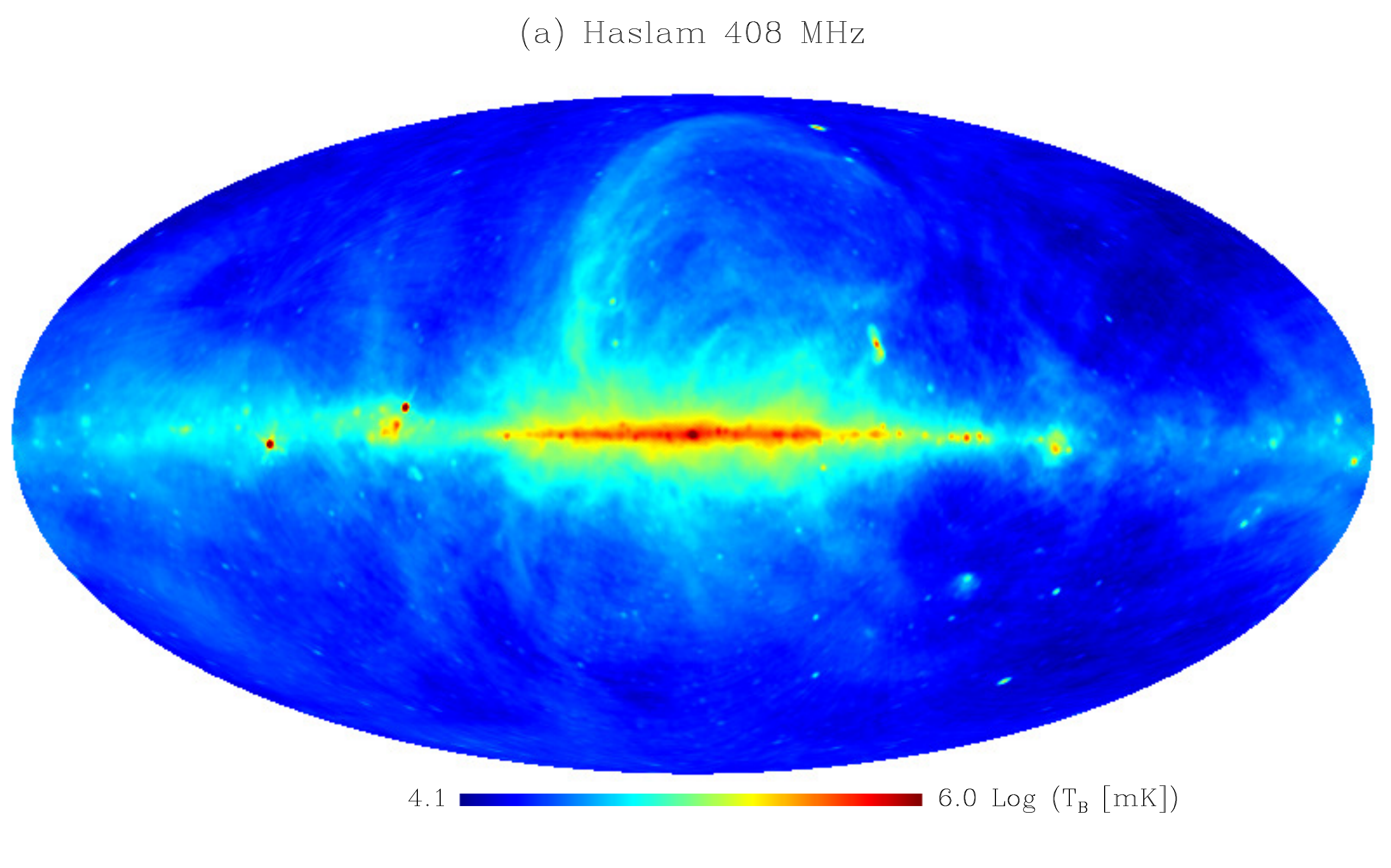}
\includegraphics[angle=90,scale=0.34]{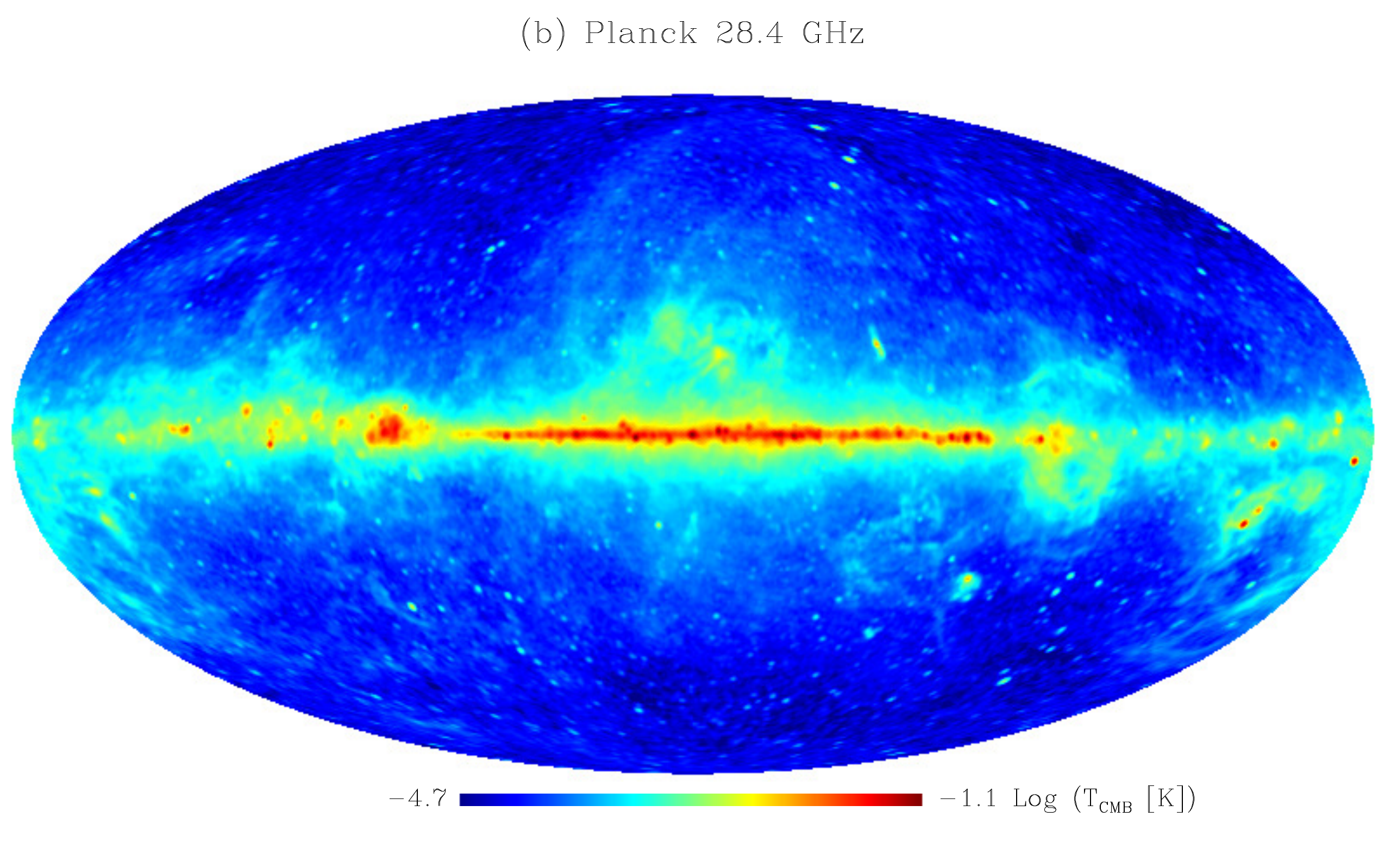}
\includegraphics[angle=90,scale=0.34]{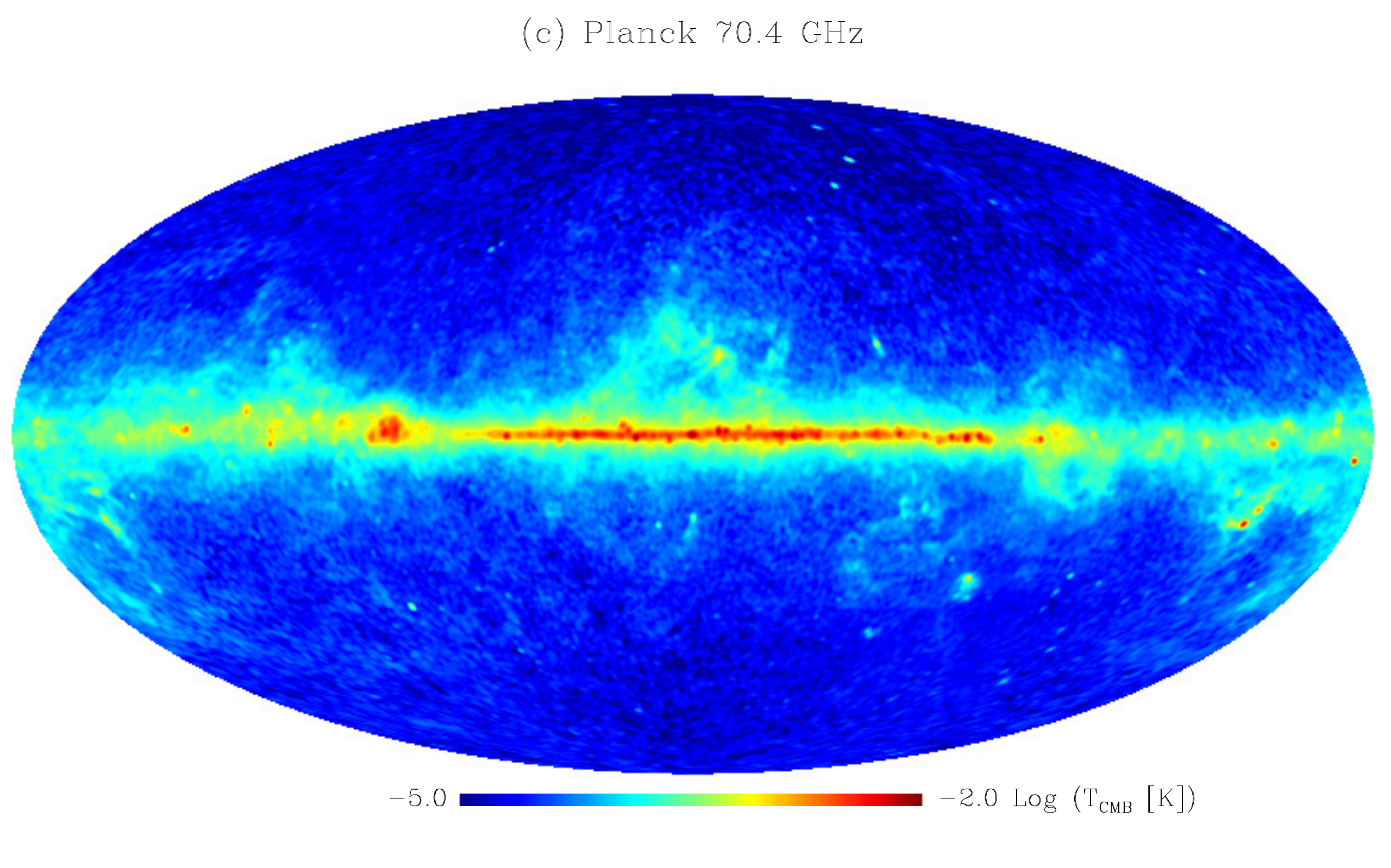}
\includegraphics[angle=90,scale=0.34]{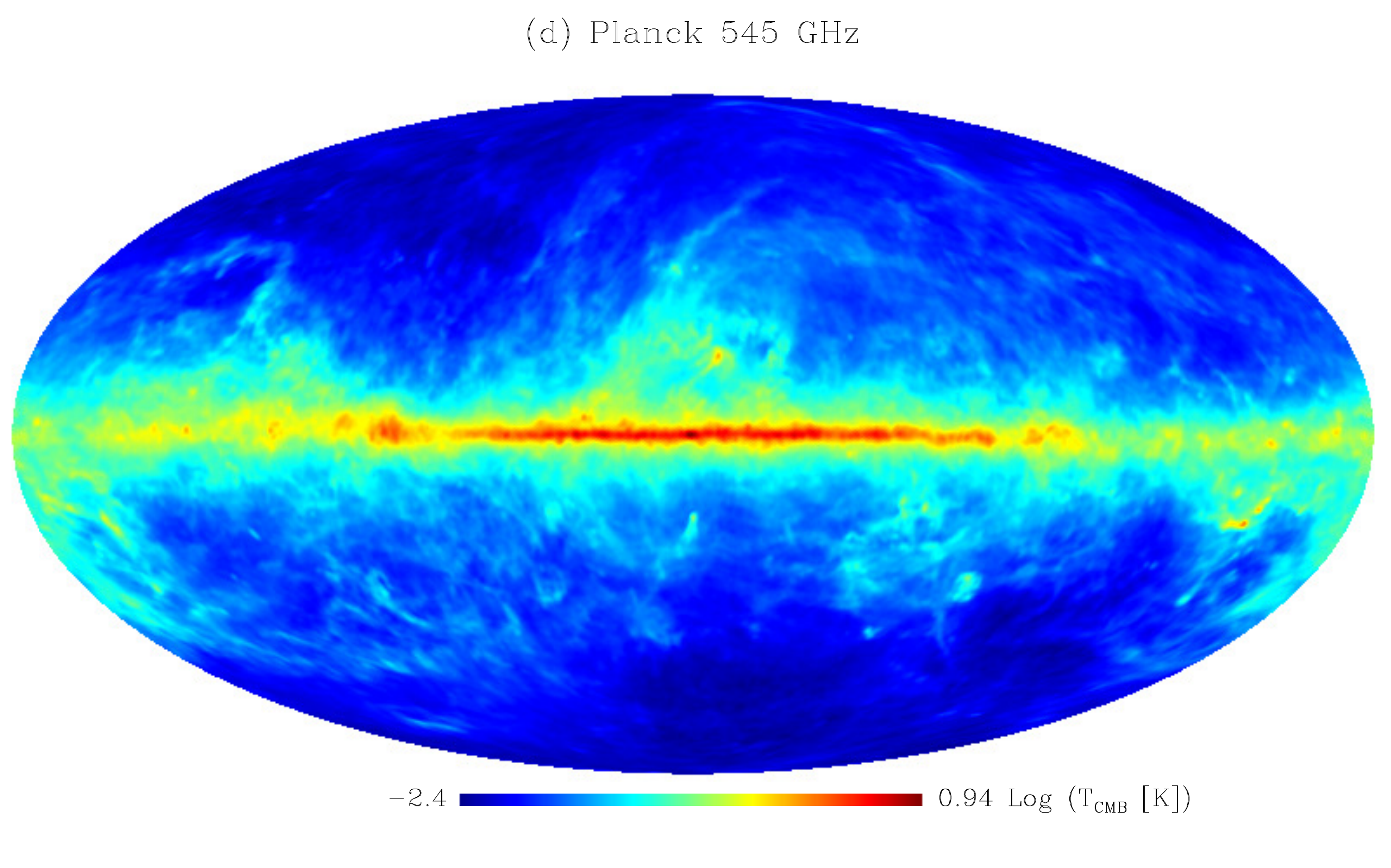}
\caption{The whole sky map at four frequencies showing the narrow inner Galaxy emission: (a) the 408\MHz\ map of \citet{Haslam:1982}, (b) the 28.4\GHz\ map from \planck, (c) the 70.4\GHz\ map from \planck, and (d) the 545\GHz\ map from \planck. The colour scales are logarithmic, and the maps are in Mollweide projection.}
\label{fig:compmaps}
\end{figure*}
\begin{figure*}
\centering
\includegraphics[angle=90,scale=1.1]{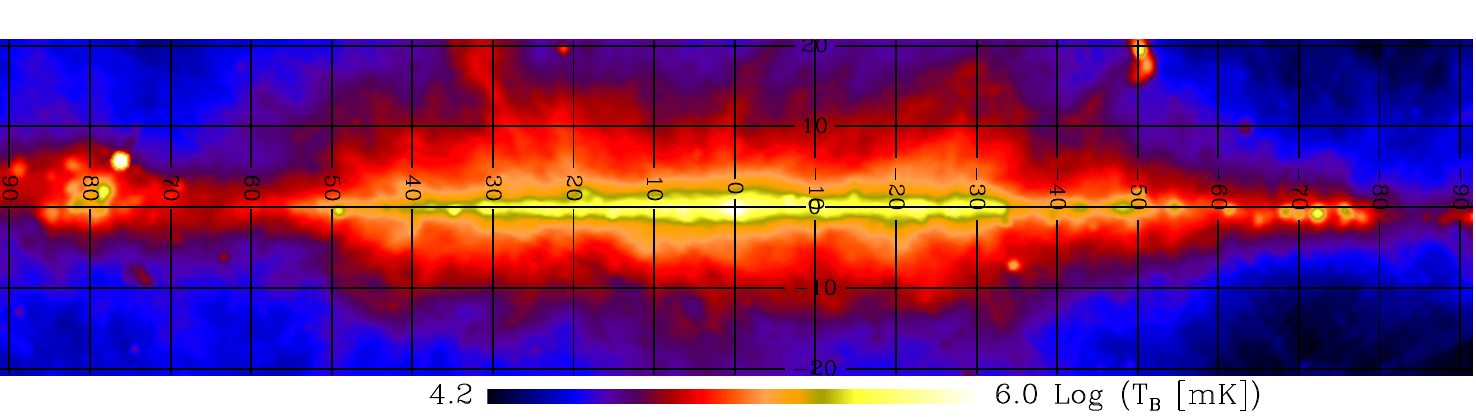}
\includegraphics[angle=90,scale=1.1]{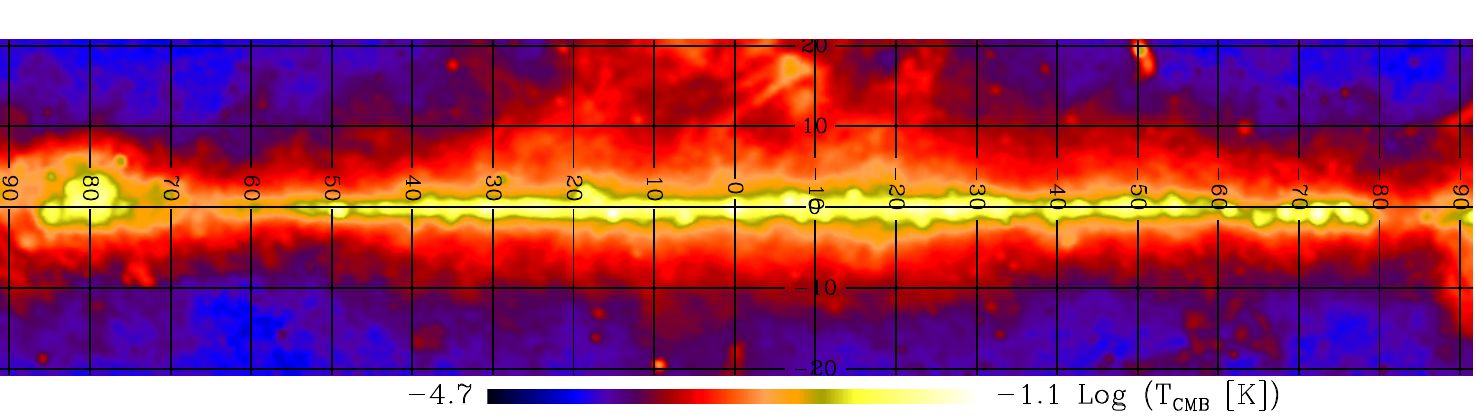}
\includegraphics[angle=90,scale=1.1]{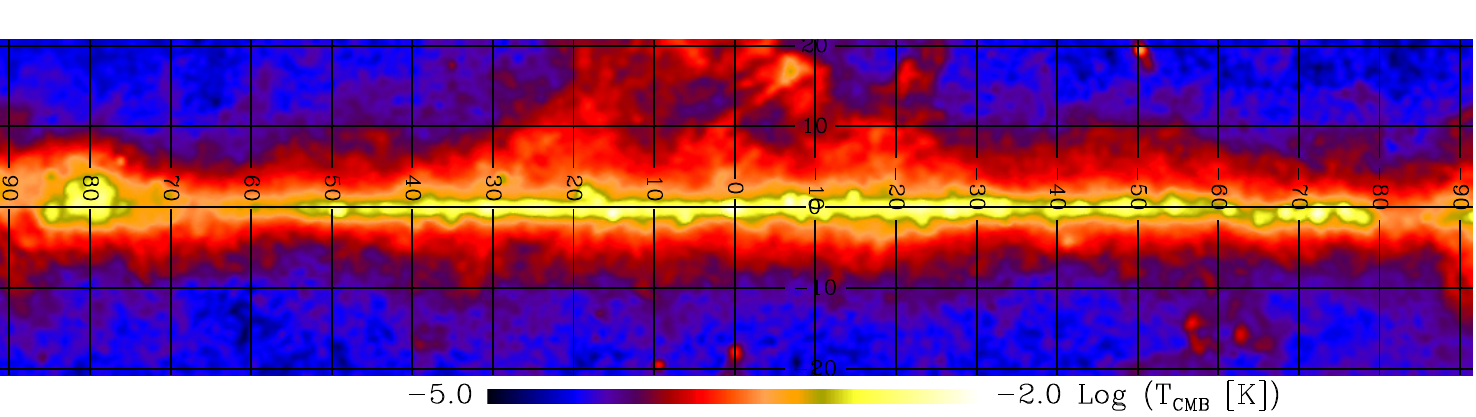}
\includegraphics[angle=90,scale=1.1]{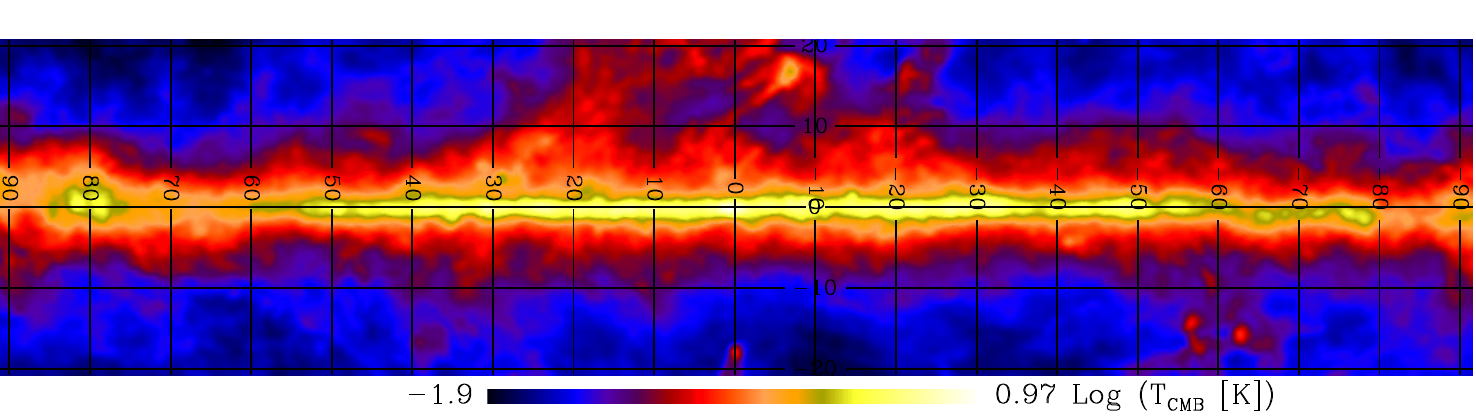}
\caption{The narrow Galactic emission at four frequencies: (a) the 408\MHz\ map of \citet{Haslam:1982}, (b) the 28.4\GHz\ map from \planck, (c) the 70.4\GHz\ map from \planck, and (d) the 545\GHz\ map from \planck. The colour scales are logarithmic. The longitude range is $l=270\deg \rightarrow 0\deg \rightarrow 90\deg$; the latitude range is $-20\deg$ to $+20\deg$. The maps are in Cartesian projection.}
\label{fig:compmaps2}
\end{figure*}

Although \planck\ observed over a wide frequency range (28.4--857\GHz), it is still important to include ancillary data to extend the frequency range into the FIR, and in particular, to lower frequencies. To supplement \planck\ data we therefore used a range of publicly available surveys and data available within the team (Table \ref{tab:data}).  In the following sections, we describe the various data sets in more detail. We used the most up-to-date versions of the data sets, and where necessary regridded the maps into the \healpix\ format \citep{Gorski:2005} by using a procedure that computes the surface intersection between individual pixels of the survey with the intersecting \healpix\ pixels. This procedure has been shown to conserve photometry \citep{Paradis2012}. We note that there are significant baseline uncertainties (i.e., offsets) in these maps, but these will not affect our results since we are subtracting a broad component from the Galactic emission, which removes any such offset. 

\subsection{Low-frequency radio data}
\label{sec:ancdata-lowfreq}

The radio data include the full-sky map at 408\MHz\ \citep{Haslam:1982}, the 
full-sky map at 1420\MHz\ \citep{Reich:1982,Reich:1986,Reich:2001}, and the 2326\MHz\ HartRAO southern survey \citep{Jonas:1998} (Table \ref{tab:data}). For this work we used a raw version of the 408\MHz\ map\footnote{Available from the Lambda webpage: \url{http://lambda.gsfc.nasa.gov/product/foreground/haslam_408.cfm}} at 51\arcm\ resolution rather than the widely used National Center for Supercomputing Applications (NCSA) version. The NCSA map has been Fourier filtered to remove large-scale striations and as a result many bright sources have also been subtracted. Since we are considering the total emission in the narrow component, including both point sources and diffuse medium, we use the raw map reprojected to the \healpix\ grid at $N_\mathrm{side}=512$, which is then smoothed from 51\arcm\ to 1\deg. Striations are visible at intermediate latitudes; however they are less than $\sim 1\,\%$ of the sky signal on the Galactic plane.

The 1420\MHz\ full-sky map was multiplied by a factor of 1.55 to account for the main beam to full beam ratio \citep{Reich:1988}. The formal uncertainty on calibration is 5\,\%, but we increase this due to the beam conversion uncertainties. We assume a 10\,\% overall calibration uncertainty for all these radio data sets as in \citet{planck2013-XV}.

\subsection{\WMAP\ data}

\WMAP\ 9-year data \citep{Bennett2013} span 23--94\GHz\ and thus complement the \planck\ data, particularly the K-band (22.8\GHz) channel. We use the 1\deg-smoothed maps available from the LAMBDA website and we apply colour corrections to the central frequencies using the recipe of \citet{Bennett2013}. We assume a conservative 3\,\% overall calibration uncertainty. The results of using the standard \WMAP\ 1\deg-smoothed maps and the deconvolved 1\deg-smoothed maps are within the uncertainties of each other.

\subsection{\DIRBE\ data}

To sample the peak of the blackbody curve for temperatures $\gtrsim 15$\,K we include the \DIRBE\ data at 240\um\ (1249\GHz), 140\um\ (2141\GHz), and 100\um\ (2997\GHz). The \DIRBE\ data are the Zodi-Subtracted Mission Average (ZSMA) maps \citep{Hauser:1998} regridded into the \healpix\ format. Colour corrections were applied as described in the \DIRBE\ explanatory supplement version 2.3. We assume a 13\,\% overall calibration uncertainty.

\subsection{IRIS data}

The IRIS data \citep{Miville-Deschenes:2005} are the reprocessed versions of the original \IRAS\ data. Colour corrections are applied as described in the \DIRBE\ and \IRAS\ explanatory supplements. We assume a 13\,\% overall calibration uncertainty \citep{IRAS1988}.

\subsection{CO data} \label{sec:co}

We use the Type 1 modified independent linear combination algorithm \citep[MILCA,][]{Hurier2013} maps of the CO emission \citep{planck2013-p03a} to derive the latitude and longitude distributions of CO $J$=1$\rightarrow$0 at 100\GHz. There is an overall agreement of 16\,\% across the whole sky between the composite survey of \citet{Dame:2001} and the \Planck\ Type~1 CO $J$=1$\rightarrow$0 map. When determining SED estimates, we use the Type~1 MILCA CO maps at 100, 217, and 353\GHz, as in \citet{planck2013-XIV}. The intensities are accurate to 10\,\%, 2\%, and 5\,\% at 100, 217, and 353\GHz\ respectively \citep{planck2013-p03a}.

\subsection{RRL data}

The RRL data from the \hi\ Parkes All-Sky Survey are used as a measure of the free-free brightness temperature at 1.4\GHz\ \citep{Alves2012}. The map covering $l=20\deg$--$44\deg$, $|b|\leq 4\deg$ is regridded into the \healpix\ format ($N_\mathrm{side}=512$) as described in the introduction of Sect. \ref{sec:ancdata} and is in units of millikelvin (Rayleigh-Jeans). These data have an overall calibration uncertainty of 15\,\% arising from systematic biases in the free-free estimate from the measured RRLs. These are due to the assumed electron temperature and to the different conversion factors of antenna to brightness temperature for point sources and for diffuse emission. These will be discussed further in Sect. \ref{sec:discussion-ff}. 

\subsection{Gamma-ray data}

The most up-to-date information on the all-sky distribution of gamma-rays is from the \Fermi-LAT, the Large Area Telescope on the \Fermi\ Gamma-ray Space Telescope \citep{Atwood:2009}, which covers the energy range 20\,MeV to 300\,GeV. The angular resolution depends strongly on energy. At 1\,GeV the 68.6\,\% containment angle is $\approx 0\pdeg8$ and varies as $\sim  E^{-0.8}$ \citep{Ackermann2012}.
\Fermi-LAT is sensitive to four components of diffuse emission; (i) gamma-rays from cosmic rays interacting with all forms of interstellar gas through \pion-production, (ii) bremsstrahlung from cosmic ray electrons (CREs), (iii) inverse Compton (IC) scattering of soft interstellar photons by CREs, and (iv) the extragalactic isotropic emission. Above 10\deg\ in latitude and 1\,GeV in energy, the extragalactic emission contributes about 20\,\% of the total \citep{Abdo:2010b}. In the Galactic plane the \pion-produced component dominates; the inverse compton (IC) emission amounts to 5--15\,\% of the total diffuse emission depending on the energy \citep{Ackermann:2011}.

In this study we use data from \citet{Dobler:2010} in the energy range 0.5--1.0\GeV, which represent \pion-decay from the total matter distribution, although bremsstrahlung is also significant in this range. Point sources have not been masked and the resolution is 88\arcm\ FWHM. This width is used when deriving the intrinsic FWHM of the gamma-ray emission.

\subsection{Pulsar data}

The ATNF catalogue of $\sim$2000 pulsars \citep{Manchester:2005} contains all the spin-powered pulsars so far detected at radio frequencies and at soft X-ray and gamma-ray energies. The latter two classes include isolated neutron stars that are spinning down in a similar manner to ordinary pulsars \citep{Lorimer:2006,Keane:2008}. The pulsar properties of interest to the present work are the positions, distances, dispersion measures (DMs), and characteristic ages estimated from the observed spin-down rates. Most high-energy pulsars are young with ages $\lsim3\times10^{5}$\,yr; they are concentrated on the Galactic plane and are most likely the counterparts of as-yet unidentified SNRs. Those associated with known SNRs have ages in the range $10^3$--$10^{5}$\,yr. By contrast, the millisecond pulsars are very old with ages up to $4 \times 10^{9}$\,yr and are more widely distributed in Galactic latitude; they are not of concern in this paper. 
The pulsar data are uniformly surveyed from $l=270\deg \rightarrow 0\deg \rightarrow 40\deg$; at $40\deg  < l < 90\deg$ the survey sensitivity is lower. The latitude widths discussed in this paper refer to the region $l=320\deg \rightarrow 0\deg  \rightarrow 40\deg$ of the highest pulsar count sensitivity.

\subsection{\hi\ data}

We use the composite \hi\ map available from the LAMBDA website to trace the spatial distribution of the neutral gas. The survey of \citet{Hartmann:1997} samples the sky for declinations $\delta \geq -30\deg$ every $0\pdeg5$, and is complemented by the lower-resolution survey of \citet{Dickey:1990} with $1\deg$ sampling. Thus the final resolution of the map is survey dependent. These data are given in units of \hi\ column density, converted from velocity-integrated \hi\ brightness by multiplying by the factor $1.8224 \times 10^{18}$\,K\,km\,s$^{-1}$\,cm$^{-2}$. On the Galactic plane this estimate of the \hi\ surface integral will be an underestimate of up to 20\,\% because of the significant optical depth of \hi\ at low Galactic latitudes \citep{Dickey:2009,Kalberla2009}.


\section{Analysis methods}
\label{sec:analysis}

A first step is to smooth all the maps to a common resolution of 1\deg\ using \healpix\ $N_\mathrm{side}=512$ \citep{Gorski:2005}. This resolution was chosen in order to include significant data sets with resolutions of $\sim$50\arcm\ (see Table \ref{tab:data}); these are 408 MHz (51\arcm), \WMAP\ K- and Ka-bands (49\arcm, 40\arcm), \DIRBE\ (40\arcm), and 0.5--1\GeV\ gamma-rays ($\sim$88\arcm).

The next step was to use a common algorithm for identifying the narrow component (1\deg--2\deg\ wide) at each frequency. This ensured that we were working with the emission from the inner Galaxy. The broad local arm and any halo emission were thereby removed. The procedure was to fit a second-order polynomial to the latitude ranges $b = -5\deg$ to $-2\deg$ and $b=+2\deg$ to $+5\deg$ and interpolate between $-2\deg$ and $+2\deg$. First, we estimated the error in the polynomial by propagating the root mean square (rms) uncertainty in the latitude profile measured at high latitudes ($|b| >20\deg$), which was then added in quadrature to the remaining narrow component.  A Gaussian fit to the narrow component provides an amplitude and FWHM that are used in the subsequent analysis. The calibration uncertainties of the various data sets given in Sect.~\ref{sec:ancdata} are added in quadrature to the Gaussian fit uncertainties for the amplitude error.  In order to have a robust estimate of the uncertainties in the latitude width, we also included the difference found between a direct measure of the width at half maximum of the narrow component and the result from the Gaussian fit. This difference is typically 0\pdeg05 and is included in the tables. The CMB contribution is small compared with the emission on the Galactic plane used in the current study. On a 1\deg\ scale the CMB is $\leq0.1$\,mK whereas the Galactic emission is $\sim2$\,mK at 60--100\GHz\ where the Galactic emission is weakest.


\subsection{Component separation methods applied to the Galactic plane}
\label{sec:compsep}

Component separation is of particular relevance to the estimation of the free-free along the Galactic plane where RRL data are not available. As indicated above, an estimate of the free-free unlocks the route to the other components. We use {\tt FastMEM} \citep{Stolyarov:2002, Stolyarov:2005, Hobson:1998}, which is a non-blind method that can be used to separate a number of components with known frequency dependence. It is a model-fitting multi-frequency technique that works in the Fourier or spherical harmonic domain. The technique uses a Bayesian approach to search for the most probable solution for components having known properties (spectral and beam transfer function). {\tt FastMEM} produces the solution, along with errors, in the spectral domain. It is particularly suitable for regions of high signal-to-noise ratio such as the Galactic plane \citep{Bennett2013}. The results of a comparison of the {\tt FastMEM} free-free estimates with those of \WMAP\ {MEM} \citep{Bennett2013} and the direct RRL free-free estimates for $l=20$\deg--44\deg\ are given in Sect. \ref{sec:discussion-ff}. A description of the {\tt FastMEM} technique is given in Appendix \ref{sec:appendix}, along with a comparison of results from other codes and from RRLs.


\subsection{SEDs along the plane}
\label{sec:seds}

The multi-frequency data at a common resolution of 1\deg\ were used to construct SEDs over appropriate longitude intervals along the Galactic plane at $b=0\pdeg0$. The units of intensity are MJy\,sr$^{-1}$. The various emission components were then identified. The RRL data were used to determine the free-free brightness temperature at each frequency with a well-determined spectral index. Where the RRL data were not available, results of  {\tt FastMEM} (see Sect.~\ref{sec:narrowcomp}  and Appendix~\ref{sec:appendix}) were used. The synchrotron component was determined from the 408\MHz\ data corrected for the free-free contribution. The thermal dust emission was estimated by fitting to the \planck\ HFI frequencies and to the \textit{IRAS} data augmented by the \DIRBE\ data. After subtracting these three components from the total emission at each frequency, the remainder was taken as the AME emission spectrum.


\section{Identification of a narrow component}
\label{sec:narrowcomp}
Figures~\ref{fig:compmaps} and \ref{fig:compmaps2} show the narrow, bright structure of the inner Galaxy at four characteristic frequencies at which the emission is dominated by just one of the four components considered in this paper. These are (a) the 408\MHz\ map of \citet{Haslam:1982} that is largely synchrotron emission \citep{Alves2010}, (b) the 28.4\GHz\ map from \planck, which has the strongest AME emission, (c) the 70.4\GHz\ map from \planck, which is mainly free-free \citep{planck2011-7.3}, and (d) the 545\GHz\ map from \planck, which is entirely thermal emission from large dust grains.

\begin{figure}
\centering
\includegraphics[scale=0.39]{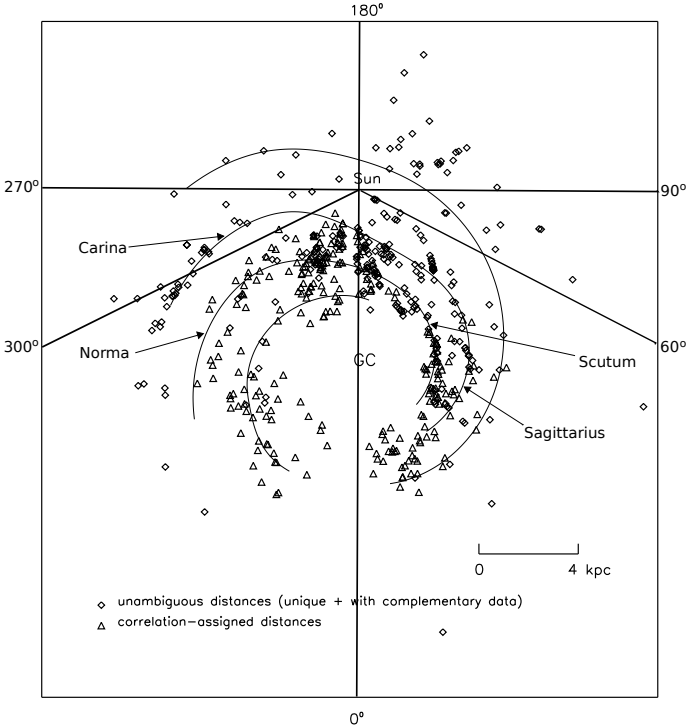}
\caption{The distribution of star-forming regions (\hii\ regions) in the inner Galaxy, from \citet{Paladini:2004} Fig. 12, with the \mbox{$l=300\deg \rightarrow 0\deg \rightarrow 60\deg$} sector indicated. The relevant spiral arms are also indicated.}
\label{fig:hiireg}
\end{figure}

A narrow component of emission, 1--2\deg wide  in Galactic latitude, is seen in the longitude range $l=270\deg \rightarrow 0\deg \rightarrow 90\deg$ at all frequencies from 408\MHz\ through \planck\ frequencies to the FIR and gamma-rays. This is the emission from the star-forming regions within the gas and dust of the inner Galaxy. The spiral arms of this most active star formation in the Galaxy are usually called the Norma-Scutum and the Carina-Sagittarius arms. Figure \ref{fig:hiireg} shows the distribution of \hii\ regions on the Galactic disk where star formation is occurring, based on the work of \citet{Paladini:2004}. In the present work we will concentrate on the longitude range $l=300\deg \rightarrow 0\deg \rightarrow 60\deg$, which encompasses the main emission from the inner arms. We will not be considering in this study the 3\kpc\ expanding arm or the Cohen-Davies bar of the innermost Galaxy \citep{Cohen:1976, Bania:2010}. We reiterate that the total emission in the narrow component being studied here includes  ``point'' (unresolved) sources such as SNRs, \hii\ regions, and dust clouds as well as the diffuse medium; the ratio of \hii\ region point sources to diffuse emission is 20--30\,\% on scales of $\simeq 15^{\prime}$ \citep{Alves2012}.

\begin{figure*}
\centering
\hspace*{-0.55cm}
\includegraphics[angle=90,scale=0.208]{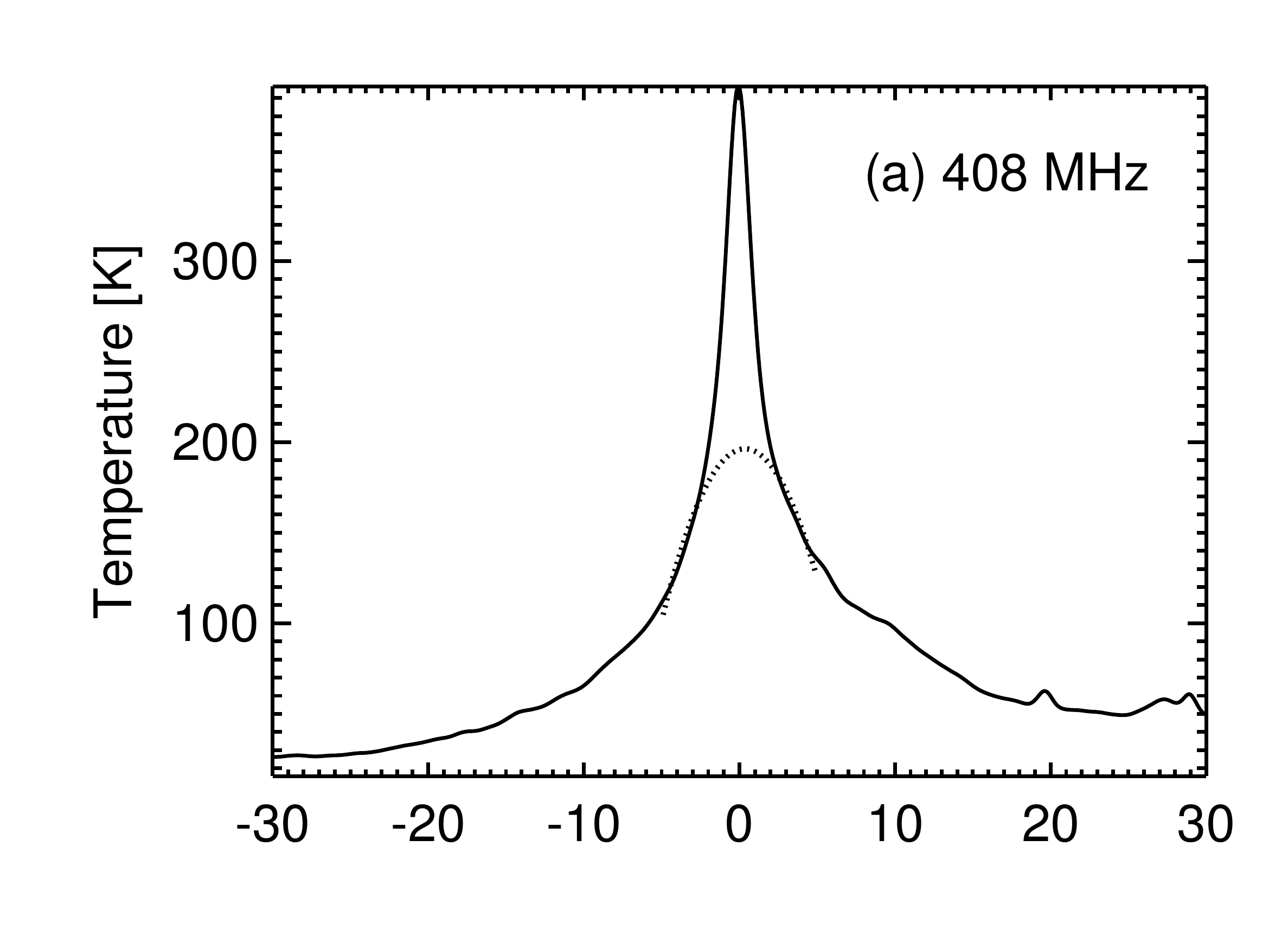}
\vspace*{-0.5cm}
\hspace*{-0.55cm}
\includegraphics[angle=90,scale=0.208]{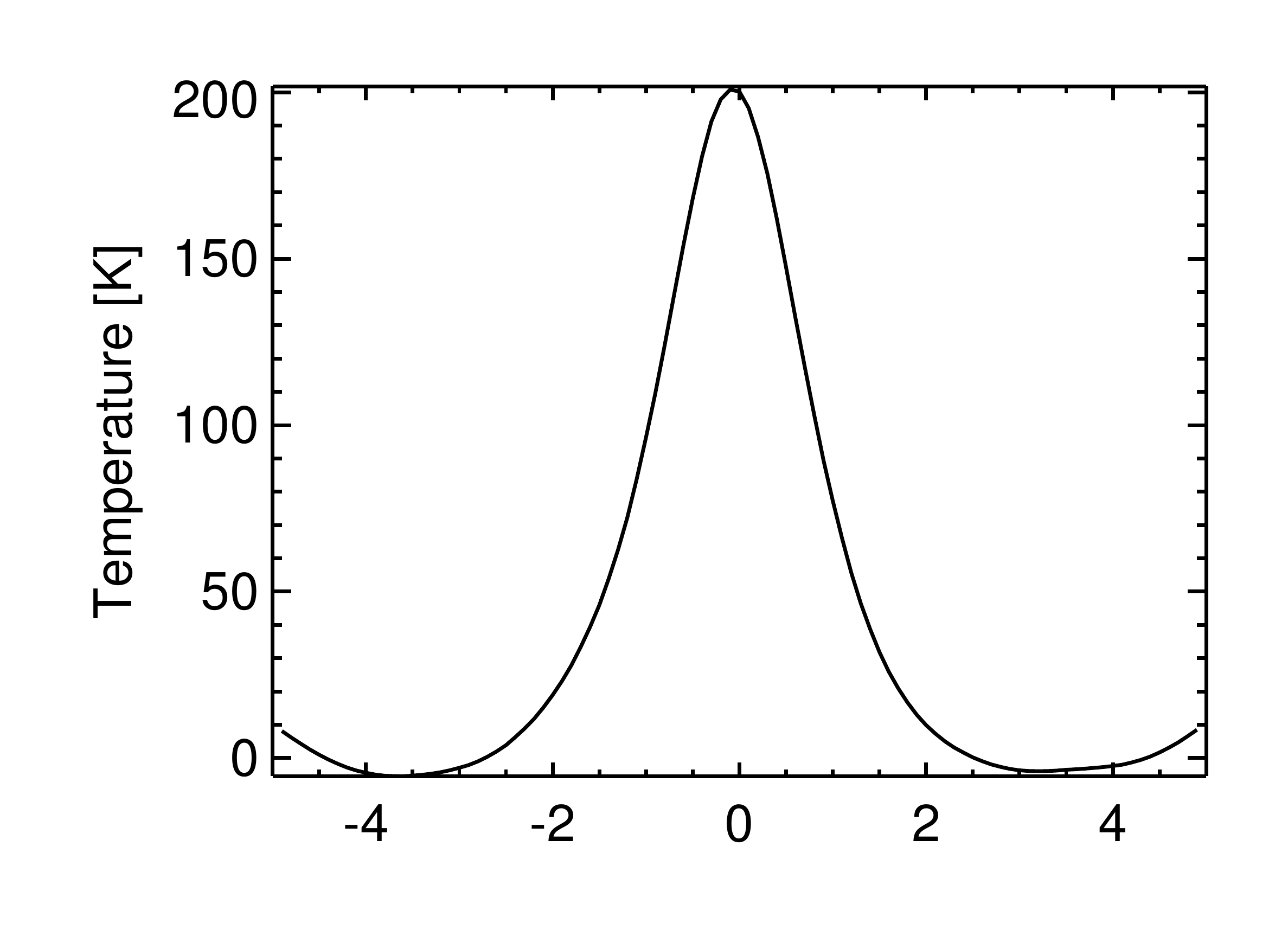}
\hspace*{-0.55cm}
\includegraphics[angle=90,scale=0.208]{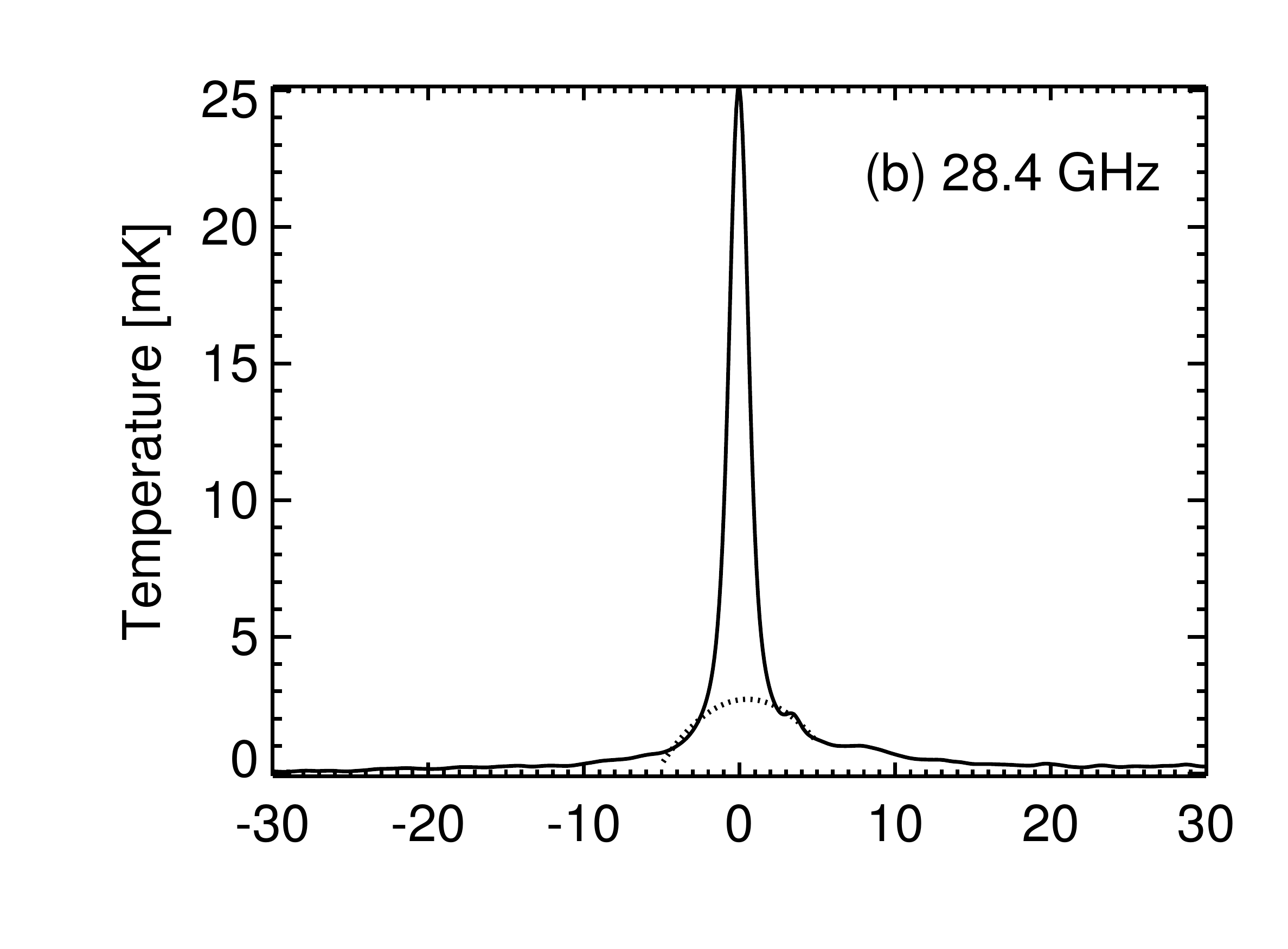}
\hspace*{-0.55cm}
\includegraphics[angle=90,scale=0.208]{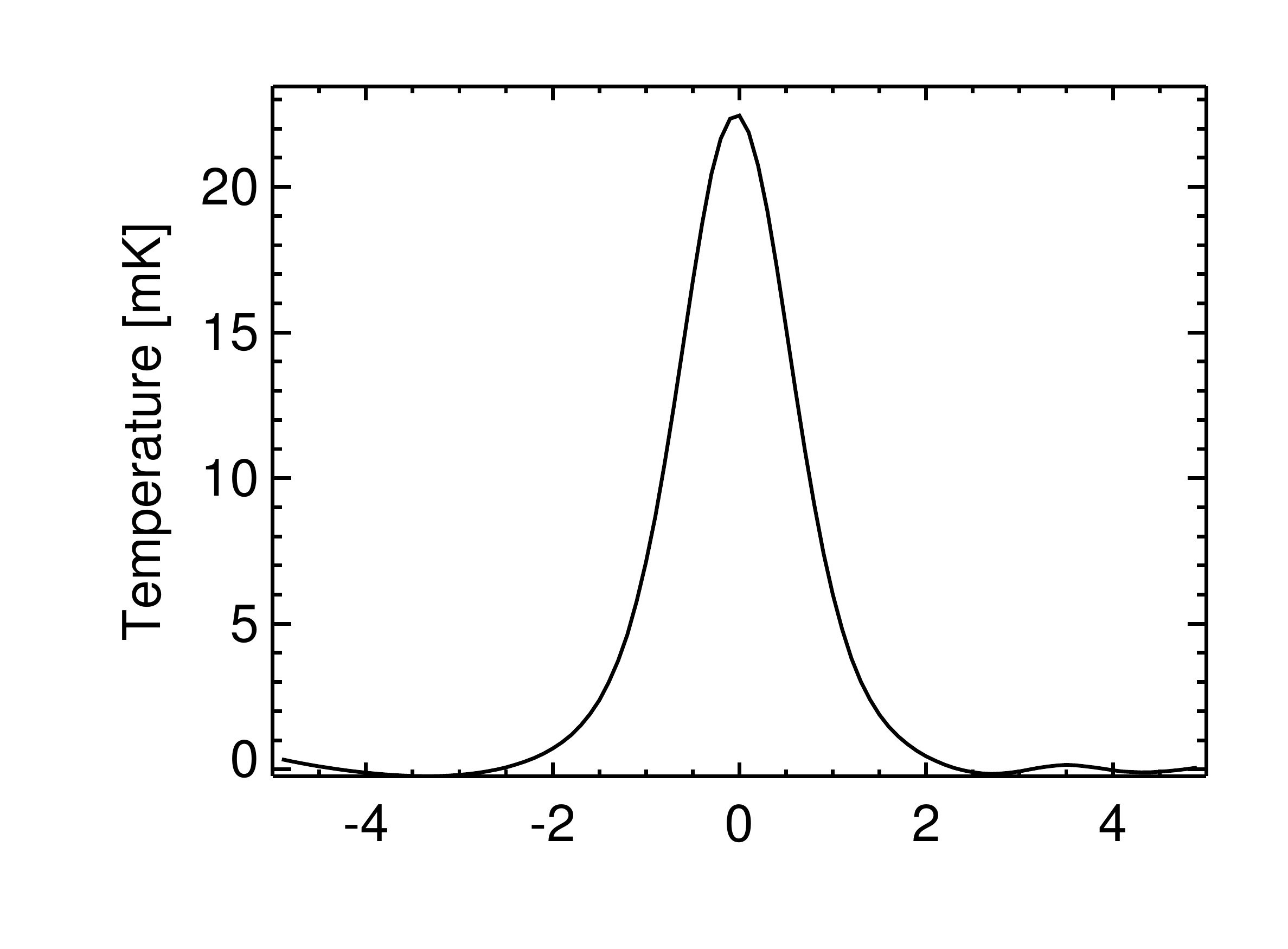}
\hspace*{-0.55cm}
\includegraphics[angle=90,scale=0.208]{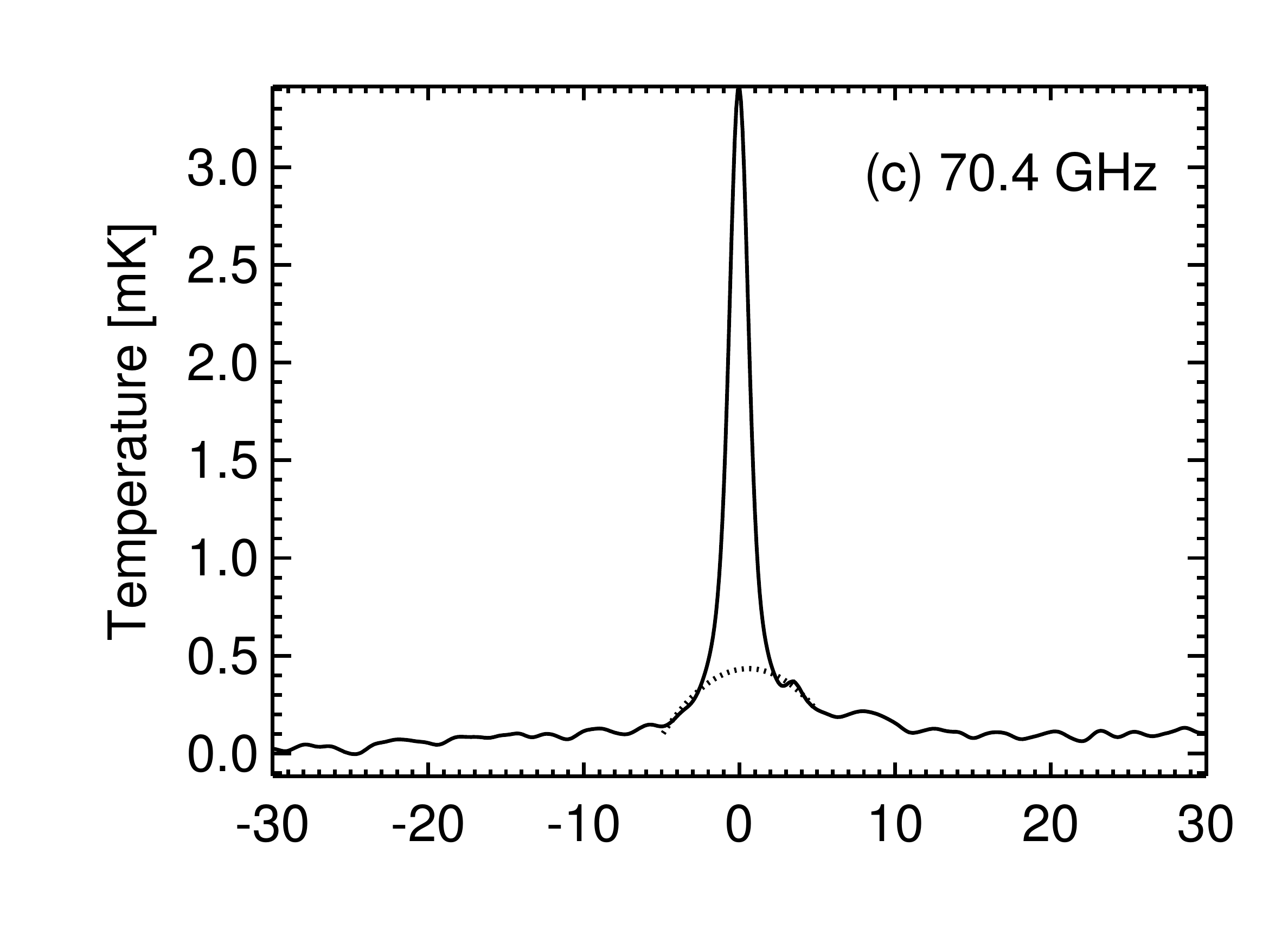}
\vspace*{-0.5cm}
\hspace*{-0.55cm}
\includegraphics[angle=90,scale=0.208]{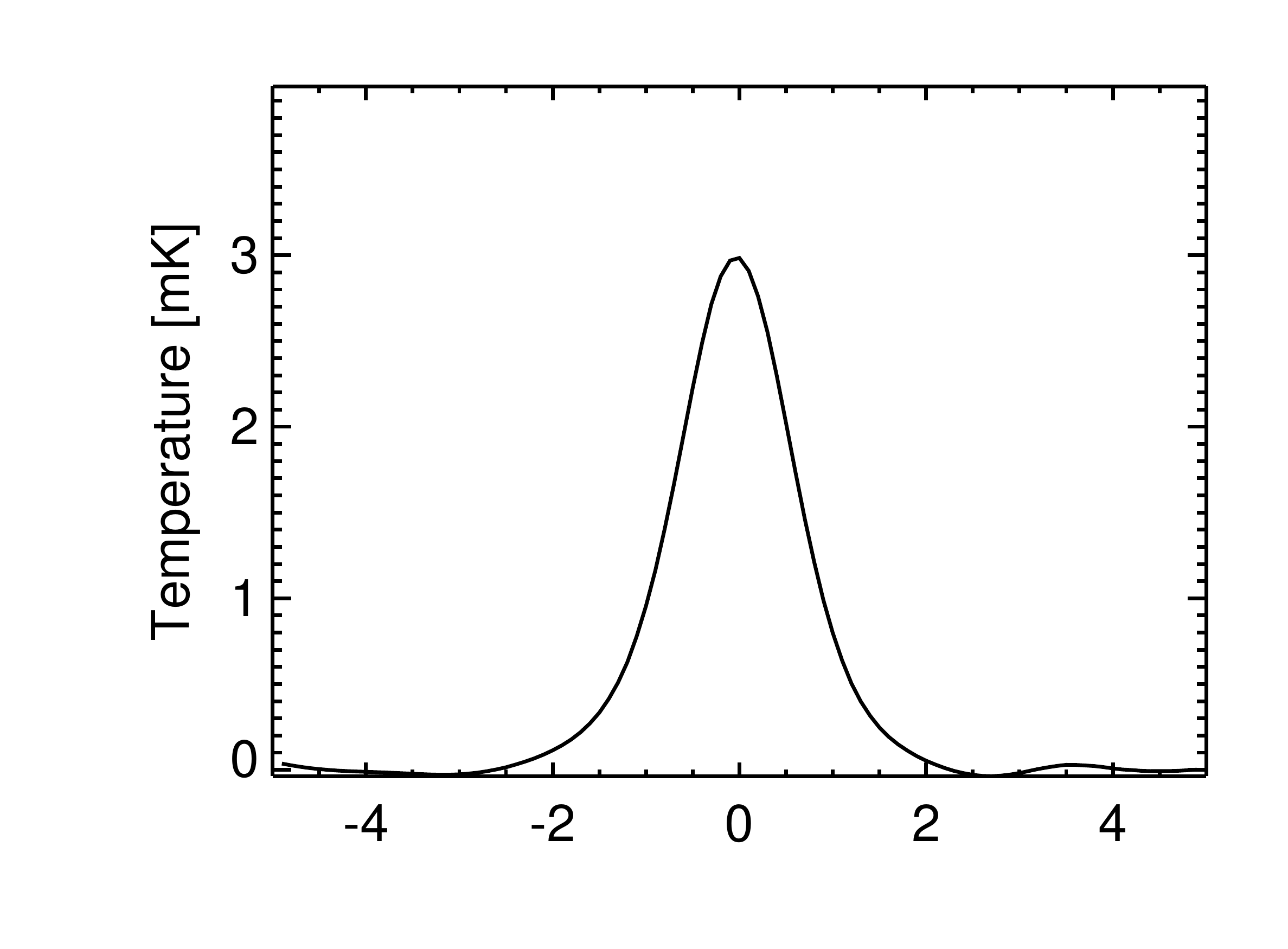}
\hspace*{-0.55cm}
\includegraphics[angle=90,scale=0.208]{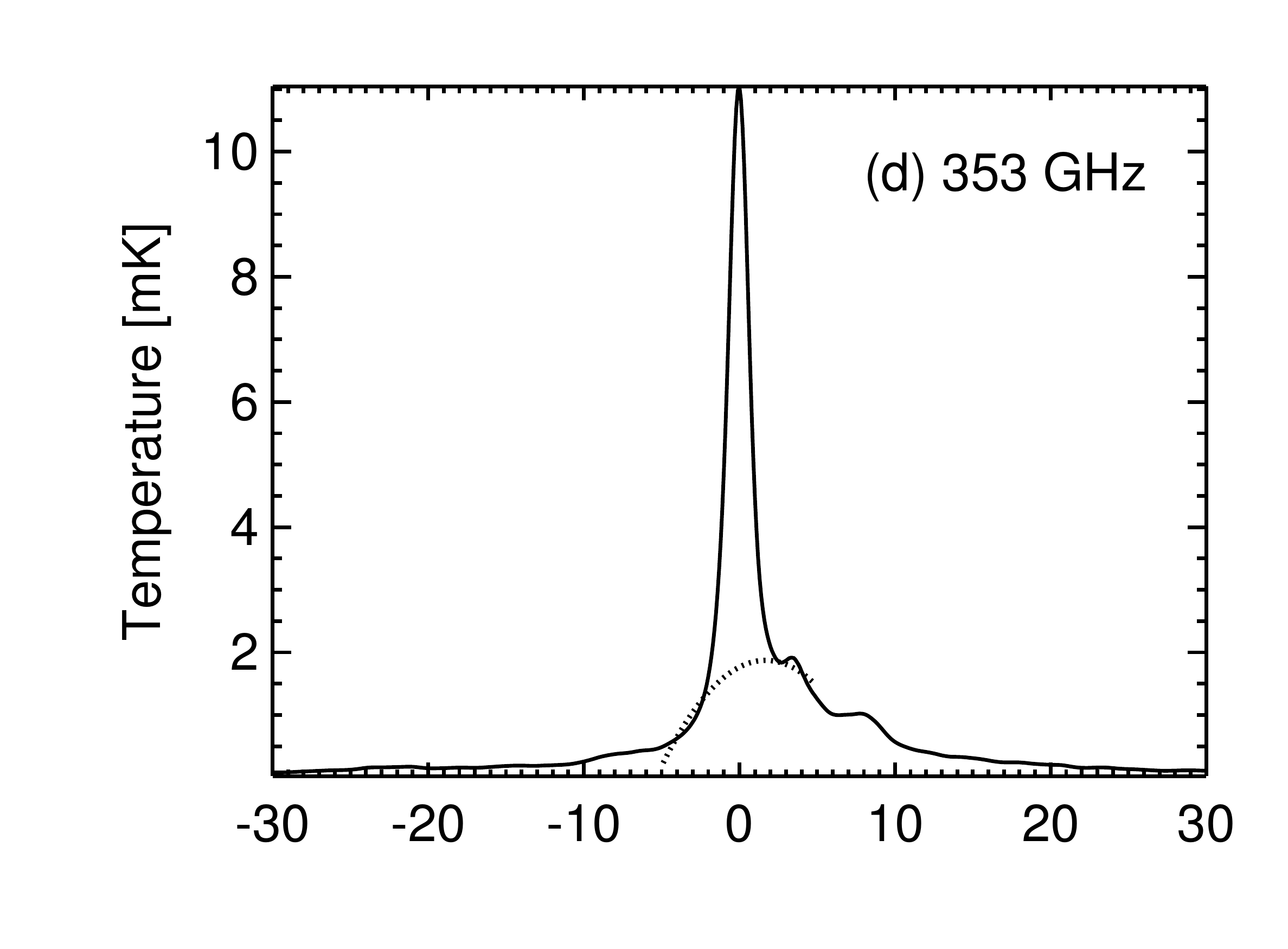}
\hspace*{-0.55cm}
\includegraphics[angle=90,scale=0.208]{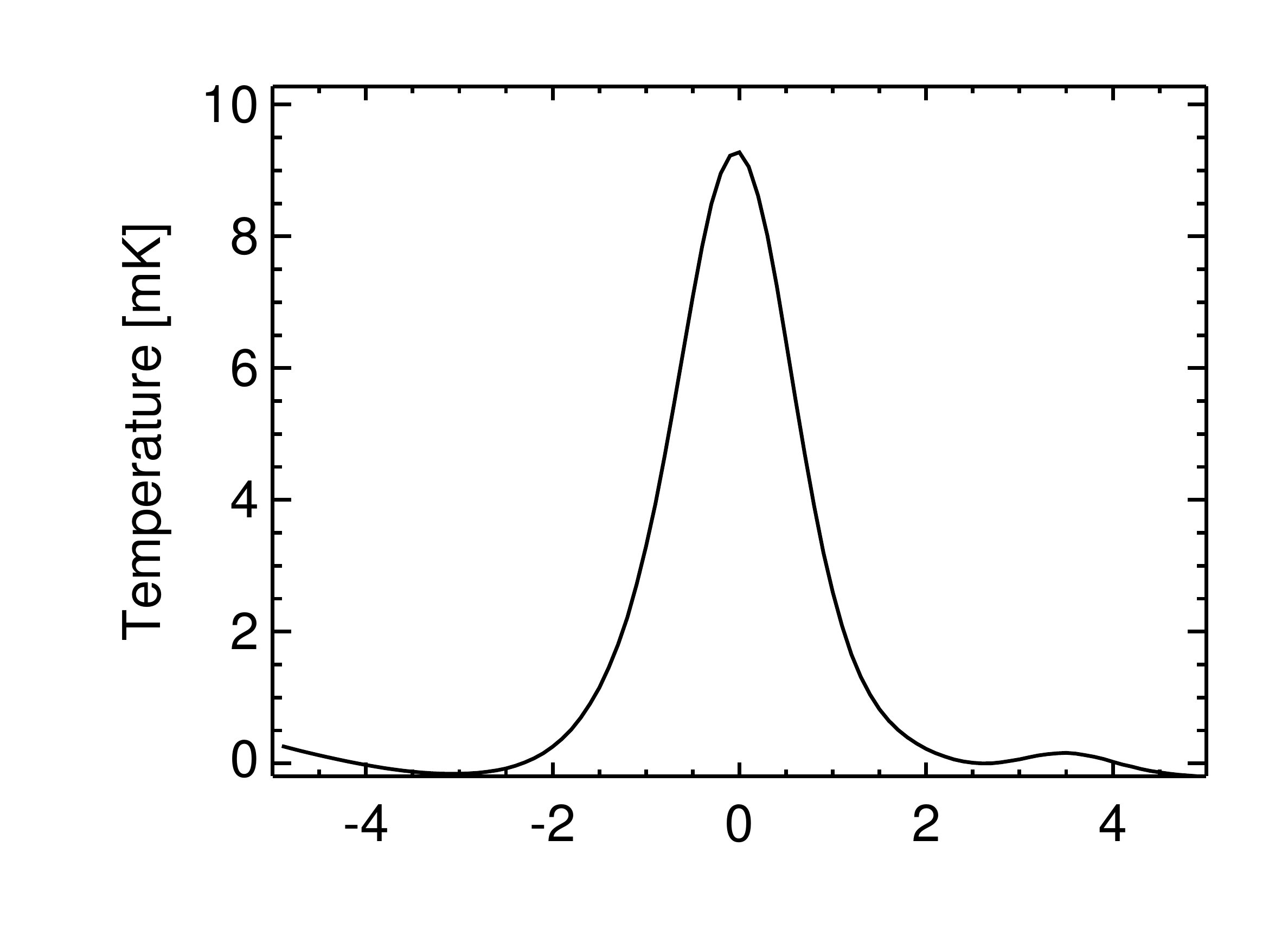}
\hspace*{-0.55cm}
\includegraphics[angle=90,scale=0.208]{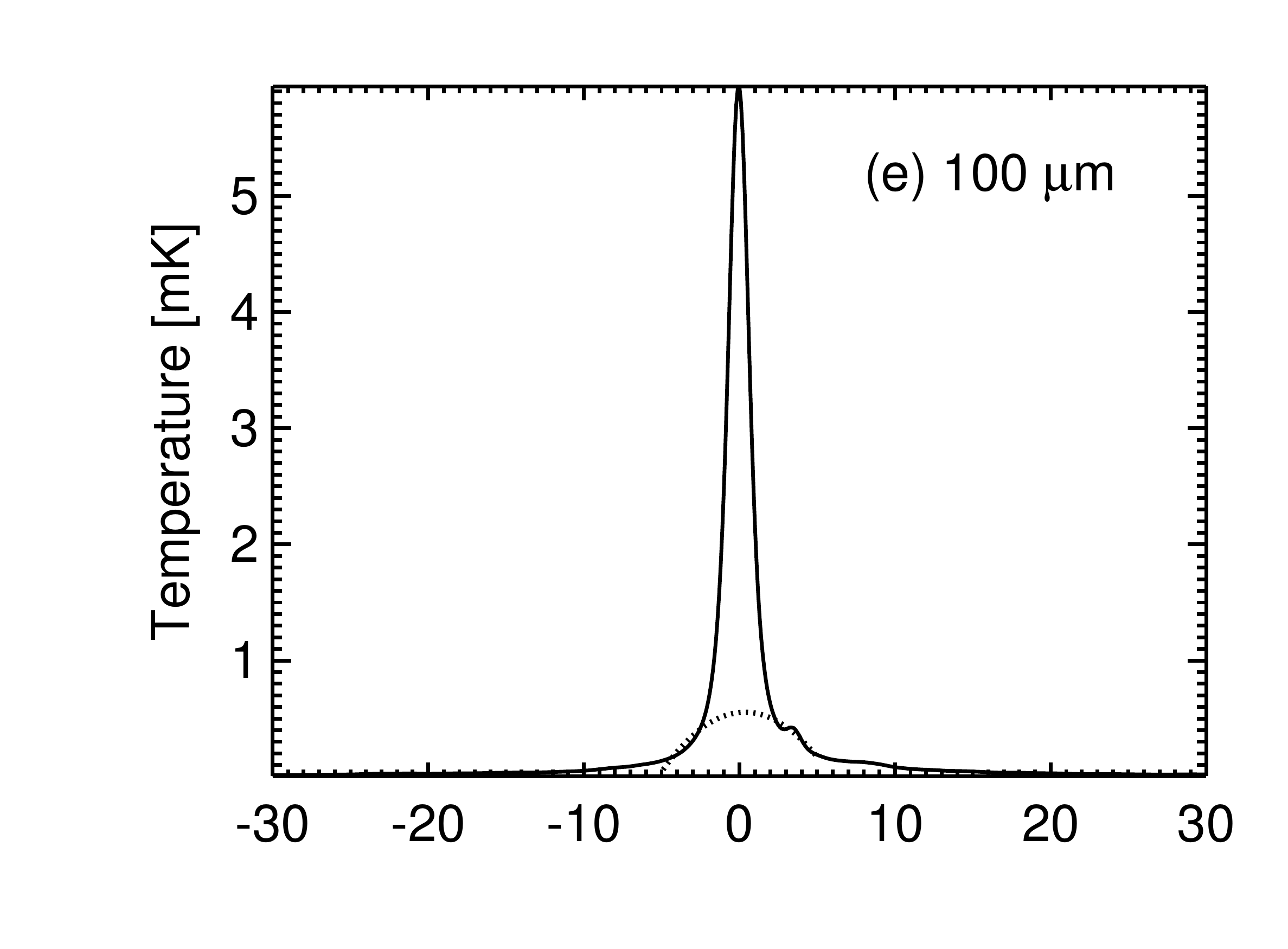}
\vspace*{-0.5cm}
\hspace*{-0.55cm}
\includegraphics[angle=90,scale=0.208]{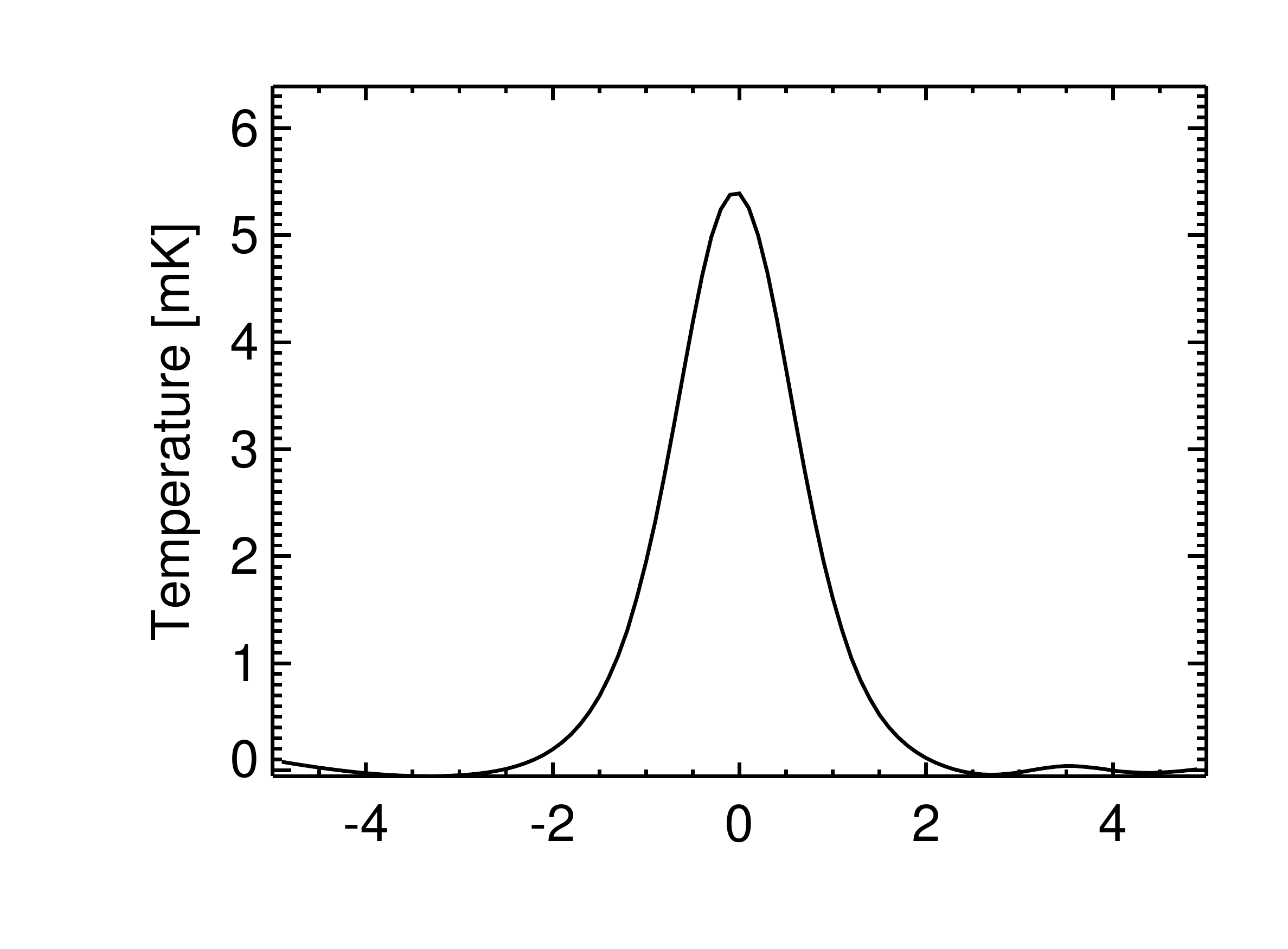}
\hspace*{-0.55cm}
\includegraphics[angle=90,scale=0.208]{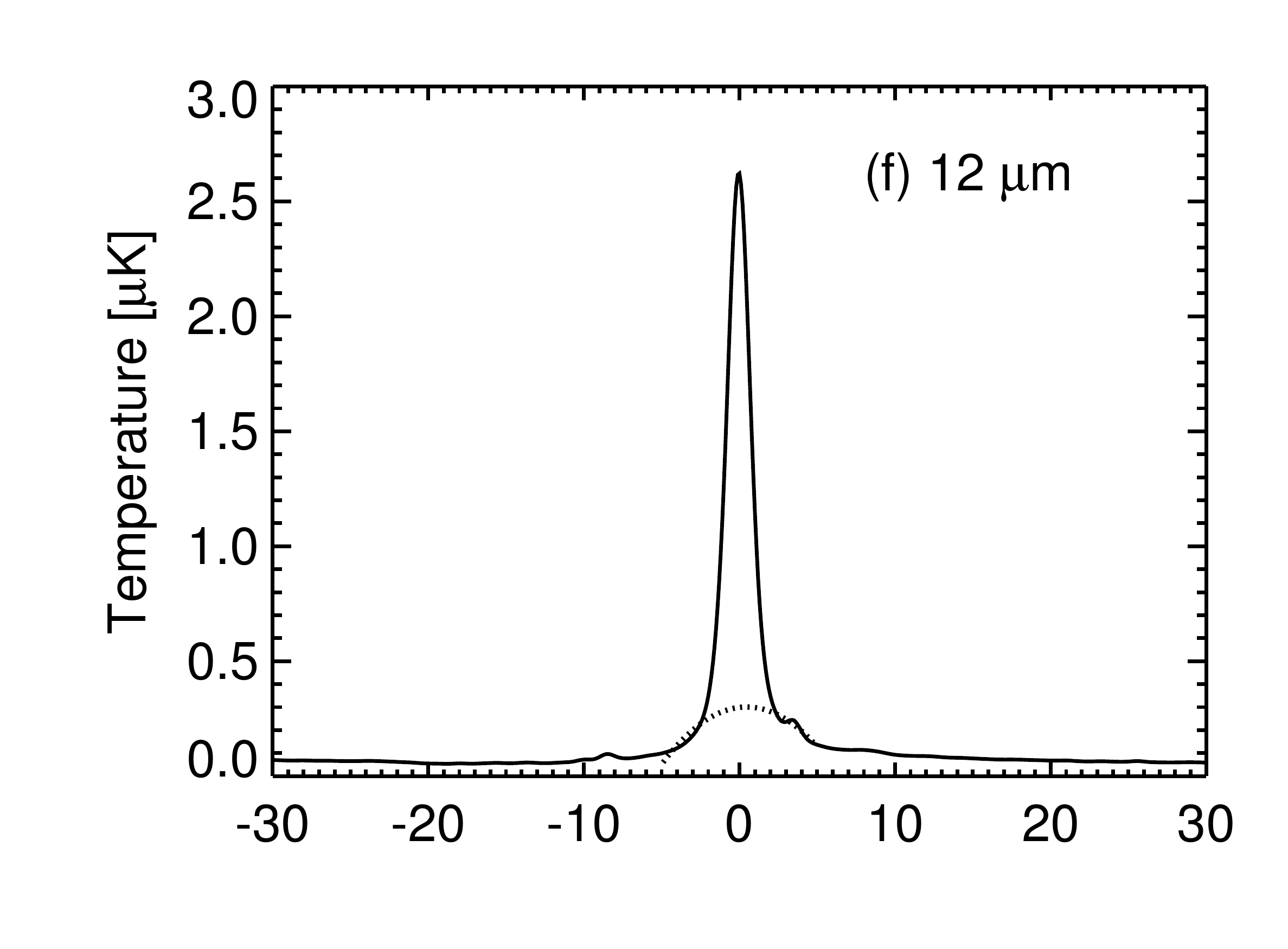}
\hspace*{-0.55cm}
\includegraphics[angle=90,scale=0.208]{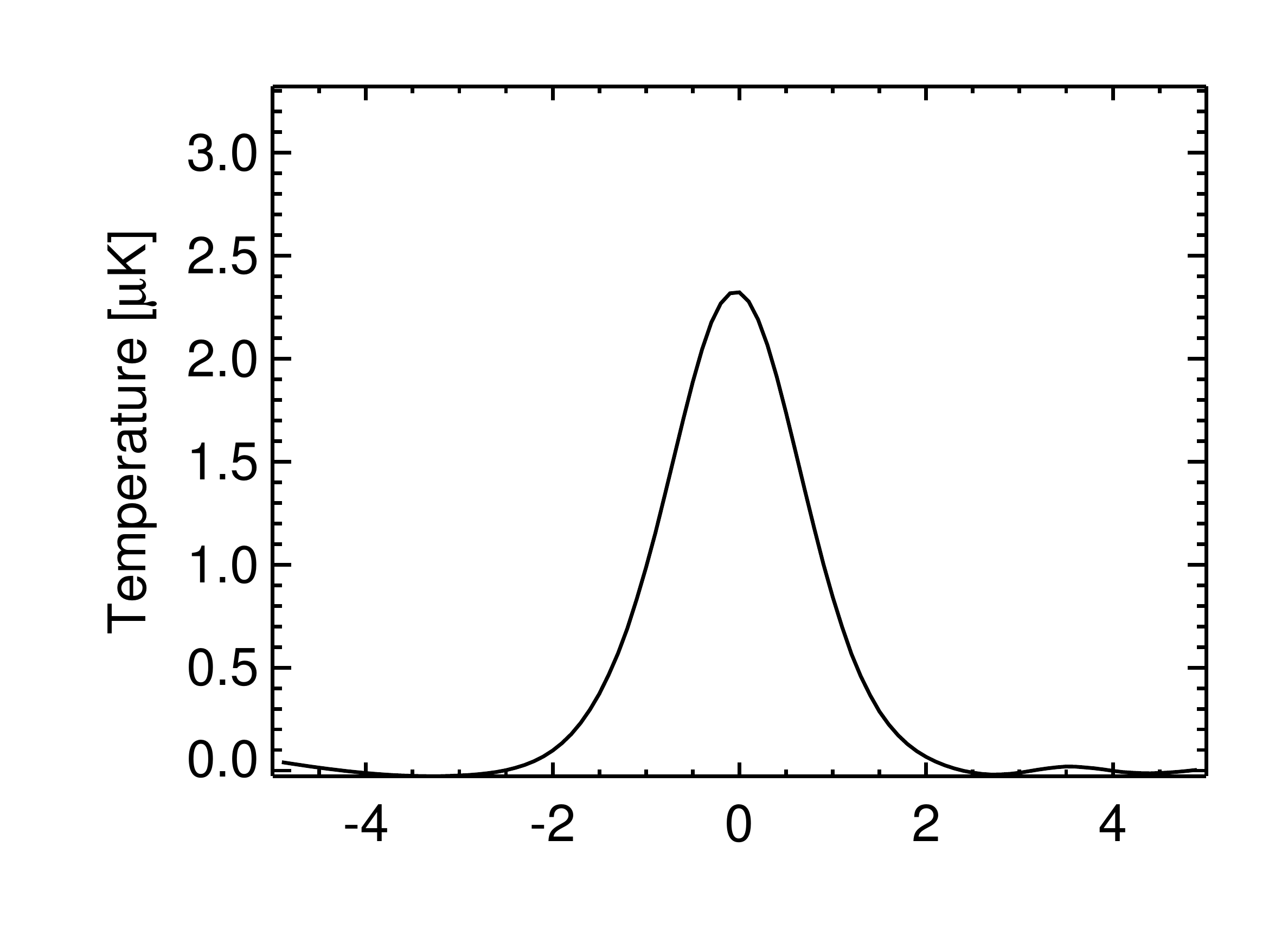}
\hspace*{-0.55cm}
\includegraphics[angle=90,scale=0.208]{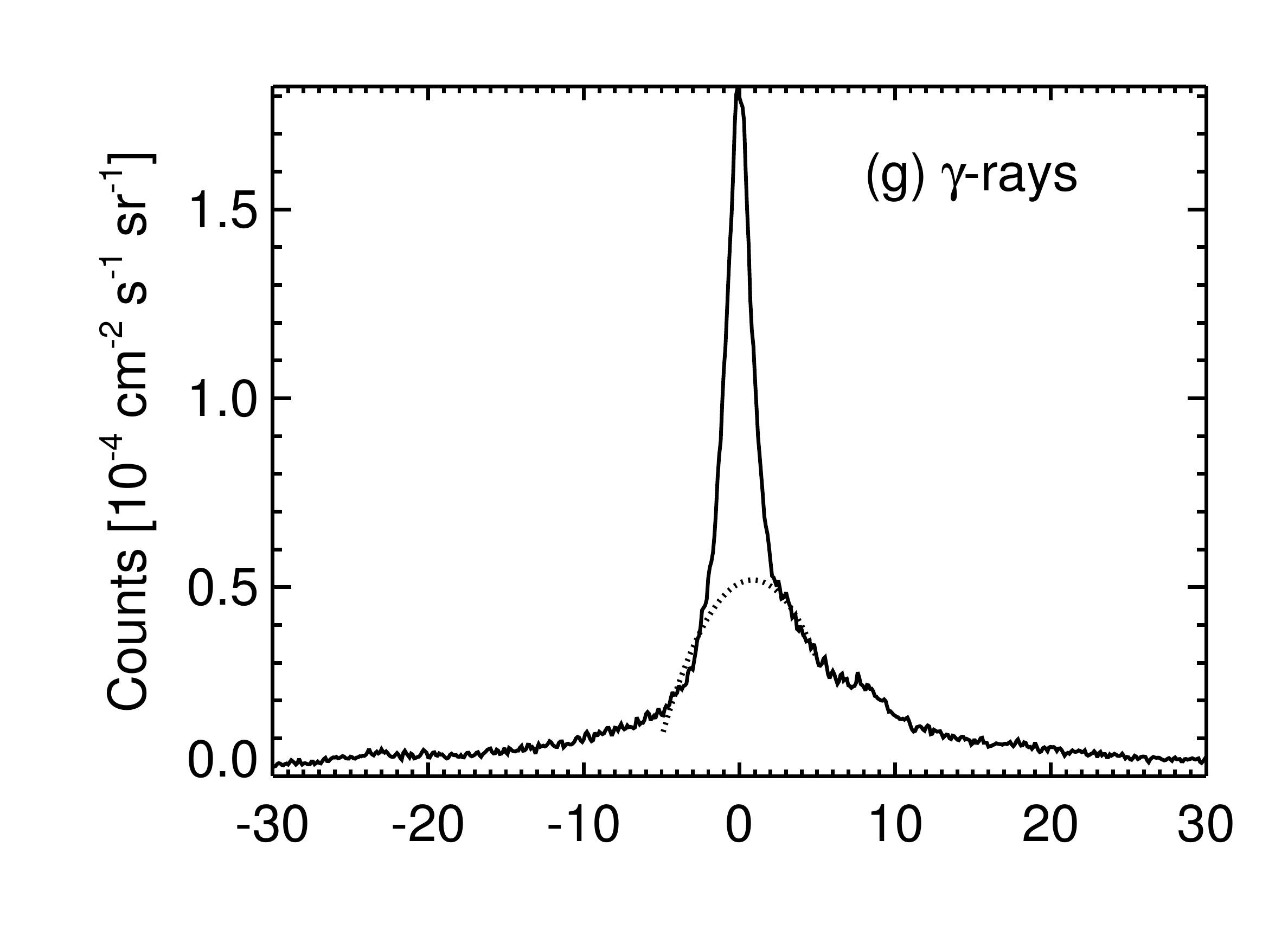}
\vspace*{-0.5cm}
\hspace*{-0.55cm}
\includegraphics[angle=90,scale=0.208]{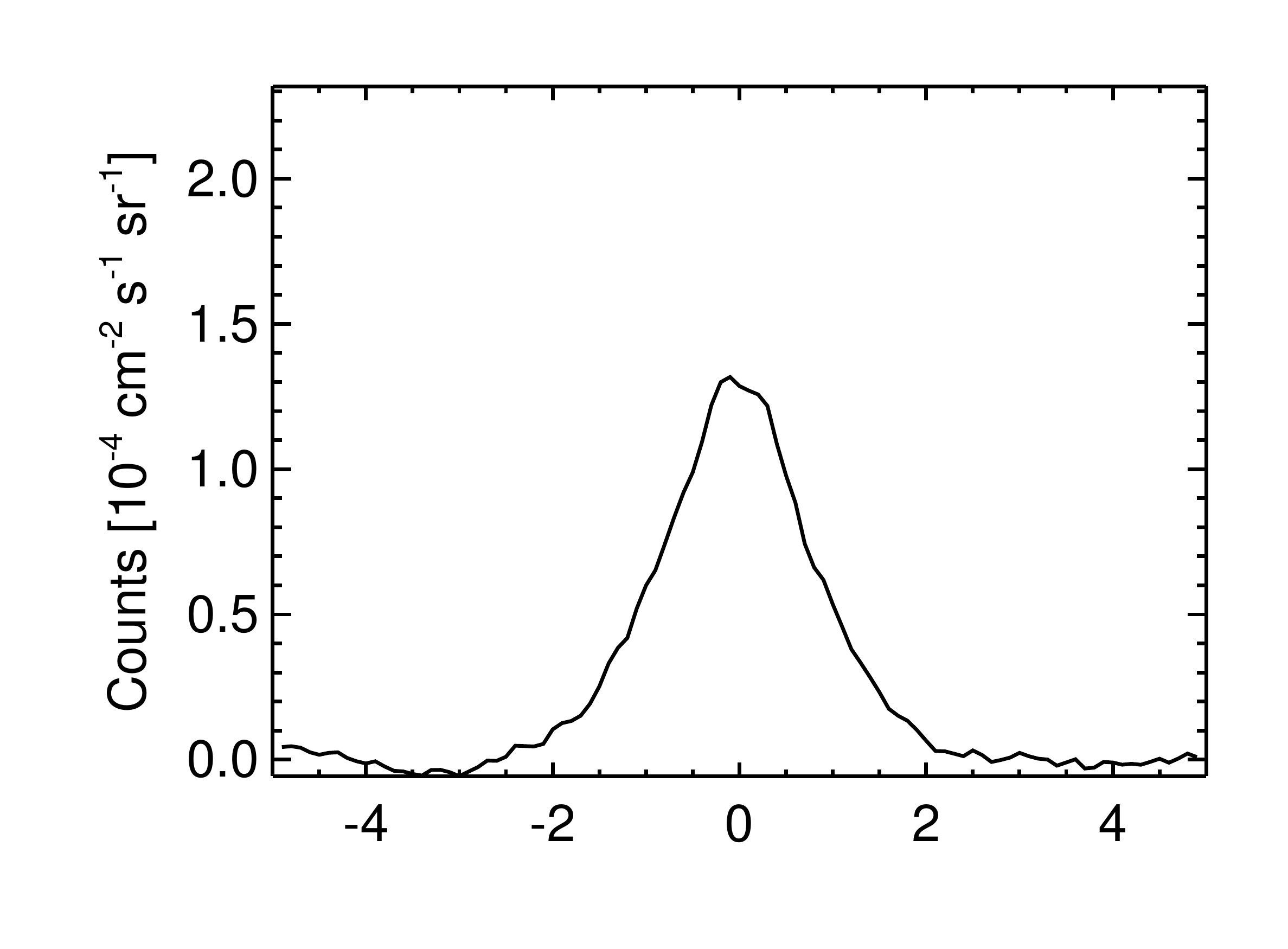}
\hspace*{-0.55cm}
\includegraphics[angle=90,scale=0.208]{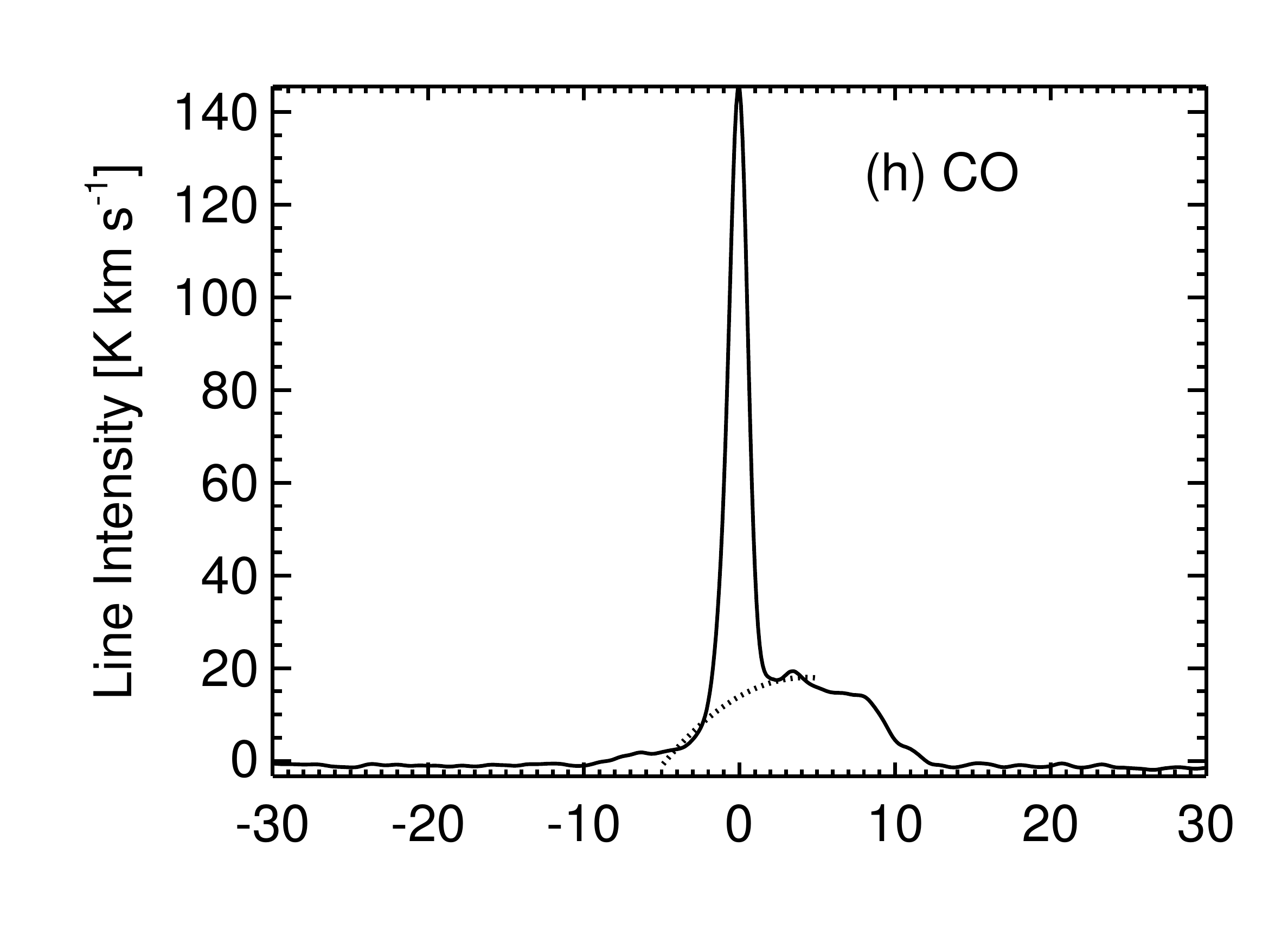}
\hspace*{-0.55cm}
\includegraphics[angle=90,scale=0.208]{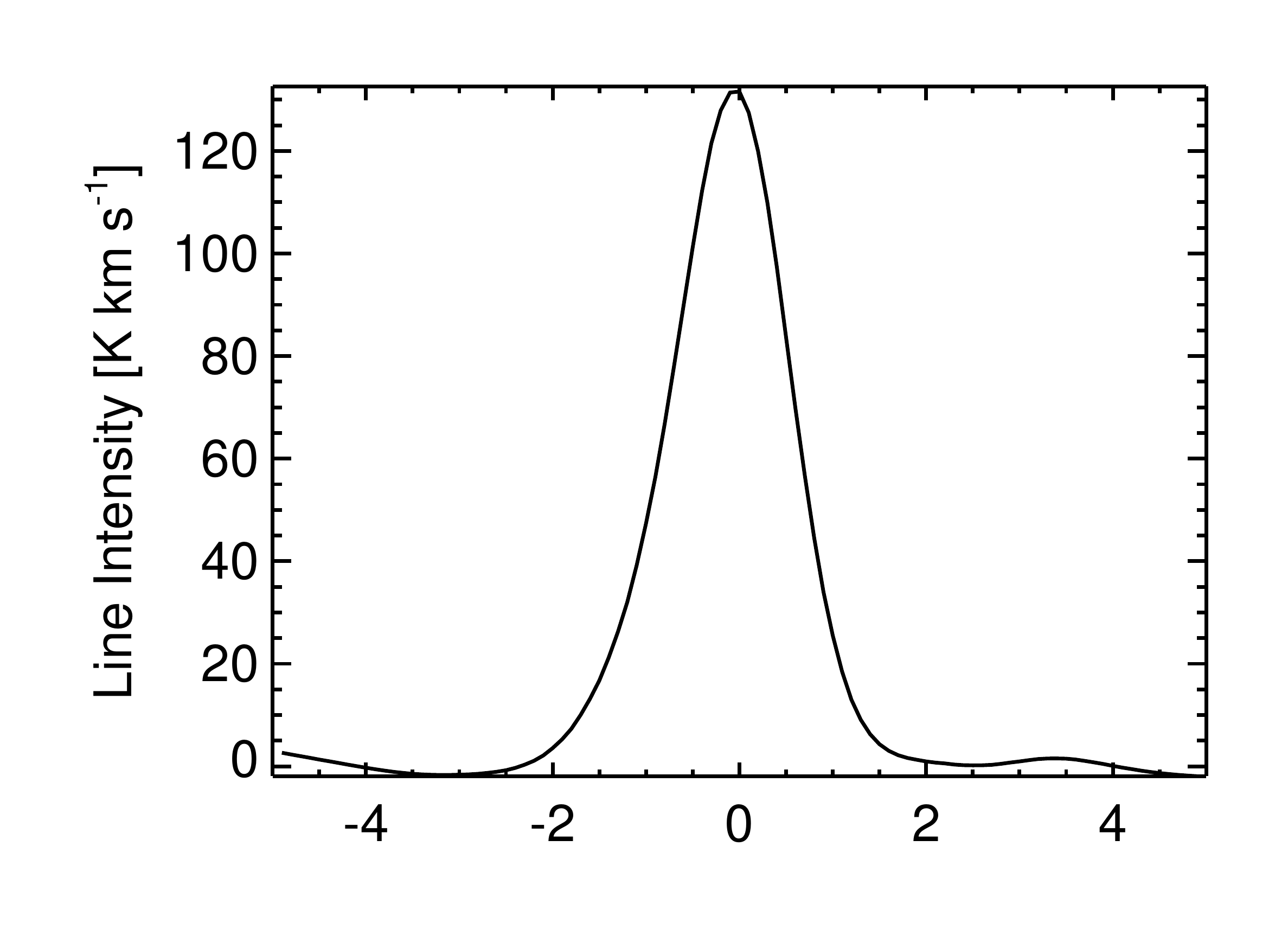}
\hspace*{-0.55cm}
\includegraphics[angle=90,scale=0.208]{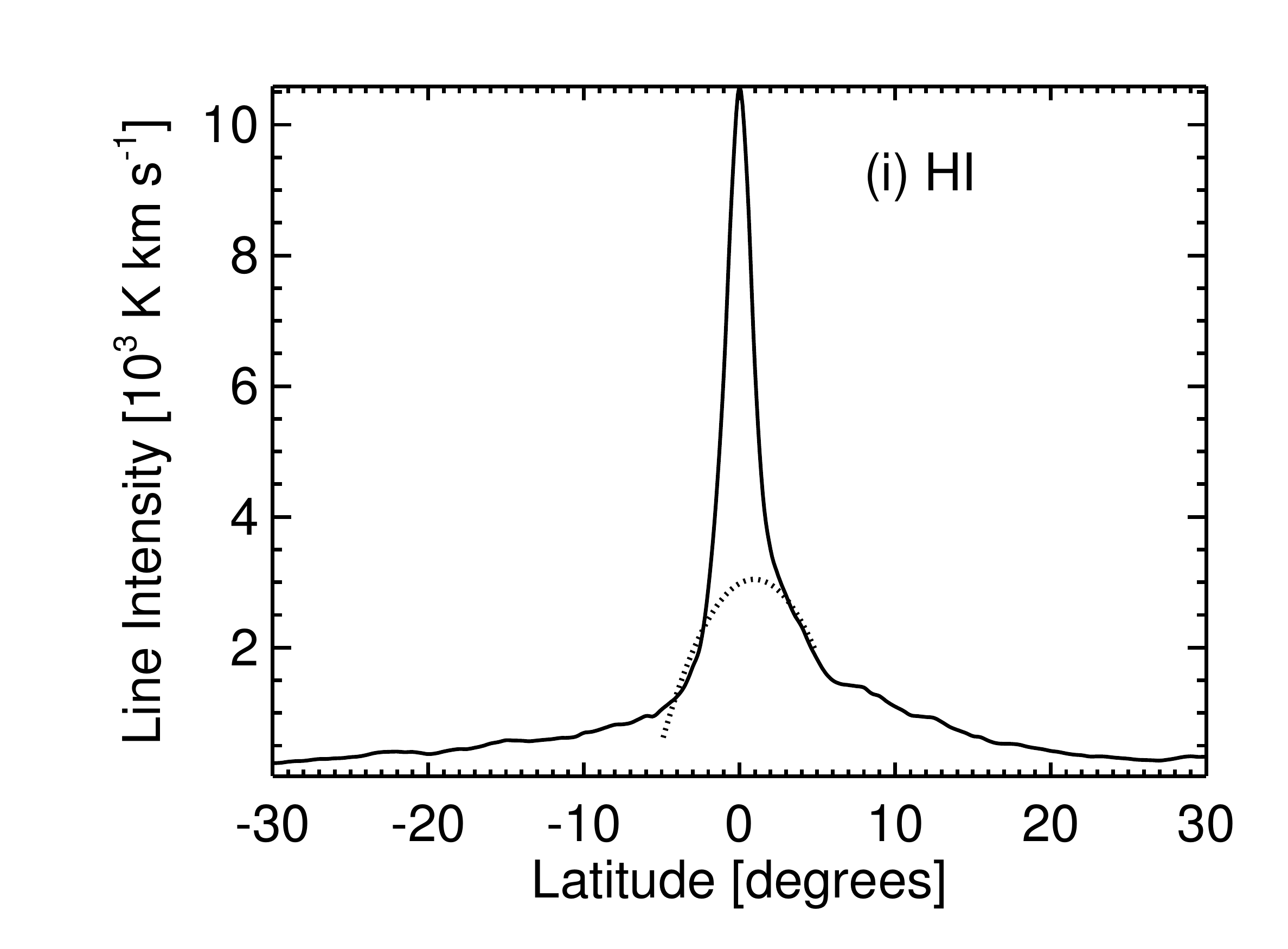}
\hspace*{-0.55cm}
\includegraphics[angle=90,scale=0.208]{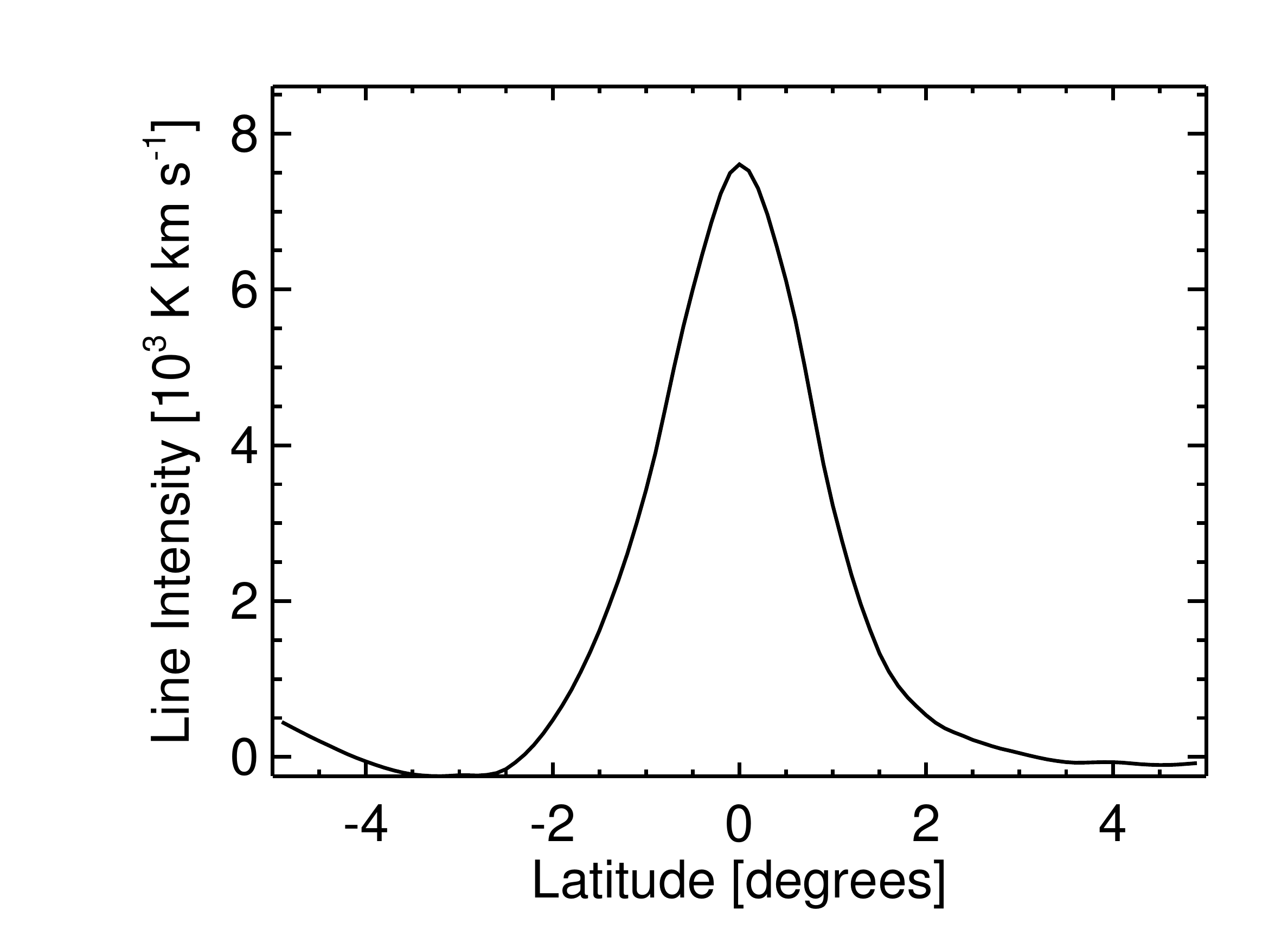}
\hspace*{-0.55cm}
\includegraphics[angle=90,scale=0.208]{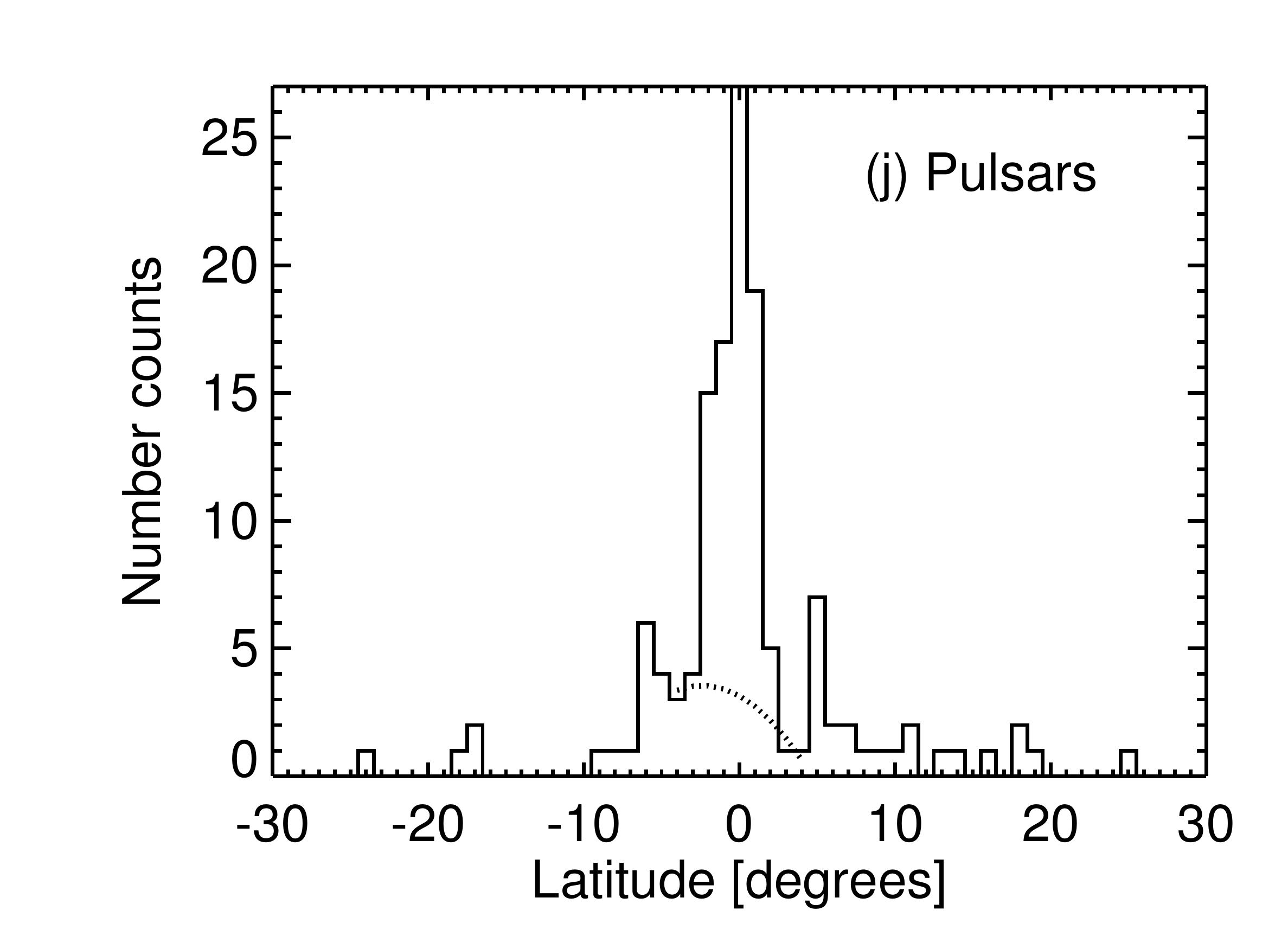}
\hspace*{-0.55cm}
\includegraphics[angle=90,scale=0.208]{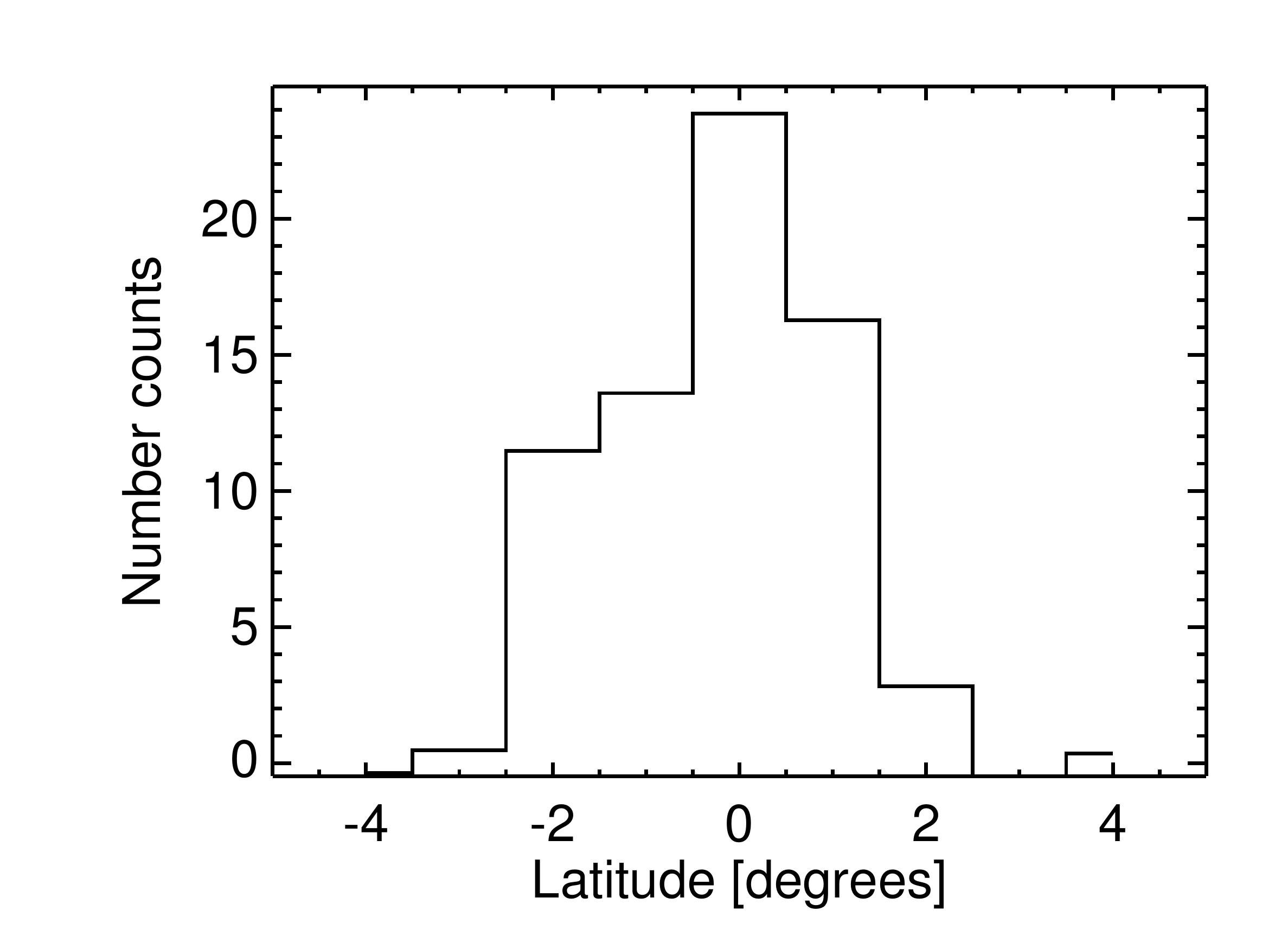}
\caption{Latitude distribution of various components averaged over the longitude range 20\deg\ to 30\deg. The first and third columns show the total emission (\textit{solid line}) with the parabolic fit (\textit{dotted line}) and the second and fourth columns show the narrow component. The pulsar data are at ${N}_\mathrm{side}=64$ to account for the low sampling and include all ages; young and old pulsars are shown separately in Fig.~\ref{fig:narrowsynch}. The northern Gould Belt system can be seen at positive latitudes at all frequencies. The broad 15\deg\ wide synchrotron component is evident in the 408\MHz\ cut.}
\label{fig:latdist}
\end{figure*}

The free-free emission measure (EM) is proportional to the square of the electron density whereas all the other emission components studied here are proportional to the density. The vertical ($z$) distribution of electron density ($n_{\rm e}$) can be derived from pulsar DM data in cases where the pulsars have reliable distances. This distribution has a narrow and a broad component. The widely used NE2001 model \citep{Cordes:2002, Cordes:2003} has a narrow distribution of scale height 140\,pc ($e^{-|z|/z_{0}}$ with $z_{0} = 140$\,pc;~$\rm FWHM = 190$\pc) and a broad component of scale height 950\,pc ($\rm FWHM = 1320$\pc). Using more recent pulsar data, \citet{Lorimer:2006} find that the thickness of the broad component is twice this value. In an attempt to reconcile the DM and EM thickness data of the broad component, \citet{Berkhuijsen:2006} introduced a clumpiness (filling) factor to the model that is a function of $z$. In a comprehensive study, \citet{Gaensler:2008} find a broad component with a scale height 1830\,pc and $n_{0} = 0.14$\,cm$^{-3}$ and a mid-plane filling factor $f_{0} = 0.04$ \citep{Berkhuijsen:2006, Gaensler:2008, Schnitzeler2012}. Although we are dealing with the narrow component in this paper, the discussion about a filling factor is relevant to our comparison of latitude distributions. If the warm ISM were simply a uniform mix of ionized and neutral gas/dust with a Gaussian latitude distribution then the FWHM of the EM would be $2^{-1/2}$ (or 71\,\%) that of the electron and the neutral gas distributions.


\subsection{Latitude distribution}
\label{sec:latdist}

Here we concentrate on the narrow component of the latitude distribution that arises in the ISM and is an integral through the Galactic disk at $\sim 2$--14\kpc\ from the Sun. The emission from the Local/Orion arm and the Gould Belt system in the longitude range being considered mostly arises within a distance of 0.5\kpc\ and produces a much broader and weaker signal that can be separated from the narrow innermost Galaxy emission. The results of this separation are illustrated in Fig. \ref{fig:latdist} for a range of frequencies with 1\deg-smoothed data averaged over the longitude range $l = 20\deg$ to $30\deg$, which includes the Scutum and Sagittarius arms.

The 408\MHz\ amplitudes of the narrow and broad components of the inner Galaxy shown in Figs. \ref{fig:compmaps}(a) and \ref{fig:compmaps2}(a) are similar. They are both closely confined to the region of high star formation in the range $l=310\deg \rightarrow 0\deg  \rightarrow 50\deg$. The FWHM of the narrow inner component is $1\pdeg8$, corresponding to a $z$-width of $\sim180$\,pc at a typical distance of 6--8\kpc. The broad inner component has a FWHM of $\sim10$\deg\ corresponding to a $z$-width of 1--2\,kpc, similar to the values given by \citet{Strong2011b} and \mbox{\citet{Mertsch2013}}. Lying outside this is a halo component $\sim 10$\kpc\ in thickness \citep{Phillipps1981,Beuermann1985,Strong2011b, Strong:2011, Ghosh:2011,Jaffe:2011,Orlando2013}.

At \Planck\ and \IRAS\ frequencies the amplitude of the broad component is 9--15\,\% of the narrow component.  This broad component arises in the emission from the Gould Belt system within 500\pc\ of the Sun, and appears strongest at positive latitudes in the longitude range $l=320\deg \rightarrow 0\deg \rightarrow 40\deg$ and at negative latitudes in the range 140\deg--220\deg \citep{Perrot2003,planck2013-XII}.  The CO and gamma-ray broad components have 12--20\,\% of the amplitude of the narrow component, and are also associated with the northern Gould Belt system.

We now compare the latitude widths, with errors, of the various components in the longitude ranges $l = 320$\deg--340\deg\ and 20\deg--40\deg\ averaged in four 10\deg\ longitude intervals. Table \ref{table:freqresults} lists the observed FWHM at 1\deg\ resolution and the intrinsic FWHM obtained by deconvolving with a 1\deg\ Gaussian for each of the \Planck\ bands. Ancillary data ranging from radio through FIR to gamma-rays are included. There are significant differences in the latitude widths in the different frequency bands.  The low-frequency synchrotron emission is broadest.  The bands associated with free-free are narrowest, while the dust emission is intermediate. These differences will be discussed in detail in Sect.~\ref{sec:discussion}.

A striking feature of Table \ref{table:freqresults} is the similarity in the widths of the emission bands in each of the longitude ranges $l = 330$\deg--340\deg, 20\deg--30\deg, and 30\deg--40\deg, while the widths at $l=320$\deg--330\deg\ are 1.35 times larger in each of the bands with a scatter of only 0.05 (at 95\,\% confidence). This suggests that all four regions have similar emission properties, but the emitting region at $l = 320$\deg--330\deg\ is closer.

\begin{table*}[tb]
\begingroup
\newdimen\tblskip \tblskip=5pt
\caption{Latitude widths and amplitudes of the narrow component in each band (\planck\ plus ancillary data) for the longitude ranges 20\deg--30\deg, 30\deg--40\deg, 320\deg--330\deg,  and 330\deg--340\deg. Note that for the spectral line (RRL, CO and \hi), gamma-ray, and pulsar data the amplitudes are in units of K\,\kms, cm$^{-2}$\,s$^{-1}$\,sr$^{-1}$ and number counts respectively. The pulsar data are at ${N}_\mathrm{side}=32$.}
\label{table:freqresults}
\nointerlineskip
\vskip -3mm
\footnotesize
\setbox\tablebox=\vbox{
 \newdimen\digitwidth 
 \setbox0=\hbox{\rm 0} 
 \digitwidth=\wd0 
 \catcode`*=\active 
 \def*{\kern\digitwidth}
 \newdimen\signwidth 
 \setbox0=\hbox{+} 
 \signwidth=\wd0 
 \catcode`!=\active 
 \def!{\kern\signwidth}
 \halign{\hbox to 1.5in{#\leaderfil}\tabskip 0.5em&
 \hfil#\hfil&
 \hfil#\hfil&
 \hfil#\hfil&
 \hfil#\hfil&
 \hfil#\hfil&
 \hfil#\hfil&
 \hfil#\hfil\tabskip 0pt\cr
 \noalign{\doubleline\vskip 2pt}
%
 \omit\hfil Data \hfil& \hfil FWHM \hfil & \hfil FWHM \hfil & \hfil Amplitude \hfil
 & \hfil FWHM \hfil & \hfil FWHM \hfil & \hfil Amplitude \hfil \cr
 \omit\hfil \hfil& \hfil ($1\deg$ reso.) \hfil& 
 \hfil (intrinsic) \hfil & \hfil [mK] \hfil & 
 \hfil ($1\deg$ reso.)\hfil & \hfil (intrinsic) \hfil & \hfil [mK] \hfil \cr
\noalign{\vskip 4pt\hrule\vskip 6pt}
\omit\hfil & \multispan3\hfil $l=20\deg-30\deg$\hfil & \multispan3\hfil $l=30\deg-40\deg$\hfil \cr
\noalign{\vskip 4pt\hrule\vskip 6pt}
0.408\,GHz & $1\pdeg91\pm0\pdeg12$ & $1\pdeg63\pm0\pdeg14$ & $(195\pm20)\times 10^{3}$ & $1\pdeg91\pm0\pdeg13$ & $1\pdeg63\pm0\pdeg16$ & $(161\pm17) \times 10^{3}$\cr
1.4\,GHz & $1\pdeg89\pm0\pdeg09$ & $1\pdeg60\pm0\pdeg11$ & $(10.5\pm1.1) \times 10^{3}$ & $1\pdeg90\pm0\pdeg11$ & $1\pdeg61\pm0\pdeg13$ & $(8.4\pm0.8)\times 10^{3}$\cr
2.3\,GHz & $1\pdeg78\pm0\pdeg08$ & $1\pdeg47\pm0\pdeg10$ & $(3.3\pm0.3) \times 10^{3}$ & $1\pdeg76\pm0\pdeg07$ & $1\pdeg45\pm0\pdeg09$ & $(2.8\pm0.3) \times 10^{3}$\cr
RRL & $1\pdeg34\pm0\pdeg04$ & $0\pdeg90\pm0\pdeg06$ & $(5.9\pm0.6) \times 10^{3} $ & $1\pdeg37\pm0\pdeg07$ & $0\pdeg93\pm0\pdeg10$ & $(4.7\pm0.5)\times 10^{3}$\cr
23\,GHz & $1\pdeg58\pm0\pdeg08$ & $1\pdeg22\pm0\pdeg10$ & $35.3*\pm1.0*$ & $1\pdeg61\pm0\pdeg11$ & $1\pdeg26\pm0\pdeg14$ & $29.8\pm0.9$\cr 
28.4\,GHz & $1\pdeg51\pm0\pdeg01$ & $1\pdeg14\pm0\pdeg01$ & $22.0*\pm0.2*$ & $1\pdeg55\pm0\pdeg04$ & $1\pdeg18\pm0\pdeg06$ & $18.3\pm0.2$\cr 
33\,GHz & $1\pdeg54\pm0\pdeg04$ & $1\pdeg18\pm0\pdeg06$ & $15.5*\pm0.5*$ & $1\pdeg57\pm0\pdeg07$ & $1\pdeg21\pm0\pdeg09$ & $12.9\pm0.4$\cr
41\,GHz& $1\pdeg52\pm0\pdeg02$ & $1\pdeg14\pm0\pdeg03$ & $*9.2*\pm0.3*$ & $1\pdeg55\pm0\pdeg05$ & $1\pdeg18\pm0\pdeg06$ & $7.7\pm0.2$\cr 
44\,GHz & $1\pdeg53\pm0\pdeg03$ & $1\pdeg16\pm0\pdeg04$ & $*7.52\pm0.08$ & $1\pdeg56\pm0\pdeg06$ & $1\pdeg20\pm0\pdeg08$ & $6.30\pm0.06$\cr 
60.7\,GHz & $1\pdeg52\pm0\pdeg02$ & $1\pdeg15\pm0\pdeg03$ & $*3.7*\pm0.1*$ & $1\pdeg57\pm0\pdeg07$ & $1\pdeg21\pm0\pdeg09$ & $3.2\pm0.1$\cr 
70.4\,GHz & $1\pdeg52\pm0\pdeg02$ & $1\pdeg15\pm0\pdeg03$ & $*2.93\pm0.03$ & $1\pdeg59\pm0\pdeg09$ & $1\pdeg23\pm0\pdeg11$ & $2.53\pm0.03$\cr
94\,GHz & $1\pdeg56\pm0\pdeg06$ & $1\pdeg19\pm0\pdeg07$ & $*2.37\pm0.07$ & $1\pdeg66\pm0\pdeg06$ & $1\pdeg32\pm0\pdeg07$ & $2.07\pm0.07$\cr 
100\,GHz & $1\pdeg52\pm0\pdeg02$ & $1\pdeg14\pm0\pdeg03$ & $*2.2*\pm0.1*$ & $1\pdeg63\pm0\pdeg03$ & $1\pdeg28\pm0\pdeg04$ & $2.0\pm0.1$\cr 
143\,GHz & $1\pdeg59\pm0\pdeg09$ & $1\pdeg24\pm0\pdeg12$ & $*2.73\pm0.05$ & $1\pdeg75\pm0\pdeg14$ & $1\pdeg43\pm0\pdeg18$ & $2.44\pm0.05$\cr 
217\,GHz & $1\pdeg57\pm0\pdeg08$ & $1\pdeg21\pm0\pdeg09$ & $*4.77\pm0.03$ & $1\pdeg72\pm0\pdeg12$ & $1\pdeg40\pm0\pdeg15$ & $4.18\pm0.02$\cr 
353\,GHz & $1\pdeg58\pm0\pdeg08$ & $1\pdeg22\pm0\pdeg10$ & $*8.7*\pm0.1*$ & $1\pdeg73\pm0\pdeg13$ & $1\pdeg41\pm0\pdeg16$ & $7.5\pm0.1$\cr 
545\,GHz & $1\pdeg58\pm0\pdeg08$ & $1\pdeg22\pm0\pdeg10$ & $15.0*\pm1.5*$ & $1\pdeg72\pm0\pdeg12$ & $1\pdeg40\pm0\pdeg14$ & $12.0\pm1.2$\cr 
857\,GHz & $1\pdeg58\pm0\pdeg08$ & $1\pdeg23\pm0\pdeg11$ & $24.0*\pm2.4*$ & $1\pdeg69\pm0\pdeg09$ & $1\pdeg36\pm0\pdeg11$ & $19.0\pm1.9$\cr 
1249\,GHz & $1\pdeg57\pm0\pdeg07$ & $1\pdeg21\pm0\pdeg09$ & $27.2*\pm3.5*$ & $1\pdeg64\pm0\pdeg04$ & $1\pdeg30\pm0\pdeg05$ & $20.9\pm2.7$\cr 
2141\,GHz & $1\pdeg58\pm0\pdeg08$ & $1\pdeg23\pm0\pdeg11$ & $16.0*\pm2.1*$ & $1\pdeg62\pm0\pdeg02$ & $1\pdeg28\pm0\pdeg03$ & $11.7\pm1.5$\cr 
2997\,GHz & $1\pdeg59\pm0\pdeg09$ & $1\pdeg23\pm0\pdeg11$ & $*4.3*\pm0.6*$ & $1\pdeg60\pm0\pdeg10$ & $1\pdeg25\pm0\pdeg13$ & $3.0\pm0.4$\cr 
12\um\ & $1\pdeg75\pm0\pdeg05$ & $1\pdeg44\pm0\pdeg06$ & $(2.3\pm0.3)\times 10^{-3}$ & $1\pdeg75\pm0\pdeg05$ & $1\pdeg44\pm0\pdeg06$ & $(1.5\pm0.2)\times 10^{-3}$\cr
CO & $1\pdeg48\pm0\pdeg08$ & $1\pdeg09\pm0\pdeg12$ & $130\pm13$ & $1\pdeg54\pm0\pdeg04$ & $1\pdeg17\pm0\pdeg05$ & $91\pm9$\cr 
\hi & $1\pdeg93\pm0\pdeg13$ & $1\pdeg65\pm0\pdeg15$ & $(7.4\pm0.7)\times10^3$ & $2\pdeg09\pm0\pdeg09$ & $1\pdeg84\pm0\pdeg10$ & $(6.1\pm0.6)\times10^3$\cr
Gamma-rays & $1\pdeg87\pm0\pdeg08$ & $1\pdeg18\pm0\pdeg11$ & $(1.4\pm0.1)\times 10^{-4}$ & $2\pdeg00\pm0\pdeg10$ & $1\pdeg37\pm0\pdeg15$ & $(1.1\pm0.1)\times 10^{-4}$\cr 
Pulsar counts & $1\pdeg53\pm0\pdeg48$ & $1\pdeg15\pm0\pdeg63$ & $15.1\pm1.6$ & $1\pdeg58\pm0\pdeg45$ & $1\pdeg23\pm0\pdeg58$ & $23.8\pm3.5$\cr 
\noalign{\vskip 4pt\hrule\vskip 6pt}
 \omit\hfil& \multispan3\hfil $l=320\deg-330\deg$\hfil & \multispan3\hfil $l=330\deg-340\deg$\hfil \cr
\noalign{\vskip 4pt\hrule\vskip 6pt}
0.408\,GHz & $2\pdeg36\pm 0\pdeg07$ & $2\pdeg14\pm 0\pdeg08$ & $(130\pm 13)\times10^3$ & $1\pdeg97\pm 0\pdeg08$ & $1\pdeg70\pm 0\pdeg09$ & $(233\pm 23)\times10^3$\cr
1.4\,GHz & $2\pdeg25\pm 0\pdeg06$ & $2\pdeg02\pm 0\pdeg07$ & $(6.7\pm0.7)\times10^3$ & $1\pdeg86\pm 0\pdeg16$ & $1\pdeg57\pm 0\pdeg19$ & $(13.1\pm  1.3)\times10^3$\cr
2.3\,GHz & $2\pdeg21\pm 0\pdeg11$ & $1\pdeg97\pm 0\pdeg12$ & $(2.1\pm0.2)\times10^3$ & $1\pdeg80\pm 0\pdeg10$ & $1\pdeg50\pm 0\pdeg12$ & $(4.4\pm 0.4)\times10^3$\cr
23\,GHz & $1\pdeg96\pm 0\pdeg06$ & $1\pdeg68\pm 0\pdeg07$ & $22.4\pm0.7$ & $1\pdeg60\pm 0\pdeg10$ & $1\pdeg25\pm 0\pdeg13$ & $46.2\pm1.4$\cr
28.4\,GHz & $1\pdeg91\pm 0\pdeg01$ & $1\pdeg63\pm 0\pdeg01$ & $13.66\pm0.14$ & $1\pdeg54\pm 0\pdeg04$ & $1\pdeg17\pm 0\pdeg06$ & $29.0\pm0.3$\cr
33\,GHz & $1\pdeg92\pm 0\pdeg02$ & $1\pdeg64\pm 0\pdeg03$ & $9.7\pm0.3$ & $1\pdeg57\pm 0\pdeg07$ & $1\pdeg21\pm 0\pdeg09$ & $20.7\pm0.6$\cr
41\,GHz & $1\pdeg91\pm 0\pdeg01$ & $1\pdeg63\pm 0\pdeg01$ & $5.82\pm0.18$ & $1\pdeg55\pm 0\pdeg05$ & $1\pdeg18\pm 0\pdeg06$ & $12.6\pm0.4$\cr
44\,GHz & $1\pdeg92\pm 0\pdeg02$ & $1\pdeg64\pm 0\pdeg02$ & $4.80\pm0.05$ & $1\pdeg56\pm 0\pdeg06$ & $1\pdeg20\pm 0\pdeg08$ & $10.3\pm0.1$\cr
60.7\,GHz & $1\pdeg92\pm 0\pdeg03$ & $1\pdeg64\pm 0\pdeg03$ & $2.50\pm0.08$ & $1\pdeg55\pm 0\pdeg05$ & $1\pdeg18\pm 0\pdeg06$ & $5.21\pm0.16$\cr
70.4\,GHz & $1\pdeg94\pm 0\pdeg05$ & $1\pdeg67\pm 0\pdeg05$ & $2.01\pm0.02$ & $1\pdeg55\pm 0\pdeg05$ & $1\pdeg19\pm 0\pdeg07$ & $4.02\pm0.04$\cr
94\,GHz & $1\pdeg99\pm 0\pdeg02$ & $1\pdeg72\pm 0\pdeg02$ & $1.70\pm0.05$ & $1\pdeg59\pm 0\pdeg09$ & $1\pdeg24\pm 0\pdeg12$ & $3.01\pm0.09$\cr
100\,GHz & $2\pdeg00\pm 0\pdeg01$ & $1\pdeg72\pm 0\pdeg02$ & $1.74\pm0.09$ & $1\pdeg60\pm 0\pdeg01$ & $1\pdeg25\pm 0\pdeg01$ & $3.0\pm0.1$\cr
143\,GHz & $2\pdeg05\pm 0\pdeg15$ & $1\pdeg79\pm 0\pdeg17$ & $2.07\pm0.04$ & $1\pdeg65\pm 0\pdeg05$ & $1\pdeg31\pm 0\pdeg06$ & $3.09\pm0.06$\cr
217\,GHz & $2\pdeg03\pm 0\pdeg03$ & $1\pdeg77\pm 0\pdeg04$ & $3.52\pm0.02$ & $1\pdeg64\pm 0\pdeg04$ & $1\pdeg30\pm 0\pdeg05$ & $5.16\pm0.04$\cr
353\,GHz & $2\pdeg03\pm 0\pdeg03$ & $1\pdeg77\pm 0\pdeg04$ & $6.33\pm0.08$ & $1\pdeg64\pm 0\pdeg04$ & $1\pdeg30\pm 0\pdeg05$ & $9.3*\pm0.1*$\cr
545\,GHz & $2\pdeg03\pm 0\pdeg03$ & $1\pdeg76\pm 0\pdeg03$ & $10.0\pm1.0$ & $1\pdeg64\pm 0\pdeg14$ & $1\pdeg30\pm 0\pdeg17$ & $16.0\pm1.6$\cr
857\,GHz & $2\pdeg01\pm 0\pdeg01$ & $1\pdeg75\pm 0\pdeg02$ & $15.0\pm1.5$ & $1\pdeg64\pm 0\pdeg14$ & $1\pdeg29\pm 0\pdeg17$ & $24.0\pm2.4$\cr
1249\,GHz & $1\pdeg98\pm 0\pdeg02$ & $1\pdeg71\pm 0\pdeg02$ & $16.4\pm2.1$ & $1\pdeg60\pm 0\pdeg10$ & $1\pdeg26\pm 0\pdeg13$ & $28\pm4$\cr
2141\,GHz & $1\pdeg96\pm 0\pdeg06$ & $1\pdeg69\pm 0\pdeg07$ & $8.8\pm1.1$ & $1\pdeg60\pm 0\pdeg01$ & $1\pdeg26\pm 0\pdeg01$ & $17\pm2$\cr
2997\,GHz & $1\pdeg93\pm 0\pdeg03$ & $1\pdeg66\pm 0\pdeg04$ & $2.2\pm0.3$ & $1\pdeg58\pm 0\pdeg02$ & $1\pdeg23\pm 0\pdeg02$ & $4.7\pm0.6$\cr
12\um\ & $2\pdeg08\pm 0\pdeg08$ & $1\pdeg83\pm 0\pdeg09$ & $(1.0\pm0.1)\times10^{-3}$ & $1\pdeg76\pm 0\pdeg04$ & $1\pdeg45\pm 0\pdeg05$ & $(2.1\pm0.3)\times10^{-3}$\cr
CO & $1\pdeg97\pm 0\pdeg03$ & $1\pdeg70\pm 0\pdeg03$ & $75\pm8$ & $1\pdeg61\pm 0\pdeg01$ & $1\pdeg26\pm 0\pdeg01$ & $108\pm11$\cr
\hi & $2\pdeg41\pm0\pdeg01$ & $2\pdeg19\pm0\pdeg01$ & $(7.1\pm0.7)\times10^3$ & $2\pdeg29\pm0\pdeg01$ & $2\pdeg06\pm0\pdeg01$ & $(7.6\pm0.8)\times10^3$\cr
Gamma-rays & $2\pdeg28\pm 0\pdeg08$ & $1\pdeg75\pm 0\pdeg11$ & $ (0.76\pm0.08)\times10^{-4}$ & $1\pdeg91\pm 0\pdeg11$ & $1\pdeg22\pm 0\pdeg17$ & $ (1.27\pm0.13)\times10^{-4}$\cr
Pulsar counts & $1\pdeg52\pm 0\pdeg48$ & $1\pdeg16\pm 0\pdeg63$ & $ 15 \pm  2$ & $1\pdeg58\pm 0\pdeg45$ & $1\pdeg23\pm 0\pdeg58$ & $ 24 \pm  4$\cr
\noalign{\vskip 3pt\hrule\vskip 4pt}}}
\endPlancktablewide
\endgroup
\end{table*}

\begin{figure*}
\centering
\includegraphics[angle=90,scale=0.33]{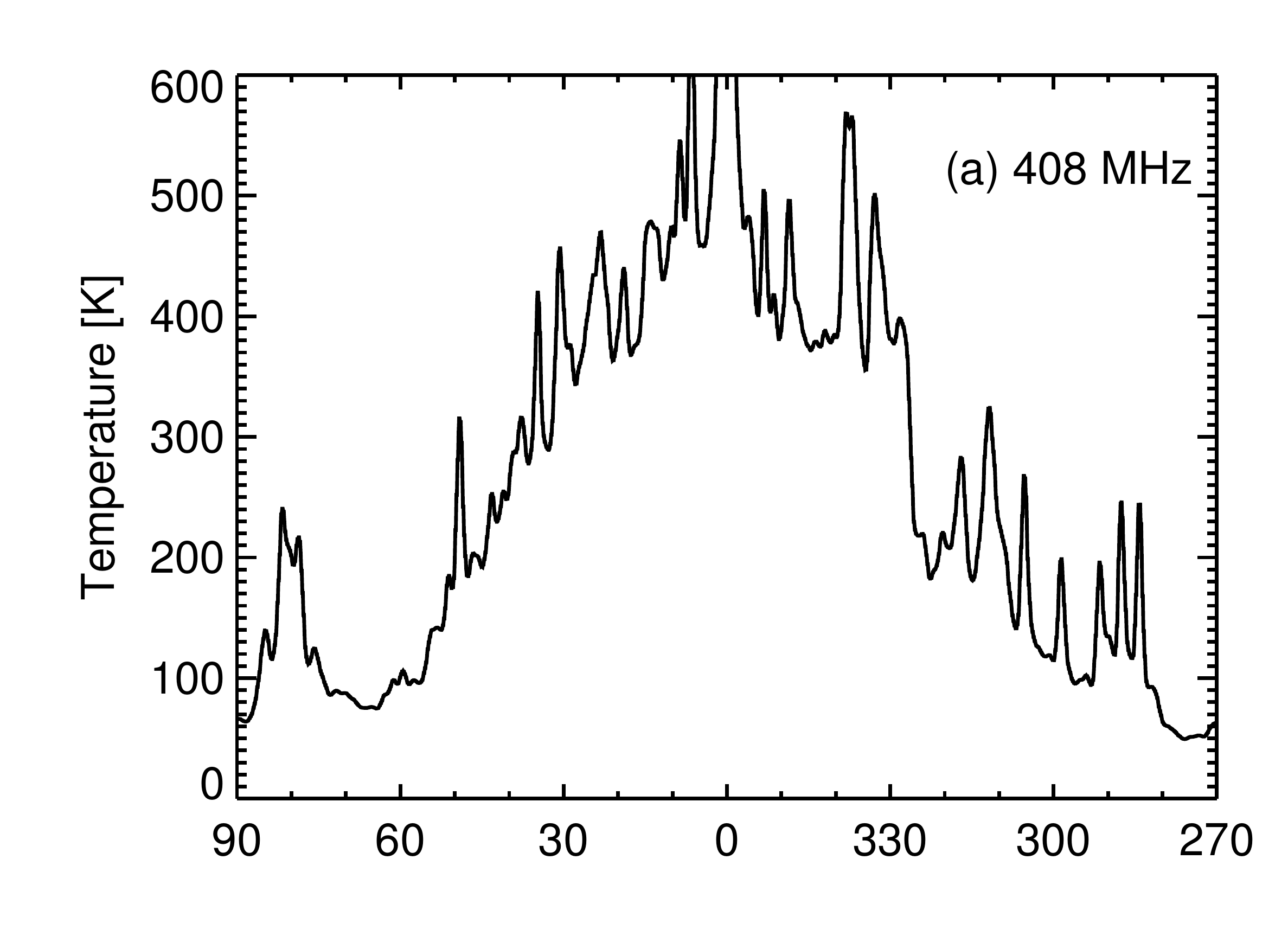}
\vspace*{-0.5cm}
\includegraphics[angle=90,scale=0.33]{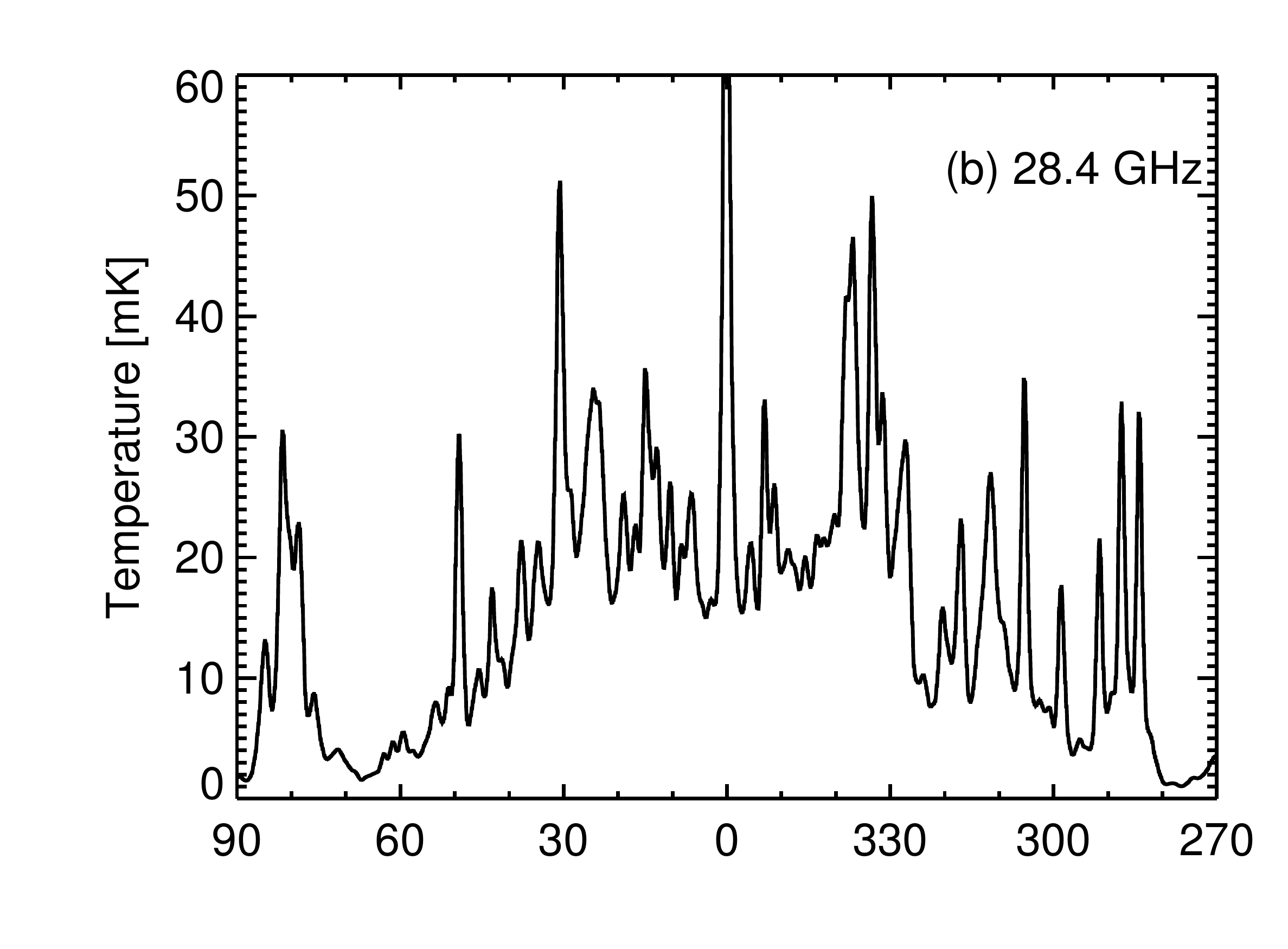}
\includegraphics[angle=90,scale=0.33]{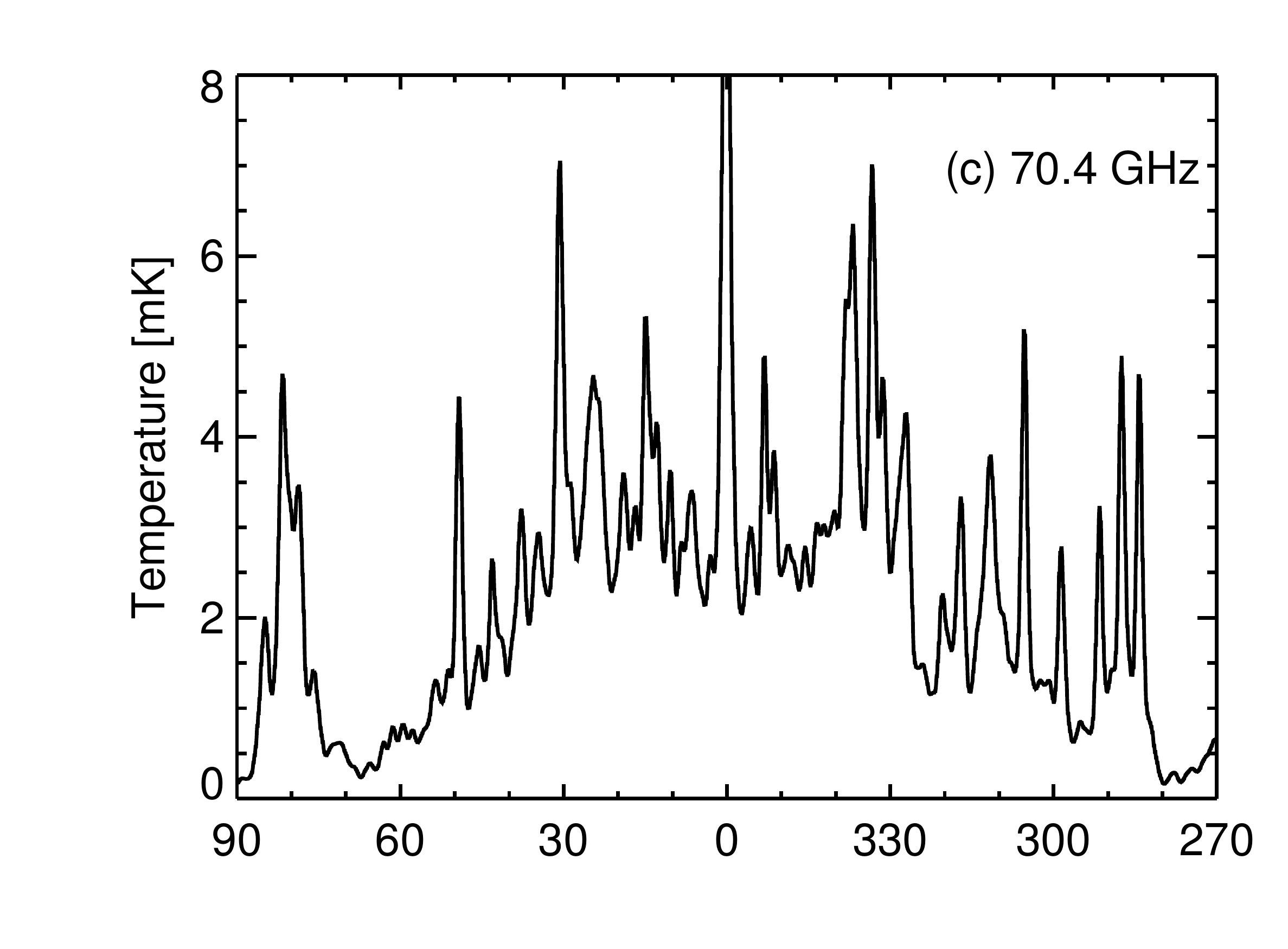}
\vspace*{-0.5cm}
\includegraphics[angle=90,scale=0.33]{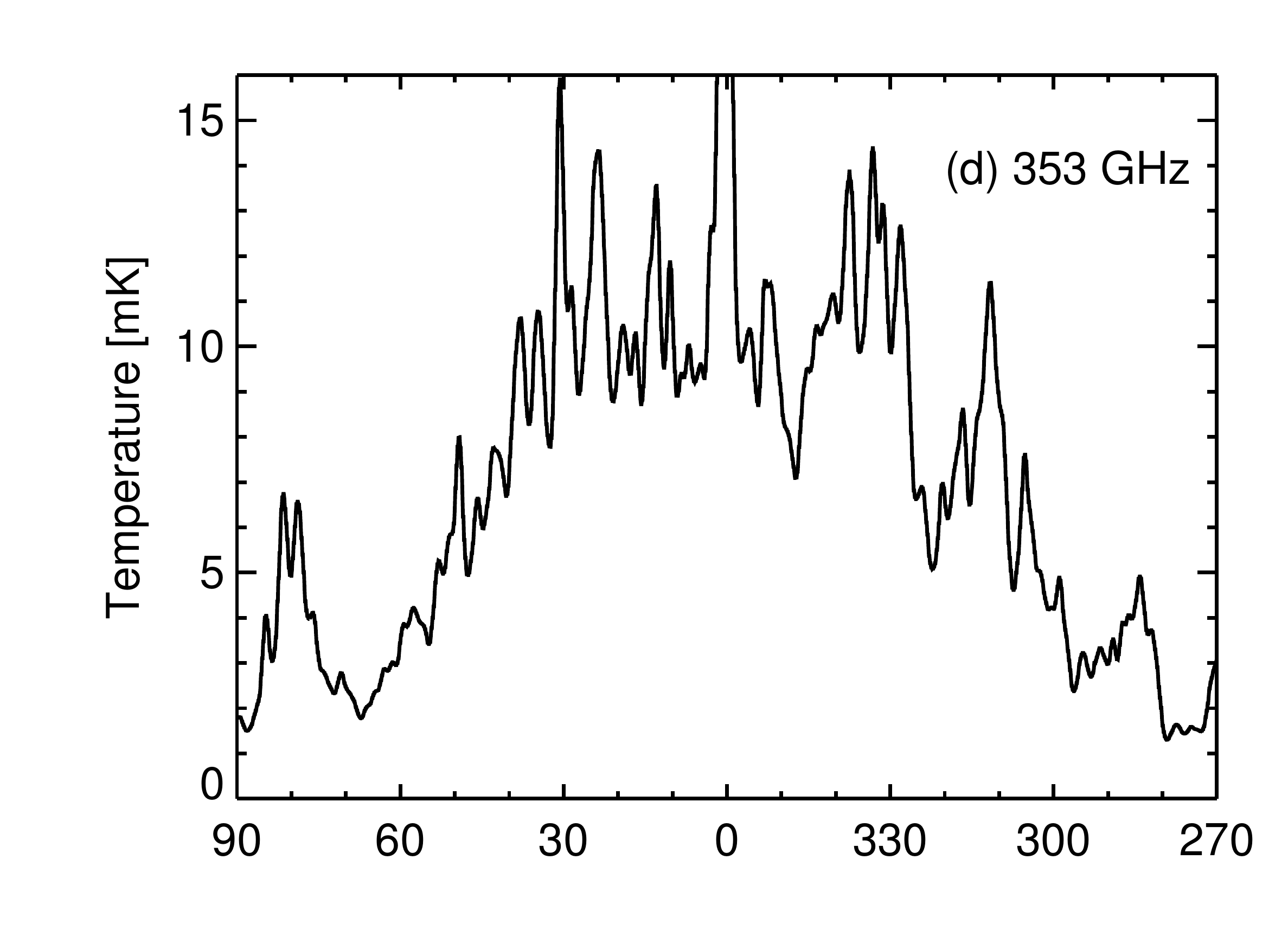}
\includegraphics[angle=90,scale=0.33]{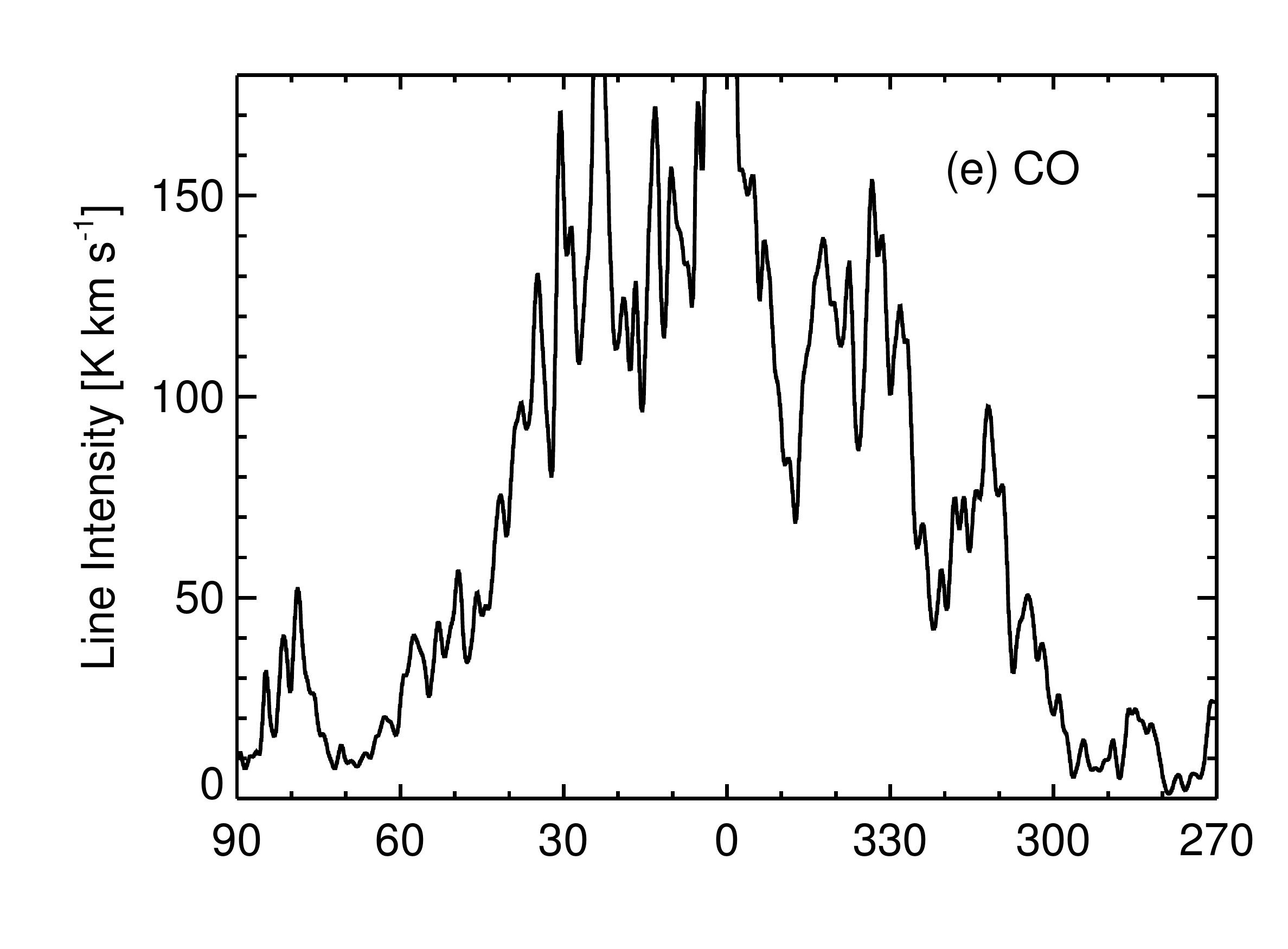}
\vspace*{-0.5cm}
\includegraphics[angle=90,scale=0.33]{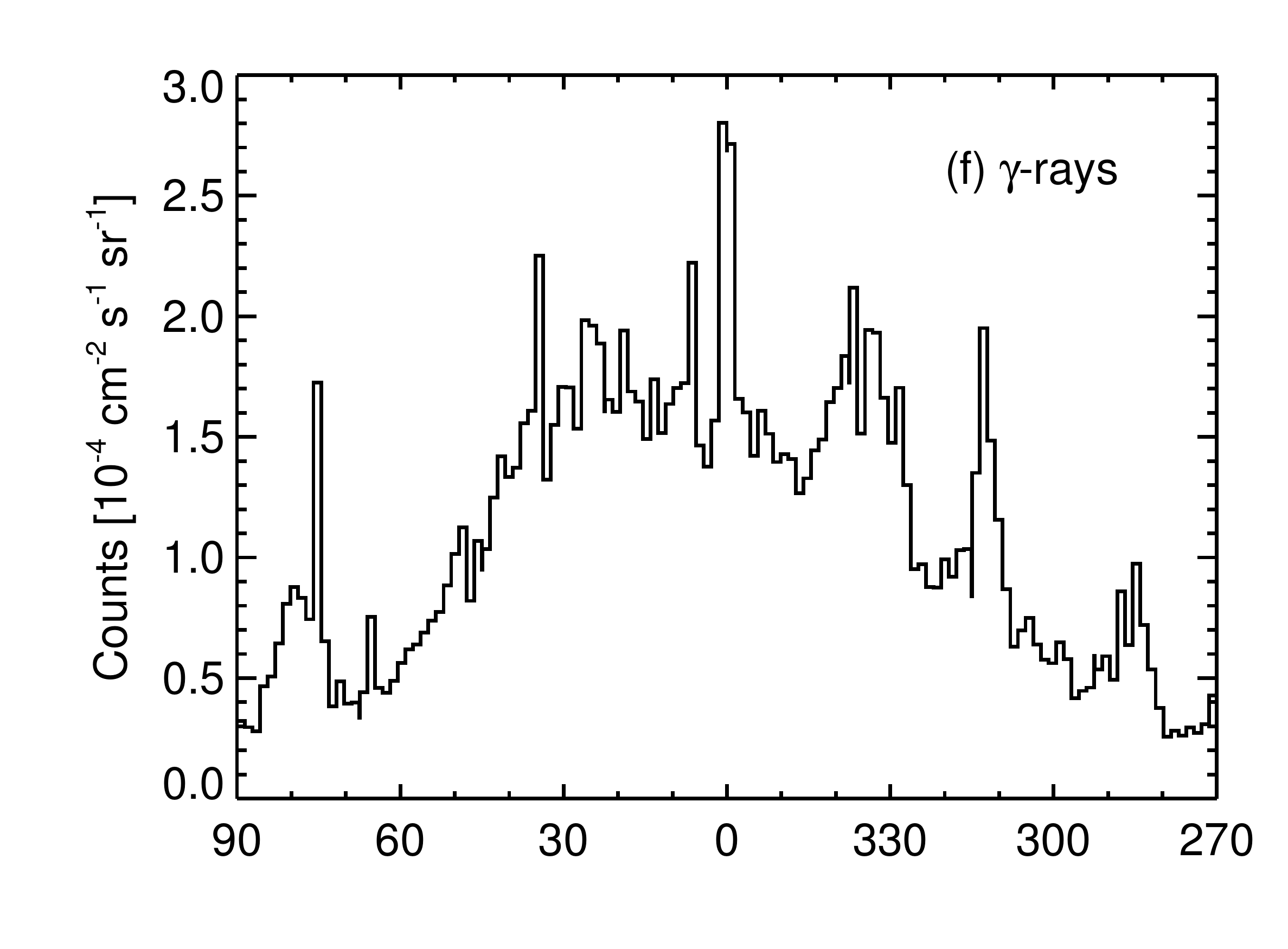}
\includegraphics[angle=90,scale=0.33]{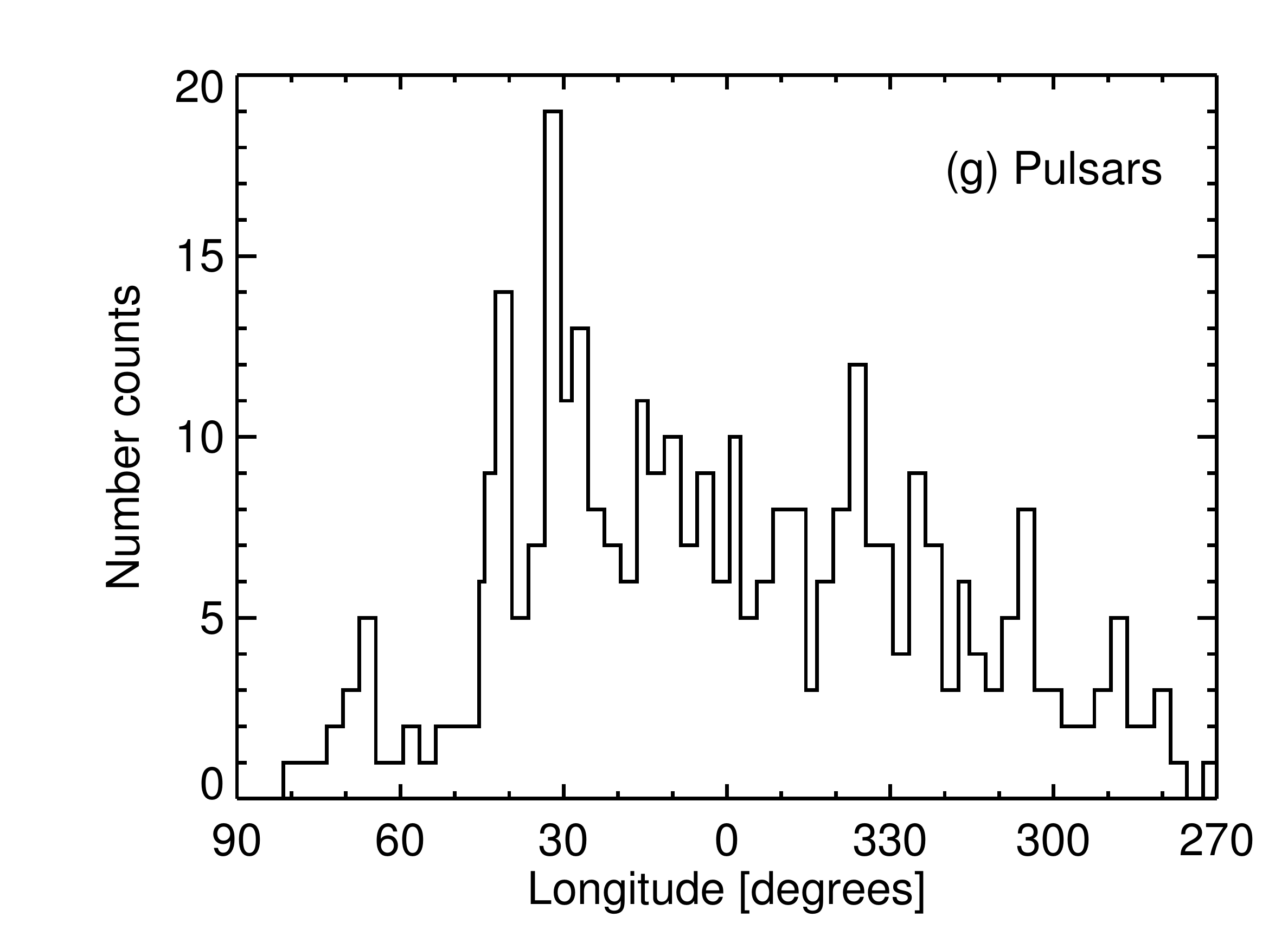}
\includegraphics[angle=90,scale=0.33]{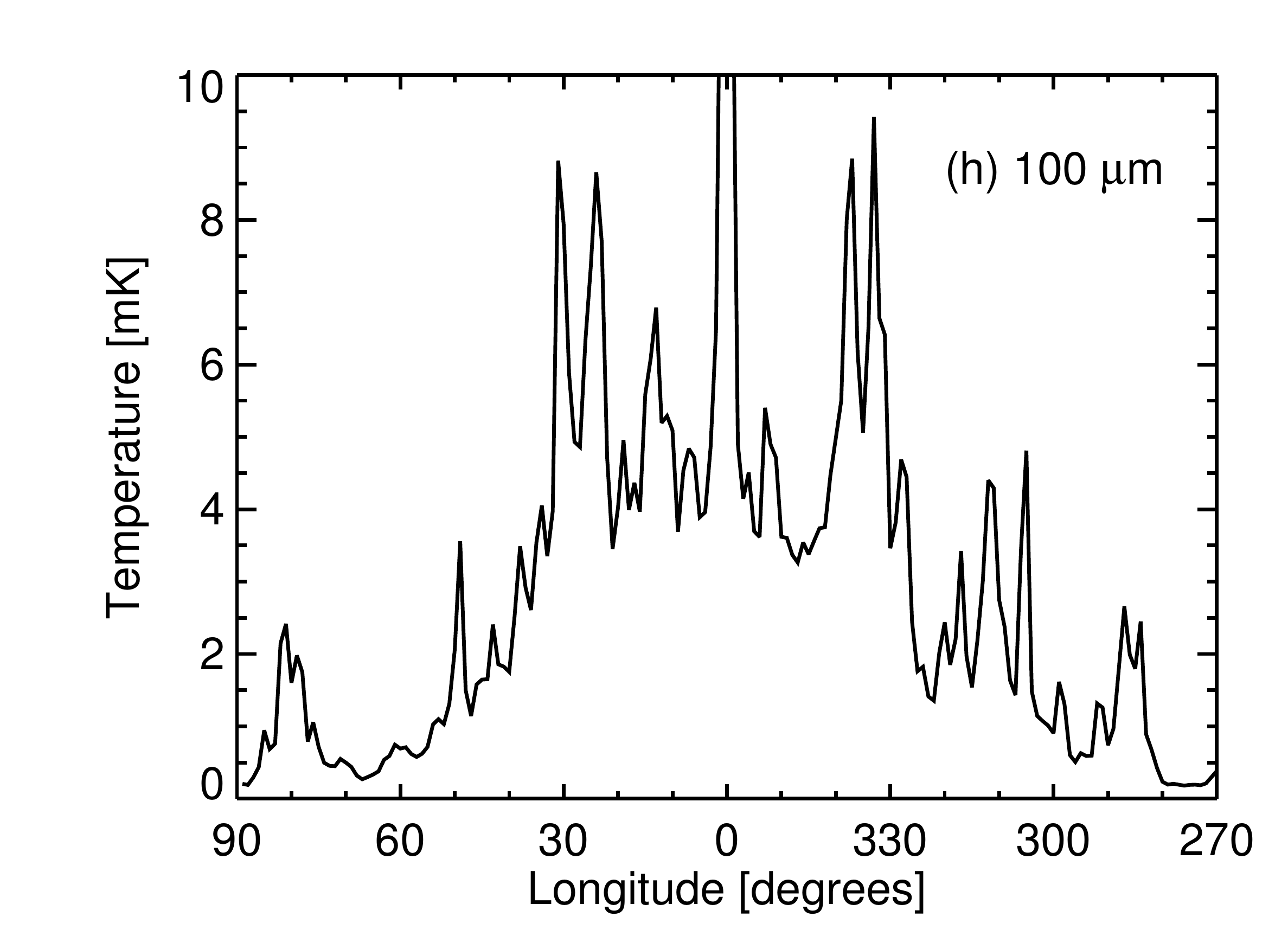}
\caption{The longitude distribution of emission components along the Galactic plane at $b = 0\deg$ in the range $l = 90\deg$ to $270\deg$. The data are all smoothed to 1\deg\ resolution except the gamma-ray data with a FWHM of 88\arcm. The components shown are 408\MHz\ (mainly synchrotron), 28.4\GHz\ (similar AME and free-free), 70.4\GHz\ (mainly free-free), 353\GHz\ (dust), CO, gamma-rays, pulsar numbers (representing supernovae), and 100\um. The gamma-ray and pulsar data are at ${N}_\mathrm{side}=64$ and ${N}_\mathrm{side}=32$ respectively, to account for the relatively low sampling. The pulsar data are uniformly surveyed from $l = 270\deg$ to 40\deg; at $l = 40\deg$ to $90\deg$ the survey sensitivity is lower. The Galactic centre is off-scale so that the diffuse emission can be displayed.}
\label{fig:londist}
\end{figure*}


\subsection{Longitude distribution}
\label{sec:londist}
We now turn to the longitude distribution of the data sets before studying the properties of each component.

The longitude distribution over the range $l=270\deg \rightarrow 0\deg \rightarrow 90\deg$ at $b = 0\deg$ of the 1\deg-smoothed data is shown in Fig. \ref{fig:londist} for selected frequencies: 408\,MHz (mainly synchrotron), 28.4\GHz\ (mainly AME and free-free), 70.4\GHz\ (mainly free-free), 353\GHz\ (mainly dust), CO, gamma-rays, pulsars (representing supernovae), and 100\um. Although each of the components represented is of a different origin, they have similar longitude distributions as a consequence of star formation in the regions of highest gas and dust density. Spiral arms can be identified through the longer emission length at their tangent points as seen from the Sun; the Carina arm tangent is seen at $l = 300\deg$, the Norma tangent at $l = 330\deg$, the Scutum tangent at $l = 25\deg$, and the Sagittarius tangent at $l = 45\deg$. These tangent points are at galactocentric distances of of 6.5, 4.5, 4.5, and 6.5\kpc\ respectively, assuming a solar galactocentric distance of 8.5\kpc\ \citep{Clemens:1985}.

Strong emission is seen in all the data sets in the longitude range $l=357\deg \rightarrow 0\deg \rightarrow 3\deg$ around the Galactic centre, but we do not consider this region in the present study.  Individual extended star-forming regions are seen in all the data sets, including  Cygnus-X at $l = 80$\deg\ and the Gum Nebula complex at $l = 286$\deg. More compact star-forming regions on a scale of 1\deg\ or less are seen at 408\MHz\ and the lower \Planck\ frequencies at $l = 283$\deg, 287\deg, 306\deg, 328\deg, 31\deg, and 50\deg.


\section{Discussion of the components}
\label{sec:discussion}
\label{sec:discussion-intro}

In Sect. \ref{sec:londist} (Fig. \ref{fig:londist}) we showed the longitude distribution of the emission at $b=0\deg$ from $l=270\deg \rightarrow 0\deg \rightarrow 90\deg$ in different frequency bands from radio to FIR. The strongest emission is clearly associated with the inner spiral arms in quadrants IV ($l=270\deg$--$360\deg$) and I ($l=0\deg$--$90\deg$). First, we consider each emission component in turn, taking account of the latitude distribution and its expected spectral index to be used in the subsequent component separation.  Then we determine SEDs at intervals along the Galactic plane at $b = 0\deg$, and the longitude distribution of each emission component between $l = 300\deg$ and 60\deg.

\subsection{The free-free component}
\label{sec:discussion-ff}

RRLs provide a  direct estimate of the free-free emission on the Galactic plane that is unaffected by the absorption that has to be taken into account for the \ha\ measures that are commonly used at intermediate latitudes.  We have a fully sampled map of the RRL emission for the region $l = 20$\deg\ to 44\deg\ and $b = -4$\deg\ to $+4$\deg\ \citep{Alves2012}, which can be used as a robust measure of the free-free emission in this range \citep{D3:2003}.  These RRL data can be used to check the calibration of the {\tt FastMEM} free-free estimate for the rest of the plane considered in this study (the parameters used are outlined in Sect. \ref{sec:analysis} and Appendix \ref{sec:appendix}).  The ratio of the {\tt FastMEM} brightness amplitude to the \WMAP\ {\tt MEM} value for the narrow component is $1.02\pm0.05$, while  a similar comparison between the {\tt FastMEM} and the RRL brightness temperatures gave a ratio of $1.19\pm0.06$. The question of systematic errors in the different datasets or analyses is discussed in Appendix \ref{sec:appendix}.

A possible systematic error in RRL free-free estimates is the uncertainty in electron temperature. We have adopted an average value of 6000\,K \citep{Alves2012}; an increase of this value by 600\,K would increase the brightness temperature by 10\,\%. No correction has been applied for non-LTE factors (Stark broadening and stimulated emission), which act in opposite senses and which in any case are small at 1.4\GHz\ in low-density regions such as the diffuse gas on the Galactic plane \citep{Shaver1980,Gordon1990}. We assign an overall error of 15\,\% in the amplitude of the free-free contribution.

In addition, there are possible systematic errors in the {\tt FastMEM} analysis, such as the contribution from the other emission components in the frequency range used in the analysis. Tests by varying the input parameters by reasonable amounts indicated that systematic effects were less than 10\,\% (see Appendix \ref{sec:appendix}). On the basis of these results we adopt a conversion factor of 1.10 for the RRL data  and use the {\tt FastMEM} data reduced by 10\,\% where the RRL data are not available.

\begin{figure}
\centering
\includegraphics[angle=90,scale=0.35]{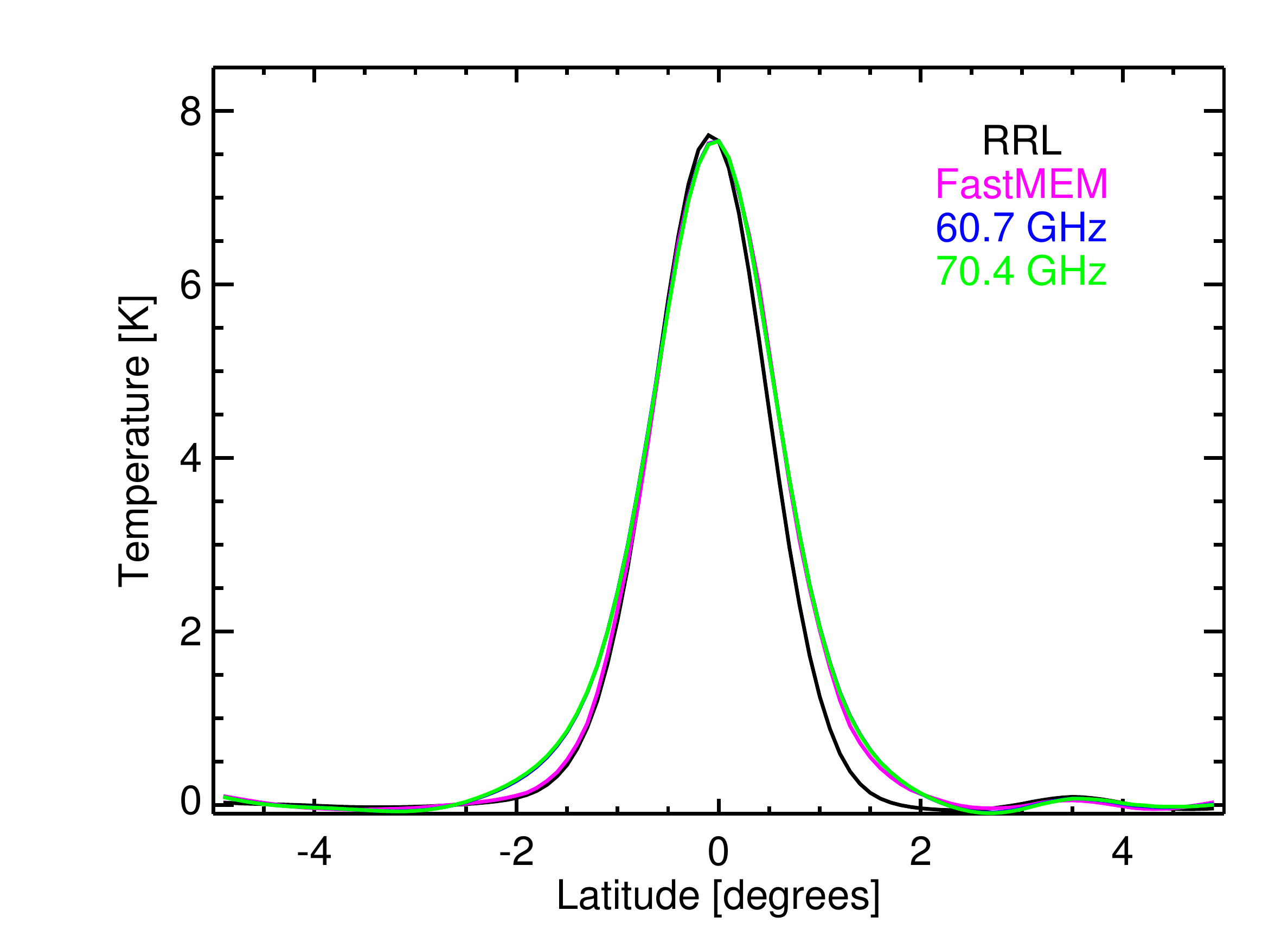}
\caption{Latitude cuts for the narrow free-free-related component in the longitude range $l=20\deg$ to $30\deg$ for RRLs, {\tt FastMEM}, \WMAP\ 60.7\GHz, and \planck\ 70.4\GHz. These are normalized to match the {\tt FastMEM} solution at $b=0\deg$ in order to show their relative widths. The re-scaled 60.7\GHz\ and 70.4\GHz\ cuts match at all latitudes and are thus not visible separately.}
\label{fig:narrowff}
\end{figure}

The latitude distribution of the $l = 20$\deg\ to 30\deg\ region of the narrow free-free component determined from RRLs at 1.4\GHz\ and from {\tt FastMEM} is compared with the \WMAP\ 60.7\GHz\ and the \Planck\ 70.4\GHz\ total emission in Fig. \ref{fig:narrowff}, where the amplitudes have been normalized at $b=0\deg$; only the narrow components are plotted.  The 60.7 and 70.4\GHz\ emission is mainly free-free (70--90\,\%) as will be shown in Fig.~\ref{fig:amesed} (see Sect. \ref{sec:discussion-ame}). The 60.7 and 70.4\GHz\ latitude profiles from the two space missions are essentially identical.

Table \ref{table:ff} gives the intrinsic width of each of the free-free related components for the four longitude ranges under study. The average intrinsic width of free-free calculated from the RRL analysis for $l=20$\deg--40\deg\ is $0\pdeg92\pm0\pdeg05$. The width of the emission at velocities corresponding to the Sagittarius and the Scutum arms is somewhat narrower at $0\pdeg85 \pm 0\pdeg03$ and $0\pdeg78 \pm 0\pdeg01$ respectively \citep{Alves2012}. The average of the 60.7 and 70.4\GHz\ FWHM is $1\pdeg18\pm0\pdeg03$. The FWHM value found from {\tt FastMEM} is $1\pdeg07\pm0\pdeg05$. We conclude that the FWHM of the free-free emission is $1\pdeg06\pm0\pdeg05$ in the longitude range $l=20$\deg--40\deg\ using all the free-free data. A similar result is found for the $l=330$\deg--340\deg\ range; the $l=320$\deg--330\deg\ range is broader in all components by a factor of $1.35\pm0.05$ due to bright structures adjacent to the plane.

\begin{table*}[tb]
\begingroup
\newdimen\tblskip \tblskip=5pt
\caption{Intrinsic latitude FWHM of free-free emission and related distributions for four longitude ranges.}
\label{table:ff}
\nointerlineskip
\vskip -3mm
\footnotesize
\setbox\tablebox=\vbox{
 \newdimen\digitwidth 
 \setbox0=\hbox{\rm 0} 
 \digitwidth=\wd0 
 \catcode`*=\active 
 \def*{\kern\digitwidth}
 \def\leaderfil{\leaders\hbox to 5pt{\hss.\hss}\hfil}
 \newdimen\signwidth 
 \setbox0=\hbox{+} 
 \signwidth=\wd0 
 \catcode`!=\active 
 \def!{\kern\signwidth}
 \halign{\hbox to 1.7in{#\leaderfil}\tabskip 1.5em&
 \hfil#\hfil&
 \hfil#\hfil&
 \hfil#\hfil&
 \hfil#\hfil&
 #\hfil\tabskip 0pt\cr
 \noalign{\doubleline\vskip 2pt}
 \omit & \multispan4\hfil Intrinsic width [degrees] \hfil & \cr
\noalign{\vskip -2pt}
\omit \hfil Data \hfil  & \multispan4{\hrulefill} & Source of data\cr
 \omit & \omit \hfil $l=20\deg$--$30\deg$ \hfil & \omit \hfil $l=30\deg$--$40\deg$ \hfil & \omit \hfil $l=320\deg$--$330\deg$ \hfil & \omit \hfil $l=330\deg$--$340\deg$ \hfil &  \hfil \cr
\noalign{\vskip 4pt\hrule\vskip 6pt}
RRL& $0\pdeg90\pm0\pdeg06$ & $0\pdeg93\pm0\pdeg10$ & \ldots & \ldots & \citet{Alves2012}\cr
60.7\GHz & $1\pdeg15\pm0\pdeg03$ & $1\pdeg21\pm0\pdeg09$ & $1\pdeg64\pm 0\pdeg03$ & $1\pdeg18\pm 0\pdeg06$ & \WMAP\cr
70.4\GHz & $1\pdeg15\pm0\pdeg03$ & $1\pdeg23\pm0\pdeg11$ & $1\pdeg67\pm 0\pdeg05$ & $1\pdeg19\pm 0\pdeg07$ & \planck\cr
{\tt FastMEM} estimate& $1\pdeg05\pm0\pdeg07$ & $1\pdeg06\pm0\pdeg07$ & $1\pdeg49\pm0\pdeg01$ & $1\pdeg10\pm0\pdeg12$ & This paper\cr
OB stars (CS dense clouds)& $0\pdeg90\pm0\pdeg05$ & $0\pdeg90\pm0\pdeg05$ & $0\pdeg90\pm0\pdeg05$ & $0\pdeg90\pm0\pdeg05$ & \citet{Bronfman:2000}\tablefootmark{a}\cr
OB stars (molecular clouds)& $0\pdeg83\pm0\pdeg05$ & $0\pdeg83\pm0\pdeg05$ & $0\pdeg83\pm0\pdeg05$ & $0\pdeg83\pm0\pdeg05$ & \citet{Wood:1989}\tablefootmark{a}\cr
100\um & $1\pdeg23\pm0\pdeg11$ & $1\pdeg25\pm0\pdeg13$ & $1\pdeg66\pm 0\pdeg04$ & $1\pdeg23\pm 0\pdeg02$ & \IRAS\cr
CO& $1\pdeg09\pm0\pdeg12$ & $1\pdeg17\pm0\pdeg05$ & $1\pdeg70\pm 0\pdeg03$ & $1\pdeg26\pm 0\pdeg01$ & \citet{planck2013-p03a}\cr
545\GHz & $1\pdeg22\pm0\pdeg10$ & $1\pdeg40\pm0\pdeg14$ & $1\pdeg76\pm 0\pdeg03$ & $1\pdeg30\pm 0\pdeg17$ & \planck\cr
\noalign{\vskip 3pt\hrule\vskip 4pt}}}
\endPlancktablewide
\tablefoot{\tablefoottext{a}{Longitude range $l = 270$\deg--90\deg.}}
\endgroup
\end{table*} 

The related data from the literature include the 99\GHz\ line of CS $J$=2$\rightarrow$1, which identifies the regions of dense gas that are the sites of massive OB star formation \citep{Bronfman:2000}. This line is integrated through the Galaxy and as published is averaged over the full longitude range; 79\,\% lies in quadrants IV and I of the inner Galaxy. The CS FWHM is $0\pdeg90\pm0\pdeg05$, which \citeauthor{Bronfman:2000} estimate as 74\pc\ within the solar circle at distances of $R_{\rm G} = 0.5 -0.8$\,$R_{\odot}$ from the Galactic centre. \citet{Wood:1989} have identified the regions of massive star formation in molecular clouds, including ultracompact \hii\ regions, by using their FIR two-colour properties. They find that the distribution of massive star candidates has a FWHM of $0\pdeg83\pm0\pdeg05$ corresponding to a scale height $0\pdeg6$ ($e^{-|b|/b_{0}}$ with $b_{0} = 0\pdeg6$); this is equivalent to a FWHM of 125\pc\ at an average distance of 8.5\kpc. We see that the OB-star latitude widths of $0\pdeg83$ are significantly narrower than those of the free-free emission as determined by the RRLs and the component separation results. This implies that the ionizing radiation from the OB stars is not fully absorbed in the dense molecular clouds in which they are born, but escapes into the surrounding lower density gas that it ionizes. Some of the broadening of the 60.7 and 70.4\GHz\ channels relative to the RRL emission is due to the contribution from AME, which will have the same width as the broader 100\um\ dust listed in Table \ref{table:freqresults}. AME in the 60--70\GHz\ range is 10--20\,\% of the free-free as given by the SEDs.

The free-free width from RRLs ($0\pdeg92\pm0\pdeg05$) is significantly less than that of the total gas density as measured by the CO ($1\pdeg13\pm0\pdeg06$), 100\um\ dust ($1\pdeg24\pm0\pdeg09$), and \Planck\ 545\GHz\ ($1\pdeg33\pm0\pdeg08$). This indicates that the fractional ionization decreases away from the OB stars on the plane.

We can compare the latitude distribution of the free-free emission, which is proportional to the emission measure (${\rm EM}=\int n_{\rm e}^{2}dl$), with the dispersion measure (${\rm DM} = \int n_{\rm e}dl$) obtained from pulsar measurements. Narrow and broad $n_{\rm e}$ distributions have been identified in modelling the Galactic electron distribution \citep{Cordes:2002, Cordes:2003, Lorimer:2006, Berkhuijsen:2006, Gaensler:2008}. These studies give a thin component with thickness FWHM of 190\pc\ and a thick disk estimated to be in the range 900--1800\pc. They are mainly based on nearer pulsars as compared whilst the free-free emission is integrated through the depth of the Galaxy. We would expect the free-free to have a narrower distribution because of its dependence on $n_{\rm e}^{2}$ \citep{Gaensler:2008}. This is indicated by the present data for the inner Galaxy where the FWHM of the free-free emission in the Sagittarius and Scutum arms has a FWHM of 100\pc\ \citep{Alves2012}. A level of clumping (filling factor) is required to bring the EM and DM into agreement. This clumping is already evident in the EM data \citep{Reynolds:1998, Berkhuijsen:2008, Alves2012}.

\subsection{Synchrotron}
\label{sec:discussion-synch}

\begin{figure*}
\centering
\includegraphics[angle=90,scale=0.27]{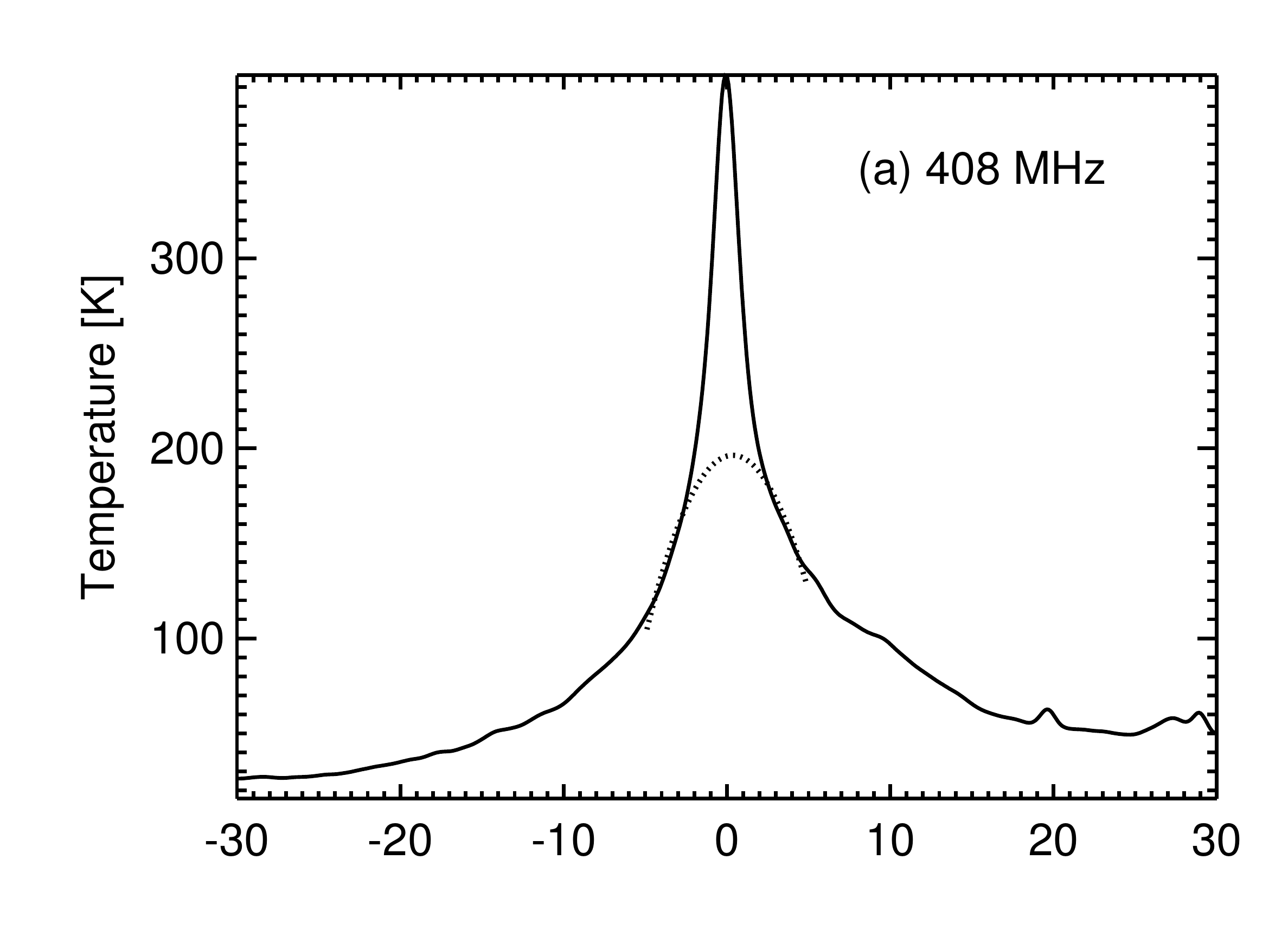}
\vspace*{-0.5cm}
\includegraphics[angle=90,scale=0.27]{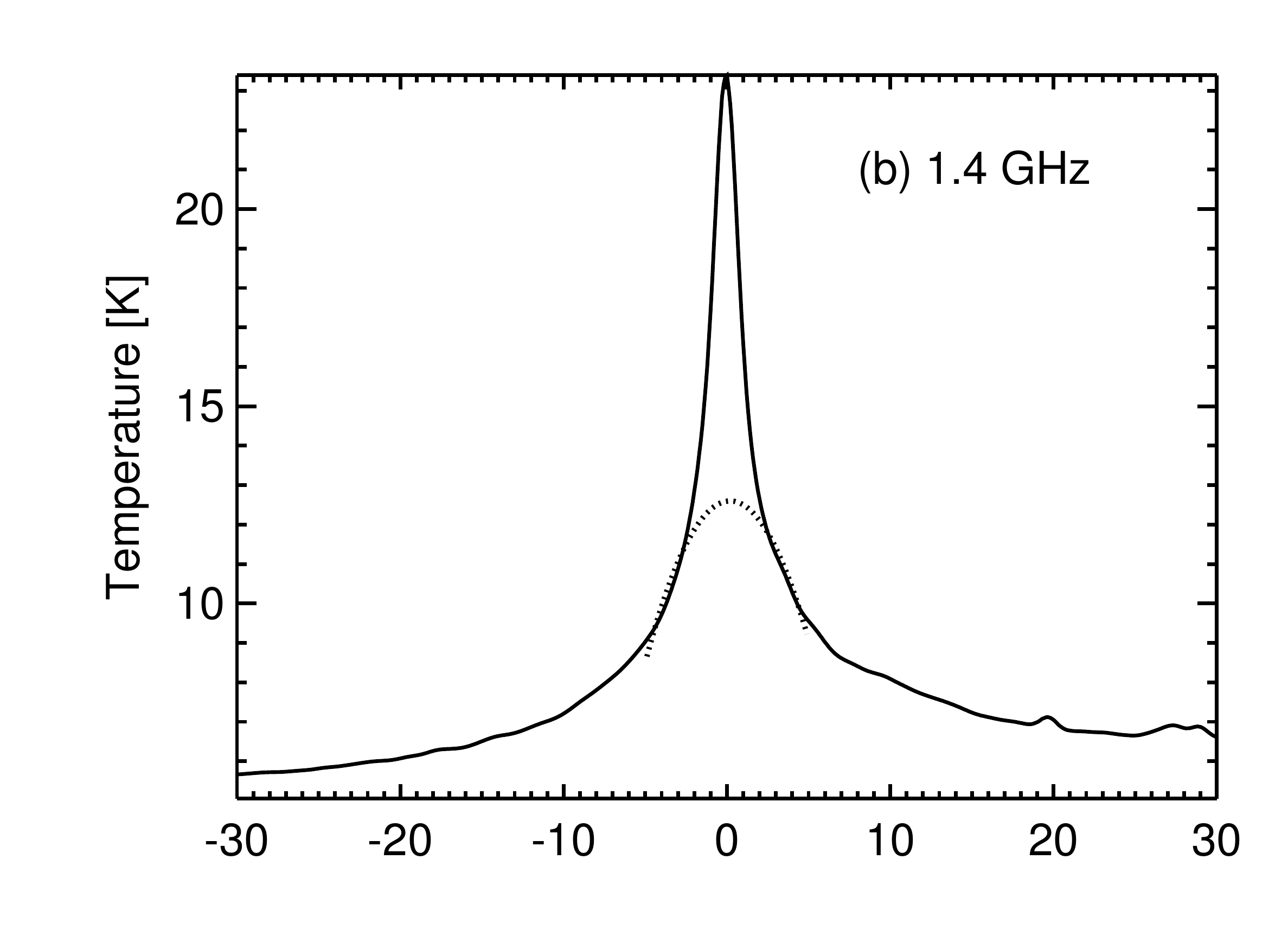}
\includegraphics[angle=90,scale=0.27]{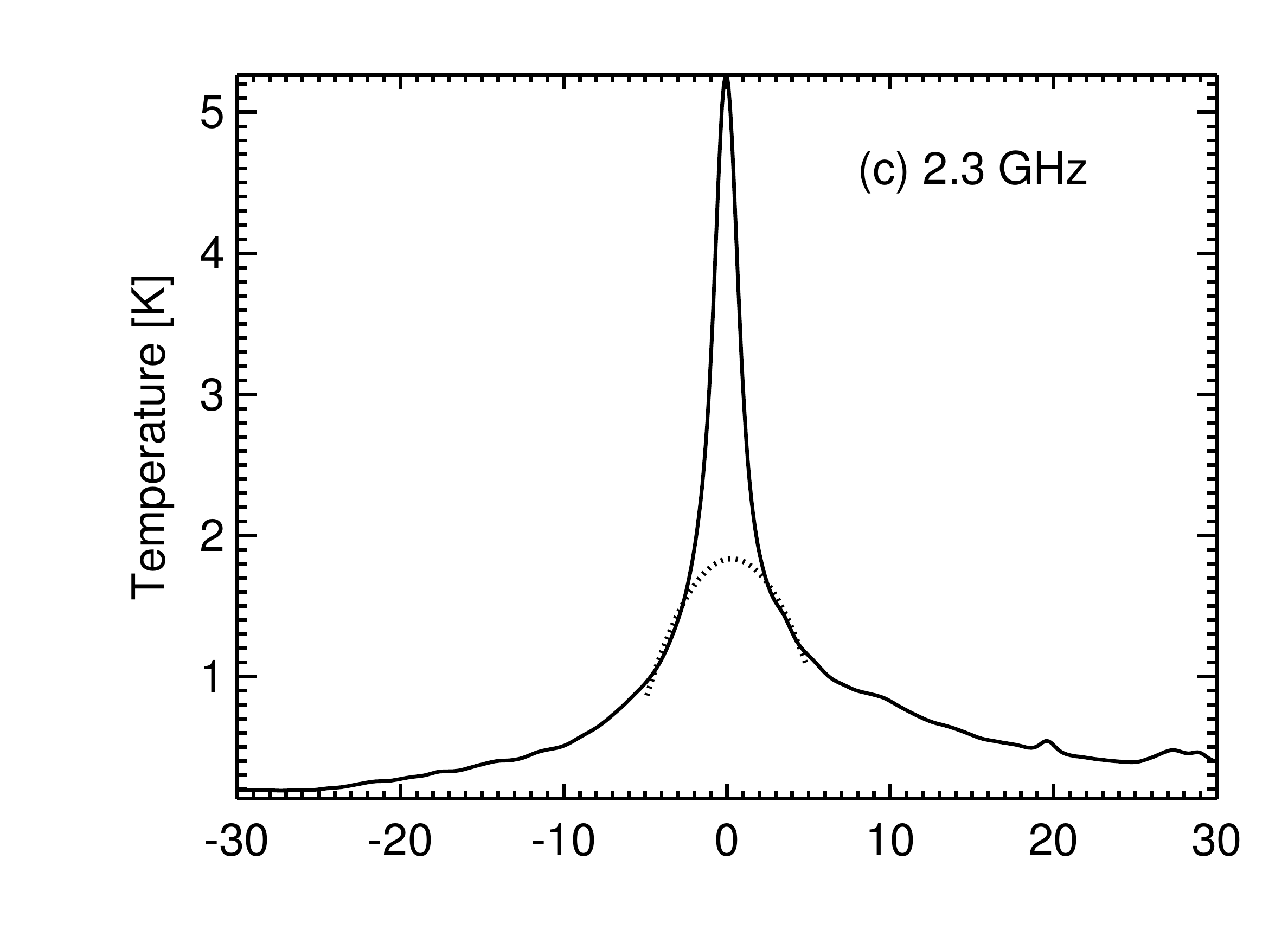}
\vspace*{-0.5cm}
\includegraphics[angle=90,scale=0.27]{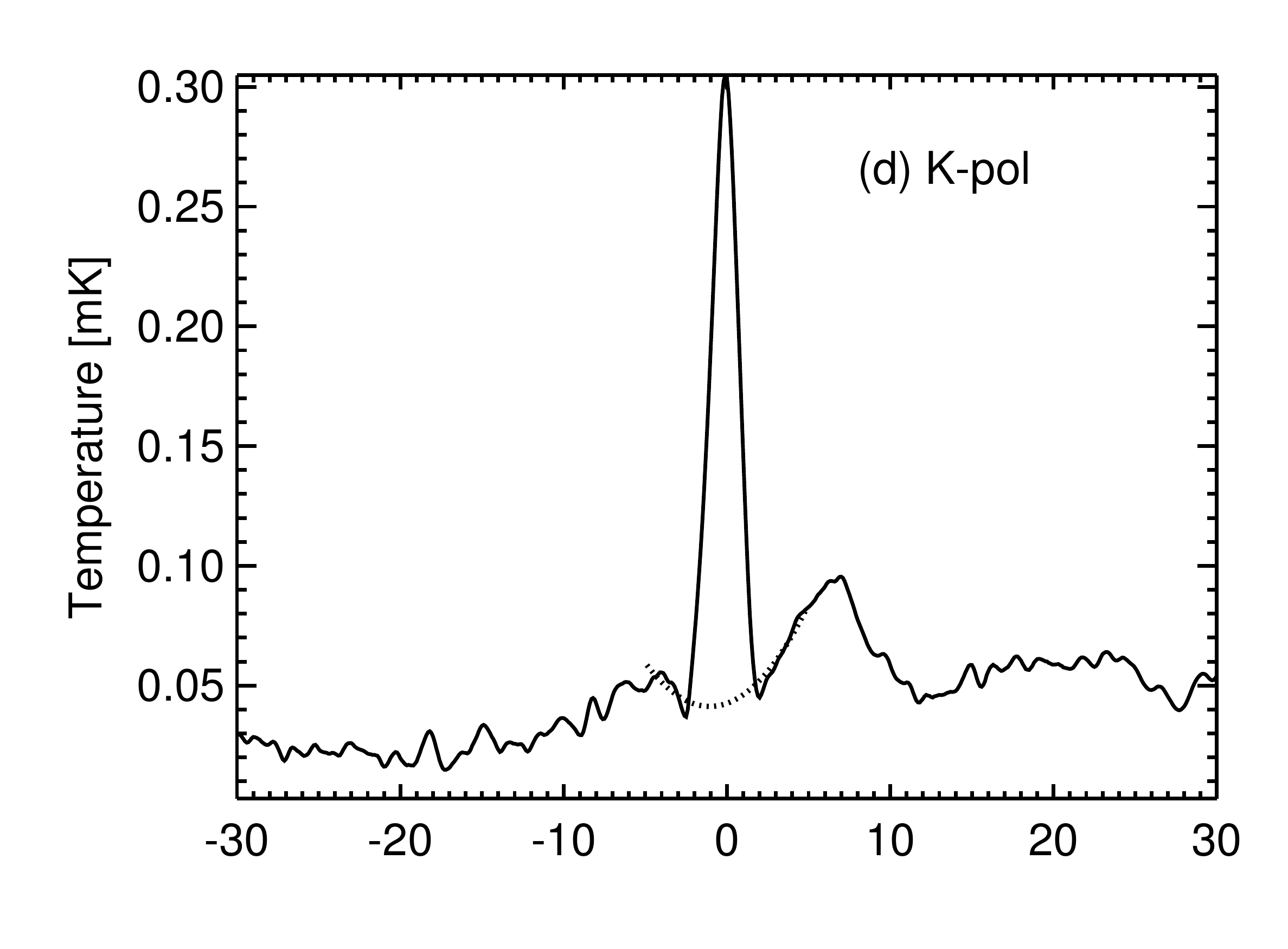}
\includegraphics[angle=90,scale=0.27]{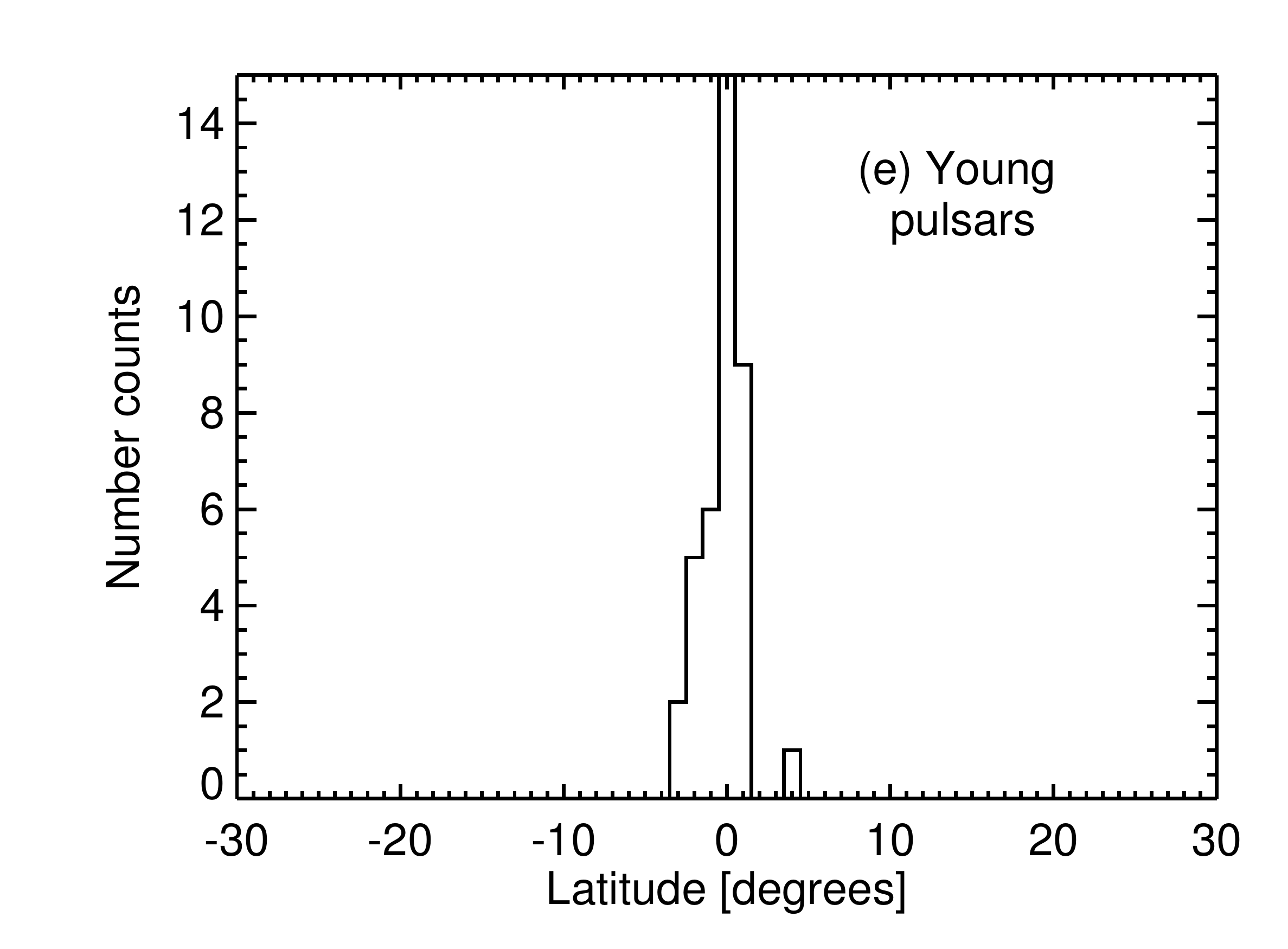}
\includegraphics[angle=90,scale=0.27]{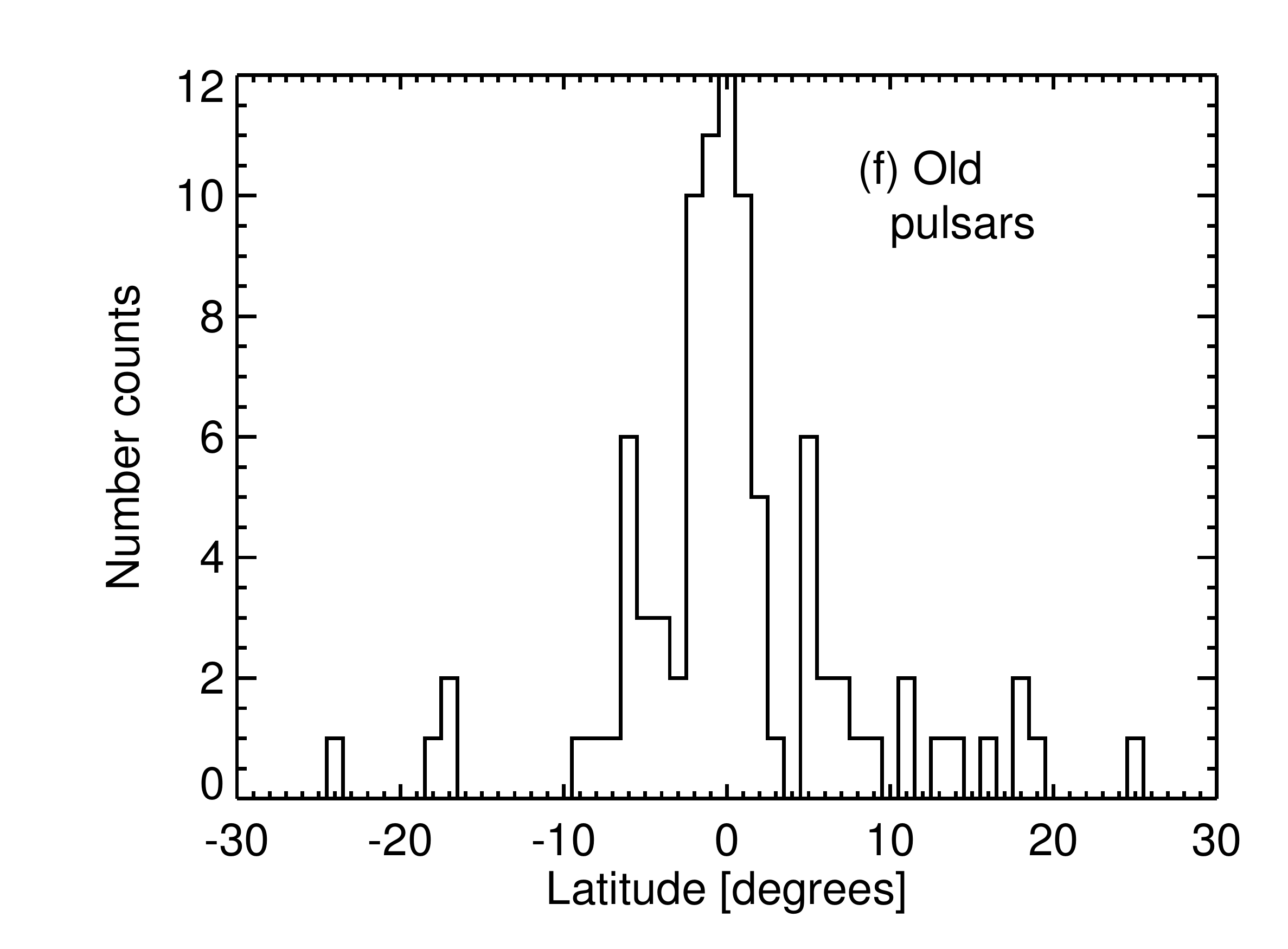}
\caption{Latitude cuts in bands relevant to synchrotron emission averaged over the longitude range 20\deg--30\deg. These are 0.408, 1.4, and 2.3\GHz, \WMAP\ K-band polarized intensity, and pulsar number counts. A reference age of $10^{6}$\,yr is chosen to define young and old pulsars. No parabolic fit is made to the separate pulsar population cuts.}
\label{fig:narrowsynch}
\end{figure*}

\begin{table*}[tb]
\begingroup
\newdimen\tblskip \tblskip=5pt
\caption{Intrinsic latitude FWHM of synchrotron and related data for four longitude ranges.}
\label{table:synch}
\nointerlineskip
\vskip -3mm
\footnotesize
\setbox\tablebox=\vbox{
 \newdimen\digitwidth 
 \setbox0=\hbox{\rm 0} 
 \digitwidth=\wd0 
 \catcode`*=\active 
 \def*{\kern\digitwidth}
 \def\leaderfil{\leaders\hbox to 5pt{\hss.\hss}\hfil}
 \newdimen\signwidth 
 \setbox0=\hbox{+} 
 \signwidth=\wd0 
 \catcode`!=\active 
 \def!{\kern\signwidth}
 \halign{\hbox to 1.7in{#\leaderfil}\tabskip 1.5em&
 \hfil#\hfil&
 \hfil#\hfil&
 \hfil#\hfil&
 \hfil#\hfil&
 #\hfil\tabskip 0pt\cr
 \noalign{\doubleline\vskip 2pt}
 \omit& \multispan4\hfil Intrinsic width [degrees] \hfil& \cr
\noalign{\vskip -2pt}
\omit \hfil Data  \hfil& \multispan4{\hrulefill}& Source of data\cr
\omit & \omit \hfil $l=20\deg$--$30\deg$ \hfil& \omit \hfil $l=30\deg$--$40\deg$ \hfil& \omit \hfil $l=320\deg$--$330\deg$ \hfil& \omit \hfil $l=330\deg$--$340\deg$ \hfil&  \hfil \cr
\noalign{\vskip 4pt\hrule\vskip 6pt}
408\MHz& $1\pdeg63\pm0\pdeg14$& $1\pdeg63\pm0\pdeg16$&  $2\pdeg14\pm 0\pdeg08$& $1\pdeg70\pm 0\pdeg09$& \citet{Haslam:1982} \cr
1.4\GHz& $1\pdeg60\pm0\pdeg11$& $1\pdeg61\pm0\pdeg13$& $2\pdeg02\pm 0\pdeg07$& $1\pdeg57\pm 0\pdeg19$& \citet{Reich:1982} \cr
2.3\GHz& $1\pdeg47\pm0\pdeg10$& $1\pdeg45\pm0\pdeg09$& $1\pdeg97\pm 0\pdeg12$& $1\pdeg50\pm 0\pdeg12$& HartRAO courtesy of J. Jonas \cr
23\GHz\ polarization& $1\pdeg66\pm0\pdeg08$& $1\pdeg18\pm0\pdeg15$& $2\pdeg94\pm0\pdeg32$& $1\pdeg84\pm0\pdeg22$& \WMAP\ Lambda \cr
545\GHz& $1\pdeg22\pm0\pdeg10$& $1\pdeg40\pm0\pdeg14$& $1\pdeg76\pm 0\pdeg03$& $1\pdeg30\pm 0\pdeg17$& \planck\cr
100\um& $1\pdeg23\pm0\pdeg11$& $1\pdeg25\pm0\pdeg13$& $1\pdeg66\pm 0\pdeg04$& $1\pdeg23\pm 0\pdeg02$& \IRAS \cr
CO& $1\pdeg09\pm0\pdeg12$& $1\pdeg17\pm0\pdeg05$& $1\pdeg70\pm 0\pdeg03$& $1\pdeg26\pm 0\pdeg01$& \citet{planck2013-p03a}\cr
Gamma-rays& $1\pdeg18\pm0\pdeg11$& $1\pdeg37\pm0\pdeg15$& $1\pdeg75\pm 0\pdeg11$& $1\pdeg22\pm 0\pdeg17$& \Fermi \cr
Pulsars\tablefootmark{a}& $1\pdeg15\pm0\pdeg63$& $1\pdeg23\pm0\pdeg58$& $1\pdeg16\pm 0\pdeg63$& $1\pdeg23\pm 0\pdeg58$& ATNF Pulsar Catalogue\cr
\noalign{\vskip 3pt\hrule\vskip 4pt}}}
\endPlancktablewide
\tablefoot{\tablefoottext{a}{Age $<10^6$\,yr}.}
\endgroup
\end{table*} 

\subsubsection{Separation of free-free and synchrotron}
\label{sec:synch-ffsep}

There is a long-standing difficulty of separating free-free and synchrotron emission on the Galactic plane (e.g., \citealt{Broadbent:1989,Bennett:2003:WMAP1foregrounds,Paladini:2003, Sun:2011}). This problem can be solved by using RRL data. Three well-calibrated maps of synchrotron continuum emission with angular resolution better than $1\pdeg0$ are available, namely those at 0.408, 1.4, and 2.3\GHz. For the longitude range $l=20$\deg--44\deg\ these maps can be corrected for free-free emission using RRL data \citep{Alves2012}. Over this longitude range at $b=0\deg$, the mean brightness temperature of the synchrotron narrow emission (5.0\,K) is similar to that of the free-free emission (4.6\,K) at 1.4\GHz. Accordingly, the free-free emission in the narrow component is 0.31 and 1.43 times the synchrotron at 0.408 and 2.3\GHz\ respectively, assuming a synchrotron spectral index of $-3.0$. For a spectral index of $-2.7$ these ratios are 0.44 and 1.22.

Figure~\ref{fig:narrowsynch} shows the raw latitude cuts for the three frequencies, 0.408, 1.4, and 2.3\GHz, that contain significant amounts of synchrotron emission, along with cuts for related emission mechanisms. The ratio of the brightness  of the broad synchrotron disk to that of the narrow disk is 0.75, 0.57, and 0.43 at 0.408, 1.4, and 2.3\GHz\ respectively, for the longitude range $l = 20$\deg--30\deg. After correcting for the contribution from free-free, these ratios are similar (1.18, 1.12, 0.98), suggesting that the spectral indices of the narrow and the broad disk synchrotron components are similar in the range 408\MHz\ to 2.3\GHz. We might expect the latitude distribution of the synchrotron emission at $\sim$1\GHz\ and \Planck\ LFI frequencies to be similar since the diffusion length in $\sim$10\,$\mu$G magnetic fields is greater than 1\kpc\ for the 4--50\,GeV electrons responsible \citep{Mertsch2010,Mertsch2013}. We see that the polarized 23\GHz\ width of the inner Galaxy is indeed similar to the $\sim$1\GHz\ widths (Fig. \ref{fig:narrowsynch} and Table \ref{table:synch}). This same is not true of the larger-scale haze emission \citep{Mertsch2010,2012Dobler,planck2012-IX}.

We note that at $|b| > 5\deg$ the free-free emission is $<5$\,\% of the value at $b=0\deg$ and accordingly we conclude that the broad synchrotron component at $|b| = 5\deg$ is essentially uncontaminated by free-free (at levels of 1.9, 3.4, 5.2\,\% at 0.408, 1.4. and 2.3\GHz\ respectively).

Table \ref{table:synch} gives the intrinsic widths of the narrow, synchrotron-dominated, low-frequency latitude distribution for the longitude ranges $l = 20$\deg--30\deg, 30\deg--40\deg, 320\deg--330\deg, and 330\deg--340\deg. Note that in going from 0.408 to 2.3\GHz\ the free-free content becomes comparable with the synchrotron emission. This may explain the slight narrowing at 2.3\GHz. Typical synchrotron FWHM intrinsic widths are $1\pdeg7$. This value may be compared with the width of the polarized emission in the \WMAP\ 23\GHz\ K-band data \citep{Bennett2013}, which is $1\pdeg73\pm0\pdeg13$; the polarized emission is the only direct measure of a synchrotron component that is uncontaminated by free-free at \WMAP\ and \Planck\ frequencies. Both these values are significantly greater than the total matter-width as represented by the dust as given by the \planck\ 545\GHz\ channel ($1\pdeg22$) or the gas as given by the CO ($1\pdeg10$) and the CS ($0\pdeg83$) distributions.

\subsubsection{Origin of the narrow component}
\label{sec:synch-pulsars}

We turn now to a comparison of the widths of the distributions of the synchrotron emission, pulsars, and gamma-rays, which all originate in the supernovae associated with the collapse of massive OB stars with FWHMs of $1\pdeg2$--$1\pdeg3$.

(a) Pulsars. Pulsars are formed during the core collapse of the OB stars with ejection velocities having a component perpendicular to the Galactic plane of 100--200\,\kms\ \citep{Chatterjee:2009}. Figure~\ref{fig:narrowsynch} shows the broader distribution of pulsars older than $10^{6}$\,yr  relative to that of the younger pulsars in the longitude range $l=20$\deg--30\deg. The width for younger pulsars is $\sim2$\deg, which is comparable to the width of the synchrotron emission and significantly broader than that of the nascent OB stars.

(b) Gamma-rays. Interstellar gamma-rays are produced by \pion-decay following the collision of relativistic cosmic ray protons with gas (principally atomic and molecular hydrogen). The relativistic cosmic ray protons and electrons themselves are accelerated by diffusive shock acceleration at collisionless shocks driven by the SN explosions \citep{Blandford1978}.  The 1--100\GeV\ CREs produce radio synchrotron emission in the enhanced magnetic fields of the SNRs while the low energy ($\lesssim$1\GeV) electrons can produce gamma-ray bremsstrahlung. The CR protons generate gamma-rays by \pion-decay in the swept-up SN gas or in any adjacent gas. IC gamma-rays may also be produced in regions of high radiation field. Good examples of young SNRs demonstrating these emission processes include Cas A, the Crab nebula, and W49B \citep{Abdo:2010d,Abdo:2010c,Abdo:2010e}. Accordingly it is expected that the distributions of CR electrons and protons and gamma-rays will follow that of the pulsars \citep{Strong:2010}. The relativistic electrons  responsible for the Galactic synchrotron emission are initially held in the magnetic field of the expanding SN shell and will escape into the ISM as its magnetic field weakens \citep{Strong:2011,Mertsch2013}.  The escaping CR protons generated in the continuing star-formation and subsequent collapse on the Galactic plane are expected to interact with the atomic and molecular hydrogen on the Galactic disk to produce a gamma-ray distribution similar to that of the gas. Table \ref{table:synch} indicates that the gamma-ray FWHM ($\sim$1\pdeg3) is indeed consistent with the higher \Planck\ frequencies, 100\um, and CO with a FWHM of 1\pdeg2--1\pdeg3. 

We now consider the possible reasons for the greater width of the synchrotron emission distribution as compared with that of the nascent OB stars. After the birth of an SNR the high energy CREs will escape, leaving the low-energy (2--100\GeV) CREs trapped by magnetic fields in the SNR emitting with a brightness temperature spectral index of $\beta_{\rm synch}\approx-2.5$; this is a mean value at 1\GHz\ although there is a spread of $\pm0.2$ \citep{Green:2009,Delahaye:2010}. Those CREs that escape will have radiative energy losses and steepen to $\beta_{\rm synch}\approx-3.0$ at tens of gigahertz in the \Planck\ frequency range.  These eventually escape into the Galactic halo at intermediate latitudes with a thickness estimated at 1 to 10\kpc\ \citep{Strong2011b,Strong:2011} and have this spectral index of $-3.0$ \citep{Davies:2006,gold2010,Jaffe:2011,Ghosh:2011}.  Assuming no reacceleration, the spectral indices will also apply to the narrow Galactic plane synchrotron component \citep{Strong:2011}. We propose that the narrow synchrotron layer is comprised of the SNRs of all ages up to the time the remnants merge into the general ISM at intermediate latitudes at a height of several kiloparsecs. On the timescale of the oldest SNRs (50\,000\,yr, \citealp{Milne:1971}) the diameters reach $\sim$$30$\pc. The observed synchrotron broadening ($1\pdeg7$) corresponds to an effective FWHM of 170\pc\ at a mean distance of 6\kpc. This extra broadening relative to the OB stars (80\pc) is most likely made up of SNR expansion and proper motions of several tens of kilometres per second. Over an effective SNR lifetime of $10^5$--$10^6$\,yr, which is also the age of the pulsars, these motions could account for this broadening. On this timescale about $10^3$--$10^4$  supernovae and (O-star) progenitors would have been born.

Strong confirmation of the identification of a narrow synchrotron component on the Galactic plane is provided by the \WMAP\ observation of 23\GHz\ polarization with the same width as the narrow component of the 408\MHz\ synchrotron total intensity (Fig. \ref{fig:narrowsynch} and Table \ref{table:synch}).  The magnetic field alignment is parallel to the Galactic plane and is the ordered field that permeates the Galactic disk. The fractional polarization on the Galactic plane at \WMAP\ frequencies, where synchrotron emission is 10\,\% of the total intensity at most, can be estimated using the 408\MHz\ data (corrected for free-free) and assuming a spectral index. If a spectral index of $-3.0$ between 408\MHz\ and \WMAP\ frequencies  is assumed, as indicated by various observations \citep{Davies:2006, kogut2007, Miville-Deschenes:2008, Dunkley2009b, gold2009, gold2010,Ghosh:2011}, the percentage polarization of the narrow synchrotron component is 50\,\%. If the spectral index were $-2.8$, the polarization would be 30\,\% in this 170\,pc-wide component. In either case the percentages for the narrow Galactic plane component are evidently higher than at intermediate latitudes where it has been found to be 10--20\,\% on average in an earlier analysis of \WMAP\ data \citep{kogut2007,Macellari2011}, however it can be up to 40--50\,\% in some regions. This reflects the fact that the aligned fraction of the field is greater on the plane and is therefore more significant in the narrow latitude component we have identified.

The lower polarization (5--10\,\% averaged over $l=20$\deg--30\deg\ in Fig. \ref{fig:narrowsynch}) in the $|b|$ range 2\deg--5\deg\ is the consequence of the superposition of filaments leaving the Galactic plane having magnetic fields with a major component perpendicular to the plane, thereby cancelling the polarization parallel to the plane.

\subsection{Thermal (vibrational) dust}
\label{sec:discussion-dust}

Two types of dust grain contribute to FIR emission. Larger grains emit in the \DIRBE/{\it Planck} bands and the \IRAS\ 60 and 100\um\ bands. Smaller warm grains radiate in the 25 and 12\um\ \IRAS\ bands and are also responsible for the AME emission at gigahertz frequencies. In this section we examine the latitude distribution and spectrum of the large grains.

\subsubsection{Thermal dust -- latitude distribution}

\begin{figure*}
\centering
\includegraphics[angle=90,scale=0.27]{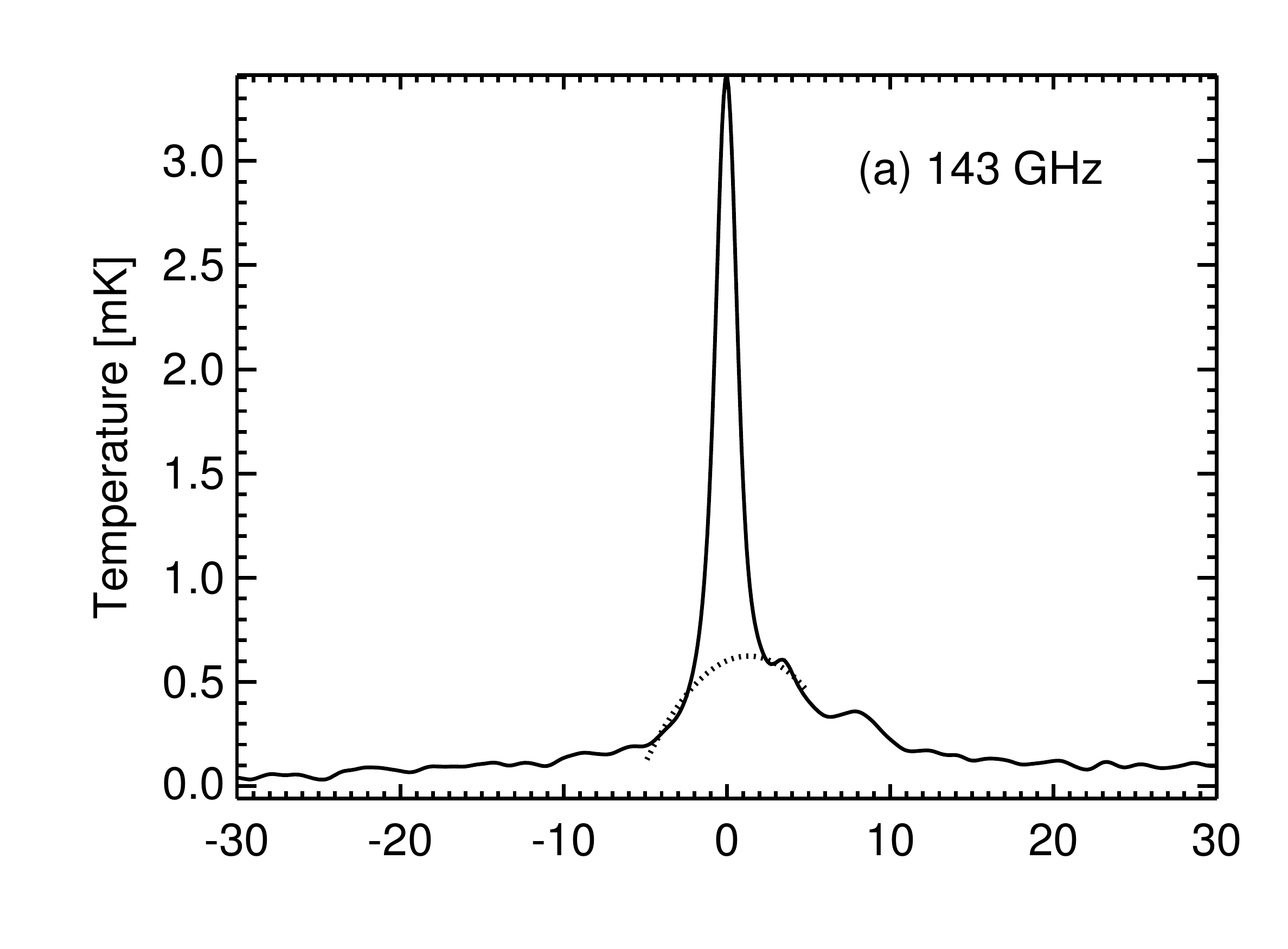}
\includegraphics[angle=90,scale=0.27]{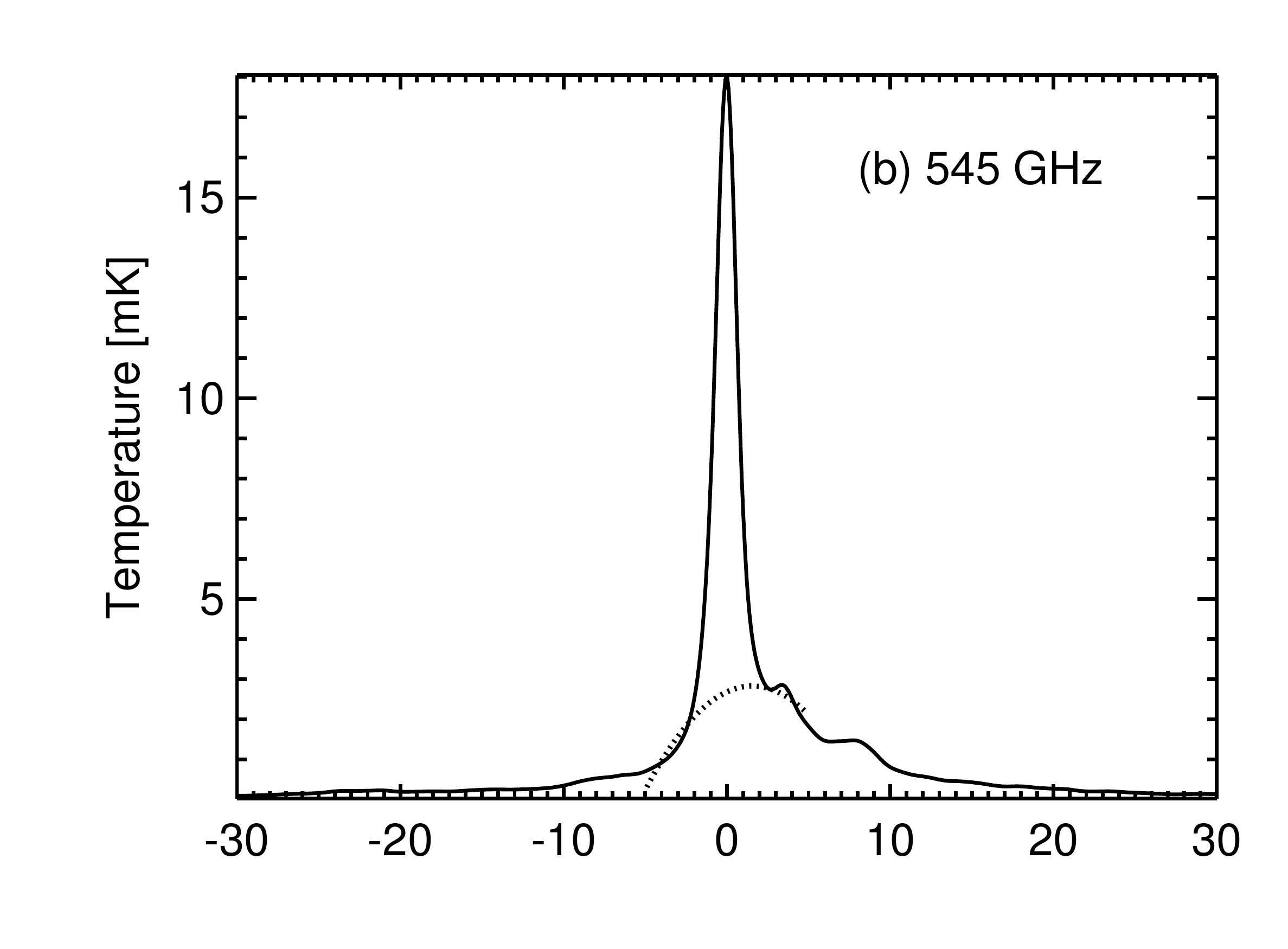}
\includegraphics[angle=90,scale=0.27]{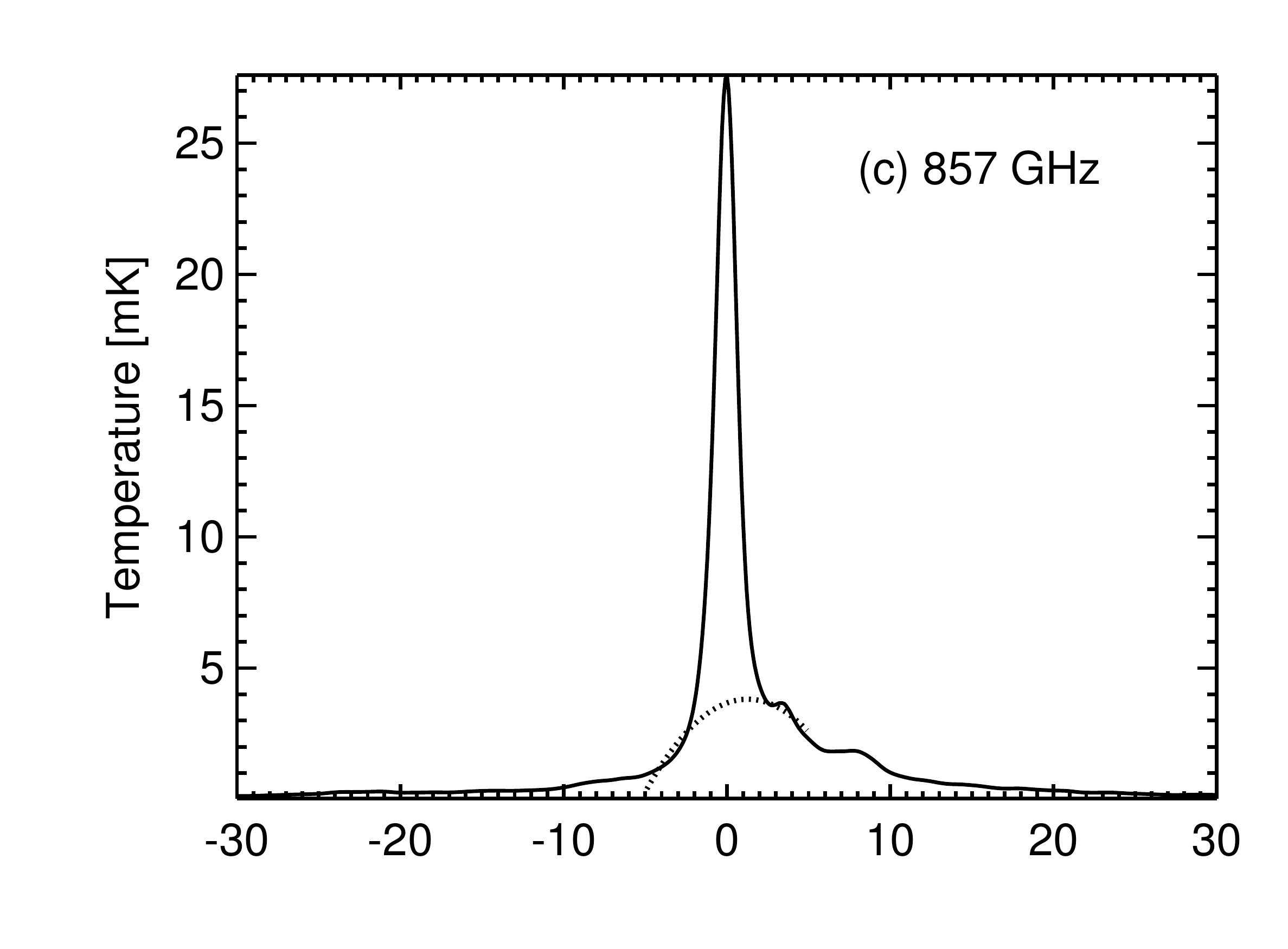}
\includegraphics[angle=90,scale=0.27]{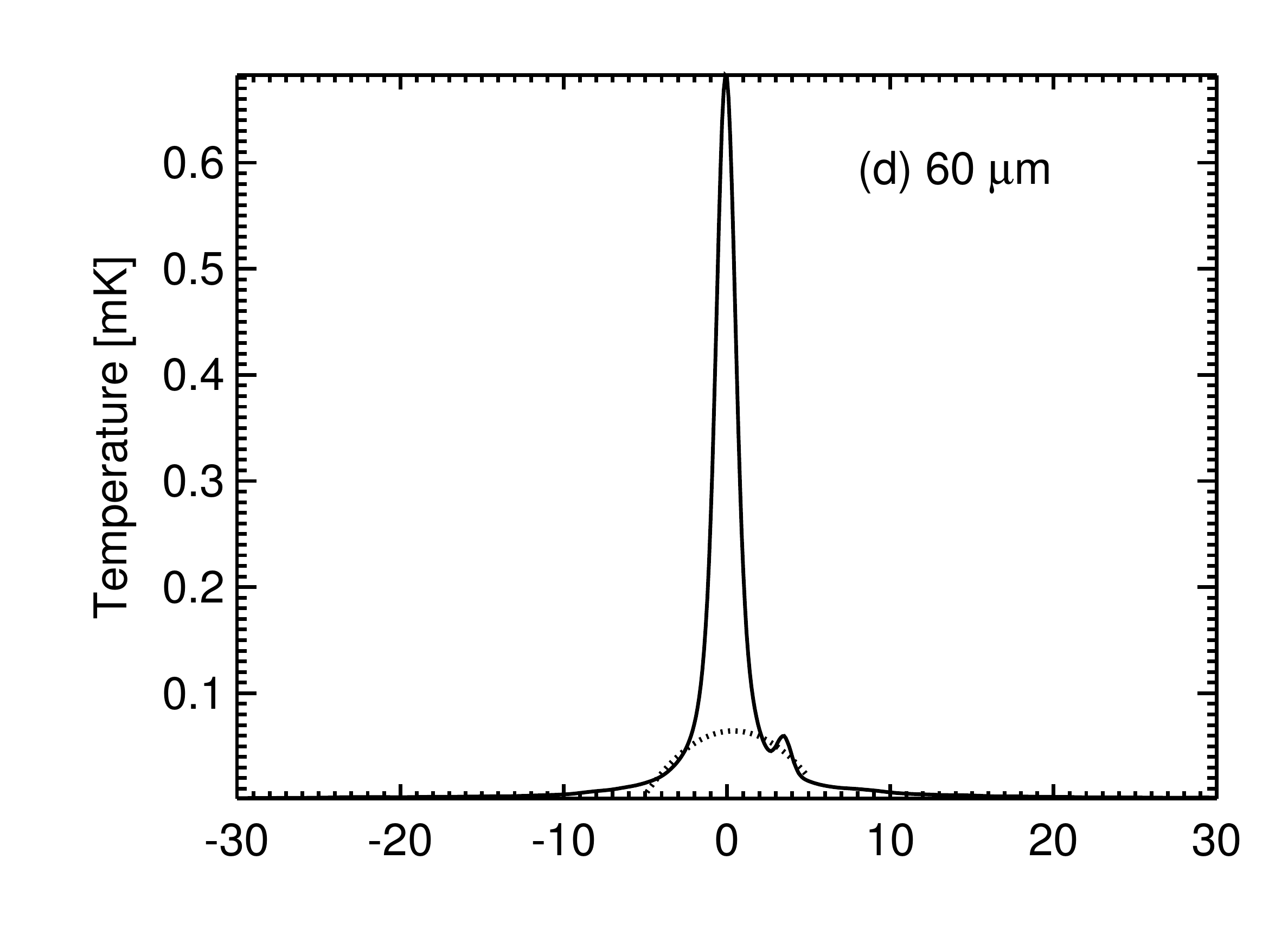}
\includegraphics[angle=90,scale=0.27]{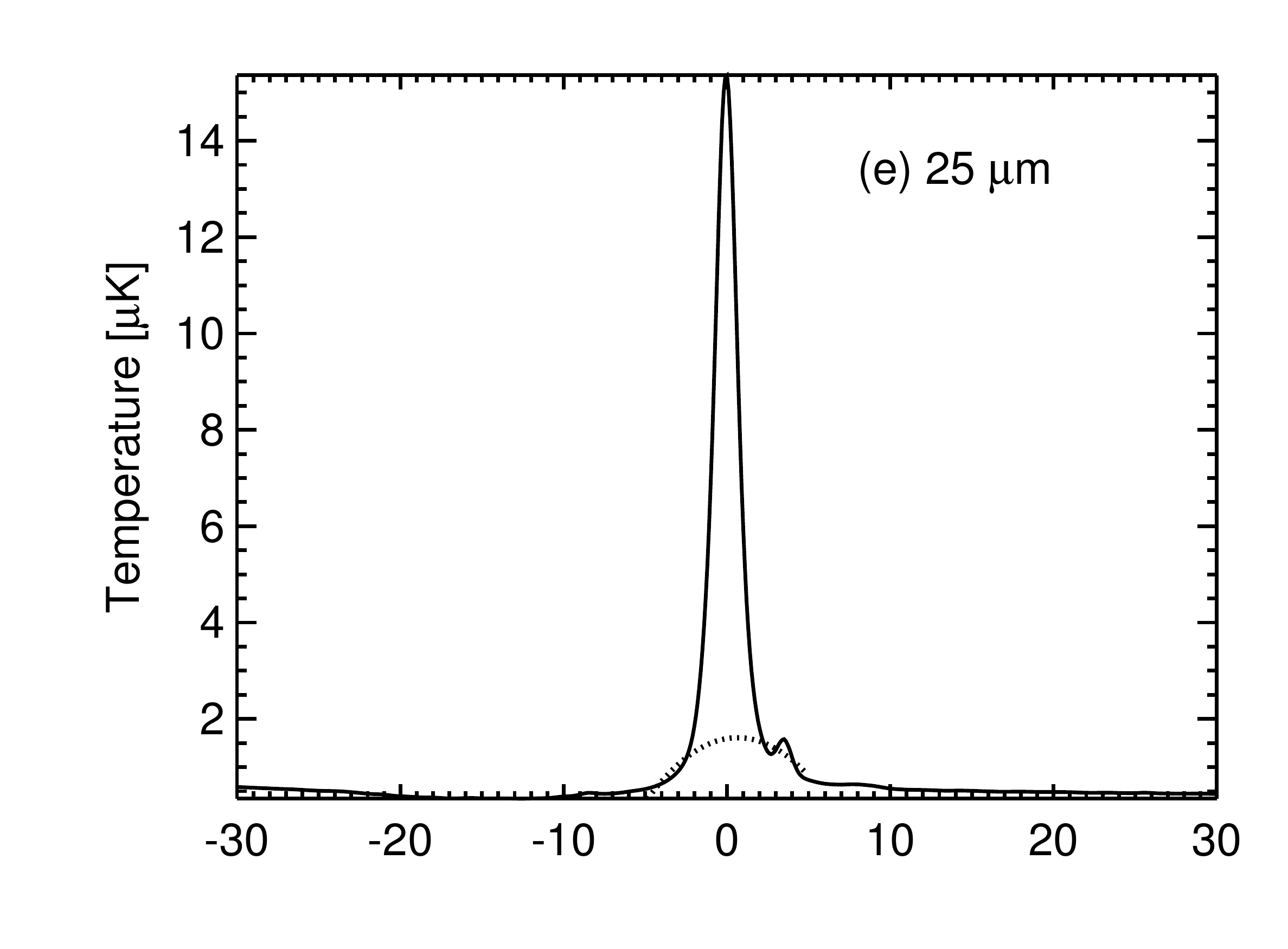}
\includegraphics[angle=90,scale=0.27]{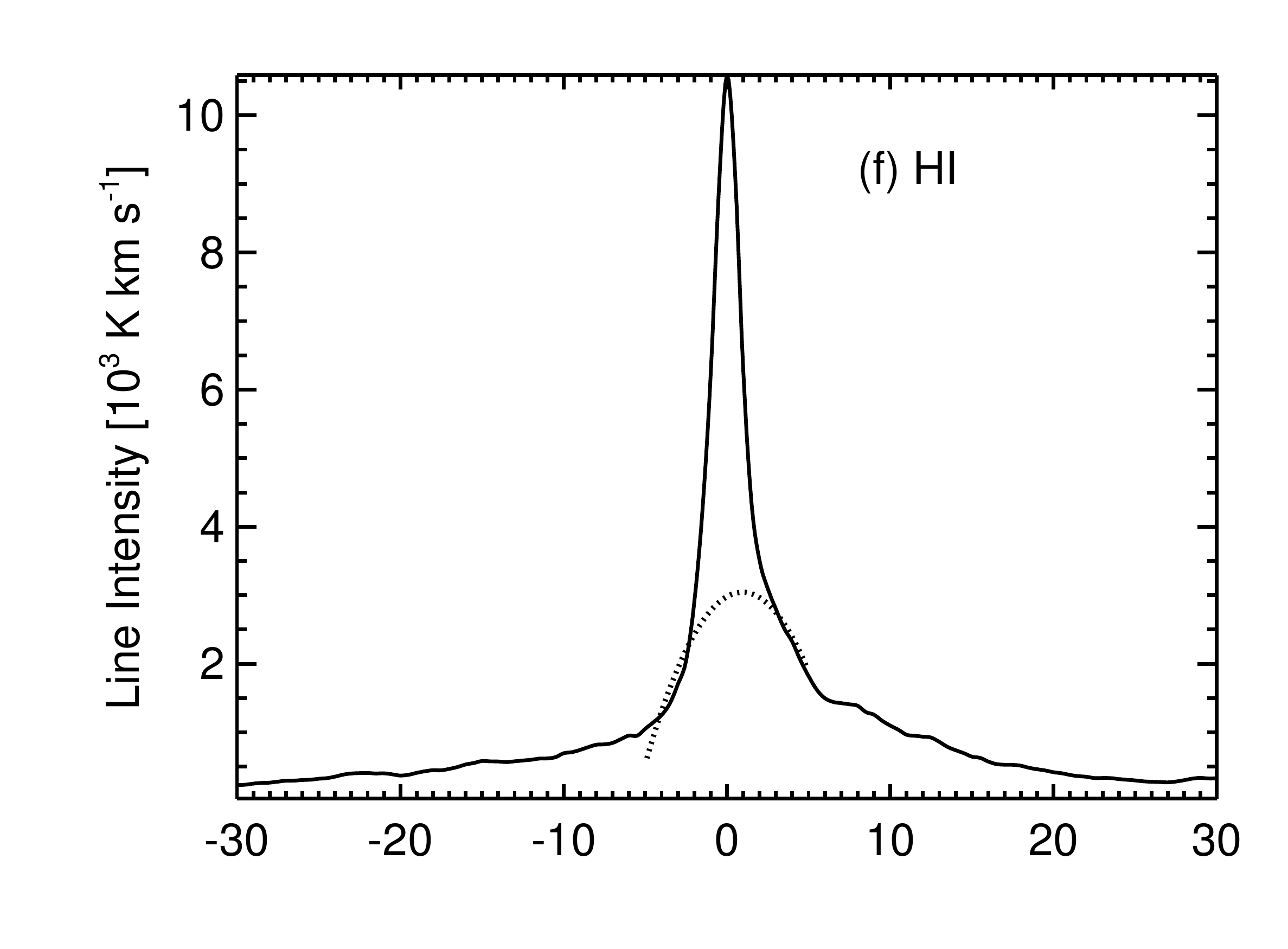}
\includegraphics[angle=90,scale=0.27]{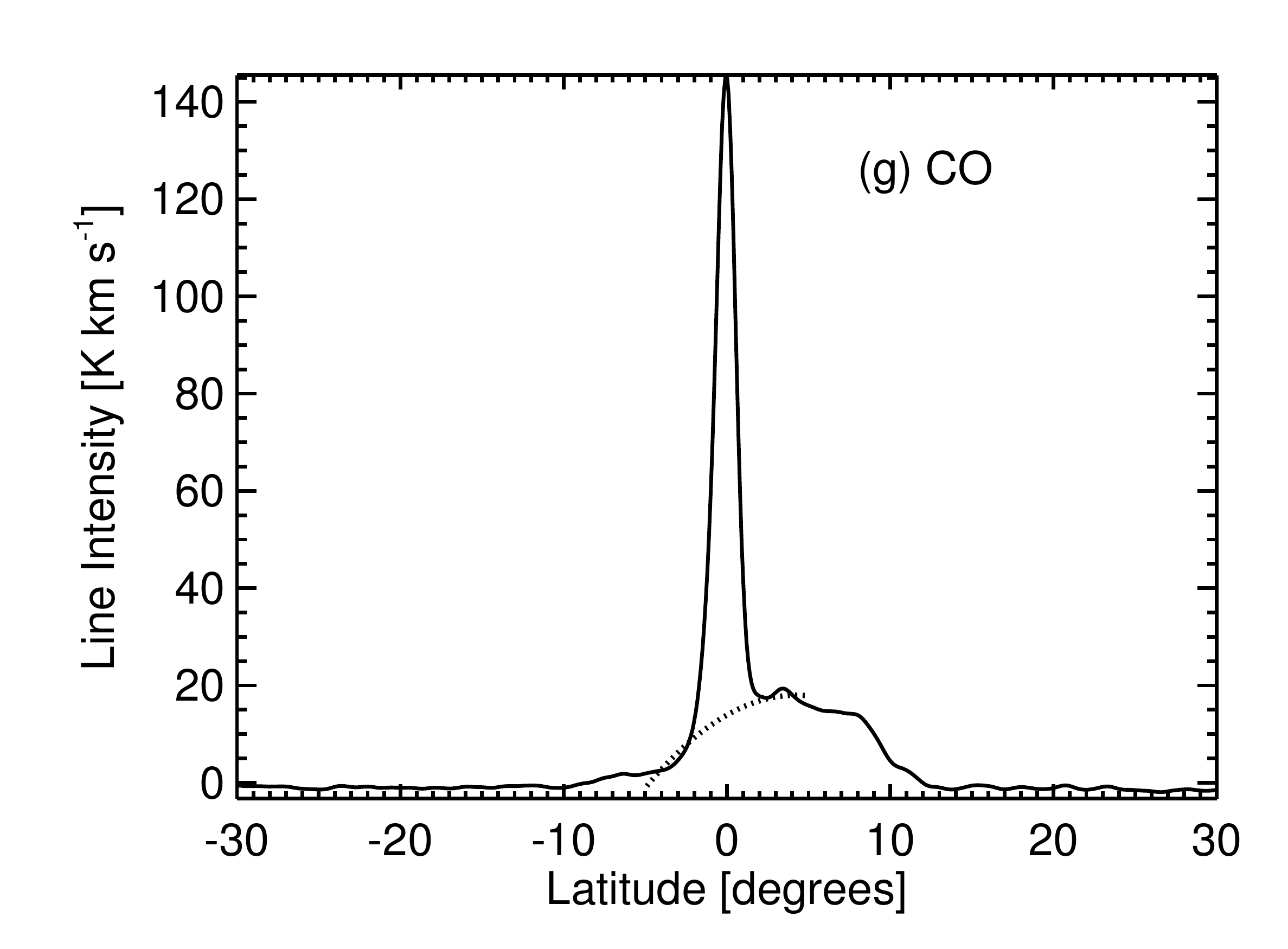}
\includegraphics[angle=90,scale=0.27]{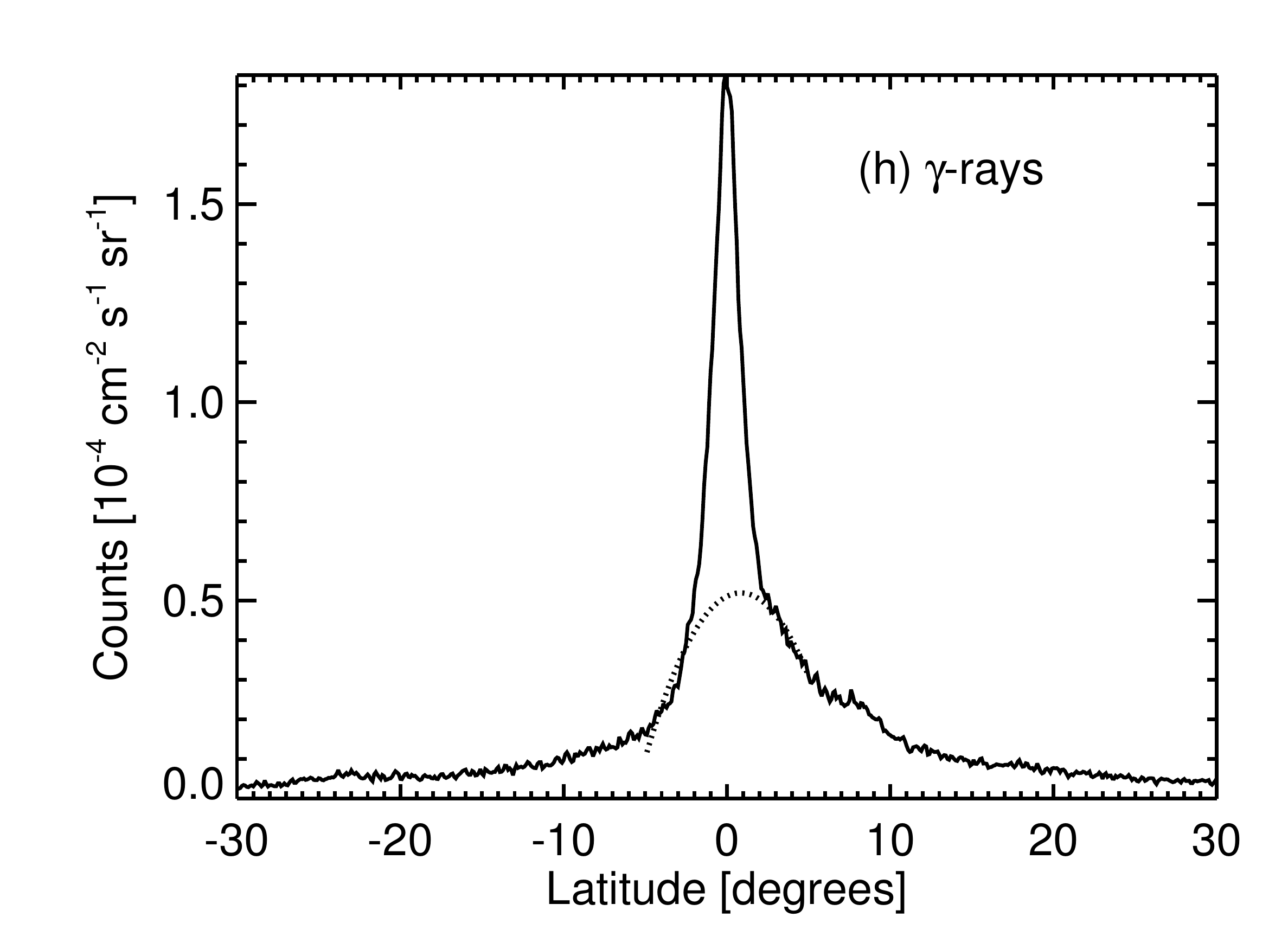}
\caption{Latitude profiles of \planck\ frequency bands and ancillary data relating principally to thermal dust, for $l = 20$\deg--30\deg.}
\label{fig:dust}
\end{figure*}
\begin{table*}[tb]
\begingroup
\newdimen\tblskip \tblskip=5pt
\caption{Intrinsic latitude FWHM of thermal dust emission and related data for  four longitude ranges.}
\label{table:dust}
\nointerlineskip
\vskip -3mm
\footnotesize
\setbox\tablebox=\vbox{
 \newdimen\digitwidth 
 \setbox0=\hbox{\rm 0} 
 \digitwidth=\wd0 
 \catcode`*=\active 
 \def*{\kern\digitwidth}
 \def\leaderfil{\leaders\hbox to 5pt{\hss.\hss}\hfil}
 \newdimen\signwidth 
 \setbox0=\hbox{+} 
 \signwidth=\wd0 
 \catcode`!=\active 
 \def!{\kern\signwidth}
 \halign{\hbox to 1.0in{#\leaderfil}\tabskip 1.0em&
 \hfil#\hfil&
 \hfil#\hfil&
 \hfil#\hfil&
 \hfil#\hfil&
 #\hfil\tabskip 0pt\cr
 \noalign{\doubleline\vskip 2pt}
 \omit& \multispan4\hfil Intrinsic width [degrees] \hfil& \cr
\noalign{\vskip -2pt}
\omit \hfil Data \hfil& \multispan4{\hrulefill}& Source of data\cr
\omit& \omit \hfil $l=20\deg$--$30\deg$ \hfil& \omit \hfil $l=30\deg$--$40\deg$ \hfil& \omit \hfil $l=320\deg$--$330\deg$ \hfil& \omit \hfil $l=330\deg$--$340\deg$ \hfil&  \hfil \cr
\noalign{\vskip 4pt\hrule\vskip 6pt}
143\GHz& $1\pdeg24\pm0\pdeg12$& $1\pdeg43\pm0\pdeg18$& $1\pdeg79\pm 0\pdeg17$& $1\pdeg31\pm 0\pdeg06$& \planck\cr
353\GHz& $1\pdeg22\pm0\pdeg10$& $1\pdeg41\pm0\pdeg16$& $1\pdeg77\pm 0\pdeg04$& $1\pdeg30\pm 0\pdeg06$& \planck\cr
545\GHz& $1\pdeg22\pm0\pdeg10$& $1\pdeg40\pm0\pdeg14$& $1\pdeg76\pm 0\pdeg03$& $1\pdeg30\pm 0\pdeg17$& \planck\cr
857\GHz& $1\pdeg23\pm0\pdeg11$& $1\pdeg36\pm0\pdeg11$& $1\pdeg75\pm 0\pdeg02$& $1\pdeg29\pm 0\pdeg17$& \planck\cr
1249\GHz& $1\pdeg21\pm0\pdeg09$& $1\pdeg30\pm0\pdeg05$& $1\pdeg71\pm 0\pdeg02$& $1\pdeg26\pm 0\pdeg13$& \DIRBE\cr
2141\GHz& $1\pdeg23\pm0\pdeg11$& $1\pdeg28\pm0\pdeg03$& $1\pdeg69\pm 0\pdeg07$& $1\pdeg26\pm 0\pdeg01$& \DIRBE\cr
100\um& $1\pdeg23\pm0\pdeg11$& $1\pdeg23\pm0\pdeg13$& $1\pdeg66\pm 0\pdeg04$& $1\pdeg23\pm 0\pdeg02$& \IRAS\cr
60\um& $1\pdeg11\pm0\pdeg01$& $1\pdeg13\pm0\pdeg02$& $1\pdeg53\pm0\pdeg03$& $1\pdeg14\pm0\pdeg02$& \IRAS\cr
25\um& $1\pdeg32\pm0\pdeg05$& $1\pdeg37\pm0\pdeg12$& $1\pdeg70\pm0\pdeg04$& $1\pdeg26\pm0\pdeg01$& \IRAS\cr
12\um& $1\pdeg44\pm0\pdeg06$& $1\pdeg44\pm0\pdeg06$& $1\pdeg83\pm 0\pdeg09$& $1\pdeg45\pm 0\pdeg05$& \IRAS\cr
Gamma-rays& $1\pdeg18\pm0\pdeg11$& $1\pdeg37\pm0\pdeg15$& $1\pdeg75\pm 0\pdeg11$& $1\pdeg22\pm 0\pdeg17$& \Fermi\cr
CO& $1\pdeg09\pm0\pdeg12$& $1\pdeg17\pm0\pdeg05$& $1\pdeg70\pm 0\pdeg03$& $1\pdeg26\pm 0\pdeg01$& \citet{planck2013-p03a}\cr
CS& $0\pdeg90\pm0\pdeg05$& $0\pdeg90\pm0\pdeg05$& $0\pdeg90\pm0\pdeg05$& $0\pdeg90\pm0\pdeg05$& \citet{Bronfman:2000}\cr
\hi& $1\pdeg86\pm0\pdeg12$& $2\pdeg03\pm0\pdeg09$& $2\pdeg19\pm0\pdeg09$& $2\pdeg06\pm0\pdeg10$& \citet{Hartmann:1997,Dickey:1990}\cr
\noalign{\vskip 3pt\hrule\vskip 4pt}}}
\endPlancktablewide
\endgroup
\end{table*} 

Figure \ref{fig:dust} shows latitude cuts at three \planck\ bands that sample vibrational dust, along with relevant ancillary data. The narrow inner Galaxy dust distribution is clear and is the major component on the plane in these bands. The weaker extended emission at positive latitudes is from the northern Gould Belt, which lies within the local spiral arm. The intrinsic latitude widths of the emission in the \planck\ bands for the four longitude ranges 20\deg--30\deg, 30\deg--40\deg, 320\deg--330\deg, 330\deg--340\deg\ are given in Table \ref{table:dust} along with widths for ancillary data of interest. The widths at the four \planck\ frequencies in the range 143--857\GHz\ are remarkably consistent at $1\pdeg23 \pm 0\pdeg02$, $1\pdeg40\pm0\pdeg03$, and $1\pdeg30\pm0\pdeg02$ for $l=20$\deg--30\deg, 30\deg--40\deg, and 330\deg--340\deg\ respectively; at $l=320$\deg--330\deg\ the widths are greater by a factor of $1.35\pm0.05$.  This agreement shows that these frequency bands are all sampling dust with the same emission properties. The widths of the FIR bands of \DIRBE\ (140\um, 240\um) and the 100\um\ band of \IRAS\ are consistent within each of the three longitude ranges with FWHM widths of $1\pdeg22\pm0\pdeg02$, $1\pdeg27\pm0\pdeg03$, and $1\pdeg25\pm0\pdeg02$, similar to those of the \Planck\ bands.  In each of the longitude ranges, on going to shorter wavelengths the width falls to a minimum of $1\pdeg13\pm0\pdeg02$ at 60\um\ and rises again to $1\pdeg44\pm0\pdeg02$ at 12\um; this is the small grain dust thought to be responsible for AME.  We consider the implication of these results further in Sect. \ref{sec:conc}.

We might expect the dust distribution at \Planck\ frequencies to follow that of the gas, as indicated by the strong correlation between the \Planck\ sky at HFI frequencies and \hi\ \citep{planck2011-7.0}.  We should therefore compare the direct dust data with other estimates of the total gas mass distribution in the Galactic plane. CO is generally considered to be a proxy for molecular hydrogen in the denser phases. The CS molecule has also been proposed as an indicator of the denser molecular clouds in which star-formation is currently occurring \citep{Bronfman:2000}.  CO is an indicator of somewhat less dense gas.  The intrinsic width of CS averaged over the inner Galaxy is $0\pdeg83$ while that of the CO in the longitude ranges 20\deg--30\deg\ and 30\deg--40\deg\ is 1\pdeg09 and 1\pdeg17 respectively, showing that denser star-forming gas has a narrower distribution than that of the dust measured in the \Planck\ HFI channels.

Another indicator of the total gas surface density is the gamma-ray brightness \citep{Grenier:2005}. The gamma-ray intrinsic width averaged over the three longitude ranges (\mbox{$l=20$\deg--30\deg}, 30\deg--40\deg, and 330\deg--340\deg) is $1\pdeg26\pm0\pdeg05$, similar to the molecular indicators. In contrast, the apparent \hi\ width is significantly greater at $1\pdeg98 \pm 0\pdeg05$ (Tables~\ref{table:dust} and ~\ref{tab:conc}). The conversion of the observed latitude profile to a true \hi\ surface density profile requires a correction for the higher optical depth on the Galactic plane, leading to a narrower FWHM. The total gas distribution (\hi\ + H$_2$) will be narrower again, approaching the gamma-ray FWHM.

\begin{table*}[tb]
\begingroup
\newdimen\tblskip \tblskip=5pt
\caption{Thermal dust properties from SEDs using MILCA for CO and fitting two $\beta_{\rm dust}$ values.}
\label{tab:dustprops}
\nointerlineskip
\vskip -3mm
\footnotesize
\setbox\tablebox=\vbox{
 \newdimen\digitwidth 
 \setbox0=\hbox{\rm 0} 
 \digitwidth=\wd0 
 \catcode`*=\active 
 \def*{\kern\digitwidth}
 \def\leaderfil{\leaders\hbox to 5pt{\hss.\hss}\hfil}
 \newdimen\signwidth 
 \setbox0=\hbox{+} 
 \signwidth=\wd0 
 \catcode`!=\active 
 \def!{\kern\signwidth}
 \halign{\hbox to 1.0in{#\leaderfil}\tabskip 1.0em&
 \hfil#\hfil&
 \hfil#\hfil&
 \hfil#\hfil&
 \hfil#\hfil&
 \hfil#\hfil\tabskip 0pt\cr
 \noalign{\doubleline\vskip 2pt}
\omit \hfil Data\hfil & \omit \hfil $l=20\deg$--$30\deg$ \hfil& \omit \hfil $l=30\deg$--$40\deg$ \hfil& \omit \hfil $l=320\deg$--$330\deg$ \hfil& \omit \hfil $l=330\deg$--$340\deg$ \hfil& \omit \hfil Average \hfil \cr
\noalign{\vskip 4pt\hrule\vskip 6pt}
$T_\mathrm{dust}$ (K)& $20.79\pm1.32$& $20.26\pm1.25$& $19.11\pm1.10$& $21.32\pm1.40$& $20.37\pm0.40$\cr
$\tau_\mathrm{353} \times 10^4$ & $*6.20\pm0.55$& $*5.51\pm0.48$& $*5.02\pm0.41$& $*6.35\pm0.58$& $*5.77\pm0.22$\cr
$\beta_\mathrm{FIR}$\tablefootmark{a}& $*1.97\pm0.14$& $*1.92\pm0.14$& $*1.96\pm0.14$& $*1.92\pm0.14$& $*1.94\pm0.03$\cr
$\beta_\mathrm{mm}$\tablefootmark{b}& $*1.71\pm0.04$& $*1.65\pm0.04$& $*1.64\pm0.04$& $*1.69\pm0.05$& $*1.67\pm0.02$\cr
Reduced $\chi^2$& 0.33& 0.30& 0.51& 0.51& 0.41\cr
\noalign{\vskip 3pt\hrule\vskip 4pt}}}
\endPlancktablewide
\tablefoot{\tablefoottext{a}{$\beta_\mathrm{FIR}$ is estimated for $\nu\ge353$\GHz.}  \tablefoottext{b}{$\beta_\mathrm{mm}$ is estimated for $\nu<353$\GHz.}}
\endgroup
\end{table*}

\subsubsection{Thermal dust spectrum via SEDs}\label{sec:dustSEDs}

The wide frequency coverage of the present data set allows a well-sampled spectral energy distribution (SED) to be determined at points along the Galactic plane.  The four emission components (synchrotron, free-free, AME, and thermal dust) all have established spectral shapes, so the observed brightness temperature $T_\mathrm{b}$ can be expressed as a sum:
\begin{equation}
T_\mathrm{b} = T_\mathrm{synch}(\nu) + T_\mathrm{ff}(\nu) + T_\mathrm{d} (\nu) + T_\mathrm{AME}(\nu) + T_\mathrm{CMB}(\nu),
\end{equation}
where
\begin{itemize}
\item $T_\mathrm{synch}$ has a brightness temperature spectral index $\beta_{\rm synch}=-2.7$ in the frequency range 0.408 to 2.3\GHz\ \citep{Platania1998,Platania2003,Peel2012}, which steepens to $-3.0$ at \WMAP\ and \Planck\ frequencies \citep{Banday:2003, Davies:2006, Kogut:2011, Ghosh:2011}.
\item $T_\mathrm{ff}$ has a spectral index of $-2.10$ to $-2.12$ over the LFI frequency range and steepens $-2.14$ at 100\GHz\ due to the Gaunt factor \citep{Draine2011Book}.  The free-free brightness can be determined directly from RRLs, which can also be used to calibrate the {\tt FastMEM} analysis. 
\item $T_\mathrm{d} = \tau_{\nu_0} (\nu/\nu_0)^{\beta_{\mathrm{dust}}} B(\nu,T_\mathrm{dust})$, where
\begin{equation}
B(\nu,T_\mathrm{dust})=2 \, h \, \frac{\nu^3}{c^2} \frac{1}{e^{h\nu/kT_\mathrm{dust}}-1}\,,
\end{equation}
 is a blackbody spectrum modified by a low frequency spectral index $\beta_\mathrm{dust}$, and $\tau_{\nu_0}$ is the dust optical depth at a reference frequency $\nu_0=353$\GHz. There is strong evidence for a break in $\beta_\mathrm{dust}$,  with a flatter value, $\beta_\mathrm{mm}$, at $\nu<353$\GHz, and a steeper value, $\beta_\mathrm{FIR}$, at $\nu \ge 353$\GHz\ \citep{Paradis2009,Gordon2010,Galliano2011,planck2013-XIV}. 
\item $T_\mathrm{AME}$ is calculated as a residual after correction for the other components but can also be fitted to models \citep{planck2011-7.2,planck2013-XII}.
\item $T_\mathrm{CMB}$ is the CMB fluctuation amplitude on scales of 1\deg\ or larger appropriate to the present study.  These signals are $<100$\,$\mu$K and can be neglected in comparison with the millikelvin Galactic plane signals.
\end{itemize}

\begin{figure}
\centering
\includegraphics[scale=0.4]{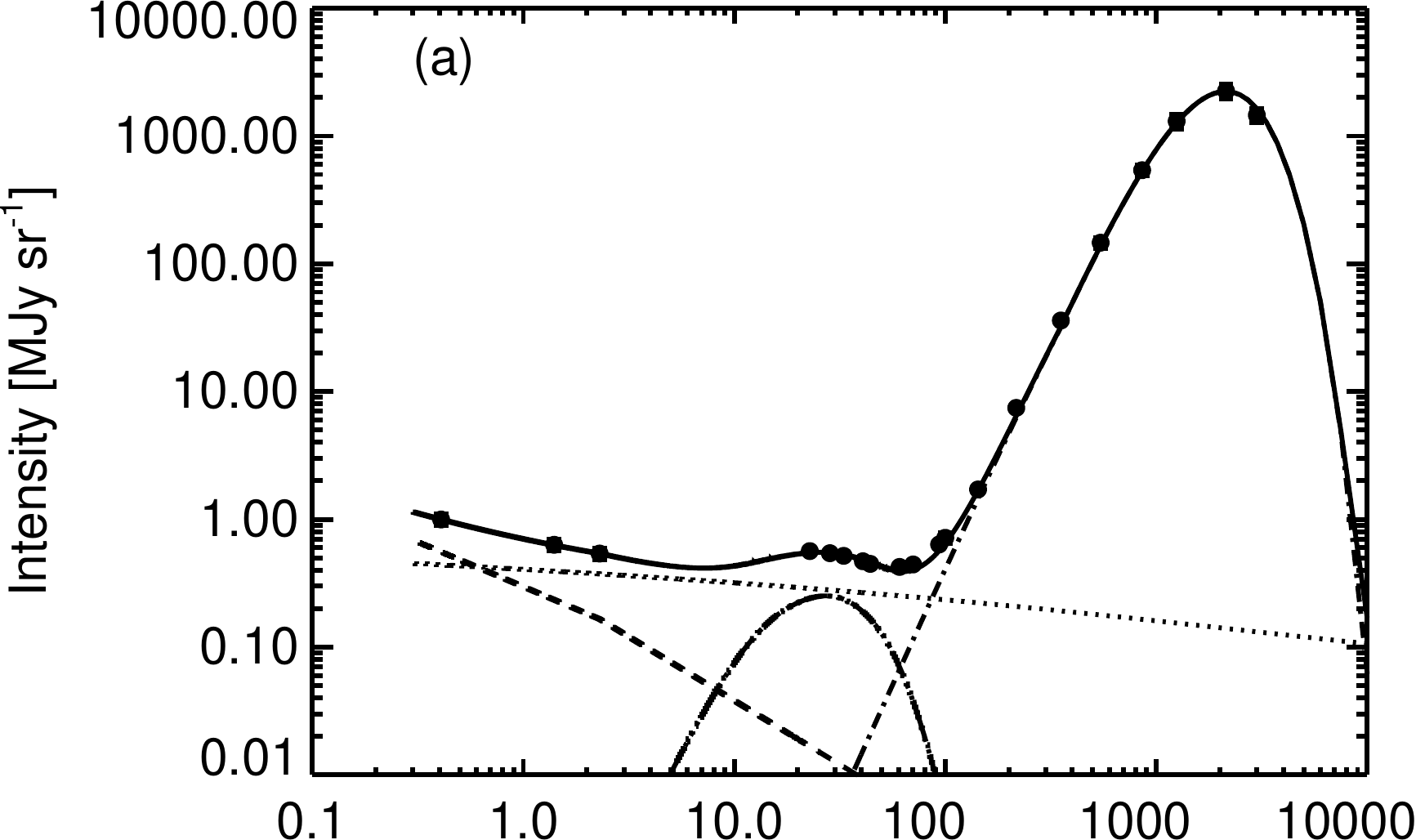}
\includegraphics[scale=0.4]{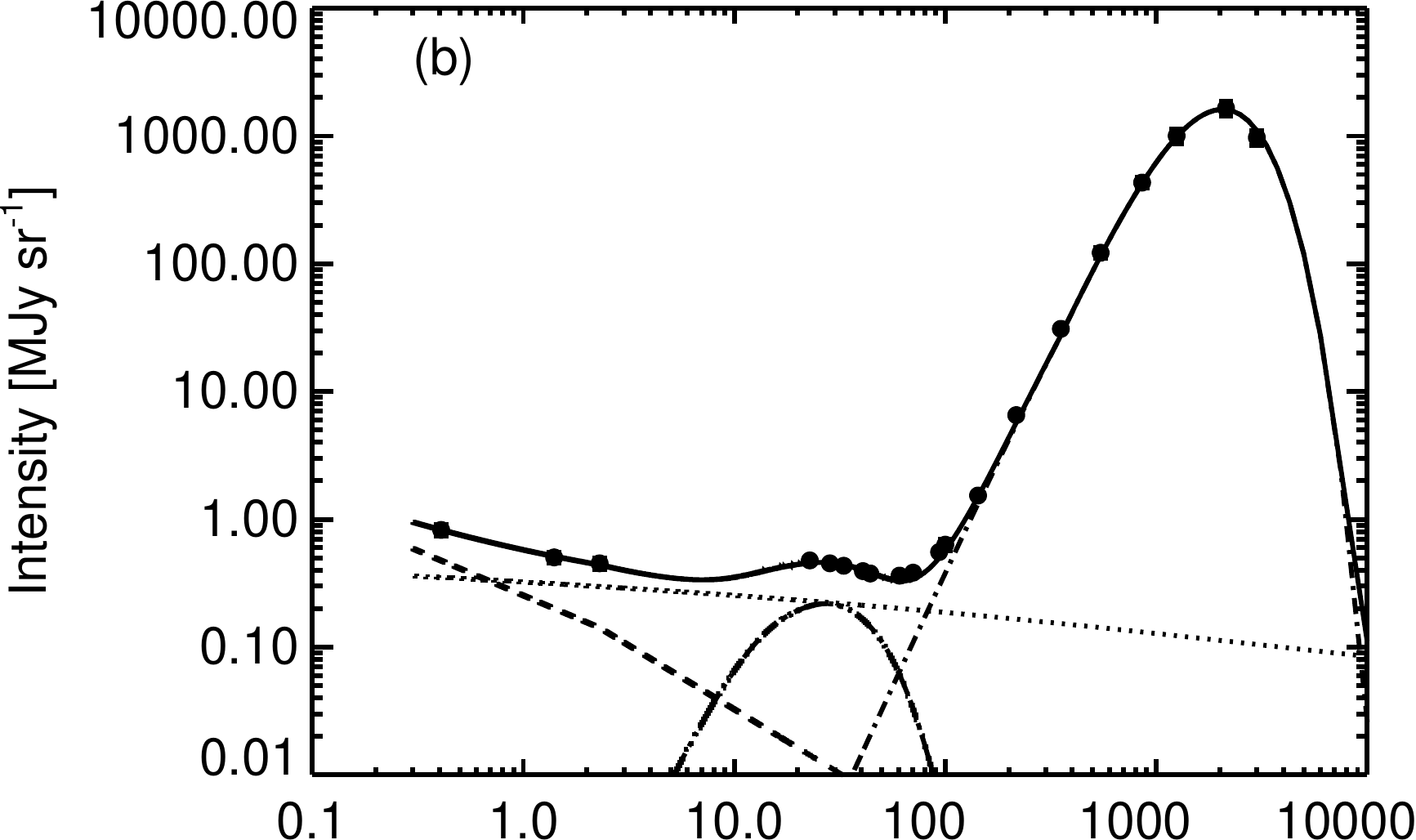}
\includegraphics[scale=0.4]{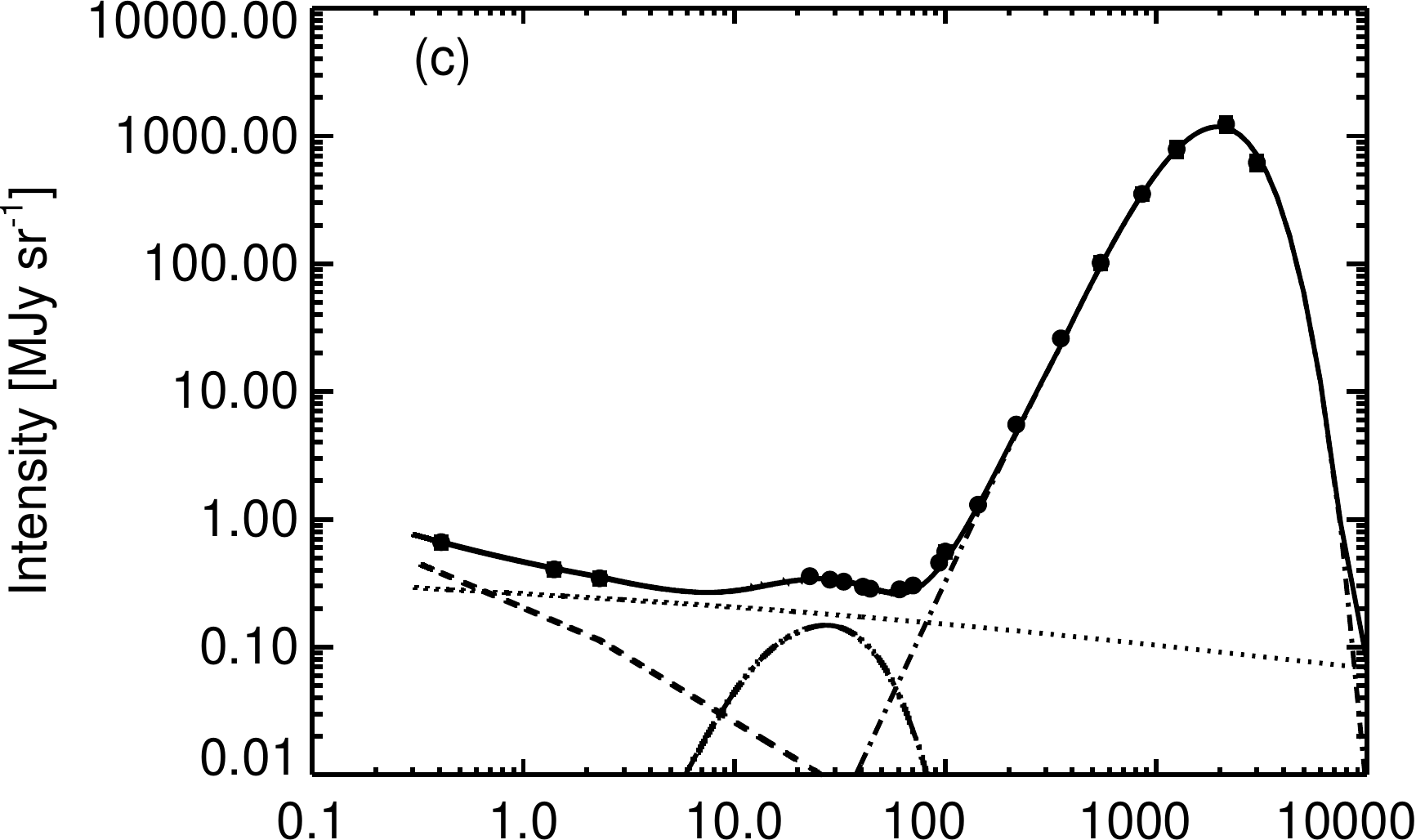}
\includegraphics[scale=0.4]{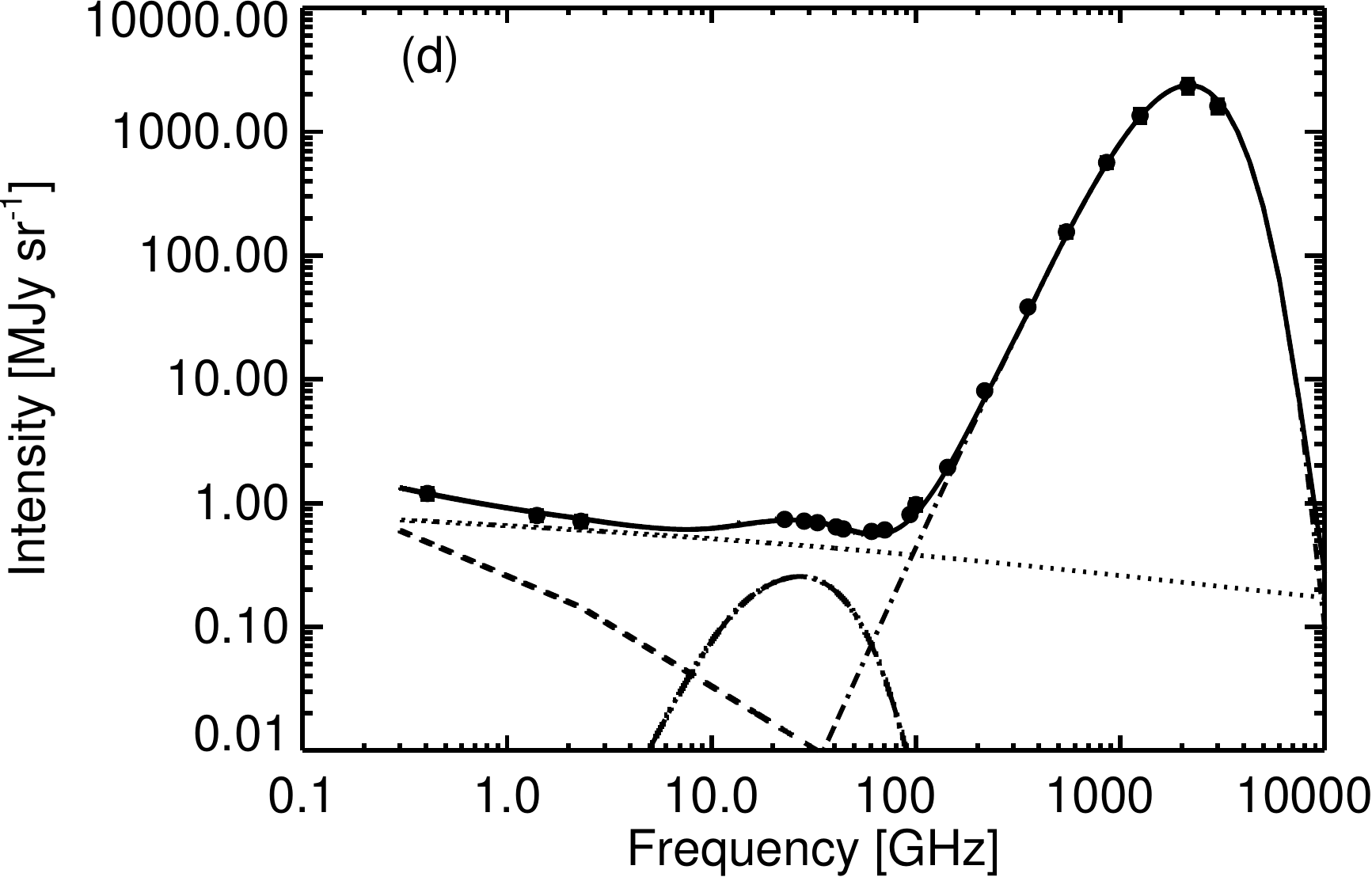}
\caption{SEDs over full frequency range 0.408\GHz\ to 3000\GHz, for the longitude ranges (a) $l=20$\deg--30\deg, (b) 30\deg--40\deg, (c) 320\deg--330\deg, and (d) 330\deg--340\deg. The points show the total emission, which is separated into free-free from the RRL and {\tt FastMEM} data (\textit{dotted line}), synchrotron from the free-free-corrected 408\MHz\ data  (\textit{dashed line}) and thermal dust (\textit{dot-dashed line}). The total of the four components is given by the full line. AME is the residual (\textit{long-dash-dotted line}).  The conversion factor from intensity to brightness temperature at 28.4\GHz\ is 40\,mK\,MJy$^{-1}$\,sr.}
\label{fig:amesed}
\end{figure}
\begin{figure}
\centering
\includegraphics[scale=0.4]{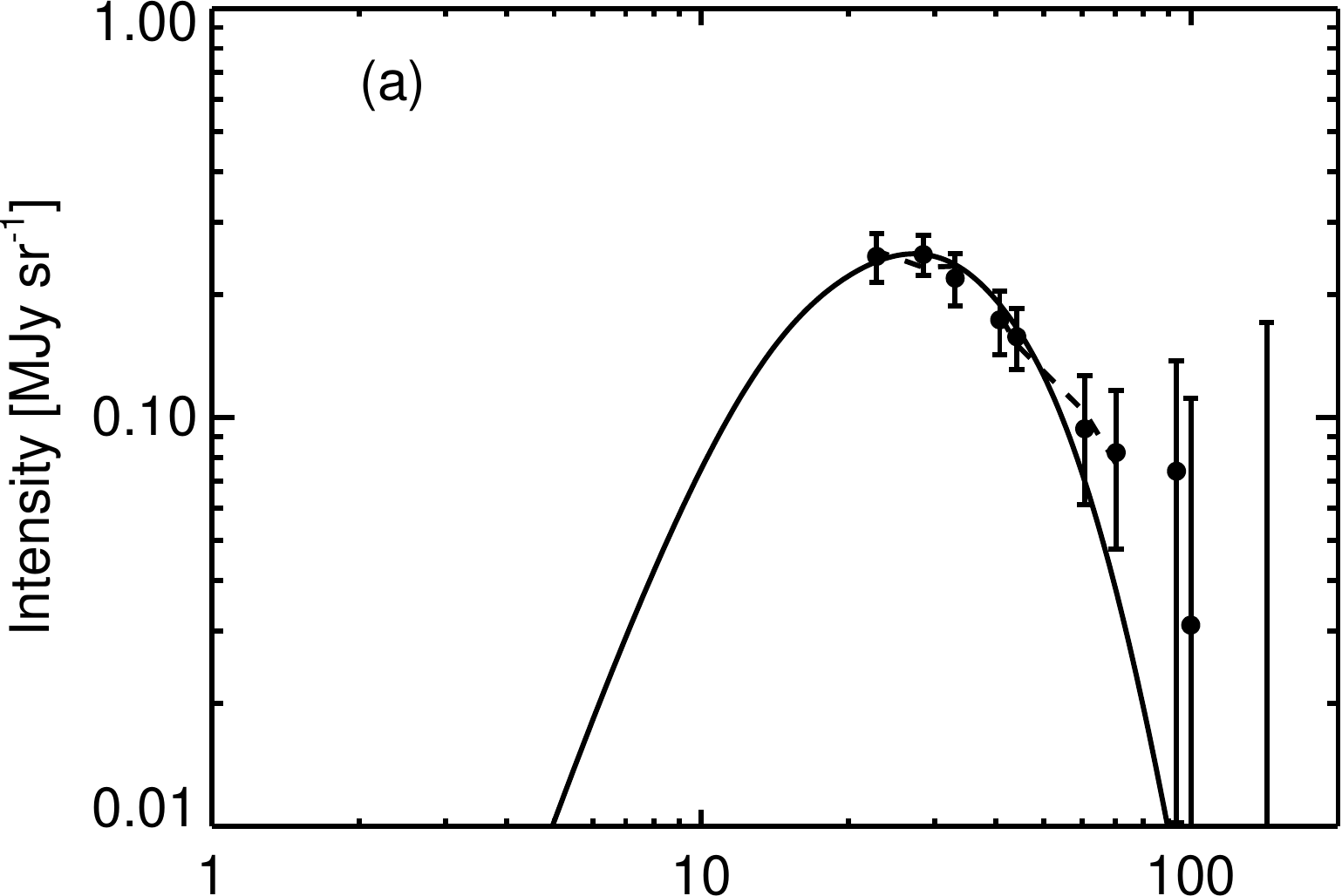}
\includegraphics[scale=0.4]{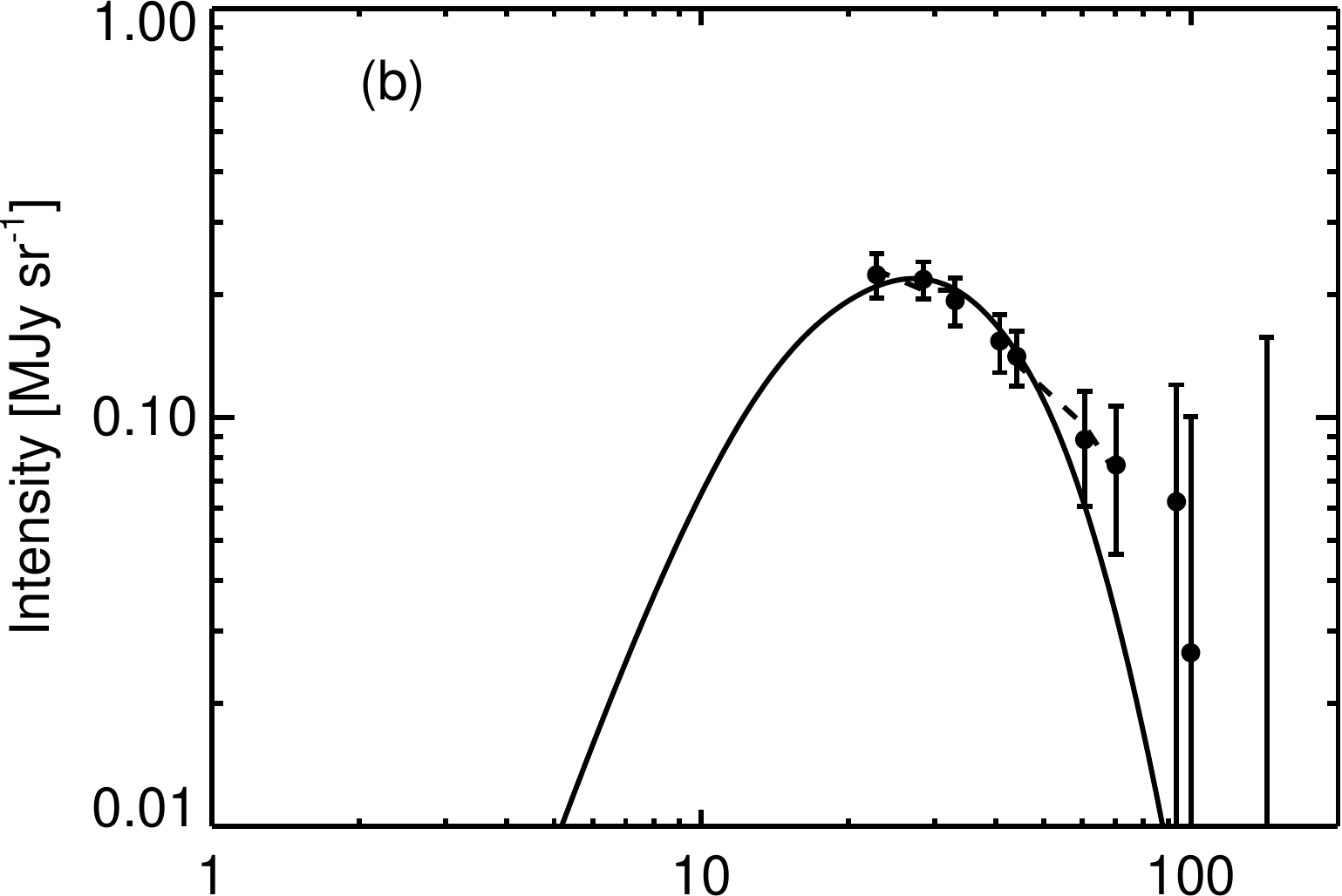}
\includegraphics[scale=0.4]{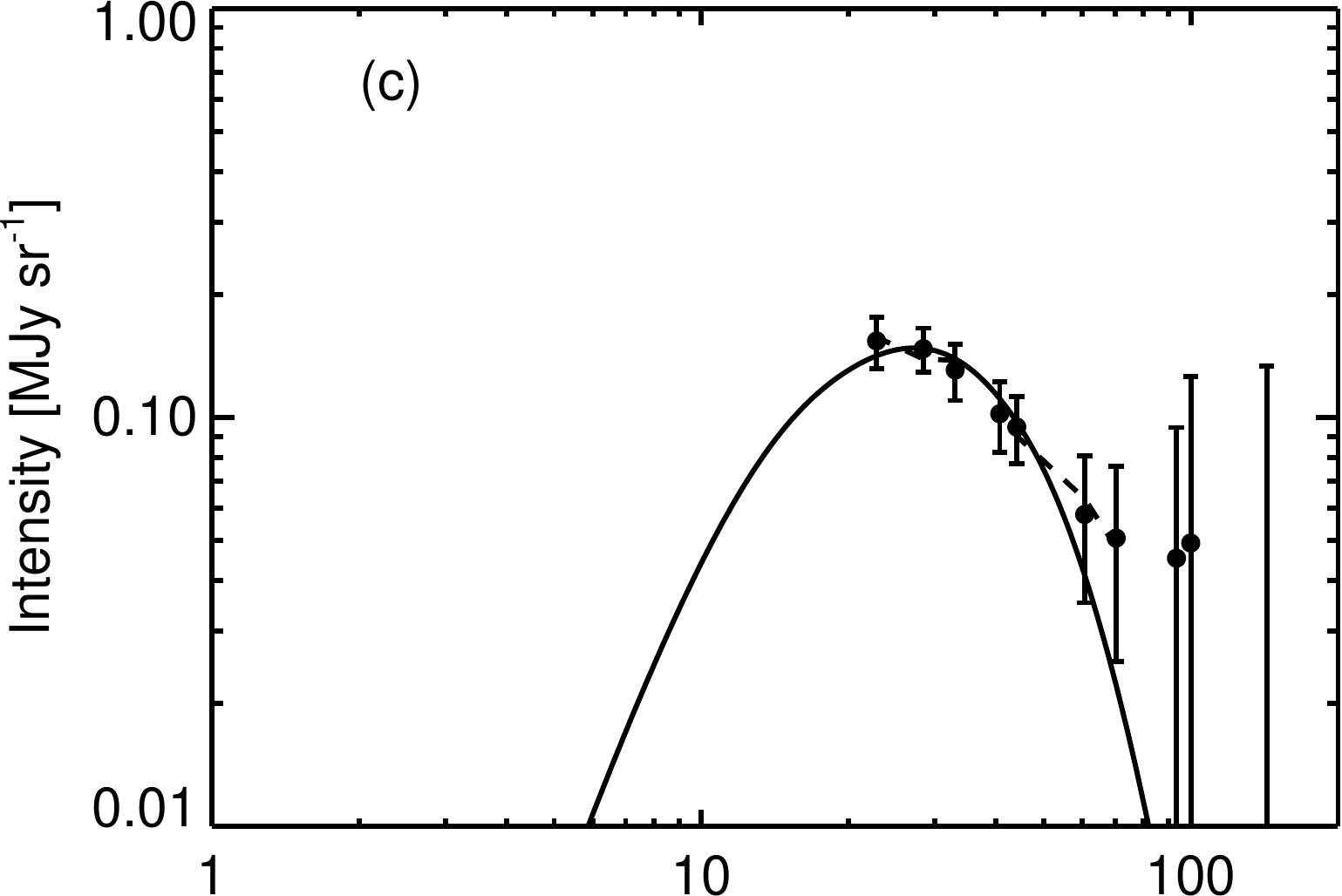}
\includegraphics[scale=0.4]{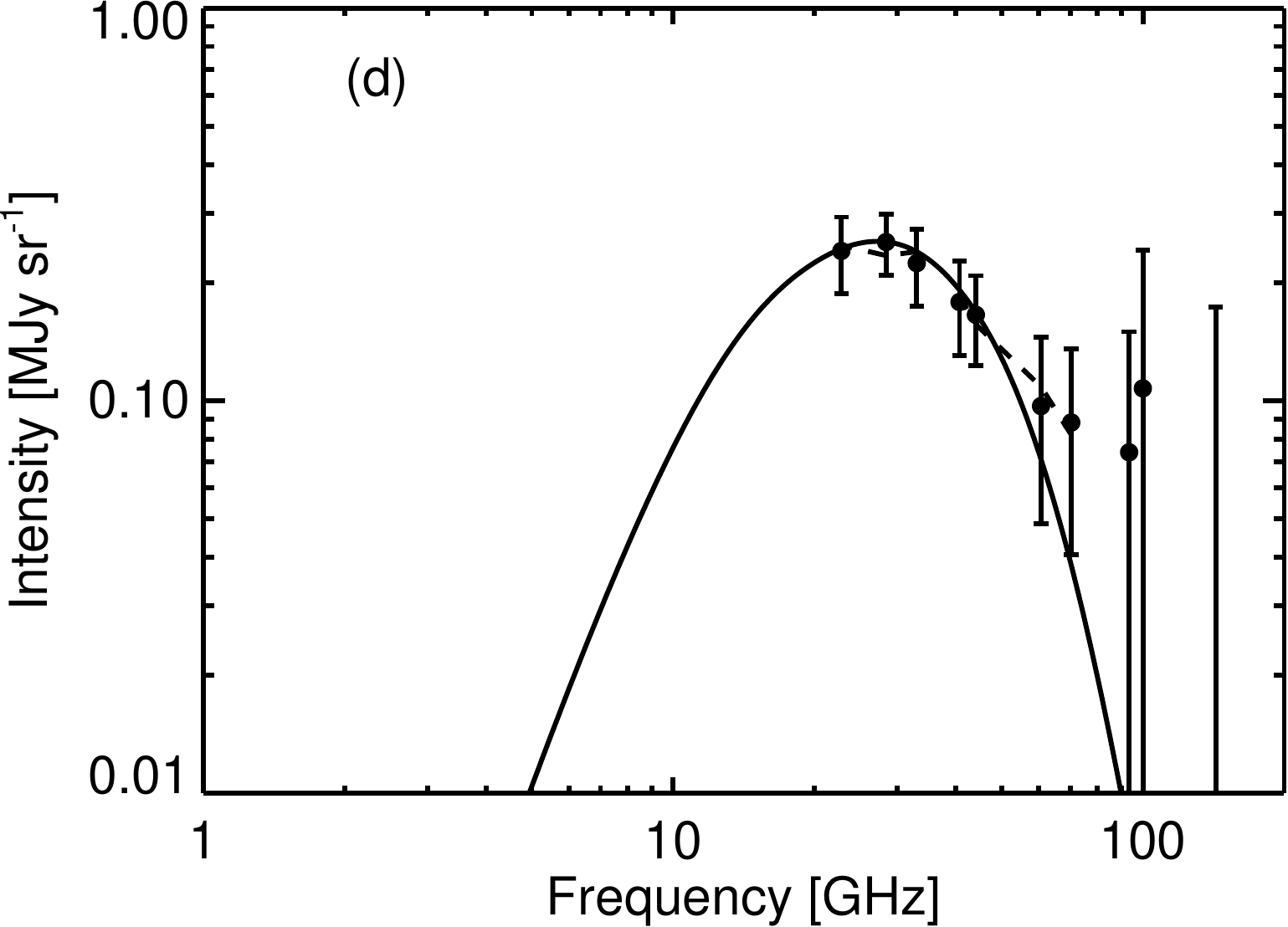}
\caption{AME spectra for the longitude ranges (a) $l = 20$\deg--30\deg, (b) 30\deg--40\deg, (c) 320\deg--330\deg, and (d) 330\deg--340\deg. 
 The spectra cover the frequency range 23--143\GHz\ and are in units of intensity. The conversion factor from intensity to brightness temperature at 28.4\GHz\ is 40\,mK\,MJy$^{-1}$\,sr.}
\label{fig:amesed2}
\end{figure}

The dust parameters $\beta_{\rm FIR}$, $\beta_{\rm mm}$, $\tau_{\nu_0}$ and $T_{\rm dust}$ are determined from the fit to the {\it Planck} HFI data supplemented with the {\it COBE}-DIRBE data, where the spectral indices $\beta_{\rm FIR}$ and $\beta_{\rm mm}$ are fitted to the data above and below 353\,GHz, respectively. The CO contribution at 100, 217 and 353\,GHz is removed using the MILCA maps (Sect.~\ref{sec:co}). We perform a least-squares fit using the IDL {\tt MPFIT} routine \citep{Markwardt2009}, taking into account the noise (both statistical and systematic) and colour corrections to the HFI and DIRBE data. 

Table \ref{tab:dustprops} shows that the values for $T_\mathrm{dust}$, $\beta_\mathrm{FIR}$, and $\beta_\mathrm{mm}$ are similar in each of the four sections of the inner Galactic plane. The low values of reduced $\chi^2$/dof with dof$=\!5$ indicate that the errors assigned to the data are overestimates. The errors given for the average values in Table~\ref{tab:dustprops} are more realistic estimates of the scatter between the four regions. Our average values for the regions $l=320$\deg--340\deg\ and $l=20$\deg--40\deg\ using a FWHM of 1\deg\ on the Galactic plane are $T_\mathrm{dust}=$ $20.4\pm0.4$\,K, $\beta_\mathrm{FIR}=1.94\pm0.03$, and $\beta_\mathrm{mm}=1.67\pm0.02$ (Table \ref{tab:dustprops}). These $\beta$ values are in agreement with a similar analysis \citep{planck2013-XIV} of the region $l=20$\deg--44\deg\ at $|b|<1$\deg, which found median values of $\beta_\mathrm{FIR}=1.88$ and $\beta_\mathrm{mm}=1.60$ with standard deviations  of $0.08$ and $0.06$ respectively, which are the spread of values across the chosen region.

Figure \ref{fig:amesed} shows the SEDs for Galactic plane emission averaged over the four 10\deg\ intervals $l=20$\deg--30\deg, $l=30$\deg--40\deg, $l=320$\deg--330\deg, and  $l=330$\deg--340\deg\ at $b=0$\deg. The synchrotron spectrum is a fit to the lower frequencies ($0.408$--$2.3$\,GHz) with $\beta_{\rm synch}=-2.7$ up to 2.3\GHz\ and $\beta_{\rm synch}=-3.0$ at higher frequencies. The free-free emission is based on the RRL data supplemented by the FastMEM analysis using a spectral index of $-2.10$ to $-2.12$ over the LFI frequency range \citep{Draine2011Book}. The two-component dust emission is derived from a least-squares fit to the HFI data as described above. The AME emission is the remainder and is shown as a fit to a equal combination of warm ionized medium (WIM) and molecular cloud (MC) models of AME emission \citep{Ali-Haimoud:2009}.

\subsection{AME estimated from SEDs}
\label{sec:discussion-ame}

\begin{table*}[tb]
\begingroup
\newdimen\tblskip \tblskip=5pt
\caption{AME fractions in the four longitude ranges shown in Figs. \ref{fig:amesed} and \ref{fig:amesed2}. The spectral indices at 28.4 and 70\GHz\ are included.}
\label{tab:amefraction}
\nointerlineskip
\vskip -3mm
\footnotesize
\setbox\tablebox=\vbox{
 \newdimen\digitwidth 
 \setbox0=\hbox{\rm 0} 
 \digitwidth=\wd0 
 \catcode`*=\active 
 \def*{\kern\digitwidth}
 \def\leaderfil{\leaders\hbox to 5pt{\hss.\hss}\hfil}
 \newdimen\signwidth 
 \setbox0=\hbox{+} 
 \signwidth=\wd0 
 \catcode`!=\active 
 \def!{\kern\signwidth}
 \halign{\hbox to 2.0in{#\leaderfil}\tabskip 2.0em&
 \hfil#\hfil&
 \hfil#\hfil&
 \hfil#\hfil&
 \hfil#\hfil&
 \hfil#\hfil\tabskip 0pt\cr
 \noalign{\doubleline\vskip 2pt}
\omit \hfil Parameter\hfil& \omit \hfil $l=20\deg$--$30\deg$ \hfil& \omit \hfil $l=30\deg$--$40\deg$ \hfil& \omit \hfil $l=320\deg$--$330\deg$ \hfil& \omit \hfil $l=330\deg$--$340\deg$ \hfil& \omit \hfil Average \hfil \cr
\noalign{\vskip 4pt\hrule\vskip 6pt}
AME / total at 28.4\GHz& $0.45\pm0.05$& $0.48\pm0.05$& $0.44\pm0.05$& $0.36\pm0.06$& $0.44\pm0.03$\cr
Significance of the above& $8.9\sigma$& $9.7\sigma$& $8.2\sigma$& $5.6\sigma$& \dots\cr
AME / free-free at 28.4\GHz& $0.9\pm0.1$& $1.0\pm0.1$& $0.8\pm0.1$& $0.6\pm0.1$& \dots\cr
AME spectral index at 28.4\GHz& $-2.2\pm0.5$& $-2.3\pm0.4$& $-2.3\pm0.5$& $-2.1\pm0.8$& $-2.2\pm0.3$\cr
AME spectral index at 70\GHz& $-3.4\pm0.6$& $-3.3\pm0.6$& $-3.3\pm0.7$& $-3.3\pm0.9$& $-3.3\pm0.4$\cr
\noalign{\vskip 3pt\hrule\vskip 4pt}}}
\endPlancktablewide
\endgroup
\end{table*}

\begin{table*}[tb]
\begingroup
\newdimen\tblskip \tblskip=5pt
\caption{Comparison of AME with free-free and 100\um\ emission in the innermost ($l=5$\deg--30\deg\ and 330\deg--355\deg) and the outer ($l=30$\deg--60\deg\ and 300\deg--330\deg) regions of the inner Galaxy as shown in Fig. \ref{fig:lonranges}.}
\label{tab:amecomparison}
\nointerlineskip
\vskip -3mm
\footnotesize
\setbox\tablebox=\vbox{
 \newdimen\digitwidth 
 \setbox0=\hbox{\rm 0} 
 \digitwidth=\wd0 
 \catcode`*=\active 
 \def*{\kern\digitwidth}
 \def\leaderfil{\leaders\hbox to 5pt{\hss.\hss}\hfil}
 \newdimen\signwidth 
 \setbox0=\hbox{+} 
 \signwidth=\wd0 
 \catcode`!=\active 
 \def!{\kern\signwidth}
 \halign{\hbox to 2.3in{#\leaderfil}\tabskip 2.0em&
 \hfil#\hfil&
 \hfil#\hfil&
 \hfil#\hfil\tabskip 0pt\cr
 \noalign{\doubleline\vskip 2pt}
\omit \hfil Parameter\hfil& \omit \hfil ~~$l=5$\deg--30\deg \hfil& \omit \hfil $l=30$\deg--60\deg \hfil& \omit \hfil Average \hfil \cr
\omit \hfil & \omit \hfil $l=330\deg$--355\deg \hfil& \omit \hfil $l=300\deg$--330\deg \hfil&  \cr
\noalign{\vskip 4pt\hrule\vskip 6pt}
AME / free-free at 28.4\GHz& $0.71\pm0.04$& $0.80\pm0.04$& $0.75\pm0.03$\cr
AME at 28.4\GHz\ / 100\um\ emission\tablefootmark{a}& $7.1\pm0.4$& $9.8\pm0.5$& $7.4\pm0.3$\cr
AME at 28.4\GHz\ / total emission& $0.42\pm0.02$& $0.48\pm0.01$& $0.45\pm0.01$\cr
\noalign{\vskip 3pt\hrule\vskip 4pt}}}
\endPlancktablewide
\tablefoot{\tablefoottext{a}{The conversion factor at 28.4\GHz\ is 40\,mK\,MJy$^{-1}$\,sr.}}
\endgroup
\end{table*}

Since AME is expected to be correlated with the dust distribution, on the plane it is not possible to use morphology to separate it from other emission components as is the case at higher latitudes. On the Galactic plane we use SEDs to separate the various emission components using their spectral shapes. Fig. \ref{fig:amesed} shows the spectra of the synchrotron, free-free and thermal dust as described in  Sect. \ref{sec:dustSEDs}. AME is the remainder of the total emission and is shown in more detail in Fig. \ref{fig:amesed2} for the four longitude ranges $l = 20$\deg--30\deg, 30\deg--40\deg, $l=320$\deg--330\deg\ and 330\deg--340\deg. The combination of WIM and MC spinning dust models \citep{Draine:1998b,Ali-Haimoud:2009} shown for comparison is a good fit to the data points. The peak emission is at $\sim$25\GHz, which is in agreement with the values of $25.5\pm1.5$\GHz\ for molecular clouds in the Gould Belt system at intermediate latitudes \citep{planck2013-XII}. A study of AME regions \citep{planck2013-XV} found a peak frequency around 28\GHz.

The AME fractions of the 28.4\GHz\ total emission in the $l=20$\deg--30\deg, 30\deg--40\deg, 320\deg--330\deg, and 330\deg--340\deg\ regions are given in Table \ref{tab:amefraction}. The errors shown in Fig. \ref{fig:amesed2} and Table \ref{tab:amefraction} are a combination of the uncertainties in the Gaussian fit of the amplitude of each latitude profile (and taking account of the error in the underlying broad component). These uncertainties are propagated in the models of free-free, synchrotron, and dust that are removed from the observations in order to derive the AME amplitude. The AME is well determined with significance in the range 5.6--9.7$\sigma$. The mean value of the AME-to-total fraction at 28.4\GHz\ is $0.44\pm0.03$.

As shown in Fig. \ref{fig:amesed2} and Table \ref{tab:amefraction} there is a significant steepening of the spectral index of the AME emission between 28.4\GHz\ ($\beta_{\rm AME}=-2.2\pm0.3$; obtained from 22.8, 28.4 and 33.0\GHz) and 70\GHz\ ($\beta_{\rm AME}=-3.3\pm0.4$; obtained from 40.7, 44.1, 60.7 and 70.4\GHz). The steepening is compatible with spinning dust models but is not well-defined due to the uncertainty in the extrapolation of the dust model below 100\GHz. 

\vspace{1mm}
A question arises as to whether the 20--60\GHz\ excess is due to a population of ultracompact \hii\ regions \citep{Wood:1989,Kurtz:2002,Dickinson2013}.  This matter is discussed in \citet{planck2013-XV} where the 100\um\ flux densities from \IRAS\ are used to determine an upper limit to the contribution from ultracompact \hii\ regions. We find that in the $l=20^{\circ}$--$30^{\circ}$ region ($|b|<0.\!^{\circ}5$) there are 907 \IRAS\ sources, of which 100 have colours that indicate them being candidate ultra compact \hii\ regions. Summing their $100\,\mu$m flux densities, and converting to an upper limit at 15\,GHz gives an average of $17$\,Jy\,deg$^{-2}$, or an upper limit of 2\,mK at 28.4\GHz. This compares with the observed value of 23\,mK in this region, as seen in Fig. \ref{fig:latdist}.  We conclude that ultracompact \hii\ regions are a minor contribution to the emission on the Galactic plane in the 20--60\GHz\ range.

\subsection{The longitude distribution of the four components}
\label{sec:discussion-lon}

\begin{figure}
\centering
\includegraphics[width=\hsize]{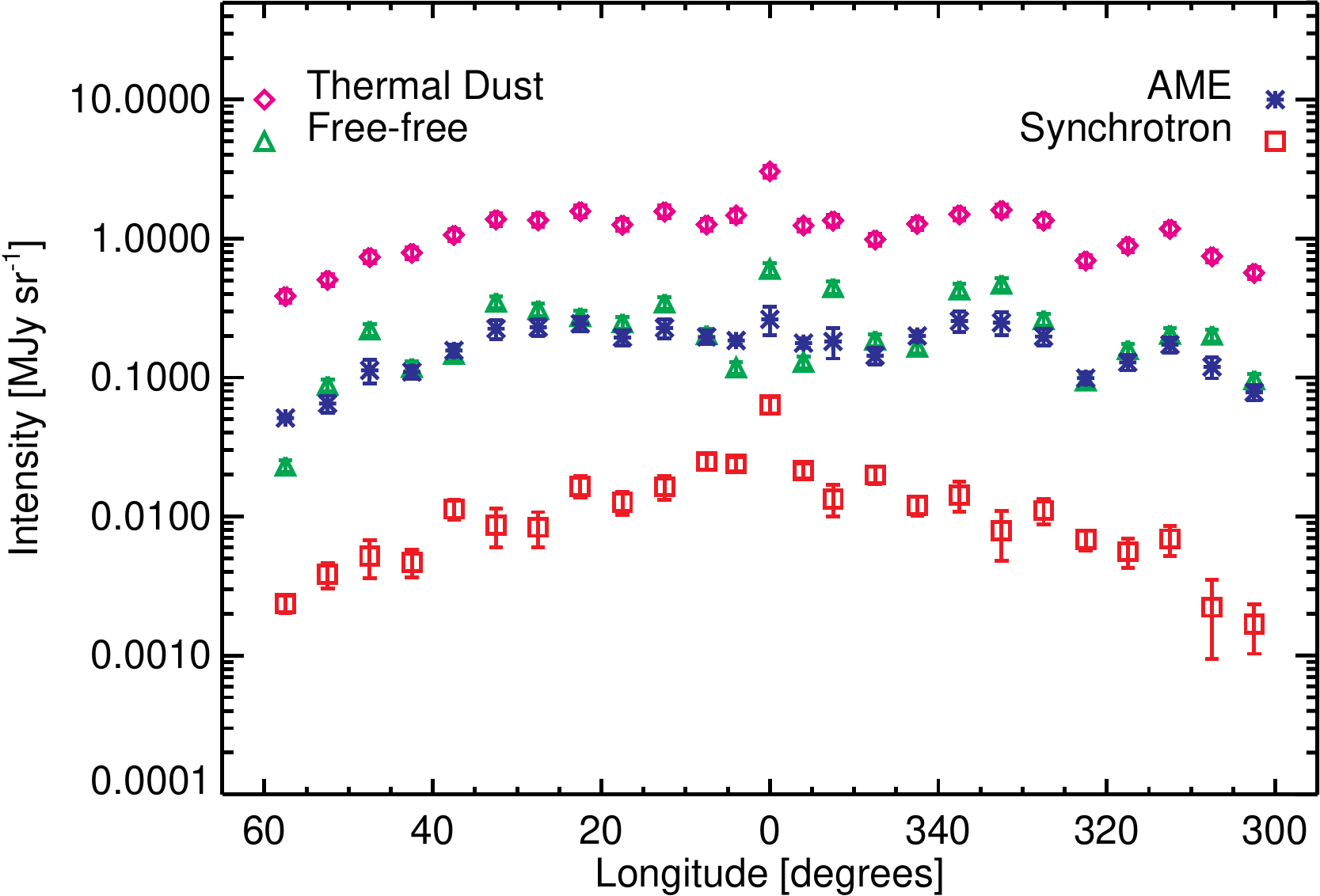}
\caption{Longitude distribution of all four components derived from SEDs and component separation, in 5\deg\ longitude intervals. The thermal dust is at 545\GHz\ scaled down by a factor of 100. The free-free, AME, and synchrotron, have been scaled to 28.4\GHz.  The conversion factor from intensity to brightness temperature at 28.4\GHz\ is 40\,mK\,MJy$^{-1}$\,sr.}
\label{fig:lonranges}
\end{figure}

Figure \ref{fig:londist} shows the longitude distribution of the total emission at $b = 0\deg$ for illustrative frequencies.  The emission is smoothed to 1\deg\ resolution and includes the diffuse, broader latitude emission as well as point sources.  The total emission is strongest in the inner Galaxy for $l=300\deg \rightarrow 0\deg \rightarrow 60\deg$ where it is 5--10 times that at $l = 80$\deg--90\deg or $l=270\deg$--280\deg.  Individual features in the local spiral arm can be seen in all four emission components at $l = 73$\deg--85\deg\ (Cygnus X) and $l = 282$\deg--295\deg.

In order to give an overview of the longitude distribution of each of the four emission components, the intensities are derived for longitude intervals of 5\deg.  This interval also gives a robust SED from which to separate the components.  Figure \ref{fig:lonranges} shows the distribution of the four emission components at $b = 0\deg$ at $l=300\deg \rightarrow 0\deg \rightarrow 60\deg$ averaged over 5\deg\ intervals in longitude.  The synchrotron emission is given by the 408\MHz\ emission corrected by free-free; this correction is typically 20\,\% on the plane \citep{Alves2012}. The free-free emission is estimated from RRL data and the {\tt FastMEM} analysis.  The thermal dust indicator is the \Planck\ 545\GHz\ emission, which is not contaminated by the other emission components.  The AME emission is the residual when the other three components are subtracted from the total emission.  For ease of comparison in Fig. \ref{fig:lonranges}, the synchrotron, free-free, and AME are evaluated at 28.4\GHz, which is close to the peak of the AME emission.

The longitude distributions of the four emission components shown in Fig. \ref{fig:lonranges} are broadly similar, as might be expected from Fig. \ref{fig:londist}. The highest rate of star formation is occurring in the inner two spiral arms of the Galaxy \citep{Bronfman:2000, Wood:1989, Alves2012}. The velocities of the gaseous components show that the inner arms located between $R_{\rm G} = 4$ to $6.5$\kpc\ are responsible. We note the increase in emission of all components towards the Galactic centre; the spiral arms are not differentiated because of the 5\deg\ smoothing and the use of a logarithmic scale. The outer arms sampled at  \mbox{$l = 270$\deg--300\deg} and at $l=60\deg$--90\deg\ are a factor of 3--4 less bright than the inner arms. The inference from Fig. \ref{fig:lonranges} is that all four components studied here are tightly related to current star formation.

%

\section{Discussion and conclusions}
\label{sec:conc}

\begin{table}[tb]
\begingroup
\newdimen\tblskip \tblskip=5pt
\caption{Intrinsic latitude widths of various Galactic components. The widths are the mean values in the longitude range $l=330\deg$--340\deg, $l=20\deg$--30\deg, and $l=30\deg$--40\deg. Values in the 320\deg--330\deg\ range are a factor of $\sim1.35\pm0.05$ greater.}
 \label{tab:conc}
\nointerlineskip
\vskip -3mm
\footnotesize
\setbox\tablebox=\vbox{
 \newdimen\digitwidth 
 \setbox0=\hbox{\rm 0} 
 \digitwidth=\wd0 
 \catcode`*=\active 
 \def*{\kern\digitwidth}
 \def\leaderfil{\leaders\hbox to 5pt{\hss.\hss}\hfil}
 \newdimen\signwidth 
 \setbox0=\hbox{+} 
 \signwidth=\wd0 
 \catcode`!=\active 
 \def!{\kern\signwidth}
 \halign{\hbox to 2.5in{#\leaderfil}\tabskip 0.5em&
 \hfil#\hfil\tabskip 0pt\cr
 \noalign{\doubleline\vskip 2pt}
 \omit \hfil Component\hfil& Width \cr
\noalign{\vskip 4pt\hrule\vskip 6pt}
OB stars \citep{Wood:1989}& $0\pdeg83\pm0\pdeg05$\cr
Dense CS clouds \citep{Bronfman:2000}& $0\pdeg90\pm0\pdeg05$\cr
Free-free via RRLs ($n_e^2$)& $0\pdeg92\pm0\pdeg05$\cr
Free-free from {\tt FastMEM}& $1\pdeg07\pm0\pdeg05$\cr
CO& $1\pdeg19\pm0\pdeg08$\cr
Dust -- FIR (100\um)& $1\pdeg24\pm0\pdeg02$\cr
\Planck\ 353, 545, 857\GHz& $1\pdeg30\pm0\pdeg04$\cr
Gamma-rays (representative of total matter)& $1\pdeg26\pm0\pdeg05$\cr
Synchrotron, GHz frequencies& $1\pdeg65\pm0\pdeg05$\cr
Synchrotron, K-band polarization& $1\pdeg67\pm0\pdeg13$\cr
Neutral hydrogen& $1\pdeg98\pm0\pdeg05$\cr
\noalign{\vskip 3pt\hrule\vskip 4pt}
}}
\endPlancktable
\endgroup
\end{table}

We have identified four components of emission within the inner Galaxy that are narrow in Galactic latitude; their emission extends over the whole frequency range of \planck\ and into the radio, FIR, and gamma-ray regions. Their longitude distribution is similar in all these bands, and indicates that this emission originates within the inner spiral arms, principally the Carina-Sagittarius arm and the Norma-Scutum arm, as shown in Figs. \ref{fig:hiireg} and \ref{fig:londist} and discussed in Sect. \ref{sec:londist}. The FWHM width in the range 1\deg\ to 2\deg\ of the various components corresponds to a FWHM \mbox{$z$-thickness} of 100 to 200\pc\ at the distance of the spiral arms. A principal result relating to Galactic structure is the difference in latitude width of the various emission components. These differences reflect the different formation processes and life-histories of the components. The FWHM data are summarized in Table \ref{tab:conc}.

\subsection{Free-free}
The free-free emission is determined by RRL observations in the range $l = 20$\deg--44\deg\ and by a {\tt FastMEM} analysis over the rest of the range, $l=300\deg \rightarrow 0\deg \rightarrow 60\deg$. Free-free is the narrowest of the four components with a typical FWHM of \mbox{$1\pdeg06\pm0\pdeg05$}, which is wider than that of the recently formed OB stars \mbox{($\rm FWHM = 0\pdeg83\pm0\pdeg05$; \citealp{Wood:1989})} and dense molecular clouds ($\rm FWHM = 0\pdeg90\pm0\pdeg05$; \citealp{Bronfman:2000}) but narrower than \hi\ ($1\pdeg98\pm0\pdeg05$), FIR dust ($1\pdeg24\pm0\pdeg02$), and the \Planck\ dust ($1\pdeg30\pm0\pdeg04$). This narrowing of the young stellar population distribution relative to that of the gas is a consequence of the star formation rate being a nonlinear function of gas density -- normally taken as proportional to $n_{\rm H}^2$.  This process naturally narrows the latitude distribution of the young stars relative to that of the gas and dust. Similarly the free-free latitude distribution will be narrower than that of the gas and dust since the free-free brightness is proportional to the EM ($n_\mathrm{e}^2 l$).

\subsection{Synchrotron}
The synchrotron study has identified a narrow latitude distribution with $\rm FWHM = 1\pdeg7$, which is superposed on a broader Galactic halo distribution of $\rm FWHM \sim 10\deg$, as shown in Figs. \ref{fig:compmaps} and \ref{fig:compmaps2}. At 408\MHz\ these two distributions have similar brightness temperatures at $b = 0\deg$ in the inner Galaxy. The synchrotron emission reflects the current CRE distribution, modified by the local magnetic field $B$ as $B^{1.7}$--$B^{2.0}$ \citep{Strong:2011} for the range of spectral indices considered here.  These observations suggest the following scenario.  The OB star population produces SNe in a narrow distribution on the plane. These SNe are the source of the CREs.  As the SNRs expand over their $10^5$--$10^6$\,yr lifetime, the CREs continue to produce synchrotron emission in the magnetic fields of the decaying remnants.  Other CREs escape from the SNRs and radiate in the ordered magnetic field of the plane.  This ordered magnetic field in the inner Galaxy produces the narrow band of highly (30--50\,\%) polarized emission on the plane seen in the \WMAP\ K, Ka, and Q bands, which is a further justification for identifying this narrow band as a separate synchrotron component.  In contrast, the synchrotron polarization at latitudes $|b|=2\deg$--5\deg\ is markedly lower at a level of 5--10\,\%, presumably as a result of  the tangled structures seen in the \WMAP\ polarization maps.  

Related information comes from the SNR and pulsar latitude distributions.  The latitude distribution of SNRs in the \citet{Green:2009} catalogue has a FWHM of  $\sim$1.0\deg, very similar to that of the parent OB star population.  Normal pulsars (neutron stars) formed in the SN event also have lifetimes of $10^5$--$10^6$\,yr and have proper motions with a component perpendicular to the plane of $\sim$200\,km s$^{-1}$.  In the pulsar lifetime they will have travelled $\sim$100\pc\ giving a FWHM of $\sim$2\deg\ at a distance of 6\kpc, a value consistent with observation (Fig. \ref{fig:narrowsynch}, Table \ref{table:synch}) again significantly broader than the OB star distribution.  Pulsar and OB star data indicate that the current SN rate for our Galaxy is 2--4 per century \citep{Keane:2008,Wood:1989}; this would imply $\sim$6000 pulsars and SNRs on or near the Galactic plane with ages up to $\sim10^5$ years.  These SNRs could make a significant contribution to the narrow distribution on the plane and provide the disordered magnetic field at intermediate latitudes seen in \WMAP\ polarization data.

\subsection{AME}
AME is clearly identified in the SED analysis where the \Planck\ data define the high frequency side of the expected peaked spectrum \citep{Draine:1998a}. The AME emission is the main emission component on the plane along with free-free emission over the frequency range 20--70\GHz. Above 70\GHz\ thermal dust becomes a significant contributor. The ratio AME/free-free for the diffuse emission on the Galactic plane is of interest for comparison with that of individual \hii\ regions.  The longitude distributions of both AME and free-free in \mbox{Fig. \ref{fig:lonranges}} indicate that there is an overall correlation between them. The weighted average value of  AME/free-free for the longitude range $l=300\deg \rightarrow 0\deg \rightarrow 60\deg$ is $0.75\pm0.03$ as measured at 28.4\GHz\ (see Table~\ref{tab:amecomparison}). There appears to be a marginally significant increase in the ratio between the innermost ($0.71\pm0.04$) and the outer ($0.80\pm0.04$) regions of the inner Galaxy. There is a similar increase in the ratio of AME to 100\um\ dust emission ($7.1\pm0.4$ to $9.8\pm0.5$\,mK\,MJy$^{-1}$\,sr; see Table~\ref{tab:amecomparison}).  These values may be compared with other estimates of AME in the total emission at 28.4\GHz:  \citet{planck2013-XV} found the average AME fraction to be $0.50\pm0.02$ for individual bright AME regions, compared with the average value of $0.45\pm0.01$ found here for the combination of $l=5$\deg--60\deg\ and 300\deg--355\deg. \cite{planck2011-7.3} found an average spinning dust fraction of $25\pm5$\,\% at 30\,GHz. This discrepancy can be explained by the fact that their value represents the average over all longitudes, with higher values found in the inner galaxy. Also, the uncertainties are likely to be under-estimated due to modelling errors.

Several investigations have shown that the dust emissivity index $\beta_{\rm dust}$ appears to vary with frequency, flattening in the millimetre range relative to the best single greybody fit, and also with environment \citep{Reach:1995,Finkbeiner:1999,Galliano:2005,Paladini:2007,planck2011-6.4b,planck2013-XIV}. We find that a single-$\beta_{\rm dust}$ fit to the dust spectrum changes the AME component at 28.4\GHz\ by $\leq$1\,\%.

\subsection{Thermal dust}
There is an indication that the dust brightness has different latitude widths in different frequency ranges. The cooler dust at HFI frequencies (143--857\GHz) has a FWHM width of $1\pdeg30\pm0\pdeg04$, while the warmer dust at the shortest \IRAS\ wavelength (12\um) has a width of $1\pdeg44\pm0\pdeg03$. Both these values of dust width are less than that of the \hi, which is $1\pdeg98\pm0\pdeg05$ (average of values in Table~\ref{table:dust}). Heating of the dust nearer the OB-stars would give a higher dust temperature and hence higher brightness temperature near the plane, leading to a narrower width. Our SED analysis gives estimates of the dust temperature and spectral index for the narrow inner Galaxy component. The dust latitude distribution is wider than that of the OB-stars heating the dust. We have determined values of dust parameters in the 1\deg\ FWHM beam on the Galactic plane for $l=320$\deg--340\deg\ and $l=20$\deg--40\deg. We find mean values of $T_\mathrm{dust}=20.4\pm0.4$\,K, $\beta_\mathrm{FIR} = 1.94\pm0.03$, and $\beta_\mathrm{mm}=1.67\pm0.02$. These values are in close agreement with those determined by \citet{planck2013-XIV} for the $l=20$\deg--44\deg\ and $|b|<1\deg$ region.

\subsection{Application to other galaxies}
The results obtained here for the inner Galactic plane can be compared with the integrated data for similar spiral galaxies. The high levels of AME and free-free relative to synchrotron found at 20--40\GHz\ for the inner Galaxy contrast with the ratios found for the integrated emission in spiral galaxies (\citealt{Niklas:1997, Peel:2011}). This could be because the equivalent of the narrow inner galaxy studied here is only a fraction of the total emission volume of the Galaxy, with the broader halo synchrotron components contributing the major part of the total emission. Data for 74 normal galaxies studied by \cite{Niklas:1997} have an average ratio of free-free to synchrotron flux density of $0.08\pm0.01$ at 1.0\GHz; this translates to a ratio at 28.4\GHz\ of 0.99, assuming a synchrotron flux density spectral index of $-0.83$ at 1--10\GHz\ and $-1.0$ at higher frequencies.  We find a free-free to synchrotron ratio of $\sim30$ at 28.4\GHz\ for the narrow component as shown in Fig. \ref{fig:amesed}. This factor of 30 difference in the ratio can be accounted for by the  larger synchrotron volume of the total galaxy as compared with the narrow inner Galaxy (a factor of $\sim$15 in latitude as shown in Figs. \ref{fig:compmaps}, \ref{fig:compmaps2} and \ref{fig:latdist} and $\sim2$ in longitude). Total-power high-resolution measurements of edge-on galaxies (tilt within 2\deg--5\deg) over a wide range of frequencies would be decisive in showing the presence or otherwise of such a narrow component at each of the four emission mechanisms in normal spiral galaxies.  

\begin{acknowledgements}
We acknowledge the use of the Legacy Archive for Microwave Background Data Analysis (LAMBDA); support for LAMBDA is provided by the NASA Office of Space Science. Some of the results in this paper have been derived using the \healpix\ package

The development of \Planck\ has been supported by: ESA; CNES and CNRS/INSU-IN2P3-INP (France); ASI, CNR, and INAF (Italy); NASA and DoE (USA); STFC and UKSA (UK); CSIC, MICINN, JA and RES (Spain); Tekes, AoF and CSC (Finland); DLR and MPG (Germany); CSA (Canada); DTU Space (Denmark); SER/SSO (Switzerland); RCN (Norway); SFI (Ireland); FCT/MCTES (Portugal); and PRACE (EU). A description of the Planck Collaboration and a list of its members, including the technical or scientific activities in which they have been involved, can be found at \url{http://www.sciops.esa.int/index.php?project=planck&page=Planck_Collaboration}.\end{acknowledgements}

\bibliographystyle{aa}
\bibliography{Planck_bib,refs1}

\appendix

\section{Validation of {\tt FastMEM}}
\label{sec:appendix}

\subsection{Reconstruction of the free-free component using {\tt FastMEM}}
\label{sec:a1}

The Maximum Entropy Method (MEM) is used to separate the signals from cosmological and astrophysical foregrounds including CMB and Galactic foregrounds such as synchrotron, free-free, anomalous microwave emission (AME) and thermal dust. The implementation 
of MEM used here works in the spherical harmonic domain, where  the separation is performed mode-by-mode allowing a huge optimization problem to be split into a number of smaller problems that can be done in parallel, saving CPU time.  This approach, called {\tt FastMEM}, is described by \citet{Hobson:1998} for Fourier modes on flat patches of the sky and by \citet{Stolyarov:2002} for the full-sky case.

The simulated data set used to check the quality of the recovery and reconstruction errors was the Full Focal Plane data set FFP4 \citep{planck2013-p28}. It includes all \Planck\ frequency maps from 30 to 857\GHz\ properly simulated with plausible noise levels, systematics due to the scanning strategy, and real beam transfer functions. The components were modelled using the Planck Sky Model \citep[PSM;][]{Delabrouille2013} v1.7.

After multiple separation tests it was found that low-frequency components are very complicated to extract, especially AME. It was decided to use only the limited number of frequency maps, from 70.4 to 353\GHz. Since there were problems with \WMAP\ modelled maps, they were not used in the analysis. However, \WMAP\ data maps are essential for low-frequency component extraction from the real data.

\begin{figure}
\centering
\includegraphics[scale=0.3]{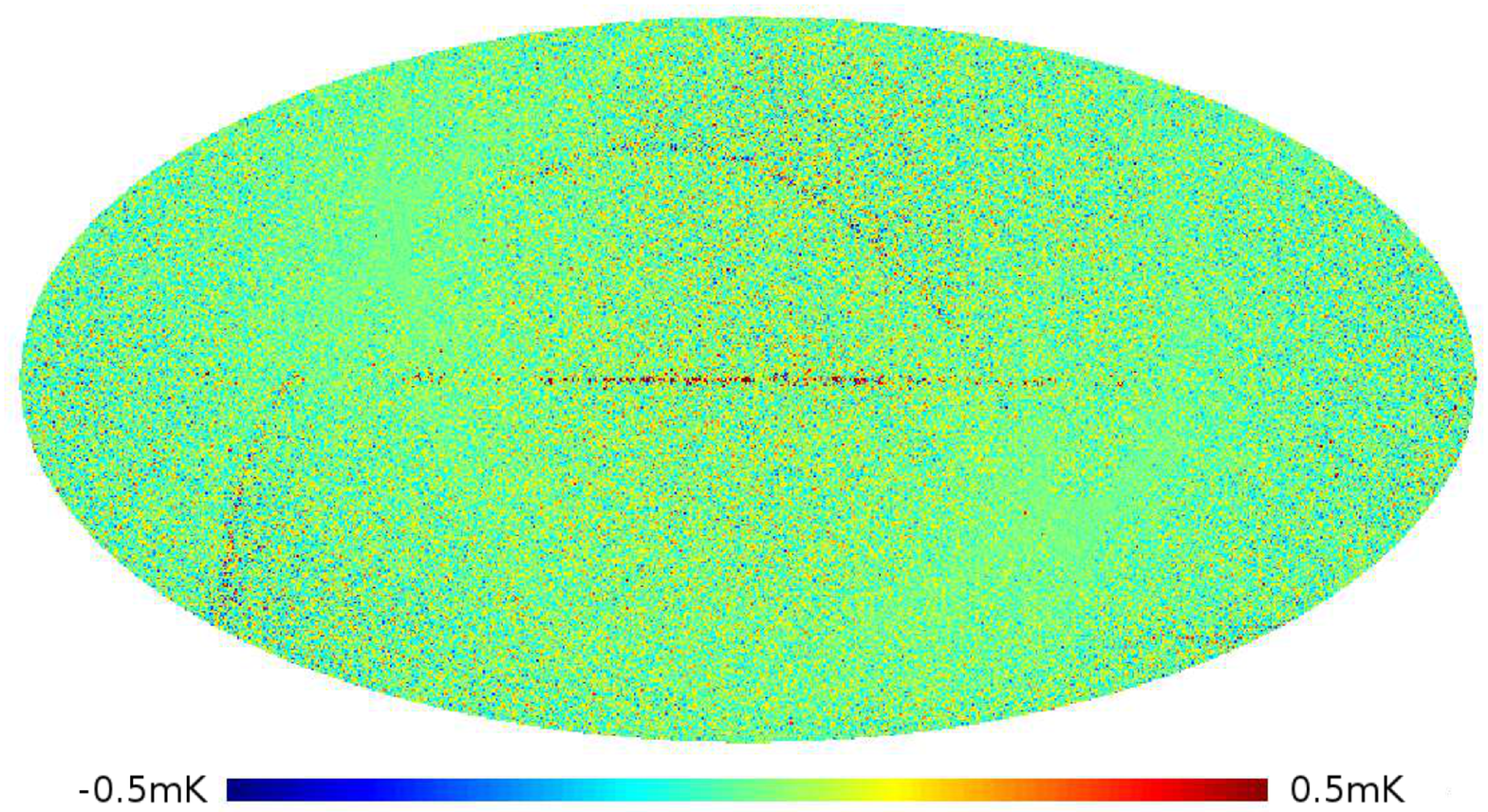}
\includegraphics[scale=0.3]{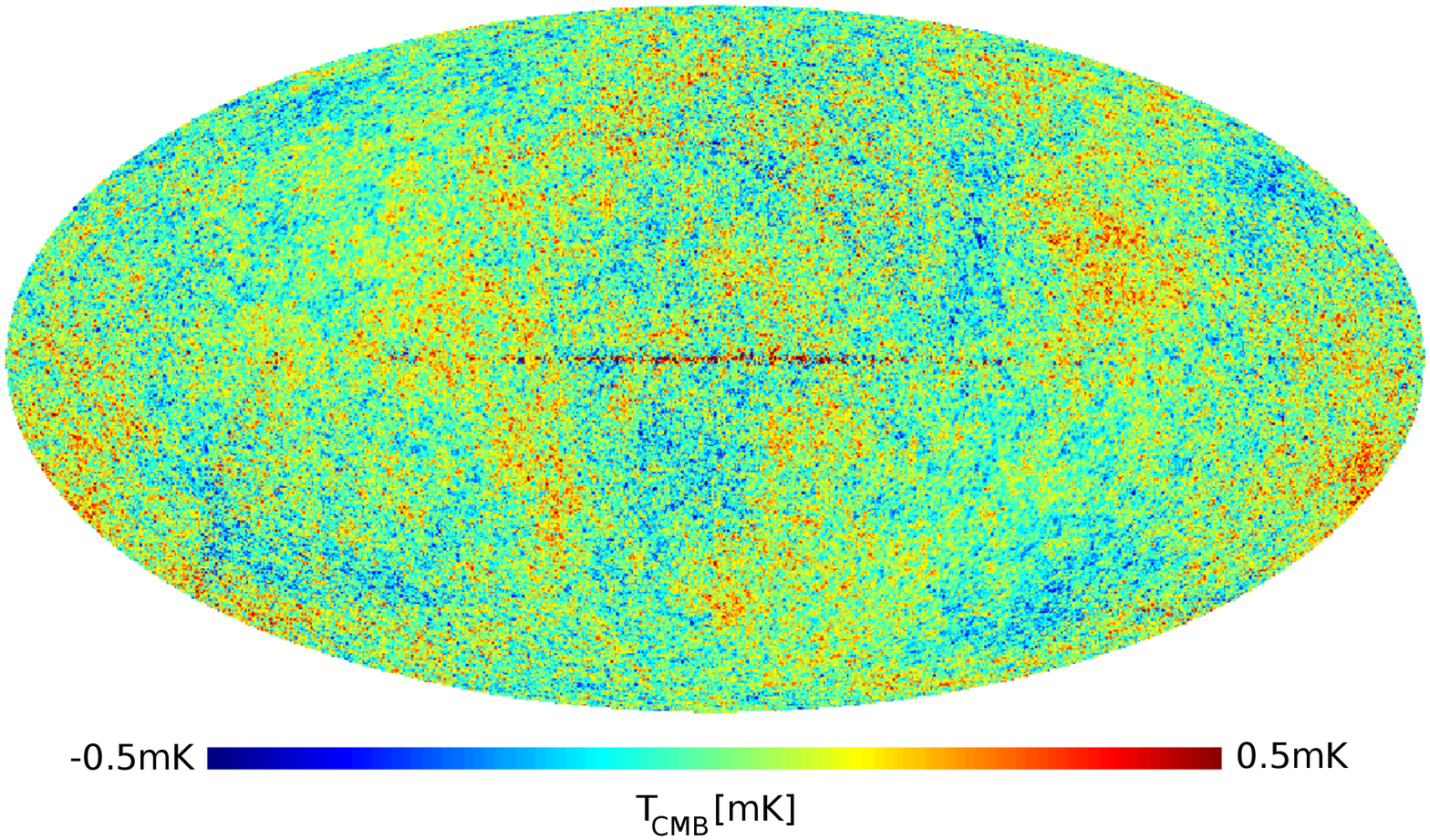}
\caption{Simulated data: residual maps at 70.4\GHz, containing only receiver noise and point sources (Eq~\ref{FastMEM_Res1}, \textit{upper panel}) and CMB component-cleaned map (Eq~\ref{FastMEM_Res2}, \textit{lower panel}). The maps are in Mollweide projection, with \mbox{$N_\mathrm{side}=1024$} and a linear temperature scale with units of T$_\mathrm{CMB}$.}
\label{fig:a1}
\end{figure}

\begin{figure}
\centering
\includegraphics[scale=0.3]{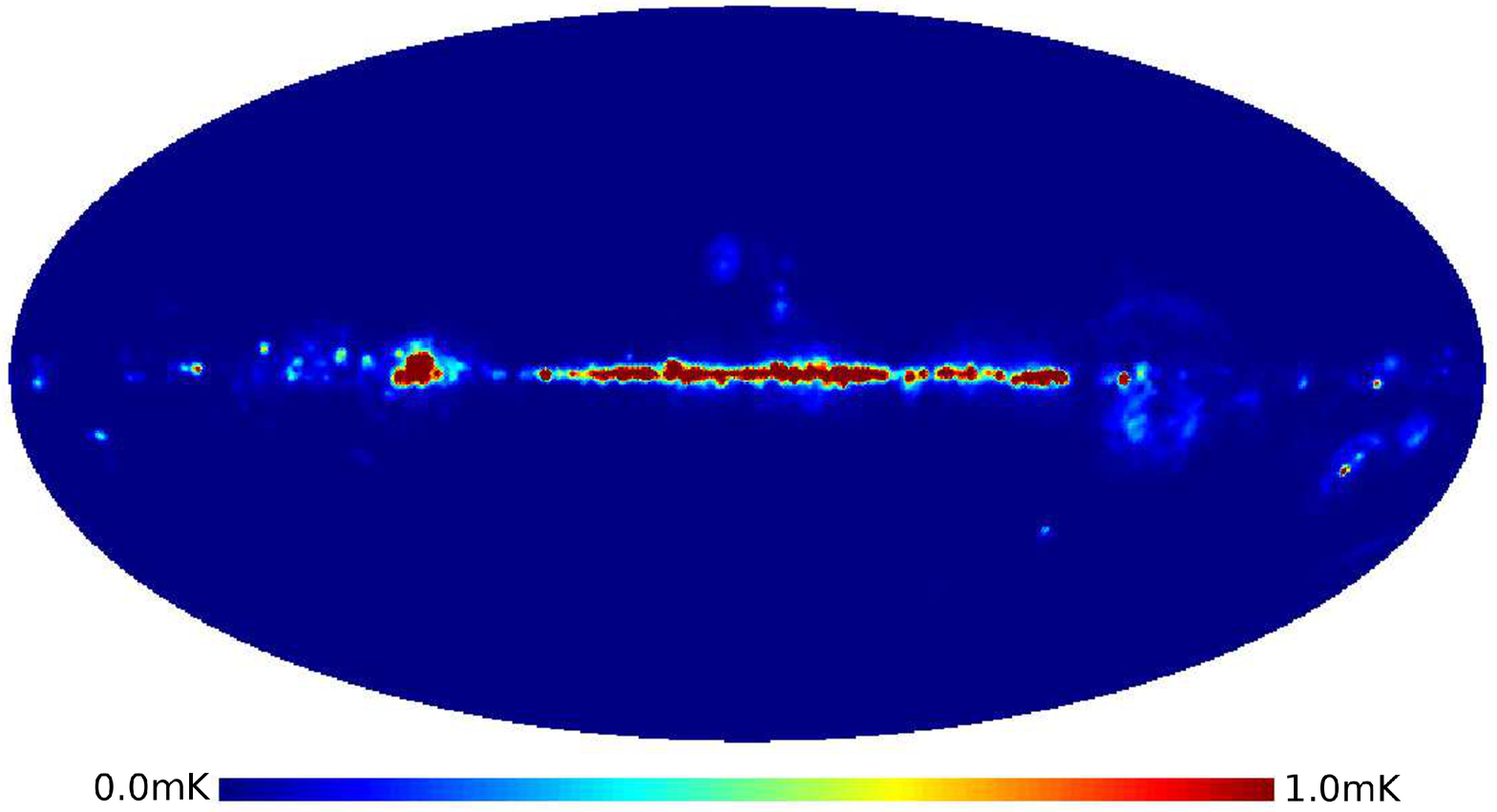}
\includegraphics[scale=0.3]{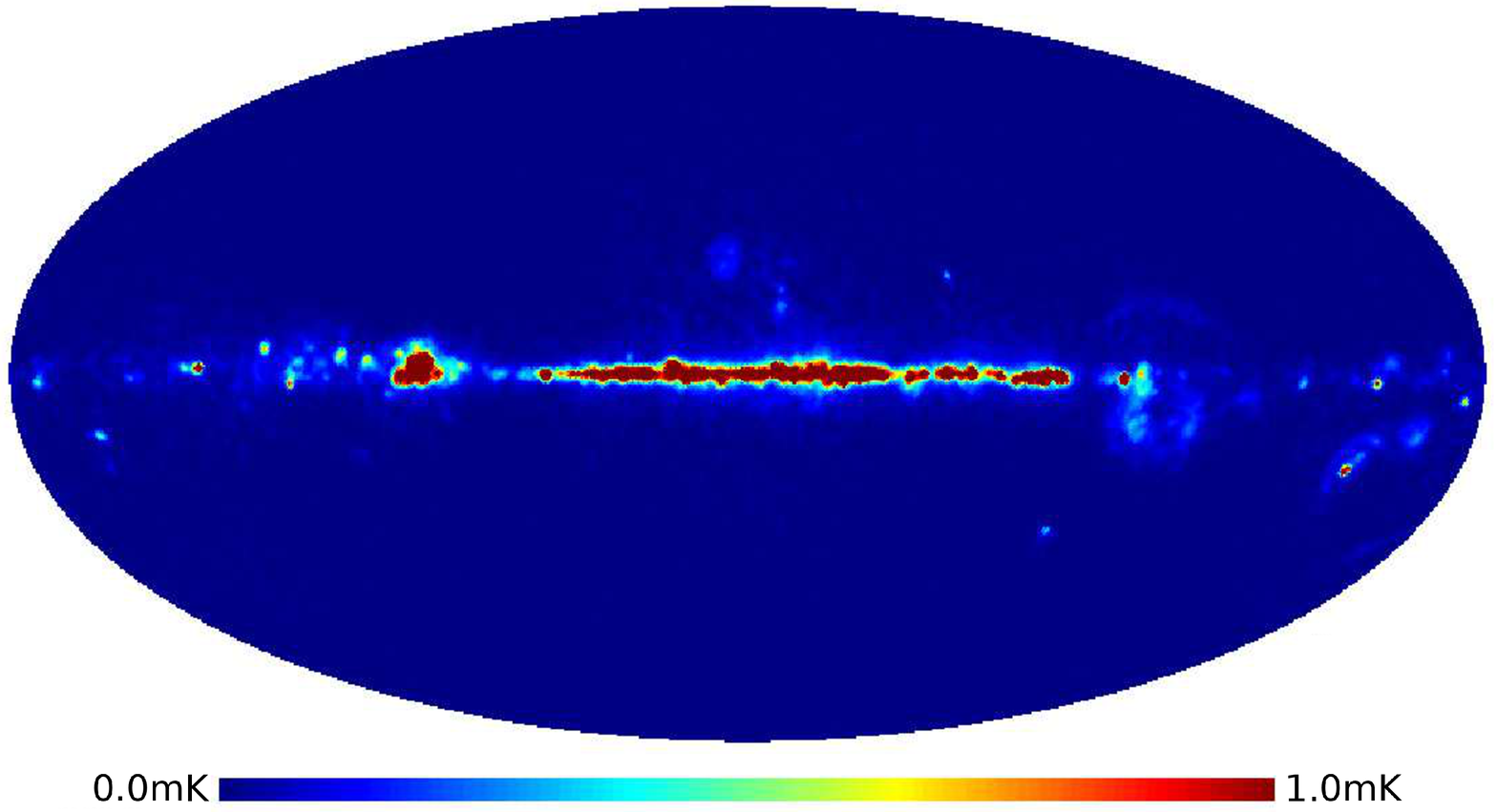}
\includegraphics[scale=0.3]{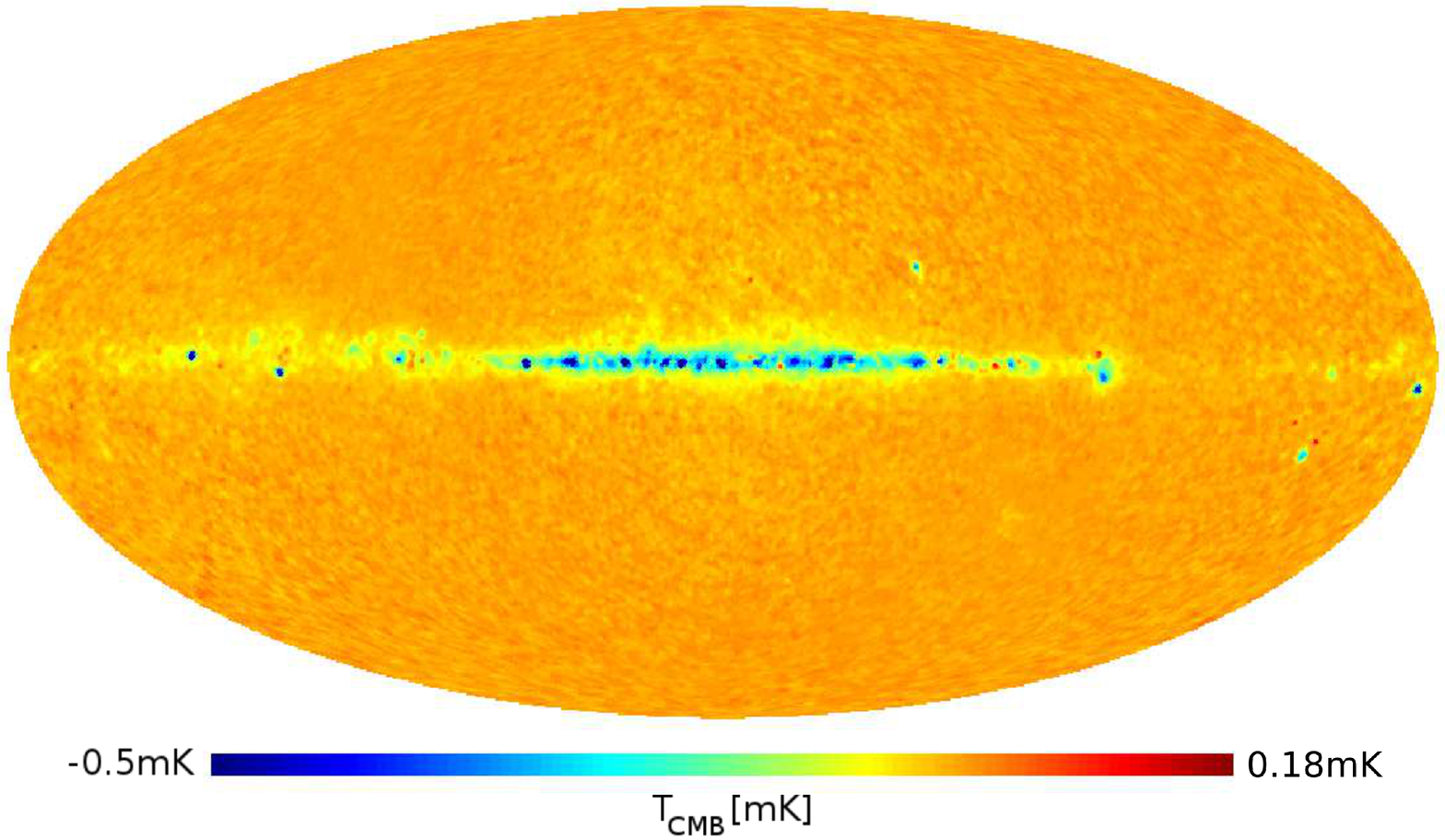}
\caption{The input, reconstructed and residual maps of the free-free emission at 70.4\GHz. All maps are in Mollweide projection, smoothed with a 1\deg\ beam, $N_\mathrm{side}=128$. The colour scale is linear, and the units are mK$_\mathrm{CMB}$.}
\label{fig:a2}
\end{figure}

The reconstructed components were CMB, thermal SZ, free-free, thermal dust, and CO $J$=1$\rightarrow$0, $J$=2$\rightarrow1$ and $J=$3$\rightarrow$2 lines. It was assumed that AME and synchrotron contributions are very small in this range of frequencies. A small bias from the synchrotron component, however, could be possible in the Galactic plane. Free-free spectral scaling was calculated assuming average electron temperature $T_\mathrm{e} = 6000$\,K \citep{Alves2012}.

The accuracy of the component separation was controlled in several ways. Firstly, the residuals between data maps and the modelled component contribution were calculated. There were component-free maps, containing only instrumental noise and point sources
\begin{equation}
\Delta T(\nu) = T(\nu)_\mathrm{data} - \hat{T}(\nu)_\mathrm{CMB} - \hat{T}(\nu)_\mathrm{dust} - \hat{T}(\nu)_\mathrm{ff} - \hat{T}(\nu)_\mathrm{SZt} - \hat{T}(\nu)_\mathrm{CO},
\label{FastMEM_Res1}
\end{equation}
and component-cleaned CMB maps at each frequency, containing CMB signal as well as receiver noise and point sources
\begin{equation}
T_\mathrm{CMB+noise+PS}(\nu) = T(\nu)_\mathrm{data} - \hat{T}(\nu)_\mathrm{dust} - \hat{T}(\nu)_\mathrm{ff} - \hat{T}(\nu)_\mathrm{SZt} - \hat{T}(\nu)_\mathrm{CO}.
\label{FastMEM_Res2}
\end{equation}
To calculate residual maps one should smooth component models with the same beam, which is taken to be a Gaussian with $\theta_\mathrm{FWHM} = 13\parcm23$ at 70.4\GHz. Residual maps are shown on the Fig.~\ref{fig:a1}.

The reconstructed free-free map can be directly compared with the input template. The residual map shows that the reconstruction slightly overestimates the real free-free contribution in the Galactic plane (Fig.~\ref{fig:a2}), where the morphology is complex and includes many bright point sources.

\begin{figure}
\centering
\includegraphics[width=\hsize]{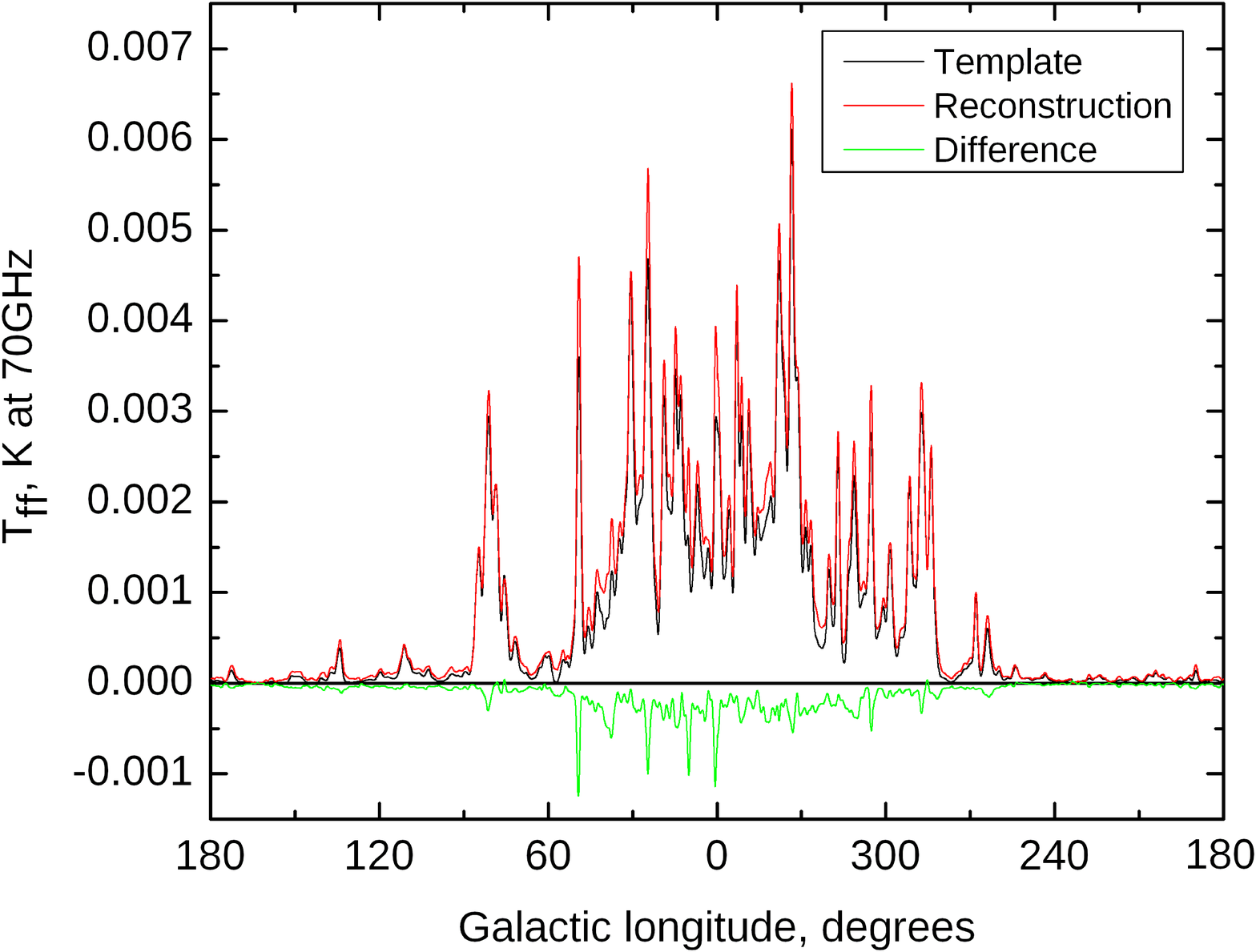}
\caption{The input, reconstructed and residual Galactic plane profiles for the free-free emission at 70.4\GHz. Profiles were calculated over the maps shown in Fig.~\ref{fig:a2};  units are K$_\mathrm{CMB}$. The plots are made for a 1\deg-resolution cut along the Galactic plane.}
\label{fig:a3}
\end{figure}

\begin{figure}
\centering
\includegraphics[width=\hsize]{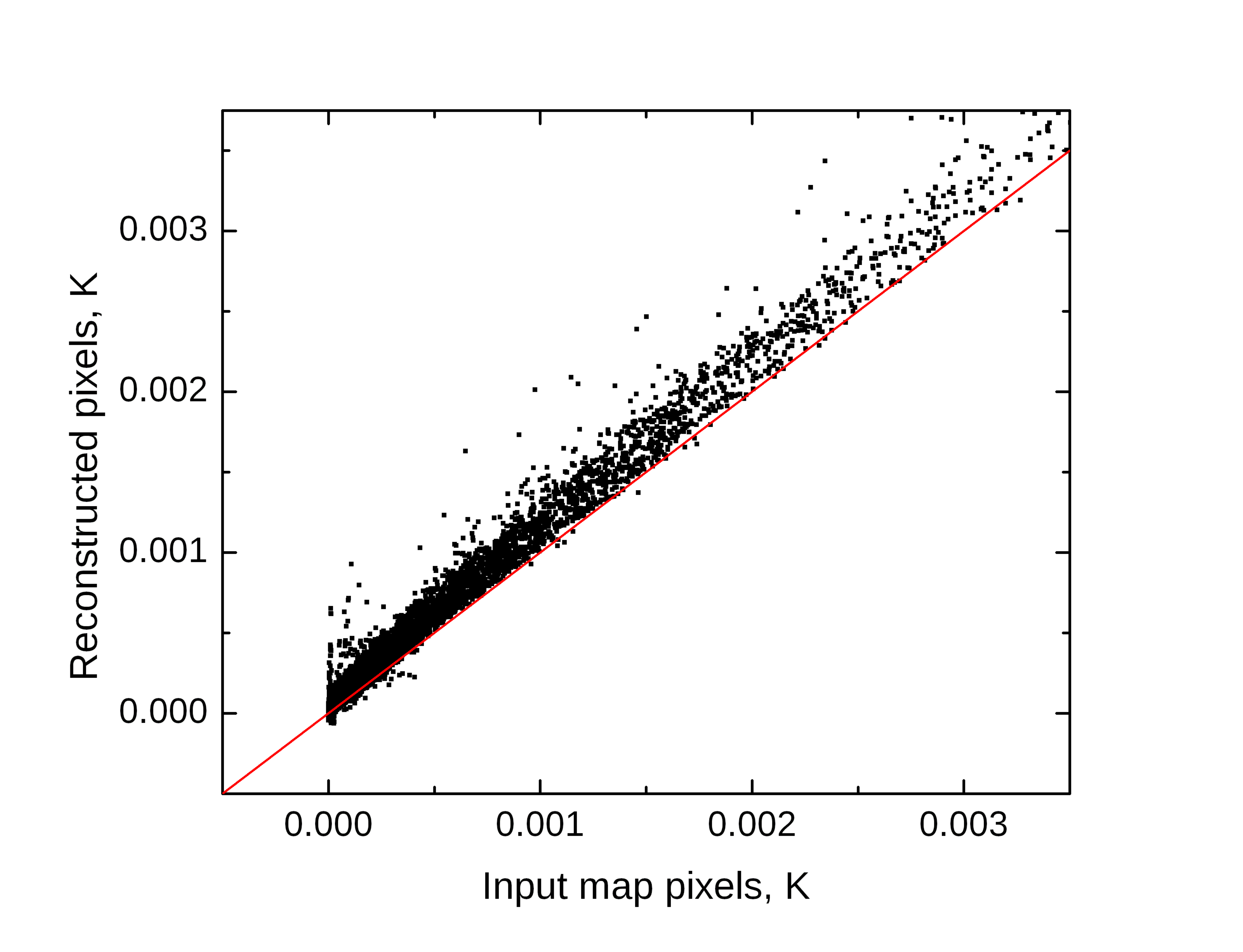}
\caption{$T$--$T$ plot comparing the pixel values in the simulated input map and in the reconstructed map. The solid red line indicates equality. The frequency is 70.4\GHz\ and the temperature scale is K$_\mathrm{CMB}$.}
\label{fig:a4}
\end{figure}

It is also useful to consider the profiles of free-free emission along the Galactic plane, which are shown on the Fig.~\ref{fig:a3}, and $T$--$T$ plots between input and reconstructed maps (Fig.~\ref{fig:a4}).

\subsection{Comparison of free-free maps from {\tt FastMEM} with other component separation methods}
\label{sec:a2}

In this section we compare the free-free emission derived from {\tt FastMEM} and other modelling methods with the direct measure of free-free using RRLs. In order to convert the RRL data to brightness temperature we have used an electron temperature, $T_\mathrm{e}$, of 6000\,K as found by \citet{Alves2012} for the Galactic plane region $l=20\deg$--$44\deg$, $|b|\leq4\deg$, which we also use for this comparison.

{\tt FastMEM} was applied to \Planck\ data in the frequency range 70.4--353\GHz\ and used to obtain the free-free, CO, and thermal dust components. The synchrotron component was assumed to be a small fraction of the total Galactic plane emission at these frequencies. This was tested by subtracting the synchrotron component found by \citet{Alves2012} with brightness temperature spectral indices in the range $\beta_{\rm synch}=-2.7$ to $-3.1$; the free-free component changed by $\leq$2\,\%. Similarly, the effect of the AME component was minimal in the 70.4--353\GHz\ range. The AME amplitude was based on the 100\um\ IRIS brightness and assuming 10\uK\ per MJy\,sr$^{-1}$ at 30\GHz\ with a spectrum peaking at 20\GHz.

\begin{figure*}
\centering
\includegraphics[scale=0.36]{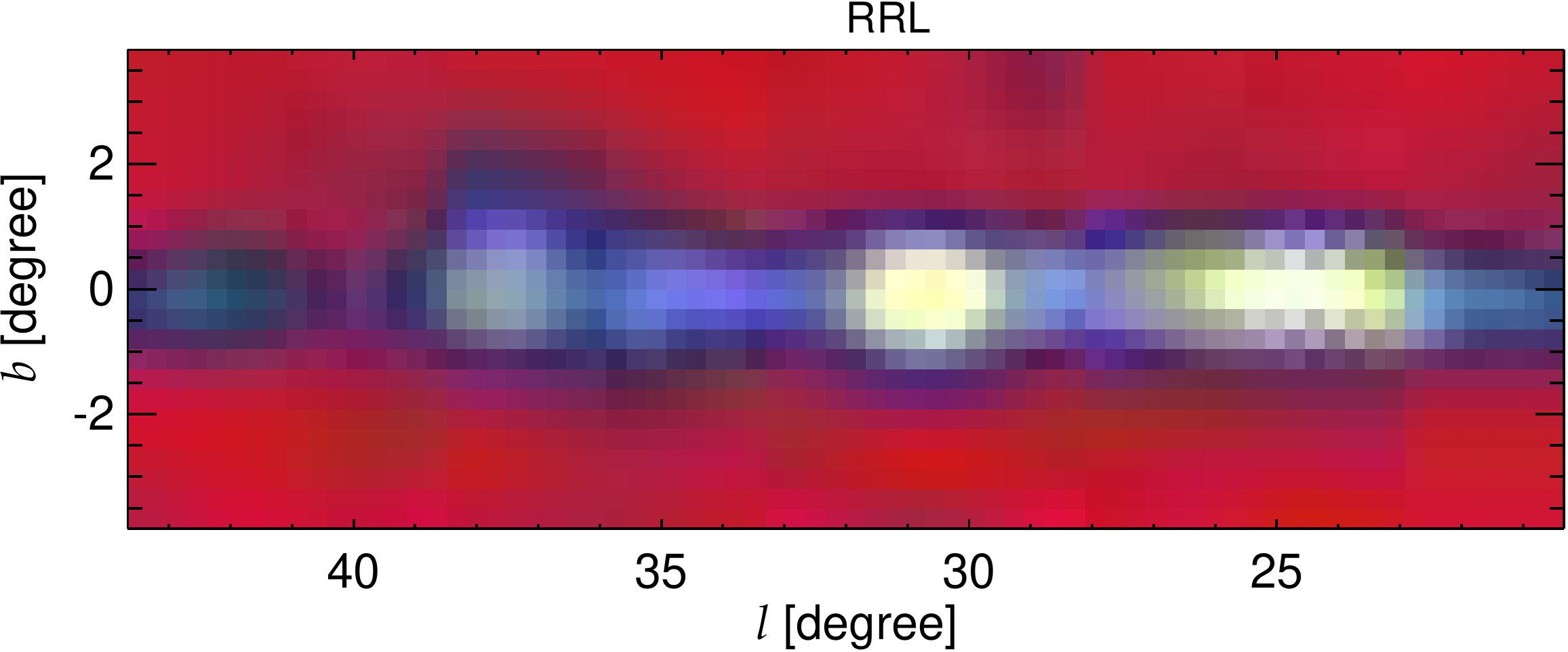}
\vspace*{0.3cm}
\includegraphics[scale=0.36]{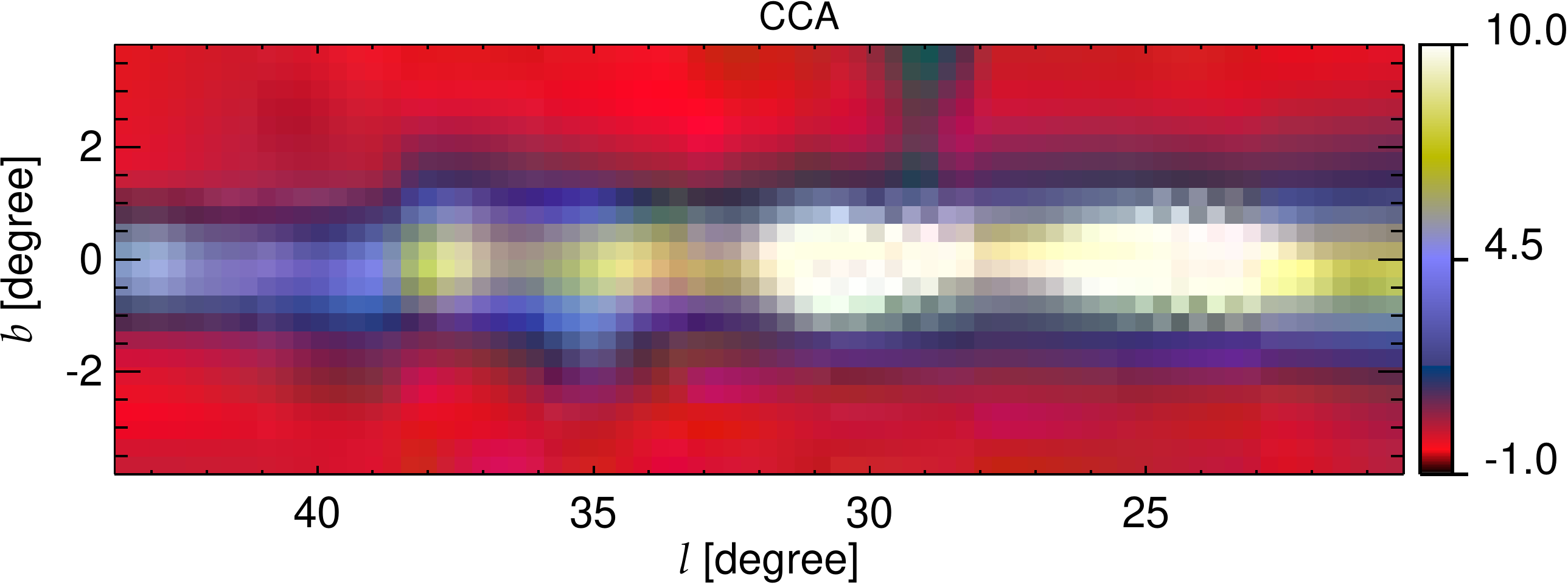}
\includegraphics[scale=0.36]{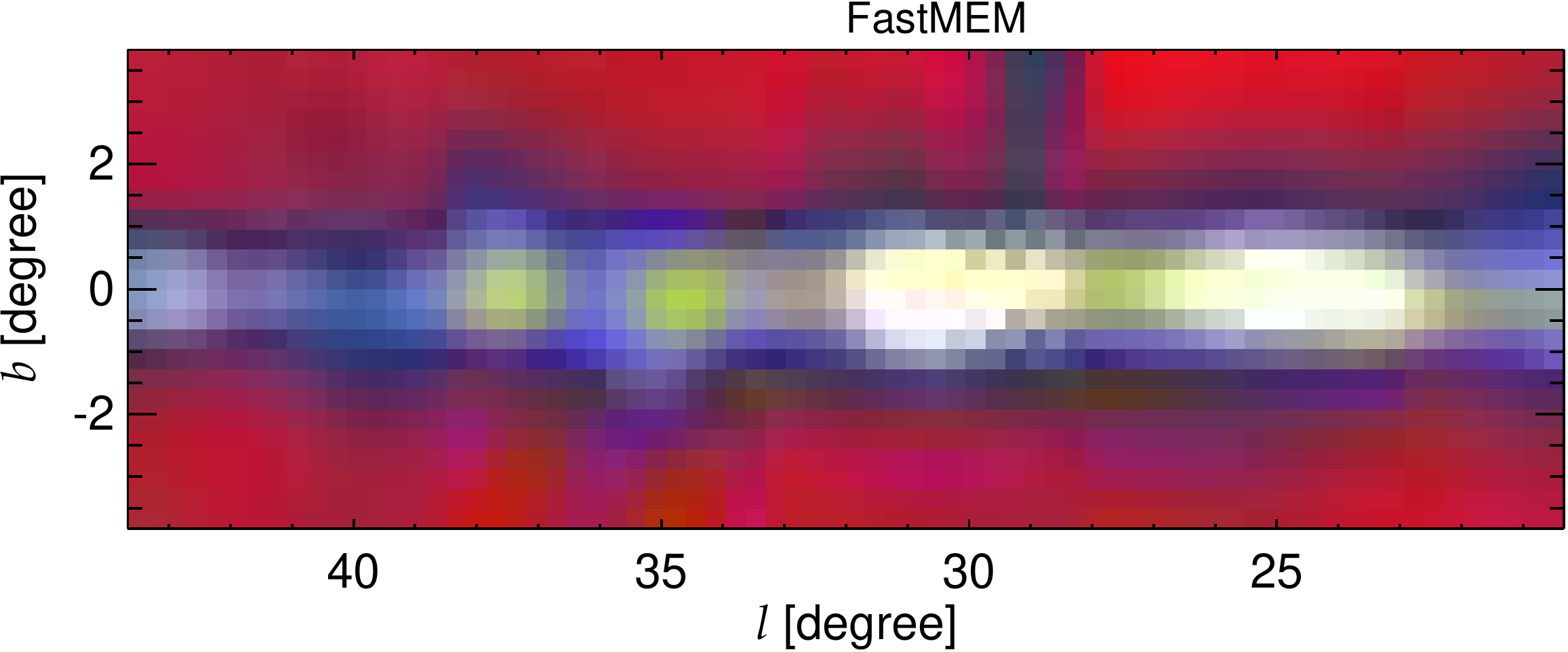}
\includegraphics[scale=0.36]{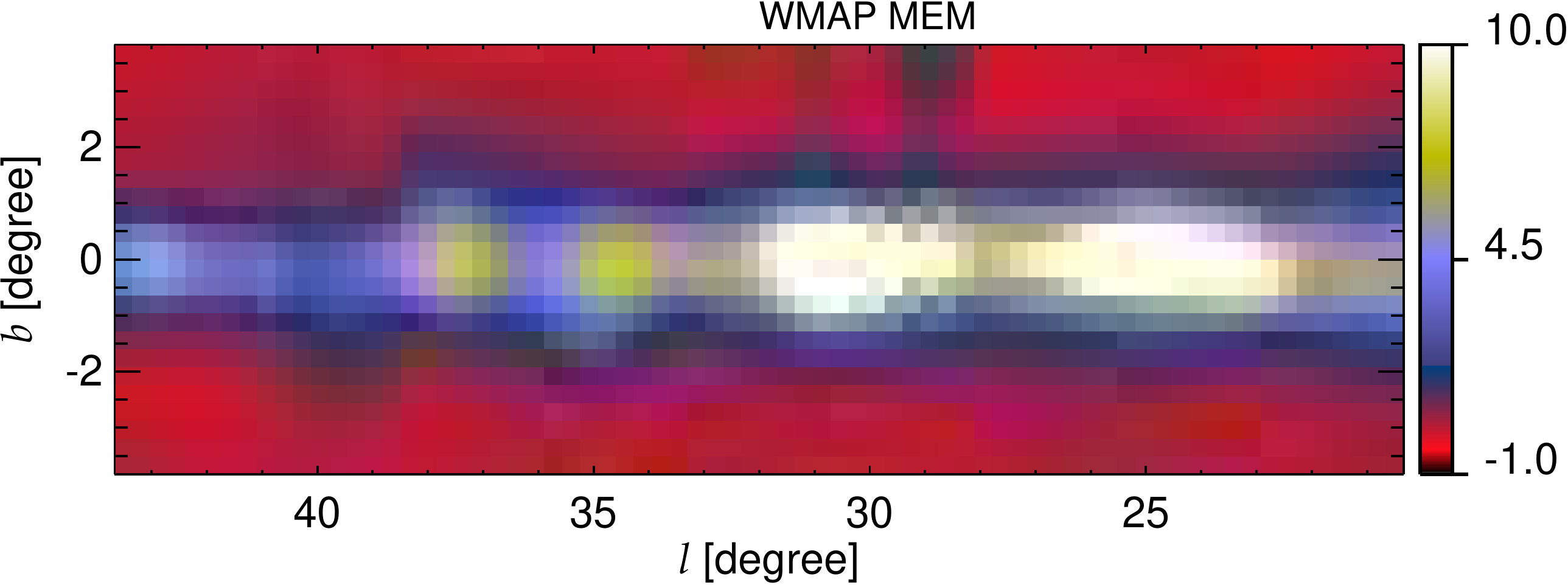}
\caption{Maps of the $l=20\deg$--$44\deg$, $b=-4\deg$--$+4\deg$ region of the inner Galaxy showing the free-free emission derived from RRLs, {\tt FastMEM}, \WMAP9 MEM, and CCA. The maps are at 33\GHz\ with 1\deg\ resolution and $N_\mathrm{side}=128$. The colour-coding is on a linear scale running from $-1$ to $10$\,mK$_\mathrm{RJ}$.}
\label{fig:a5}
\end{figure*}
\begin{figure*}
\centering
\includegraphics[angle=0,scale=0.5]{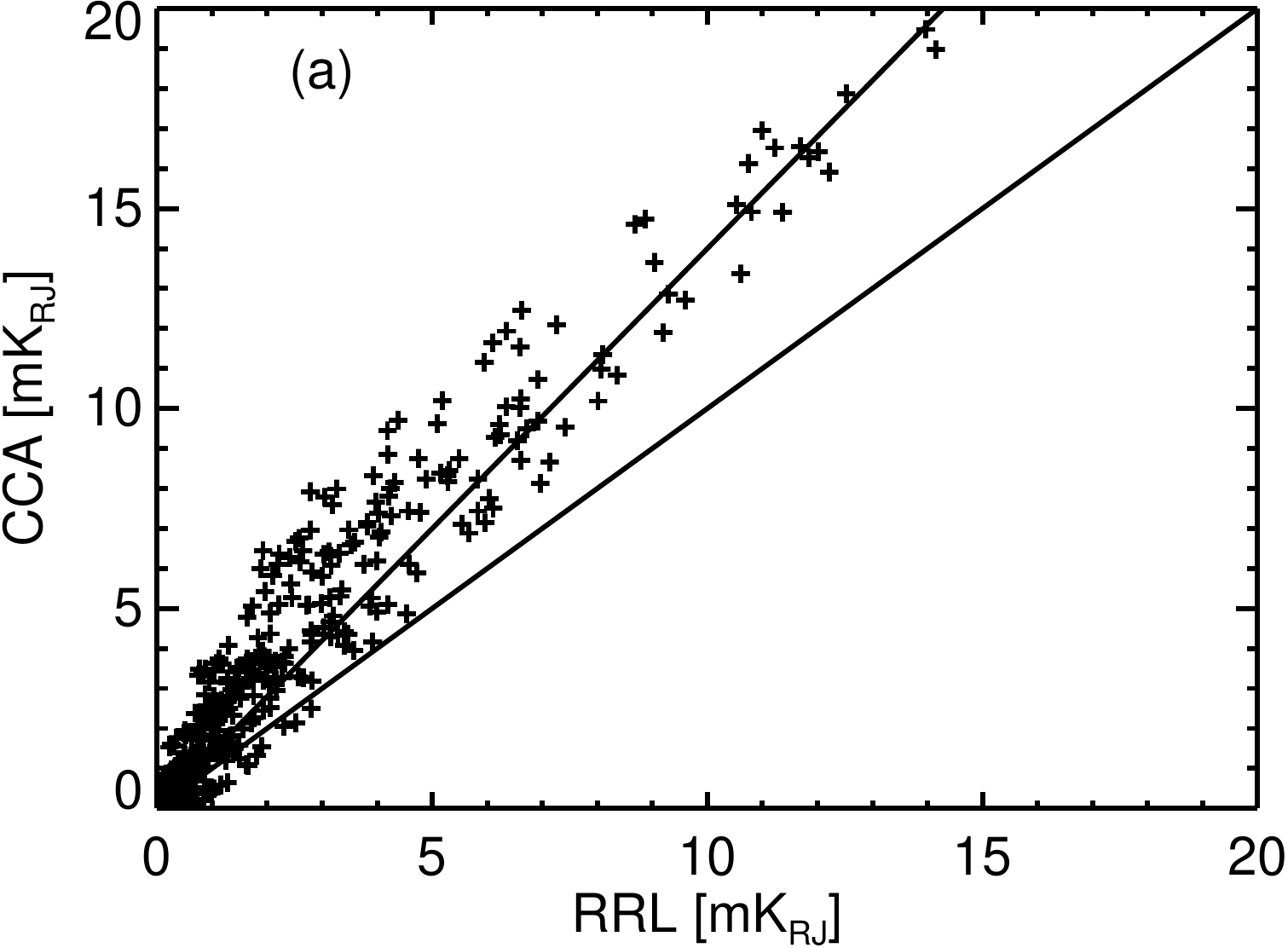}
\includegraphics[angle=0,scale=0.5]{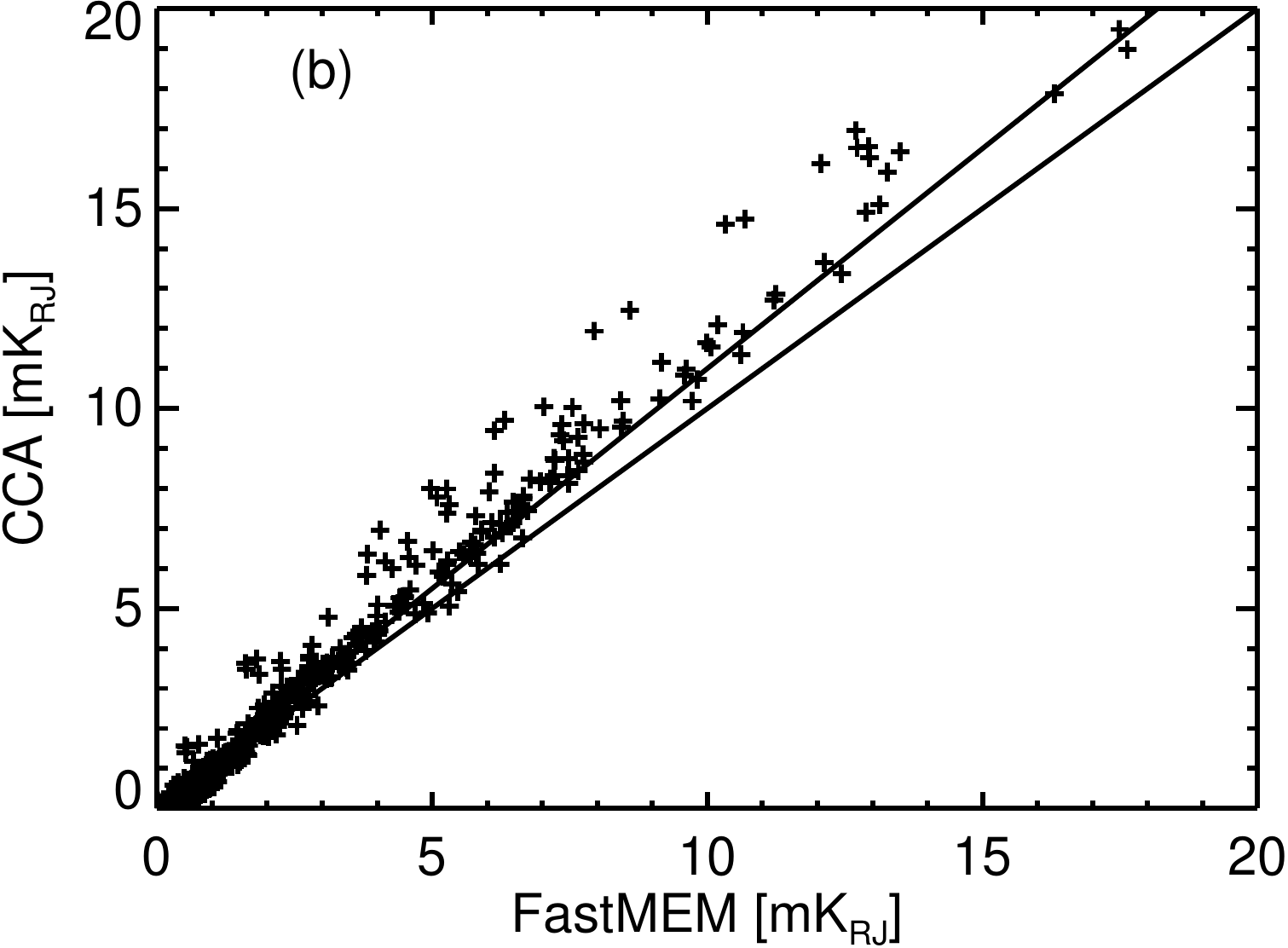}
\includegraphics[angle=0,scale=0.5]{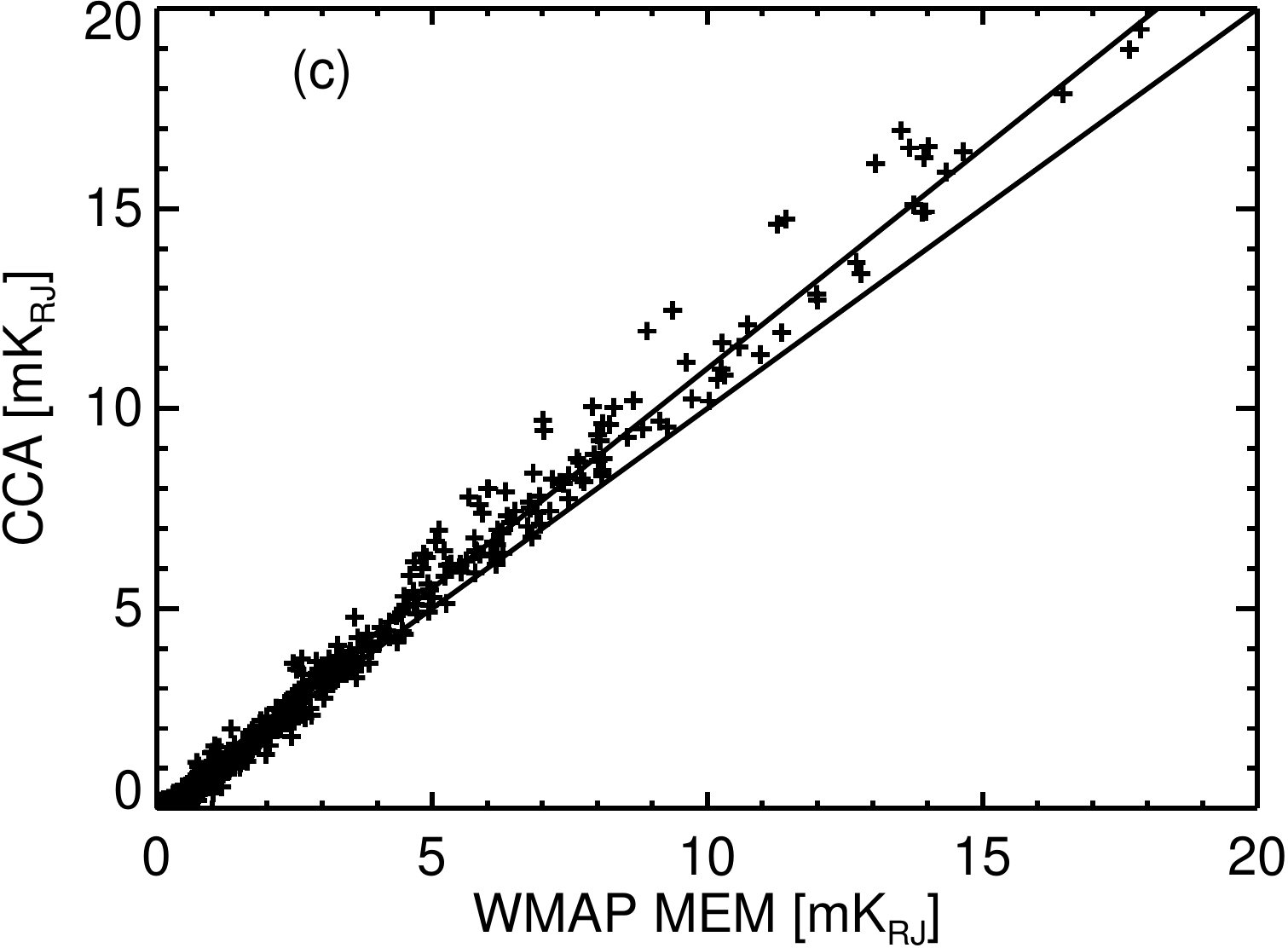}
\includegraphics[angle=0,scale=0.5]{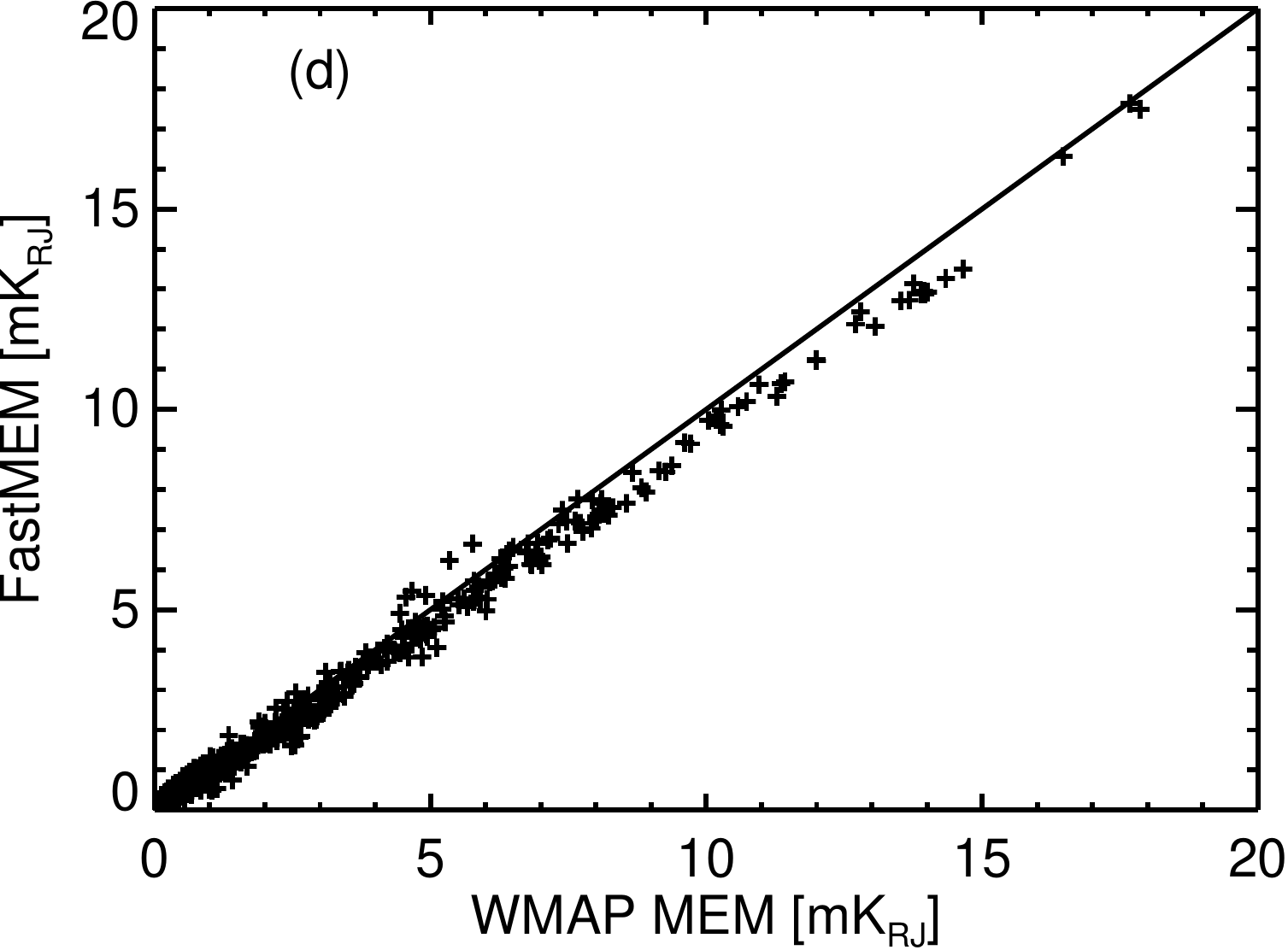}
\caption{$T$--$T$ plots comparing the free-free estimates from RRLs, \WMAP9 MEM, and \Planck\ data using {\tt FastMEM} and CCA. The lower line indicates equality while the other line (where present) indicates the best fit. The frequency is 33\GHz\ and the resolution is 1\deg, with $N_\mathrm{side}=128$.}
\label{fig:a6}
\end{figure*}

The \WMAP\ 9-yr MEM free-free all-sky fit \citep{Bennett2013} was used to map the $l=20\deg$--44\deg, $|b|\leq4\deg$ region of the \cite{Alves2012} RRL survey, which includes emission from the Sagittarius and Scutum spiral arms of the inner Galaxy. This fit is based on the \WMAP\ frequency range 23--94\,GHz and includes significant synchrotron and AME emission that could also be estimated separately. 

The Correlated Component Analysis (CCA) component separation method \citep{Bonaldi2006}, as used in \citet{planck2013-XII}, was also applied to the RRL region. The map of the free-free from RRLs is compared with those from {\tt FastMEM}, \WMAP9 MEM and CCA in Fig.~\ref{fig:a5}. The two MEM models and CCA are remarkably similar. The RRLs show the same structure but are somewhat weaker as we now show.

The $T$--$T$ plots in Fig.~\ref{fig:a6} show the relationships between the amplitudes of the RRL measurements and the three models. It can be  seen that {\tt FastMEM} from \Planck\ and \WMAP9 MEM agree closely. The CCA amplitudes are somewhat larger than the MEM solutions. The corresponding RRL amplitudes calculated for $T_\mathrm{e}=6000$\,K are lower by $\sim$30\,\%. The best fitting ratios of the slopes after removing offsets and including the brighter data only ($T_\mathrm{b}\ge0.3$\,mK) are:
\begin{eqnarray*}
&\textrm{{\tt FastMEM} / \WMAP9 MEM}&= 1.02\\
&\textrm{CCA / 0.5 ({\tt FastMEM}+\WMAP9 MEM)}&= 1.05\\
&\textrm{CCA / RRLs}&= 1.35\\
&\textrm{0.5 ({\tt FastMEM} + \WMAP9 MEM)} / \textrm{RRLs}&= 1.30
\end{eqnarray*}

We now consider what systematics could account for the difference between the RRL measurement and the three models, all of which agree within $\sim$5\,\%. The assumption of $T_\mathrm{e}=6000$\,K for the RRLs could be the source of part of the 30\,\% difference. An increase of $T_\mathrm{e}$ from 6000\,K to 7000\,K for example would increase the free-free brightness temperature by 19\,\%. On the other hand, the three models considered here each need to take account of the other foregrounds, which they do in different ways. This could possibly lead to an overestimate of the free-free emission if the other three spectral components were underestimates. We suggest that this difference may be resolved by using $T_\mathrm{e}=7000$\,K and reducing the free-free component map by 10\,\%.

\clearpage
\raggedright
\end{document}